\documentclass[11pt,a4paper,twoside]{report}
\def\papertype{4}
\usepackage{lmodern}
\usepackage{times}
\usepackage{color}
\usepackage{graphicx}
\usepackage{epsfig}
\usepackage[pdf]{pstricks}

\if4\papertype
\else
\fi
\catcode`@=11
\definecolor{chcolor}{cmyk}{0.1,0.9,0.0,0.0}
\definecolor{hrcolor}{cmyk}{0.1,0.9,0.0,0.0}
\definecolor{airesblue}{cmyk}{0.85,0.85,0.0,0.0}
\def\chtcolor{\color{chcolor}}
\def\head@rule{{\color{hrcolor}\hrule height 0.8pt depth 0.8pt}}
\renewcommand\@pnumwidth{1.85em}
\if@twoside
  \def\ps@headings{%
      \let\@oddfoot\@empty\let\@evenfoot\@empty
      \def\@evenhead{\vbox{\hsize=\textwidth
                            \makebox[\textwidth]{\sffamily\upshape%
                            \thepage\hfill\leftmark
                            \vrule depth0.21ex width 0pt}\head@rule}}%
      \def\@oddhead{\vbox{\hsize=\textwidth%
                           \makebox[\textwidth]{\sffamily\upshape%
                           \rightmark\hfill\thepage%
                            \vrule depth0.21ex width 0pt}\head@rule}}%
      \let\@mkboth\markboth
    \def\chaptermark##1{%
      \markboth {{\scshape
        \ifnum \c@secnumdepth >\m@ne
            {\bfseries\@chapapp\ \thechapter.} \ %
        \fi
        ##1}}{}%
      \markright {{\scshape
        \ifnum \c@secnumdepth >\m@ne
            {\bfseries\@chapapp\ \thechapter.} \ %
        \fi
        ##1}}}
}
\else
  \def\ps@headings{%
    \let\@oddfoot\@empty
    \def\@oddhead{\vbox{\hsize=\textwidth%
                         \makebox[\textwidth]{\sffamily\upshape%
                         \rightmark\hfill\thepage%
                         \vrule depth0.21ex width 0pt}\head@rule}}%
    \let\@mkboth\markboth
    \def\chaptermark##1{%
      \markright {{\scshape
        \ifnum \c@secnumdepth >\m@ne
            {\bfseries \@chapapp\ \thechapter.} \ %
        \fi
        ##1}}}}
\fi
\def\ps@plain{\let\@mkboth\@gobbletwo
     \let\@oddhead\@empty\def\@oddfoot{\reset@font\hfil\sffamily\thepage
     \hfil}\let\@evenhead\@empty\let\@evenfoot\@oddfoot}
\def\@chapter[#1]#2{\ifnum \c@secnumdepth >\m@ne
                         \refstepcounter{chapter}%
                         \typeout{\@chapapp\space\thechapter.}%
                         \addcontentsline{toc}{chapter}%
                           {\protect\numberline{{\chtcolor\thechapter}}#1}%
                    \else
                      \addcontentsline{toc}{chapter}{#1}%
                    \fi
                    \chaptermark{#1}%
                    \addtocontents{lof}{\protect\addvspace{10\p@}}%
                    \addtocontents{lot}{\protect\addvspace{10\p@}}%
                    \if@twocolumn
                      \@topnewpage[\@makechapterhead{#2}]%
                    \else
                      \@makechapterhead{#2}%
                      \@afterheading
                    \fi}
\def\@makechapterhead#1{%
  \vspace*{50\p@}%
  {\parindent \z@ \raggedright \normalfont
    \ifnum \c@secnumdepth >\m@ne
        \huge\sffamily\bfseries{\chtcolor \@chapapp\space \thechapter}%
        \par\nobreak
        \vskip 20\p@
    \fi
    \interlinepenalty\@M
    \Huge \sffamily\bfseries #1\par\nobreak
    \vskip 40\p@
  }}
\def\@makeschapterhead#1{%
  \vspace*{50\p@}%
  {\parindent \z@ \raggedright
    \normalfont
    \interlinepenalty\@M
    \Huge \sffamily\bfseries  #1\par\nobreak
    \vskip 40\p@
  }}
\renewcommand\section{\@startsection {section}{1}{\z@}%
                                   {-3.5ex \@plus -1ex \@minus -.2ex}%
                                   {2.3ex \@plus.2ex}%
                                   {\normalfont\Large\sffamily\bfseries}}
\renewcommand\subsection{\@startsection{subsection}{2}{\z@}%
                                     {-3.25ex\@plus -1ex \@minus -.2ex}%
                                     {1.5ex \@plus .2ex}%
                                     {\normalfont\large\sffamily\bfseries}}
\renewcommand\subsubsection{\@startsection{subsubsection}{3}{\z@}%
                                  {-3.25ex\@plus -1ex \@minus -.2ex}%
                                  {1.5ex \@plus .2ex}%
                                  {\normalfont\normalsize\sffamily\bfseries}}
\renewcommand\paragraph{\@startsection{paragraph}{4}{\z@}%
                                    {3.25ex \@plus1ex \@minus.2ex}%
                                    {-1em}%
                                    {\normalfont\normalsize\bfseries}}
\renewcommand\subparagraph{\@startsection{subparagraph}{5}{\parindent}%
                                  {3.25ex \@plus1ex \@minus .2ex}%
                                  {-1em}%
                                  {\normalfont\normalsize\sffamily\bfseries}}
\def\footnoterule{\kern-3\p@
  {\color{hrcolor}\hrule \@width 2in}\kern 2.6\p@} 
\long\def\@caption#1[#2]#3{%
  \par
  \addcontentsline{\csname ext@#1\endcsname}{#1}%
    {\protect\numberline{\csname the#1\endcsname}{\ignorespaces #2}}%
  \begingroup
    \@parboxrestore
    \if@minipage
      \@setminipage
    \fi
    \normalsize\itshape
    \@makecaption{\bfseries\csname fnum@#1\endcsname}{\ignorespaces #3}\par
  \endgroup}
\long\def\@makecaption#1#2{%
  \vskip\abovecaptionskip
  \sbox\@tempboxa{\raggedright\parindent=0pt{\bfseries #1.} #2}%
  \ifdim \wd\@tempboxa >\hsize
    \raggedright\parindent=0pt{\bfseries #1.} #2\par
  \else
    \global \@minipagefalse
    \hb@xt@\hsize{\hfil\box\@tempboxa\hfil}%
  \fi
  \vskip\belowcaptionskip}
\def\rcaption{\refstepcounter\@captype \@dblarg{\@rcaption\@captype}}
\long\def\@rcaption#1[#2]#3{%
  \par
  \addcontentsline{\csname ext@#1\endcsname}{#1}%
    {\protect\numberline{\csname the#1\endcsname}{\ignorespaces #2}}%
  \begingroup
    \@parboxrestore
    \if@minipage
      \@setminipage
    \fi
    \normalsize\itshape
    \@rmakecaption{\bfseries\csname fnum@#1\endcsname}{\ignorespaces #3}\par
  \endgroup}
\long\def\@rmakecaption#1#2{%
  \vskip\abovecaptionskip
  \sbox\@tempboxa{\raggedleft\parindent=0pt{\bfseries #1.} #2}%
  \ifdim \wd\@tempboxa >\hsize
    \raggedleft\parindent=0pt{\bfseries #1.} #2\par
  \else
    \global \@minipagefalse
    \hb@xt@\hsize{\hfil\box\@tempboxa\hfil}%
  \fi
  \vskip\belowcaptionskip}
\def\@biblabel#1{{\bfseries[#1]}}

\catcode`@=12
\def\thpop{$\vphantom{|^{|^|}_{|_|}}$}
\long\def\tabcaption#1#2#3{\caption[{#2}]{\label{TAB:#1}#3}}
\long\def\figcaption#1#2#3{\caption[{#2}]{\label{FIG:#1}#3}}
\long\def\rfigcaption#1#2#3{\rcaption[{#2}]{\label{FIG:#1}#3}}
\long\def\lowfig#1#2#3#4#5{%
\begin{figure}#1
\begin{center}
\makebox{#3}
\end{center}
\figcaption{#2}{#4}{#5}
\end{figure}}
\long\def\latfig#1#2#3#4#5{%
\begin{figure}#1
\begin{flushright}
\makebox{\makebox{#3}\kern 1em
\parbox[b]{0.3\textwidth}{\figcaption{#2}{#4}{\strut #5\strut}}}
\end{flushright}
\end{figure}}
\long\def\rlatfig#1#2#3#4#5{%
\begin{figure}#1
\begin{flushleft}
\makebox{\parbox[b]{0.3\textwidth}%
{\rfigcaption{#2}{#4}{\strut #5\strut}\kern 1em}\makebox{#3}}
\end{flushleft}
\end{figure}}
\long\def\lowtable#1#2#3#4#5{%
\begin{table}#1
\begin{center}
\makebox{#3}
\end{center}
\tabcaption{#2}{#4}{#5}
\end{table}}
\long\def\lattable#1#2#3#4#5{%
\begin{table}#1
\begin{flushright}
\makebox{\makebox{#3}\kern 1em
\parbox[b]{0.3\textwidth}{\tabcaption{#2}{#4}{\strut #5\strut}}}
\end{flushright}
\end{table}}
\newcounter{iicounter}
\def\iiset{\setcounter{iicounter}{0}}
\def\ii{\addtocounter{iicounter}{1}{\rm(\roman{iicounter})\/ }}
\newcounter{aacounter}
\def\aaset{\setcounter{aacounter}{0}}
\def\aa{\addtocounter{aacounter}{1}{\rm(\alph{aacounter})\/ }}
\def\`#1{\if#1i{\accent18 \i}\else{\accent18 #1}\fi}
\def\'#1{\if#1i{\accent19 \i}\else{\accent19 #1}\fi}

\def\ppref#1{\ref{#1} (page \pageref{#1})}
\def\ttbf{\ttfamily\bfseries\upshape}
\def\itbf{\rmfamily\bfseries\itshape}
\def\kw;#1;{{\ttbf #1}}\def\kwbf;#1;{{\bf #1}}
\def\kwt;#1;{{\ttfamily\mdseries\upshape #1}}
\def\akw;#1;#2;{$\underline{\smash{\hbox{\ttbf #1}}}\hbox{\ttbf #2}$}
\def\akwbf;#1;#2;{$\underline{\smash{\hbox{\bf #1}}}\hbox{\bf #2}$}
\def\vkw;#1;{{\rmfamily\itshape\bfseries #1\/}}
\def\kwdisplay;#1;{\begin{displaymath}
\makebox[11cm][l]{\kw;#1;}
\end{displaymath}}
\def\kwexample#1{\begin{displaymath}
\makebox[11cm][l]{\parbox{\textwidth}{{\ttbf\parindent=0pt
\frenchspacing\obeyspaces\obeylines#1}}\hss}
\end{displaymath}}
\def\idlidx;#1;{\index{Input Directive Language!directives!#1@\kw;#1;}}
\def\subidlidx;#1;#2;{%
\index{Input Directive Language!directives!#1!#2@\kw;#2;}}
\def\idlbfpx;#1;{%
\index{Input Directive Language!directives!#1@\kw;#1;|bfpage{ }}}
\def\obsidlidx;#1;{\idlidx;#1 {\rm(obsolete directive)};%
\index{#1 (obsolete IDL directive)@\kw;#1; (obsolete IDL directive)}}
\def\libidx;#1;{\index{AIRES object library!#1@\kw;#1;}}
\def\arsidx;#1;{\index{AIRES Runner System!commands!#1@\kw;#1;}}
\def\idlbfx;#1;{\kwbf;#1;\idlidx;#1;}
\def\obsidlbfx;#1;{\kwbf;#1;\obsidlidx;#1;}
\def\libbfx;#1;{\kwbf;#1;\libidx;#1;}
\def\arsbfx;#1;{\kwbf;#1;\arsidx;#1;}
\def\idlref;#1;{IDL0:#1}
\def\idlidl;#1;{\idlbfx;#1; (\pageref{\idlref;#1;})}
\def\libref;#1;{rou0:#1}
\def\liblib;#1;{\libbfx;#1; (\pageref{\idlref;#1;})}
\def\mrm#1{\hbox{\rm #1}}
\def\gcmsq{$\mrm{g}/\mrm{cm}^2$}

\def\vr#1{\mathbf{#1}}
\def\versionday{24}%
\def\versionmonth{April}%
\def\versionyear{2019}%
\def\currairesversion{19.04.00}  
\def\prevairesversion{18.09.00}  

\date{\sffamily {\versionmonth} \versionday, \versionyear}
\makeindex
\begin{document}
%
%
\baselineskip=15pt
\setlength{\textfloatsep}{22pt plus10pt minus5pt}
\setlength{\headsep}{22pt}
\pagenumbering{roman}
%
%
\definecolor{versiongrey}{cmyk}{0.0,0.0,0.0,0.55}
\newfont{\atitle}{phvb at 48pt}
\title{%
\scalebox{1.0}[1.3]{\color{airesblue}\atitle
                                   AIRES%
}\\*[2.8ex]
\scalebox{1.0}[1.3]{\Huge\sffamily\bfseries
                    A system for air shower simulations%
}\\*[4ex]
{\huge\sffamily\bfseries
                     User's guide and reference manual%
}\\*[3ex]
{\Huge\sffamily\bfseries\color{versiongrey} Version \currairesversion}}
\author{{\vrule depth 2.5\baselineskip width 0pt}\\
    {\Large\sffamily\upshape S.~J.~Sciutto
    \vrule depth \baselineskip width 0pt}\\
  {\large\sffamily\itshape Departamento de F\'isica and IFLP (CONICET)}\\
  {\large\sffamily\itshape Universidad Nacional de La Plata}\\
  {\large\sffamily\itshape C. C. 67 - 1900 La Plata}\\
  {\large\sffamily\itshape Argentina}\\*[0.2ex]
  {\large\ttbf sciutto@fisica.unlp.edu.ar}}
\maketitle
\setcounter{page}{2}
\thispagestyle{empty}
\hbox{}
\clearpage
\chapter*{\ }
\begin{center}
\raisebox{-0.55\textheight}[0pt][0pt]{\kern1cm\relax%
\includegraphics[width=13cm]{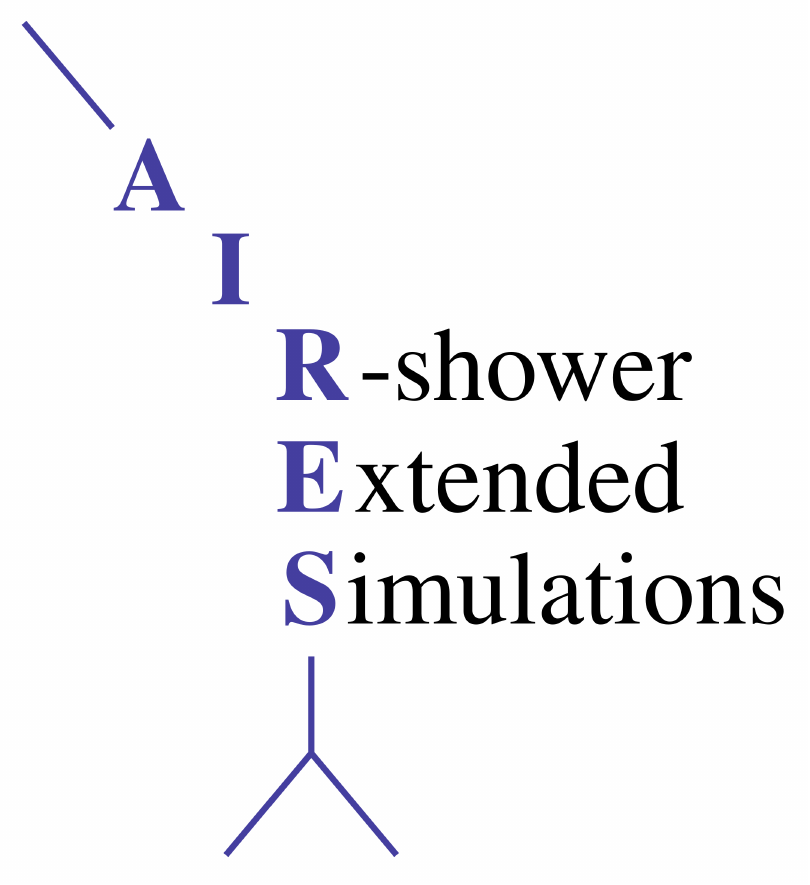}}
\end{center}
\clearpage
\hbox{ }
\vfill
{\small\parindent=0pt
\begin{center}
AIRES user's guide and reference manual\\
Version {\currairesversion} (\versionyear)\\
S. J. Sciutto, La Plata, Argentina.
\end{center}
\bigskip
\begin{quotation}\parindent=0pt\noindent
This manual is part of the AIRES {\currairesversion} distribution.
The AIRES system is distributed worldwide as ``free software'' for all
scientists working in educational/research non-profit
institutions. Users from commercial or non-educational institutions
must obtain the author's written permission before using the software
and/or its related documentation.

\smallskip
The present document makes obsolete all the previous versions of the
AIRES user's manual and reference guide.

\smallskip
{\bf NO WARRANTY.{}} The AIRES system is provided in an {\bfseries\itshape
``as is''\/} basis, without warranty of any kind, either expressed or
implied, including, but not limited to, the implied warranties of
merchantability and fitness for a particular purpose. The entire risk as to
the quality and performance of the program is with the user. Should the
program prove defective, the user assumes the cost of all necessary
servicing, repair or correction. In no event will the AIRES author(s), be
liable to any user for damages, including any general, special, incidental or
consequential damages arising out of the use or inability to use the
Simulation System (including, but not limited to, loss of data or data being
rendered inaccurate or losses sustained by the user or third parties or a
failure of the System to operate with any other programs), even if the
author(s) have been advised of the possibility of such damages.

\smallskip
Product and company names mentioned in this manual are trademarks or trade
names of their respective companies.
\end{quotation}}
%
\chapter*{Summary}%
\addcontentsline{toc}{chapter}{\protect\numberline{ }%
{\large\bfseries Summary}}
\markboth {{\scshape SUMMARY}}{{\scshape SUMMARY}}%
The name {\bf AIRES}
($\underline{\hbox{\bf AIR}}$-shower
$\underline{\hbox{\bf E}}$xtended
$\underline{\hbox{\bf S}}$imulations)
identifies a set of programs
and subroutines to simulate particle showers produced after the
incidence of high energy cosmic rays on the Earth's atmosphere, and to
manage and analyze all the related output data.

AIRES provides full space-time particle propagation in a realistic
environment, where the characteristics of the atmosphere, the
geomagnetic field and the Earth's curvature are taken into account
adequately. A statistical sampling procedure (the so-called {\em
thinning\index{thinning}\/}) is used when the number of particles in
the showers is exceedingly large. The thinning algorithms used in
AIRES are unbiased, that is, the statistical sampling never alters the
average values of output observables.

The particles taken into account by AIRES in the simulations are:
Gammas; electrons; positrons; muons; taus; pions; kaons; eta, rho,
$D$, $D_s$, and $B$ mesons; lambda, sigma, xi, and omega
baryons; nucleons; antinucleons; and nuclei up to $Z=36$. Electron and
muon neutrinos are generated in certain processes (decays) and
accounted for their energy, but not propagated. The primary particle
can be any one of the already mentioned particles, with energy ranging
from less than 1 GeV up to more than 1 ZeV ($10^{21}$ eV). It is also
possible to simulate showers initiated by ``special'' primary
particles\index{special primary particles} via a call to a
user-written module capable of processing the ``first interaction'' of
the primary and returning a list of standard particles suitable for
being processed by AIRES.

Among all the physical processes that may undergo the shower
particles, the most important from the probabilistic point of view are
taken into account in the simulations. Such processes are: (i) {\em
Electrodynamical processes:\/} Pair production\index{pair production}
and electron-positron annihilation\index{positron annihilation},
bremsstrahlung\index{bremsstrahlung} (electrons, positrons and
muons\index{muon bremsstrahlung}),
muonic pair production, knock-on electrons\index{knock-on electrons}
 ($\delta$ rays), Compton\index{Compton effect}
and photoelectric\index{photoelectric effect} effects,
Lan\-dau-Pomeranchuk-Migdal (LPM)
effect\index{LPM effect} and dielectric suppression\index{dielectric
suppression}. (ii) {\em Unstable particle decays,\/} pions and muons,
for instance. (iii) {\em Hadronic processes:\/} Inelastic collisions
hadron-nucleus and photon-nu\-cleus, sometimes simulated using an
external package\index{external packages} which implements a given
hadronic interaction model, like the well-known EPOS\index{EPOS},
QGSJET\index{QGSJET} or SIBYLL\index{SIBYLL} models. Photonuclear
reactions\index{photonuclear reactions}. Nuclear
fragmentation, elastic and inelastic. (iv) {\em Propagation of charged
particles:\/} Losses of energy in the medium (ionization), multiple
Coulomb scattering and geomagnetic deflections.

The AIRES simulation system provides a comfortable environment where
to perform reliable simulations: The {\em Input Directive Language
(IDL)\index{Input Directive Language}\/} is a set of simple directives
which allow for an efficient control of the input parameters of the
simulation. The {\em AIRES Runner System\index{AIRES Runner System}\/}
is a powerful tool to manage long simulation tasks in UNIX
environments, allowing the user to coordinate several tasks running
concurrently, controlling the evolution of a given job while running,
etc. The {\em AIRES summary program\index{AIRES summary program}\/}
processes the internal dump files\index{internal dump file} generated
by the main simulation program, and allows to obtain data related with
physical observables either after or during the simulations. Finally,
the {\em AIRES object library\/} provides a series of auxiliary
routines to process the data generated by the simulation program, in
particular the data contained in the {\em compressed output
files\index{compressed output files},\/} the detailed particle data
files containing per-particle information for particles reaching the
ground, crossing different observing levels during their evolution,
etc.

The present version of AIRES (\currairesversion) represents a new
release of the Air Shower Simulation System where many new technical
features have been added to it. In particular, the present version
includes the possibility of installing {\em extensions\index{AIRES
extensions},\/} additional modules that are incorporated to the
main simulation system that enlarge its capabilities. This is the case
of the {\em ZHAireS\index{ZHAireS}\/} extension, that allows to
simulate the radio waves emitted during the shower development.

The present version of AIRES (\currairesversion) represents a new
release of the Air Shower Simulation System where many new features
and algorithm improvements have been added to it.
 {\bfseries The most important additions for this new version are
   summarized in next section {\em Release Notes.}}

It is worthwhile mentioning that some of the modules that that make up
the AIRES simulation engine have been developed with the help of
various colleagues: The routines that simulate the geomagnetic
deflections, LPM effect and muon bremsstrhlung and pair production
were developed with the help of A. Cillis (La Plata, 1998-2001). The
atmospheric model GAMMA was designed and developed in collaboration
with J. C. Moreno (La Plata, 2008-2012). The HQIP preprocessor was
designed and developed in collaboration with J. I. Illana and M. Masip
(Universidad de Granada, Spain, 2010-2014). The updated cross sections
for photonuclear interactions have been investigated in collaboration
with C. A. Garc{\'\i}a Canal (La Plata), G. Pancheri (INFN Frascati,
Italy), and F. Cornet and A. Grau (Universidad de Granada, Spain,
2008-2015). Additionally, many of the developments presented for the
current release were performed taking into account users' suggestions
and remarks. The author is indebted to everybody that have contacted
him (a very long list of persons indeed), either to report a bug or to
make a comment on the program.

\vskip1.2cm minus1em
\begin{flushright}
La Plata, {\versionmonth} {\versionyear}.
\end{flushright}

%
%
\chapter*{Release notes}\index{release notes}
\addcontentsline{toc}{chapter}{\protect\numberline{ }%
{\large\bfseries Release notes}}
\markboth {{\scshape RELEASE NOTES}}{{\scshape RELEASE NOTES}}%
\label{A:relnotes}

We include here a brief summary of the new developments that
are included in the current version of AIRES (\currairesversion), as
well as a description of the differences with the previous release of the
system (\prevairesversion).

The number in brackets placed after directive names or library
routines indicates the page where the corresponding directive/routine
is described. Example: \kwbf;TaskName; (\pageref{IDL0:TaskName}).

\section*{Differences between AIRES {\currairesversion} and AIRES
{\prevairesversion}}
\index{differences between AIRES {\currairesversion} and AIRES
{\prevairesversion}}
\subsection*{Algorithm modifications}
\begin{description}
\item[Atmosphere.] All the code related with the atmospheric model
  management has been carefully reviewed. New atmospheric profile
  models, namely \kwbf;SouthPoleAvg; and \kwbf;MalargueAvg;, are now
  available (see directive \idlidl;Atmosphere;), corresponding to
  annual averages of atmospheric profiles of the South Pole and
  Malargue sites, respectively. Additionally, the new directive
  \idlidl;AddAtmosModel; allows the user to define custom atmospheric
  models that can be used alternatively to the built-in models.
\item[Neutrinos.] New tables available (6296, 6496, and 6796) that
  record the number and energy of created secondary neutrinos as a
  function of atmospheric depth.
\item[Dynamic setting of primary energy.] The special primary
  modules\index{special primary particles} are now capable of setting
  the shower primary energy dynamically at the moment they are invoked
  to inject the special primary particles (the \idlidl;PrimaryEnergy;
  IDL directive is no more mandatory in such cases).
\end{description}
\subsection*{Installation procedure}\index{installing AIRES}
\begin{itemize}
\item The installation procedure now supports the installation of
  AIRES extensions\index{AIRES extensions}.
\end{itemize}
\subsection*{Input Directive Language}\index{Input Directive Language}
The number in brackets placed after directive names indicate the page
where the corresponding directive is described.
\clearpage
\begin{description}

\item[New features.] \ \ \ 
\begin{itemize}\setlength{\itemsep}{0pt}
  \item The site \vkw;TelescArray; was included to the default site
    library, and can be selected by means of the \idlidl;Site; directive.
  \item New atmospheric models \vkw;SouthPoleAvg; and
    \vkw;MalargueAvg; are available. They
    can be selected using the \idlidl;Atmosphere; directive.
\end{itemize}

\item[New directives.] \ \ \ 
\begin{itemize}\setlength{\itemsep}{0pt}
  \item \idlidl;AddAtmosModel;.
\end{itemize}


\item[Directives that changed their syntax or behavior.] \ \ \ 
\begin{itemize}\setlength{\itemsep}{0pt}
  \item \idlidl;AddSpecialParticle;.
  \item \idlidl;Atmosphere;.
  \item \idlidl;Input;.
  \item \idlidl;InputPath;.
  \item \idlidl;Shell;.
\end{itemize}


\end{description}
\subsection*{Output data}

\begin{description}\setlength{\itemsep}{0pt}
\item[New output tables.] The current AIRES version includes 4 new
tables\index{output data tables}:
\begin{itemize}
\item 0100. Atmospheric profile.
\item 6296. Number of created particles: All neutrinos.
\item 6496. Number of created entries: All neutrinos.
\item 6796. Energy of created particles: All neutrinos.
\end{itemize}
\end{description}
%
%


%
%

%
%
\subsection*{Bugs}

Several problems with AIRES \prevairesversion simulation program were detected
or reported by several users. All the errors in the program's
code --mostly minor bugs-- were fixed and are no longer present in the
current version of AIRES. Some of those bugs are:
\begin{itemize}
\item Fixed a problem installing the AIRES Runner System\index{AIRES
  Runner System} on Mac OS platforms generated by the fact that in
  those platforms the file identifications are case independent.
\item Fixed a problem appearing with the installation of external
  input data files\index{External input data file} in cases of very
  long directory paths.
\end{itemize}

%
%
{\sffamily
\renewcommand{\normalcolor}{\sffamily\color{black}}
\tableofcontents}
\listoffigures
\listoftables
\markboth {\ }{}%
\cleardoublepage
\pagenumbering{arabic}
\setcounter{page}{1}
%
\chapter{Introduction}

Cosmic rays with energies larger than 100 TeV must be studied --at
present-- using experimental devices located on the surface of the
Earth. This implies that such kind of cosmic rays cannot be detected
directly; it is necessary instead to measure the products of the
atmospheric cascades of particles initiated by the incident astroparticle.
An atmospheric particle shower begins when the primary cosmic particle
interacts with the Earth's atmosphere. This is, in general, an inelastic
nuclear collision that generates a number of secondary particles. Those
particles continue interacting and generating more secondary particles
which in turn interact again similarly as their predecessors. This
multiplication process continues until a maximum is reached. Then the
shower attenuates as far as more and more particles fall below the
threshold for further particle production.

A detailed knowledge of the physics involved is thus necessary to
interpret adequately the measured observables and be able to infer the
properties of the primary particles. This is a complex problem
involving many aspects: Interactions of high energy particles,
properties of the atmosphere and the geomagnetic field, etc. Computer
simulation is one of the most convenient tools to quantitatively
analyze such particle showers.

In the case of air showers initiated by ultra-high energy
astroparticles ($E\ge 10^{19}$ eV), the primary particles have
energies that are several orders of magnitude larger than the maximum
energies attainable in experimental colliders. This means that the
models used to rule the behavior of such energetic particles must
necessarily make extrapolations from the data available at much lower
energies, and there is still no definitive agreement about what is the
most convenient model to accept among the several available ones.

The {\bf AIRES system}%
\footnote{{\bf AIRES} is an acronym for
$\underline{\hbox{\bf AIR}}$-shower
$\underline{\hbox{\bf E}}$xtended
$\underline{\hbox{\bf S}}$imulations.}
is a set of programs to simulate such air showers. One of the basic
objectives considered during the development of the software is that of
designing the program modularly, in order to make it easier to switch
among the different models that are available, without having to get
attached to a particular one.

Several simulation programs that were developed in the past were
studied in detail in order to gain experience and improve the new
design. Among such programs, the MOCCA\index{MOCCA} code created by
A. M. Hillas\index{Hillas, A. M.}  \cite{MOCCA} has been extensively
used as the primary reference when developing the first version of
AIRES \cite{Aires120} released in May 1997. Needless to say, the
present version of AIRES (\currairesversion), released more that 20
years after the first one (1.2.0) does include modifications,
sometimes massive, of the original algorithms, and in consequence
both programs are no longer equivalent.

Another characteristic of ultra-high energy simulations that was taken
into account when developing AIRES is the large number of particles
involved. For example, a $10^{20}$ eV shower contains about $10^{11}$
secondary particles. From the computational point of view, this fact
has two main consequences that were specially considered at the moment
of designing AIRES: (i) With present day computers, it is virtually
impossible to follow all the generated particles, and therefore a
suitable sampling technique must be used to reduce the number of
particles actually simulated. The so-called thinning\index{thinning}
algorithm introduced by Hillas \cite{splithin} or the sampling
algorithm of Kobal, Filip\v{c}i\v{c} and Zavrtanik \cite{slthin}
represent examples of such sampling methods. (ii) The simulation
algorithm is CPU intensive, and therefore it is necessary to develop a
series of special procedures that will provide an adequate environment
to process computationally long tasks.

There are many quantities that define the initial or environmental
conditions for an air shower, for example, the identity of the primary
particle and its energy, the position of the ground surface, the
minimum energy a particle must have to be taken into account in the
simulation, the intensity and orientation of the geomagnetic
field\index{geomagnetic field}, etc. Additionally, it is possible to
define many observables that are useful to characterize the particle
shower, namely, longitudinal\index{longitudinal development} and
lateral distribution\index{lateral distributions} of particles, energy
distributions\index{energy distributions}, position of the shower
maximum and so on.

A comfortable environment is provided by AIRES to manage all the input
and output data: The {\em Input Directive Language (IDL)\index{Input
Directive Language}\/} is a set of human-readable input directives
that allow the user to efficiently steer the simulations. The {\em
AIRES summary program\index{AIRES summary program}\/} and the {\em
AIRES object library\index{AIRES object library}\/} represent a set of
tools to manage the output data after the simulations are finished,
and even during them, allowing to control their evolution. Data
associated with particles reaching ground or crossing predetermined
observing levels can be recorded into {\em compressed output
files\index{compressed output files}.\/} A special data compression
procedure is used to reduce as much as possible the size of the files,
which tends to be very large in certain circumstances. The compressed
files can be processed with the help of some auxiliary routines that
are included in the AIRES library. The machine and operating system
used to generate such files may be different that the ones used to
read them.

The particles taken into account by AIRES in the simulations are:
Gammas; electrons; positrons; taus; muons; pions; kaons; eta, rho,
$D$, $D_s$, and $B$ mesons; lambda, sigma; xi, and omega
baryons; nucleons; antinucleons; and nuclei up to $Z=36$. Electron and
muon neutrinos are generated in certain processes (decays) and
accounted for their energy, but not propagated. The primary particle
can be any one of the already mentioned particles, with energy ranging
from less than 1 GeV up to more than 1 ZeV ($10^{21}$ eV). It is also
possible to simulate showers initiated by ``special'' primary
particles\index{special primary particles} via a call to a
user-written module capable of processing the ``first interaction'' of
the primary and returning a list of standard particles suitable for
being processed by AIRES. A detailed description on how to define and
use special primaries is placed in section \ref{S:specialprim}.

Among all the physical processes that may undergo the shower
particles, the most important from the probabilistic point of view are
taken into account in the simulations. Such processes are: (i) {\em
Electrodynamical processes:\/} Pair production\index{pair production}
and electron-positron annihilation\index{positron annihilation},
bremsstrahlung\index{bremsstrahlung} (electrons, positrons and
muons\index{muon bremsstrahlung}),
muonic pair production, knock-on electrons\index{knock-on electrons}
 ($\delta$ rays), Compton\index{Compton effect}
and photoelectric\index{photoelectric effect} effects,
Lan\-dau-Pomeranchuk-Migdal (LPM) 
effect\index{LPM effect} and dielectric suppression\index{dielectric
suppression}. (ii) {\em Unstable particle decays,\/} pions and muons,
for instance. (iii) {\em Hadronic processes:\/} Inelastic collisions
hadron-nucleus and photon-nu\-cleus, generally simulated using an
external package\index{external packages} which implements a given
hadronic interaction model like the well-known EPOS\index{EPOS}
\cite{EPOSLHC}, QGSJET\index{QGSJET} \cite{QGSJET2R4}, or SIBYLL\index{SIBYLL}
\cite{SIBYLL23c} models; or by a
built-in algorithm called {\em extended Hillas splitting algorithm}%
\index{extended Hillas splitting algorithm}
(EHSA)\index{EHSA|see{extended Hillas splitting algorithm}}%
\index{splitting algorithm!extended|see{extended Hillas splitting algorithm}}.
Photonuclear reactions\index{photonuclear reactions}. Nuclear
fragmentation, elastic and inelastic. (iv) {\em Propagation of charged
particles:\/} Losses of energy in the medium (ionization), multiple
Coulomb scattering and geomagnetic deflections.

All the general characteristics of AIRES and the physics involved in air
shower simulations are summarized in table \ref{TAB:airesfeatures};
they are described in more detail in chapter \ref{C:genchar}.
%
\lowtable{[p]}{airesfeatures}{%
\def\rbeg{\vrule height 11pt depth 0pt width 0pt\relax}
\def\rend{\vrule height 0pt depth 5pt width 0pt\relax}
{\small
\begin{tabular}{|p{4.3cm}|p{10cm}|}
\hline
{\bf Propagated particles}%
&\rbeg
Gammas. Leptons: $e^{\pm}$, $\mu^{\pm}$, $\tau^{\pm}$.\par
Mesons: $\pi^0$, $\pi^{\pm}$; $\eta$, $K^0_{L,S}$, $K^{\pm}$,
        $\rho^0$, $\rho^{\pm}$, $D^0$, $D^{\pm}$,
        $D_s^{\pm}$, $B^0$, $B^{\pm}$.\par
Baryons: $p$, $\bar p$, $n$, $\bar n$, $\Lambda$, $\Lambda_c$,
         $\Sigma^0$, $\Sigma^{\pm}$, $\Xi^0$, $\Xi^-$,
         $\Omega^-$.\par
Nuclei up to $Z=36$.\par
Neutrinos are generated (in decays) and accounted for their
number and energy, but not propagated.
\rend\\ \hline

{\bf Primary particles}
&\rbeg
All propagated particles can be injected as primary
particles.\par
Multiple and/or ``exotic'' primaries can be
injected using the {\em special primary\/}
feature\index{special primary particles}.
\rend\\ \hline

{\bf Primary energy range}
&\rbeg
From 500 MeV to 3 ZeV ($3 \times 10^{21}$ eV).
\rend\\ \hline

{\bf Geometry and environment}
&\rbeg
Incidence angles from vertical to horizontal showers.\par
The Earth's curvature\index{Earth's curvature}
is taken into account for all inclinations.\par
Realistic atmosphere.\par
Geomagnetic deflections: The geomagnetic
field\index{geomagnetic field} can be calculated using the
IGRF\index{International Geomagnetic Reference Field} model \cite{IGRF}.
\rend\\ \hline

{\bfseries Propagation (general)}
&\rbeg
Medium energy losses (ionization).\par
Scattering of all charged particles including corrections for finite
nuclear size.\par
Geomagnetic deflections.
\rend\\ \hline

{\bfseries Propagation: {\itshape Electrons and\par gammas}}
&\rbeg
Compton\index{Compton effect} and
photoelectric\index{photoelectric effect} effects,
$e^+e^-$ pair production\index{pair production}.\par
Bremsstrahlung\index{bremsstrahlung},
emission of knock-on electrons\index{knock-on electrons}, and
$e^+$ annihilation\index{positron annihilation}.\par
LPM effect\index{LPM effect}, and
dielectric suppression\index{dielectric suppression}.\par
Photonuclear reactions\index{photonuclear reactions}.
\rend\\ \hline

{\bfseries Propagation: {\itshape Muons}}
&\rbeg
Bremsstrahlung\index{muon bremsstrahlung} and
muonic pair production\index{muonic pair production}.\par
Emission of knock-on electrons.\par
Decay.
\rend\\ \hline

{\bfseries Propagation: {\itshape Hadrons and\par nuclei}}
&\rbeg
Hadronic collisions using the
EHSA\index{extended Hillas splitting algorithm} (low energy)
and EPOS\index{EPOS}, QGSJET\index{QGSJET} or SIBYLL\index{SIBYLL}
(high energy).\par
Nuleus-nucleus collisions\index{nucleus-nucleus collisions}
via EPOS, QGSJET or SIBYLL, or using a built-in
nuclear fragmentation algorithm.\par
Hadronic cross sections\index{hadronic cross sections} are evaluated
from fits to
experimental data (low energy), or to EPOS, QGSJET or SIBYLL predictions
(high energy).\par
Emission of knock-on electrons.\par
Decay of unstable hadrons.
\rend\\ \hline

{\bfseries Statistical sampling}
&\rbeg
Particles are sampled by means of the Hillas thinning\index{thinning}
algorithm \cite{splithin}, extended to allow control of maximum weights.
\rend\\ \hline

{\bfseries Main observables}
&\rbeg
Longitudinal development\index{longitudinal development}
of all particles recorded in up to 510 observing levels.\par
Energy deposited in the atmosphere.\par
Lateral\index{lateral distributions}, energy\index{energy distributions}
and time distributions at ground level.\par
Detailed list of particles reaching ground, and/or crossing
predetermined observing levels.
\rend\\ \hline
\end{tabular}}
}%
{Main characteristics of the AIRES air shower simulation system.}%
{Main characteristics of the AIRES air shower simulation system.%
\index{AIRES!table of features}%
}

AIRES is completely written in standard FORTRAN (using a few
extensions that are, to the best of our knowledge, accepted by all
FORTRAN compilers). The complete AIRES {\currairesversion} source
code, which includes the EPOS LHC \cite{EPOSLHC}, EPOS 1.99
\cite{EPOSLHC}, QGSJET-II-03 \cite{QGSJET2R3}, QGSJET-II-04
\cite{QGSJET2R4}, SIBYLL 2.3c \cite{SIBYLL23c}, SIBYLL 2.3
\cite{SIBYLL23}, and SIBYLL 2.1 \cite{SIBYLL21}
 hadronic collisions packages%
\index{external packages}\index{hadronic models}\label{P:AiresModel},
the IGRF \cite{IGRF} routines\index{external packages} to evaluate geomagnetic
data and Netlib/minpack/lmder nonlinear least squares fitting
package\index{external packages}\index{Netlib} \cite{netlib}, consists
of more than 2000 routines, adding up to more than 300,000 source lines
extensively commented.

In the present version, the AIRES simulation system consists of the
following:
\begin{itemize}
\item The main air shower simulation programs:
  \begin{itemize}
     \item \kwbf;AiresEPLHC; and \kwbf;AiresEP199;;
     \item \kwbf;AiresQIIr04; and \kwbf;AiresQIIr03;;
     \item  \kwbf;AiresS23;, \kwbf;AiresS23c;, and \kwbf;AiresS21;;
  \end{itemize}
  containing
  the interfaces with the hadronic collision packages\index{hadronic
    models}\index{external packages} EPOS LHC \cite{EPOSLHC}, EPOS
  1.99 \cite{EPOSLHC}, QGSJET-II-04 \cite{QGSJET2R4}, QGSJET-II-03
  \cite{QGSJET2R3}, SIBYLL 2.3 \cite{SIBYLL23}, SIBYLL 2.3c
  \cite{SIBYLL23c}, and SIBYLL 2.1 \cite{SIBYLL21}, respectively. The
  default simulation program, \kwbf;Aires;, is equivalent to
  \kwbf;AiresS23;.
\item The summary program\index{AIRES summary program}
(\kwbf;AiresSry;) designed to process a part of the data generated by
the simulation programs, allowing the user to analyze the results of
the simulation after completing it, or even while it is being run.
\item The IDF to ADF file format converting program
\kwbf;AiresIDF2ADF;\index{AIRES IDF to ADF converting program}.
\item A library of utilities to help the user to process the
compressed output data files\index{compressed output files} generated
by the simulation program, write external modules to process special
primaries\index{special primary particles}, etc. In LINUX environments
this library is implemented as an object library called
\kwbf;libAires.a;, or a shared object library called \kwbf;libAires.so;.
\item The AIRES runner system\index{AIRES Runner System}: A set of
shell scripts to ease working with AIRES in UNIX environments.
\end{itemize}
\index{AiresSry@\kwbf;AiresSry;|see{AIRES summary program}}%
\index{AiresIDF2ADF@\kwbf;AiresIDF2ADF;|see{AIRES IDF to ADF converting
program}}%
\section{Structure of the main simulation programs}
\label{S:airesstruct}

An air shower starts when a cosmic particle reaches the Earth's
atmosphere and interacts with it. In most cases the first interaction
is an inelastic collision of the (high energy) primary particle with
an air nucleus. The product of this collision is a set of secondary
particles carrying a fraction of the primary's energy. These
secondaries begin to move through the atmosphere and will eventually
interact similarly as the primary did, generating new sets of
secondaries. This multiplication process continues until a maximum is
reached. After that moment the shower begins to attenuate because an
increasing number of secondaries are produced with energies too low
for further particle generation.

This phenomenon is simulated in AIRES in the following way:
\begin{enumerate}
\item
Several data arrays or {\em stacks\/} are defined. Every record within
any stacks is a particle entry, and represents a physical
particle. The data contained in every record are related to the
characteristics of the corresponding particle: Identity, position,
energy, etc.
\item The particles can move inside a volume within the atmosphere where
the shower takes place. This volume is limited by the {\em ground,\/}
and {\em injection\/} surfaces, and by vertical planes which limit the
region of interest.
\item Before starting the simulations all the stacks are empty. The
first action is to add the first stack entry, which corresponds to the
primary particle. The primary is initially located at the injection
surface, and its downwards direction of motion defines the sower axis.
\item The stack entries are repeatedly processed sequentially. Every
particle entry is updated analyzing first all the possible
interactions it can have, and evaluating the corresponding
probabilities for each possibility, taking into account the physics involved.
\item Using a stochastic method, the mentioned probabilities are used to
select one of the possible interactions. This selection defines what
is going to happen with the corresponding particle at that moment.
\item The interaction is processed: First the particle is moved a
certain distance (which comes out from the mentioned stochastic
method), then the products of the interaction are generated. New stack
entries are appended to the existing lists for every one of the
secondary particles that are created. Depending on the particular
interaction that is being processed, the original particle may survive
(the corresponding entry remains in the stack for further processing)
or not (the entry is deleted).
\item When a charged particle is moved, its energy is modified to take
into account the energy losses in the medium (ionization).
\item Particle entries can also be removed when one of the following
events happens: \aaset\aa The
energy of the particle is lower than a certain threshold energy called
{\em cut energy.\/} The cut energies may be different for different
particle kinds. \aa The particle reaches ground level. \aa A particle
going upwards reaches the injection surface. \aa A particle with
quasi-horizontal motion exits the region of interest.
\item After having scanned all the stacks, it is checked whether or
not there are remaining particle entries pending further
processing. If the answer is positive, then all the stacks are
re-scanned once more; otherwise the simulation of the shower is
complete.
\end{enumerate}

The group of algorithms related with interaction selection and
processing, as well as calculation of energy losses is the group of
{\em physical algorithms.\/}

The most important air shower observables are those related with
statistical distributions of particle properties. To evaluate such
quantities the simulation engine of AIRES also possesses internal
monitoring procedures that constantly check and record particles
reaching ground and/or passing across predetermined observing surfaces
located between the ground and injection levels.

From this description, it shows up clearly that the air shower
simulation programs consist of various interacting procedures that
operate on a data set with a variable number of records, modifying its
contests, increasing or decreasing its size accordingly with
predetermined rules.

It is necessary to do a modular design of such a program to make it
more manageable; and this is particularly relevant for the case of the
algorithms related with the physical laws that rule the interactions
where --as mentioned-- there are still open problems requiring
continuous change and testing of procedures.

\def\idioma{2}
\lowfig{[p]}{aires_struct}{\includegraphics[width=14cm]{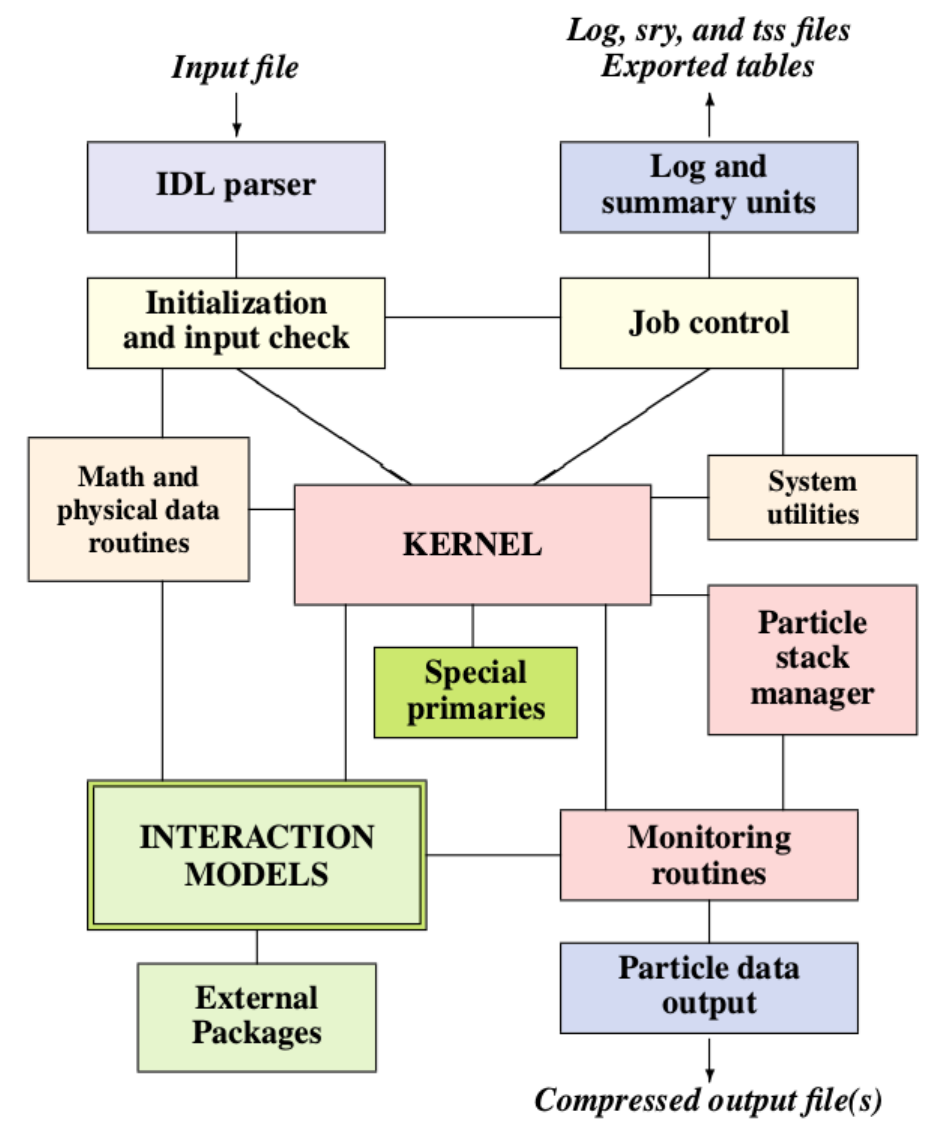}}%
{Structure of AIRES simulation program}%
{The structure of AIRES main simulation program.}
Figure \ref{FIG:aires_struct} contains a schematic representation of
the modular structure of the main simulation programs. Every unit
consists of a set of subroutines performing the tasks assigned to the
corresponding unit. In general, every unit can be replaced virtually
without altering the other ones. In the case of the external
interaction models where complete packages\index{external packages}
developed by other groups are linked to the simulation program via a
few interface routines, the modularity acquires particular importance
since it makes it possible to easily switch among the various packages
available.

The user controls the simulation parameters by means of input
directives. The {\em Input Directive Language\index{Input Directive
Language}\/} (IDL) is a set of human-readable directives than provides
a comfortable environment for task control. After the input data is
processed and checked, control is transferred to the program's kernel.
During the simulations the particles of the cascade are generated and
processed by several packages. The interactions model package contains
the ``physics'' of the problem.

The job control unit is responsible (among other tasks) of updating
the {\em internal dump file (IDF)\index{internal dump file}.\/} This
file contains all the relevant internal data used during the
simulation, and is the key for system fault tolerant
processing\index{fault tolerant processing} since it makes it possible
to restart a broken simulation process from the last update of the
IDF.

The kernel interacts also with other modules that generate the output
data, namely, log, summary\index{summary file}, and task summary
script\index{task summary script file} files, internal dump
file --in either binary or ASCII (portable) format-- and compressed output
files\index{compressed output files} generated by the monitoring routines
and the particle data output unit.

In the current version of AIRES, there are two compressed output files
implemented: The ground particle file and the longitudinal tracking
particle file. Records within the ground particle file (longitudinal
tracking file) contain data related with particles reaching ground
level (passing across observing levels).

Since the number of data records contained in such files can be enormous,
a special compression mechanism has been developed to reduce file size
requirements. The compressing algorithm is part of the particle data
output module. To give an idea of the space needed to store the particle
records, let us consider the case of the ground particle file with its
default settings: For each particle reaching ground and fulfilling certain
(user settable) conditions, a 18 byte long record is written. The record
data items are: particle identity, statistical weight, position, time of
arrival and direction of motion.  Leading and trailing records are written
before and after an individual shower is completely
simulated. Considering, for instance, a ``hard'' simulation regime where
$2\times10^{19}$ eV primary energy showers (proton or iron) are simulated
with $10^{-7}$ relative thinning\index{thinning} level using the standard
Hillas algorithm (see section \ref{S:thinning}), generate a compressed
ground particle file of size less than 11 MB/shower when storing all the
particles whose distance from the shower core is larger than 50 m and less
than 12 km.

The green unit named ``special primaries''\index{special primary
particles} consists basically in a kernel-operated interface with
user-provided external modules capable of generating lists of
particles that will be used to initiate a shower. This feature allows
the user to start showers initiated by non-conventional (exotic)
primary particles like neutrinos, for example.

The math and physical data routines are called from several units
within the program and provide many utility calculations. In
particular, they contain the atmospheric model\index{atmospheric
model} (used to account for the varying density of the Earth's
atmosphere) and the geomagnetic field\index{geomagnetic field}
auxiliary routines that can evaluate the geomagnetic field in any
place around the world.
\section{Getting and installing AIRES}
\index{AIRES!installation}\index{installing AIRES}%
\label{S:downloading}

AIRES is distributed worldwide as ``free software'' for all scientists working
in educational/research non-profit institutions. Users from commercial or
non-educational institutions must obtain the author's written permission
before using the software.

The present version of AIRES (\currairesversion) can be obtained from
the World Wide Web, at the following site:
\begin{displaymath}
\hbox{\ttbf
  aires.fisica.unlp.edu.ar}
\end{displaymath}

AIRES is distributed in the form of compressed UNIX tar files. The
installation is automatic for LINUX and Mac OS systems. For other operating
systems some adaptive work may be needed. Appendix \ppref{A:installing}
contains detailed instructions on how to install AIRES and/or maintain
an existing installation.

\chapter{General characteristics of AIRES}\label{C:genchar}

The aim of this chapter is to introduce the basic concepts needed to
adequately define the problem being considered.

\section{The environment of an air shower}

\subsection{Coordinate system}\label{S:coordinates}

The AIRES coordinate system\index{AIRES coordinate system} is a
Cartesian system whose origin is placed at sea level at a
user-specified geographical location. The $xy$ plane is located
horizontally at sea level and the positive $z$-axis points
upwards. The $x$-axis points to the ``local'' magnetic North, that is,
the local direction of the horizontal component of the geomagnetic
field\index{geomagnetic field} (see section \ref{S:geomag} for
details). The $y$-axis points to the West.

Figure \ref{FIG:coor1} shows an schematic representation of the
coordinate system used by AIRES. The $xy$ plane is tangent to the sea
level surface, here taken as a spherical surface of radius $R_e =
6370949$ m\label{def:Re} centered at the Earth's center. The {\em
ground level,\/} and the {\em injection level,\/} refer to spherical
surfaces concentric to the sea level surface and intersecting the
$z$-axis at $z=z_g$ ($z_g\ge0$) and $z=z_i$ ($z_i>z_g$) respectively.
\lowfig{[htb]}{coor1}{\includegraphics[width=13cm]{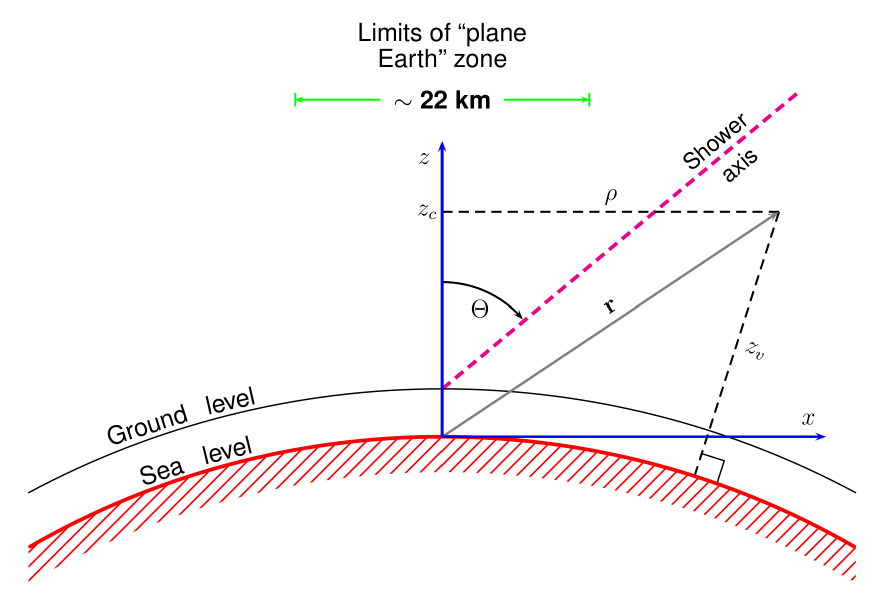}}%
{AIRES coordinate system}{AIRES coordinate system.\index{AIRES
coordinate system}}

The {\em shower axis\/} of a shower with zenith angle $\Theta$ is
defined as the straight line that passes by the intersection point
between the ground level and the $z$-axis, and makes an angle $\Theta$
with the $z$-axis ($0\le\Theta<90^\circ$). The azimuth angle $\Phi$ is
the angle between the horizontal projection of the shower axis and the
$x$-axis ($0\le\Phi<360^\circ$).

In AIRES version 1.2.0, all the spherical surfaces mentioned in the
preceding paragraphs were approximated as planes. This approximation
is justified every time the horizontal distances involved are
negligible in comparison with the Earth's radius, $R_e$. This is the
case for showers whose zenith angle is small, but certainly not for
those with large zenith angles, especially for quasi-horizontal
showers.

For AIRES version 1.4.0 or later the curvature of the
Earth\index{Earth's curvature} is taken into account to make it
possible to reliably simulate showers with zenith angles in the full
range $0\le\Theta<90^\circ$. Since full spherical calculations are
computationally expensive, an effort was made to optimize the
corresponding algorithms. These optimizations are based on two key
concepts: \iiset\ii Even if a non-vertical shower can start in a very
distant point, most of the shower development takes place relatively
near the $z$-axis where the ``plane Earth'' approximation is
acceptable.  \ii Many calculations that employ spherical geometry can
be substantially simplified if the coordinate system is temporarily
rotated so the involved point lies near the new $z$-axis, and plane
geometry is used in the rotated system. If necessary, an inverse
rotation is applied to express results in the original coordinate
system.

In order to apply the first concept, a zone where the Earth can be
acceptably approximated as plane must be defined. As it will be justified
later in this chapter (see section \ref{S:planeok}), the Earth's spherical
shape can be ignored in a conic region region centered at the $z$-axis,
with a varying diameter ranging from 8 km at sea level to 45 km at an
altitude of 100 km.a.s.l. The average limits of that region (about 22 km
diameter) are indicated in figure \ref{FIG:coor1}.

To fastly perform the rotation operations needed to express
coordinates and vector in a temporary local coordinate system, it
results convenient to use a redundant set of coordinates, defined as
follows: Let $\vr{r}$ be the position vector of a point with
coordinates $(x,y,z)$. We define the {\em vertical altitude,\/} $z_v$,
of the point as the minimum distance between the point and the sea
level surface. It is straightforward to demonstrate that
\begin{equation}                                \label{eq:zv}
(R_e +z_v)^2 = (R_e + z_c)^2 + \rho^2
\end{equation}
where $\rho^2=x^2 + y^2$ and $z_c=z$ denotes the point's {\em central
altitude,\/} an alternative way to express the $z$-coordinate which
stresses the fact that this coordinate is always measured along the
same central axis. The redundant set of coordinates
\begin{equation}                                 \label{eq:redundant}
(x, y, z_c, z_v)
\end{equation}
is used by AIRES to define the position of a point. The difference
between $z_c$ and $z_v$ gives information about how far from the
$z$-axis is the point, and in the ``plane Earth'' zone $z_v$ is set
equal to $z_c$.

This way of taking into account the Earth's shape in the simulations
proved to be accurate enough when compared with exact procedures
while being economic from the computational point of view.

\subsection{Atmosphere}
\label{S:atmosphere}

The Earth's atmosphere is the medium where the particles of the shower
propagate and their evolution depends strongly on its
characteristics. The simulations must therefore be based on realistic
models of the relevant atmospheric quantities.\index{atmospheric model}

The atmosphere has been extensively measured and studied during the
last decades. As a result, a variety of models and parameterizations
of measured data have been published. Among them, the so-called {\em
US standard atmosphere\index{US standard atmosphere}\/} \cite{usatm76}
is a widely used model based on experimental data\footnote{The US
standard atmosphere is sometimes referred as the US extension of the
ICAO (International Civil Aviation Organization) standard
atmosphere.}.  We have selected it as a convenient default model to use in
AIRES which gives an acceptably realistic approximation of the average
atmosphere.

Besides the default model, AIRES accepts other atmospheric models,
including the possibility of adding user-defined custom models, as
explained with more detail in section ....

An evident characteristic of the atmospheric medium is that of being
inhomogeneous. Its density, for instance, diminishes six orders of
magnitude when the altitude above sea level passes from zero to 100
km, and another additional six orders for the range 100 km to 300 km
\cite{usatmCRC}. This fact is taken into account in the model we have
selected, where most of the relevant observables are regarded as
functions of the altitude above sea level, or vertical altitude, $h$:
The atmosphere is thus a spherically symmetric ``layer'' a few
hundreds kilometers thick, whose internal radius is the Earth's radius
(6370 km).

For a variety of processes that the particles can undergo during the
development of the shower, it is essential to know the chemical
composition as well as the density of the medium they are passing
through \cite{rossi}. For this reason, we have studied the behavior of
these two quantities, especially their dependence with the vertical
altitude.

\latfig{[htb]}{Mvsh}{\epsfig{file=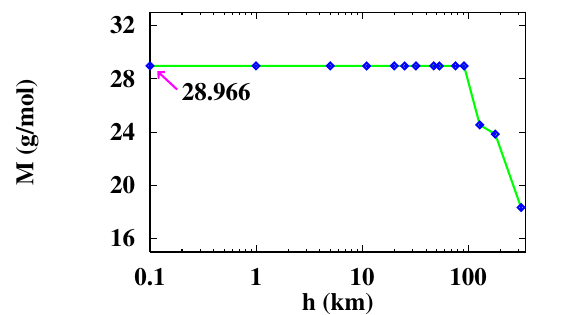}}%
{Mean molecular weight of the atmosphere versus altitude}%
{Mean molecular weight of the atmosphere as a function of the vertical
altitude (US standard atmosphere\index{US standard atmosphere}
\cite{usatmCRC}). The line is only to guide the eye.}
The chemical composition of the air, as given by the mean molecular weight,
remains virtually unchanged in all the region $0\le h\le 90$ km, and
diminishes progressively for larger values of $h$. This clearly shows up in
figure \ref{FIG:Mvsh}, where the US standard atmosphere mean molecular weight
\cite{usatmCRC} has been plotted versus the vertical altitude. The constant
value $M=28.966$ is the mean molecular weight corresponding to an atomic
mixture of 78.47\% N, 21.05\% O, 0.47\% Ar and 0.03\% other elements. The
corresponding mean atomic weight (atomic number) is 14.555 (7.265). The ratio
between mean atomic number and weight is 0.499.

On the other hand, the density of the air does change considerably
with the vertical altitude, as shown in figure \ref{FIG:rhovsh}. The
dots are the US standard atmosphere\index{US standard atmosphere}
data, taken from reference \cite{usatmCRC}.  The green full line
corresponds to Linsley's parameterization of the US standard
atmosphere \cite{Linsley1}, also called Linsley's atmospheric model or
Linsley's model, which effectively reproduces very accurately the US
standard atmosphere data. The isothermal atmosphere
\begin{equation}                                \label{eq:isotermic}
\rho(h)=\rho_0\,e^{-gMh/RT}
\end{equation}
was also plotted (dotted red line) for comparison. $\rho_0$ and $T$
match the corresponding US standard atmosphere values at sea level.
\latfig{[htb]}{rhovsh}{\epsfig{file=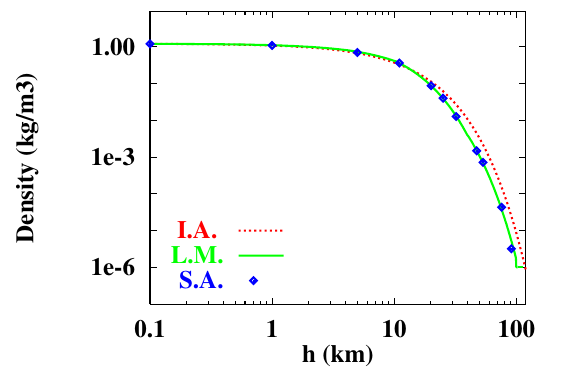}}%
{Density of air versus altitude}%
{Density of the air as a function of the vertical altitude. The dots
represent the US standard atmosphere data\index{US standard
atmosphere} \cite{usatmCRC}, while the full green line corresponds to
Linsley's model \cite{Linsley1} and the dashed red one to the
isothermal atmosphere $\rho(h) = \rho_0\, e^{-gMh/RT}$ with
$\rho_0=1.225$ $\mrm{kg}/\mrm{m}^3$, $M=28.966$ and $T=288$ K.}

It is worthwhile mentioning that Linsley's model is limited to
altitudes less than 100 km. Currently, in AIRES the model has been
extended up to $h_{\mathrm{max}}\sim 420$ km, approximately; the
density is considered to be zero for $h>h_{\mathrm{max}}$. This
approximation helps very much to simplify different algorithms used in
air shower simulations while being absolutely justified since only
affects an atmospheric zone placed much above the region where the air
showers take place, which at most extends up to 50 vertical kilometers
above sea level.

For the same reason, the chemical composition of the air can be
assumed to be constant in the full range of non-vanishing density
($0\le h\le h_{\mathrm{max}}$). As shown in figure \ref{FIG:Mvsh},
this only affect the very upper layer of the atmosphere, with
altitudes larger than 90 km.

A further approximation that will be made when necessary is to assume
that the air is a ``pure'' substance made with ``air'' atoms whose
nuclei possess charge $Z_{\mathrm{eff}}$ and mass number
$A_{\mathrm{eff}}$. To match the actual molecular weight, it is
necessary to set $Z_{\mathrm{eff}} =7.3$ and
$\langle Z_{\mathrm{eff}}/A_{\mathrm{eff}}\rangle=0.5$ \cite{MOCCA}.

The density of the air is not directly used by the related algorithms:
The quantity that naturally describes the varying density of the
atmospheric medium is the so called {\em vertical atmospheric
depth,\/} $X_v$, defined as follows\index{atmospheric
depth}\index{atmospheric depth!vertical}:
\begin{equation}                               \label{eq:Xvdef}
X_v(h) = \int_h^\infty \rho(z)\>dz.
\end{equation}
The integration path is the vertical line that goes from the given
altitude, $h$, up to infinity. The usual unit to express $X_v$ is
\gcmsq. In figures \ref{FIG:Xvsh} and \ref{FIG:Xvsh2}, $X_v(h)$
(Linsley's model) is plotted against $h$. Notice that $X_v(0) \approx
1000$ {\gcmsq} and $X_v(h)\to 0$ for $h\to\infty$ as expected.
\lowfig{[t]}{Xvsh}{\epsfig{file=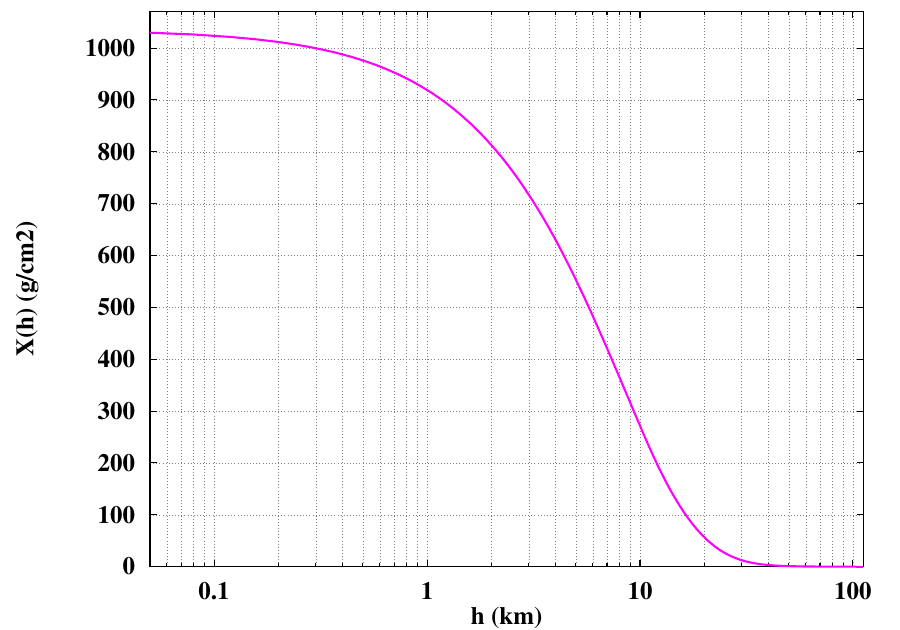}}%
{Vertical atmospheric depth versus altitude}%
{Vertical atmospheric depth\index{atmospheric depth!vertical}, $X_v$,
versus vertical altitude over sea level, $h$, accordingly with
Linsley's model \cite{Linsley1}.}
\latfig{[htb]}{Xvsh2}{\epsfig{file=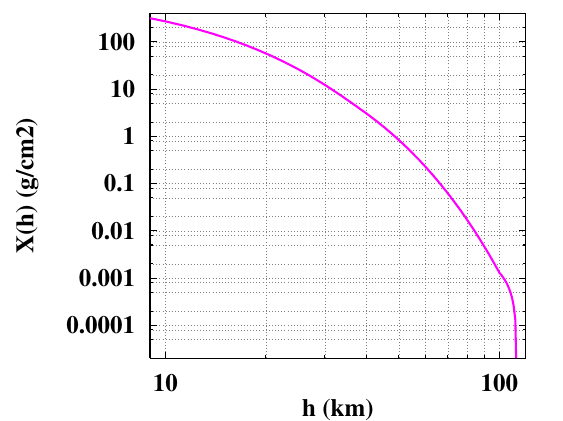}}%
{Vertical atmospheric depth for altitudes larger than 10 km}%
{Same as figure \ref{FIG:Xvsh}, but for altitudes larger than 10 km.}

$\rho(h)$ can be obtained from $X_v(h)$ via
\begin{equation}                               \label{eq:rhofromX}
\rho(h)= -{{dX_v(h)}\over{dh}}.
\end{equation}

Linsley's parameterization of $X_v(h)$ \cite{Linsley1}, is done as
follows: \iiset\ii The atmosphere is divided in $L$ layers. For
$l=1,\ldots,L$ layer $l$ starts (ends) at altitude $h_l$ ($h_{l+1}$).
It is clear that $h_1 = 0$ and $h_{L+1}=h_{\mathrm{max}}$.
\ii $X_v(h)$ is given by:
\begin{equation}                             \label{eq:Xfromh}
X_v(h) = \left\{\begin{array}{ll}
a_l + b_l e^{-h/c_l} & h_l \le h < h_{l+1}, \quad l=1,\ldots,L-1 \\
a_L - b_L (h/c_L)    & h_L \le h < h_{L+1} \\
0                    & h \ge h_{L+1}.
\end{array}\right.
\end{equation}
Where the coefficients $a_l$, $b_l$ and $c_l$, $l=1,\ldots,L$ are
adjusted to fit the corresponding experimental data. The coefficients
used in AIRES, which correspond to a model with $L=5$
layers, are listed in table \ref{TAB:Linsley0},
and are the ones that come out from a fit to the US standard
atmosphere data. The Linsley's model\index{atmospheric model}
prediction for $\rho(h)$ plotted
in figure \ref{FIG:rhovsh} was obtained using this coefficient set and
equations (\ref{eq:Xfromh}) and (\ref{eq:rhofromX}).
%
\lowtable{[t]}{Linsley0}{%
\def\phb{$\phantom{0}$}
\begin{tabular}{c|rr|rrc}
{\bfseries\itshape Layer} & \multicolumn{2}{c|}{\bfseries\itshape
 Layer limits\/ {\rm (km)}} &
\multicolumn{1}{c}{$a_l$} & \multicolumn{1}{c}{$b_l$} & $c_l$ \\
$l$ & \multicolumn{1}{c}{\bfseries\itshape From} &
  \multicolumn{1}{c|}{\bfseries\itshape\kern1em To} &
\multicolumn{1}{c}{(\gcmsq)} & \multicolumn{1}{c}{(\gcmsq)} & (m) \\ \hline
1 &   0 &   4 & $-$186.5562\phb & 1222.6562 & 9941.8638 \\
2 &   4 &  10 &  $-$94.9199\phb & 1144.9069 & 8781.5355 \\
3 &  10 &  40 &      0.61289    & 1305.5948 & 6361.4304 \\
4 &  40 & 100 &      0.0\phb\phb\phb\phb    
                                &  540.1778 & 7721.7016 \\
5 & 100 & $\sim$113
              & \multicolumn{1}{c}{0.01128292} & \multicolumn{1}{c}{1}
                                            & $10^7$
\end{tabular}
}{%
Original Linsley's model coefficients for the US standard atmosphere}%
{Linsley's model coefficients for the US standard atmosphere\index{US
standard atmosphere} \cite{Linsley1}. The number of layers is $L=5$.
}
%

Another important property of Linsley's parameterization is that
$X_v(h)$ can easily be inverted to obtain $h=X_v^{-1}(X)$ ($X>0$): Let
$X_l= X_v(h_l)$, $l=1,\ldots,L$, then
\begin{equation}                             \label{eq:hfromX}
h=\left\{\begin{array}{ll}
{\displaystyle -c_l\, \ln\!\left({{X-a_l}\over{b_l}}\right)} &
X_{l+1} < X \le X_l, \quad l =1,\ldots, L-1\\*[10pt]
c_L(a_L-X)/b_L & 0 < X \le X_L ,
\end{array}\right.
\end{equation}
where the replacement $X_v(h_{L+1})=0$ has been made.

A quantity related to the vertical depth that appears frequently in air
shower calculations is the {\em slant atmospheric depth\index{atmospheric
depth!slant|bfpage{ }},\/} $X_s$, defined similarly as $X_v$ (equation
(\ref{eq:Xvdef})) but using a non-vertical integration path. In most
applications the integration path is a straight line going along the
shower axis, from the given point to infinity. In this case $X_s$ takes
the form:
\begin{equation}                               \label{eq:Xsdef}
X_s(z) = \int_z^{'\,\infty} \rho(z_v)\>dl,
\end{equation}
where the prime in the integral indicates that the path is along a
non-vertical line and $z_v$ is the vertical altitude defined in
equation (\ref{eq:zv}).

The integral in equation (\ref{eq:Xsdef}) cannot be solved
analytically in the general case of an arbitrary geometry (see page
\pageref{rou0:xslant}).
If the Earth's curvature\index{Earth's curvature} is not taken into
account (plane Earth), then it is straightforward to prove that
\begin{equation}                         \label{eq:Xsplane}
X_s(h) = {{X_v(h)}\over{\cos\Theta}},
\end{equation}
where $\Theta$ is the zenith angle of the shower axis (see section
\ref{S:coordinates}). From this equation it comes out that $X_s$
depends not only on $h$ but also on $\Theta$ and the location of the
ground surface.

Unless otherwise specified, any reference to atmospheric depth, or
depth, is assumed to be a reference to $X_v$ which may also be
noted simply $X$.\footnote{Notice that in some publications the symbol
$X$ is used to represent the slant depth.}
\subsection{The slant depth and the Earth's curvature.}
\label{S:curvature}\index{atmospheric depth!slant}
\index{slant atmospheric depth|see{atmospheric depth, slant}}
\index{vertical atmospheric depth|see{atmospheric depth, vertical}}
\index{Earth's curvature}
Many air shower observables, especially the ground level
distributions, depend on
the thickness of the air layer that separates the starting point of an
air shower from the ground level. For non-vertical showers starting at
the top of the atmosphere, this thickness is measured in terms of the
slant depth evaluated at ground level, $X_s(z_g)$. The plane Earth
approximation given by equation (\ref{eq:Xsplane}) is usually employed
to evaluate that quantity. However, this approximate equation can give
inaccurate estimations for large zenith angles, and in fact it is
divergent for $\Theta=90^\circ$.

%
\lowtable{[htb]}{Xsvszen}{%
{\def\pz{$\phantom{0}$}
\begin{tabular}{|c|c|c|c|c|}\hline
\multicolumn{1}{|c}{\thpop{\bfseries\itshape Zenith angle}}&
\multicolumn{2}{|c}{\bfseries\itshape Curved Earth}&
\multicolumn{2}{|c|}{\bfseries\itshape Plane Earth}\\\cline{2-5}
\multicolumn{1}{|c}{\thpop(deg)}&
\multicolumn{1}{|c}{{\bfseries\itshape length\/} (km)}&
\multicolumn{1}{|c}{{\bfseries\itshape path\/} (\gcmsq)}&
\multicolumn{1}{|c}{{\bfseries\itshape length\/} (km)}&
\multicolumn{1}{|c|}{{\bfseries\itshape path\/} (\gcmsq)}\\\hline
 \pz0 & \pz110 & \pz1036.1  & \pz110  &  \pz1036.1 \\ 
   30 & \pz127 & \pz1195.9  & \pz127  &  \pz1196.4 \\
   45 & \pz154 & \pz1463.6  & \pz156  &  \pz1465.3 \\
   60 & \pz215 & \pz2065.1  & \pz220  &  \pz2072.2 \\
   70 & \pz303 & \pz3003.7  & \pz322  &  \pz3029.4 \\
   80 & \pz518 & \pz5765.5  & \pz633  &  \pz5966.7 \\
   85 & \pz757 &   10571.7  &   1262  &    11887.9 \\
   89 &   1083 &   25919.3  &   6303  &    59367.2 \\
   90 &   1189 &   36479.9  &$\infty$ & $\infty$   \\\hline
\end{tabular}}
}{%
Total shower axis length and slant path versus zenith angle}%
{Total shower axis length ($\mathrm{m}$) and slant
path\index{atmospheric depth!slant} (\gcmsq) measured from the top of
the atmosphere (110 km.a.s.l) down to sea level, tabulated versus the
zenith angle.  }
%
To precisely estimate $X_s(z_g)$ we have evaluated numerically the
integral of equation (\ref{eq:Xsdef}) for various representative cases.
In table \ref{TAB:Xsvszen} the results corresponding to $z_g=0$
(ground level located at sea level) are tabulated for different zenith
angles. The top of the atmosphere is located at an altitude of 110
km.a.s.l, and Linsley's parameterization is used in the
calculations. The respective data corresponding to the plane Earth
model are also tabulated for comparison purposes.

The tabulated quantities indicate that the plane and curved Earth
estimations differ in less than 4\% for all zenith angles
$\Theta\le80^\circ$, and the differences increase notably as long as
the zenith angle approaches $90^\circ$.

The geometrical length of the shower axis, $a$, is also tabulated for both
models. In the plane Earth approximation this length is given by
\begin{equation}                            \label{eq:planelaxis}
a = {{z_{\mathrm{max}}-z_g}\over{\cos\Theta}},
\end{equation}
where $z_{\mathrm{max}}$ is the (vertical) altitude of the top of the
atmosphere (110 km in the present case).
On the other hand, if the Earth's curvature\index{Earth's curvature}
is taken into account, the expression for $a$ becomes
\begin{equation}                            \label{eq:curvedlaxis}
a=\sqrt{(R_e+{z_v}_{\mathrm{max}})^2-(R_e+z_g)^2\sin^2\Theta}
-(R_e+z_g)\cos\Theta,
\end{equation}
where ${z_v}_{\mathrm{max}}$ stands for the vertical altitude of the
injection point (110 km). Equation (\ref{eq:planelaxis}) is the
$R_e\to\infty$ limit of equation (\ref{eq:curvedlaxis}).
\subsection{Range of validity of the ``plane Earth'' approximation}
\label{S:planeok}
In section \ppref{S:coordinates} it is specified that the limit of the
``plane Earth'' zone is located at a certain distance from the central
$z$-axis. This distance varies linearly with the altitude and goes from 4
km at sea level up to 22.5 km at 100 km above sea level.

To determine the boundaries of that zone, that is, a region
where plane geometry can safely be used in the involved procedures,
the requirement of expressing the vertical depth of a given point with
enough precision was taken into account. The condition actually
imposed can be defined in the following terms: Let $d$ be the
horizontal distance of a certain point to the $z$-axis, and let $z$
and $z_v$ be the point's central and vertical altitudes. Let
\begin{equation}                             \label{eq:hdeltaX}
\Delta X(d) = X_v(z_v) - X_v(z).
\end{equation}
In a plane geometry, $\Delta X$ is zero for all $d$ provided $z$ is
kept fixed. We can use this quantity to determine a safe ``plane
zone'' imposing a bound on $\Delta X$. After a series of technical
considerations, too many to be explained in detail here, we concluded
that the geometry can be acceptably taken as plane for all points
whose distances to the $z$-axis are less than $d_{\mathrm{max}}$
defined by the condition\footnote{This requirement is more stringent that
the one used for AIRES version 1.4.2a or earlier. The original equations
\cite{Aires142} were not adequate in certain particular conditions,
namely, quasi-horizontal showers, and were thus modified.}:
\begin{equation}                             \label{eq:dmax}
\Delta X(d_{\mathrm{max}}) < 0.25\;\hbox{\gcmsq}
\quad {\bf AND}\quad
2\,\Delta X(d_{\mathrm{max}}) < 1\,\hbox{\%} \times X_v(z).
\end{equation}

Using equations (\ref{eq:zv}) and (\ref{eq:dmax}), and taking into account
that $\Delta X \cong (z_v -z) \rho(z)$, it is simple to obtain estimations
for $d_{\mathrm{max}}$ at different altitudes.  At sea level, for example,
where the vertical depth is approximately 1030 {\gcmsq}, and the density
of the air is 1.22 $\mathrm{kg/m^3}$, we obtain
\begin{equation}                             \label{eq:dmax2a}
d_{\mathrm{max}}\> \widetilde<\> 5.5\;\mathrm{km}.
\end{equation}
The same calculation for 100 km above sea level yields
\begin{equation}                             \label{eq:dmax2b}
d_{\mathrm{max}}\> \widetilde<\> 24\;\mathrm{km}.
\end{equation}
The boundaries of used by AIRES (see section \ref{S:coordinates}) agree
with these results.
\subsection{Geomagnetic field}%
\index{geomagnetic field}%
\index{Earth's magnetic field|see{geomagnetic field}}
\label{S:geomag}
All charged particles that move near the Earth are deflected by the
geomagnetic field. Such deflections are taken into account in the
internal algorithms of AIRES.

The Earth's magnetic field, $\vr{B}$, is described by its strength, F,
$\mathrm{F}= \|\vr{B}\|$; its inclination, I, defined as the angle
between the local horizontal plane and the field vector; and its
declination, D, defined as the angle between the horizontal component
of $\vr{B}$, H, and the geographical North (direction of the local
meridian). The angle I is positive when $\vr{B}$ points downwards and
D is positive when H is inclined towards the East.

Let $(B_x, B_y, B_z)$ be the Cartesian components of $\vr{B}$ {\em
with respect to the AIRES coordinate system\index{AIRES coordinate
system}\/} (section \ref{S:coordinates}). They can be obtained from
the field's strength and inclination via
\begin{equation}                                   \label{eq:Bxyz}
B_x = F\cos\mathrm{I},\quad B_y=0,\quad B_z=-F\sin\mathrm{I}.
\end{equation}
$B_y$ is always zero by construction, since in the AIRES coordinate
system the $x$-axis points to the local {\em magnetic\/} north,
defined as the direction of the H component of the geomagnetic field.

There are two alternatives for specifying the geomagnetic field in AIRES:
(i) Manually, entering F, I and D. (ii) Giving the geographical
coordinates, altitude and date of a given event. In the later case, the
magnetic field is evaluated using the International Geomagnetic Reference
Field (IGRF)
\cite{IGRF}\index{International Geomagnetic Reference Field}%
\index{IGRF|see{International Geomagnetic Reference Field}}, a
widely used model based on experimental data that gives accurate
estimations of all the components of the Earth's magnetic
field\index{external packages}.

We are not going to place here any further analysis of the geomagnetic
field and its implementation in an air shower simulation program. The
interested reader can consult
reference \cite{geomag1} which contains
a detailed description of general aspects of the geomagnetic
field and the IGRF, together with a discussion about the practical
implementation of the deflection procedure and an analysis of the
effect of the geomagnetic field on several air shower observables.

\section{Air showers and particle physics}

We are going to describe here how the particles of an air shower are
identified and processed and which interactions are taken into account.

\subsection{Particle codes.}\label{S:acodes}%
\index{particle codes}\index{AIRES particle codes|bfpage{ }}

AIRES recognizes all the particles commonly taken into account in air
shower simulations plus additional ones included for completeness.
Each particle is internally identified by a particle code. It is
important to notice, however, that user level particle specifications
are made by means of {\em particle names\/} instead of numeric codes.

Table \ref{TAB:eqcode} lists AIRES particle codes, together with the
corresponding particle names and synonyms.
%
\lowtable{[p]}{eqcode}{%
\def\phigh{$\vphantom{|^{|^|}}$}\def\plow{$\vphantom{|_{|_|}}$}
\begin{tabular}{|crl||crl|}
\hline
{\bfseries\itshape Particle}\phigh\plow&
{\bfseries\itshape Code}&
{\bfseries\itshape Name and synonyms}~~~~ &
{\bfseries\itshape Particle}&
{\bfseries\itshape Code}&
{\bfseries\itshape Name and synonyms}\\ \hline
$\gamma$\phigh &     1 & {Gamma gamma} &
                         $n$       &  30 & {n Neutron neutron}\\*[1pt]
$e^+$          &     2 & {e$+$ Positron positron} &
                         $\bar{n}$ & $-$30 &{nbar AntiNeutron antineutron}\\*[1pt]
$e^-$          &  $-$2 & {e$-$ Electron electron} &
                         $p$         &    31 & {p Proton proton}\\*[1pt]
$\mu^+$        &     3 & {mu$+$ Muon$+$ muon$+$} &
                          $\bar{p}$  & $-$31 & {pbar AntiProton antiproton}\\*[1pt]
$\mu^-$        &  $-$3 & {mu$-$ Muon$-$ muon$-$} &
                          $\Lambda$        &    40 & {Lambda}\\*[1pt]
$\tau^+$       &     4 & {tau$+$} &
                          $\bar{\Lambda}$ & $-$40 & {Lambdab}\\*[1pt]
$\tau^-$       & $-$4 & {tau$-$} &
                          $\Sigma^0$       &    41 & {Sigma0}\\*[1pt]
$\nu_e$        &     6 & {nu(e)} &
                          $\bar\Sigma^0$   & $-$41 & {Sigma0b}\\*[1pt]
$\bar\nu_e$    &  $-$6 & {nubar(e)} &
                          $\Sigma^+$       &    42 & {Sigma$+$}\\*[1pt]
$\nu_\mu$  &     7 & {nu(m)} &
                           $\bar\Sigma^+$   & $-$42 & {Sigma$+$b}\\*[1pt]
$\bar\nu_\mu$  &  $-$7 & {nubar(m)} &
                           $\bar\Sigma^-$   &    43 & {Sigma$-$b}\\*[1pt]
$\nu_\tau$ &     8 & {nu(t)} &
                           $\Sigma^-$       & $-$43 & {Sigma$-$}\\*[1pt]
$\bar\nu_\tau$ &  $-$8 & {nubar(t)} &
                           $\Xi^0$          &    44 & {Xi0}\\*[1pt]
$\pi^0$        &    10 & {pi0} &
                           $\bar\Xi^0$      & $-$44 & {Xi0b}\\*[1pt]
$\pi^+$        &    11 & {pi$+$} &
                           $\bar\Xi^-$      &    46 & {Xi$-$b}\\*[1pt]
$\pi^-$        & $-$11 & {pi$-$} &
                           $\bar\Omega^-$   &    47 & {Omega$-$b}\\*[1pt]
$K^0_S$        &    12 & {K0S} &
                           $\Omega^-$       & $-$47 & {Omega$-$}\\*[1pt]
$K^0_L$        &    13 & {K0L} &
                           $\Lambda_c^+$    &    48 & {Lambdac$+$}\\*[1pt]
$K^+$          &    14 & {K$+$} &
                           $\Lambda_c^-$    & $-$48 & {Lambdac$-$}\\*[1pt]
$K^-$          & $-$14 & {K$-$} &&& \\*[1pt]
$D^0$          &    16 & {D0} &&& \\*[1pt]
$\eta$         &    15 & {eta} &&& \\*[1pt]
$\bar D^0$     & $-$16 & {D0bar} &&& \\*[1pt]
$D^+$          &    17 & {D$+$} &&& \\*[1pt]
$D^-$          & $-$17 & {D$-$} &&& \\*[1pt]
$D_s^+$        &    18 & {Ds$+$} &&& \\*[1pt]
$D_s^-$        & $-$18 & {Ds$-$} &&& \\*[1pt]
$B^0$          &    20 & {B0} &&& \\*[1pt]
$\bar B^0$     & $-$20 & {B0bar} &&& \\*[1pt]
$B^+$          &    21 & {B+} &&& \\*[1pt]
$B^-$          & $-$21 & {B-} &&& \\*[1pt]
$\rho^0$       &    25 & {rho0} &&& \\*[1pt]
$\rho^+$       &    26 & {rho$+$} &&& \\*[1pt]
$\rho^-$       & $-$26 & {rho$-$} &&& \\*[1pt]
\hline
\end{tabular}
}{AIRES particle codes and names}%
{AIRES particle codes\index{AIRES particle codes} and names. The
  nuclear coding system and nuclear names are explained in the text.
}
%

Nuclear codes are set taking into account $Z$ (atomic number), $N$
(number of neutrons) and $A= Z + N$ (mass number), in a
computationally convenient codification formula:
\begin{equation}                                     \label{eq:nuclcode}
\hbox{code} = 100 + 32\, Z + (N - Z + 8),
\end{equation}
with $0\le N-Z+8\le 31$.  Taking $1\le Z\le 26$ (from hydrogen to
iron), this coding system allows to uniquely identify
all known isotopes.

Regarding the names of nuclei, they can be specified in several ways:
(i) By their chemical names, for example \kwbf;Fe\^{}56; (56 refers to
the mass number $A$, which defaults to the most abundant isotope's mass
 number when not specified).
(ii) By special names, as \kwbf;Deuterium; for $\mrm{H}^2$ or
\kwbf;Iron; for $\mrm{Fe}^{56}$. (iii) By direct specification of $Z$,
$N$ and/or $A$, for example \kwbf;NZ 2 2; ($\mrm{He}^4$), \kwbf;ZA 26
54; ($\mrm{Fe}^{54}$), etc.

In certain cases it may be needed to refer to {\em groups\/} of
particles having some properties in common. There are several particle
groups defined in the AIRES system which can be useful in such
situations. The most important groups of particles are listed in table
\ref{TAB:pclegrp}.
%
\lowtable{[htb]}{pclegrp}{%
\begin{tabular}{lcl}
{\bfseries\itshape Group name and synonyms}&\ &
{\bfseries\itshape Particles in the group}\\*[5pt]
{NoParticles~~None} && {Empty group}\\
{AllParticles~~All} && {Universal group containing all
                        particles}\\
{AllCharged}        && {All charged particles, including all
                        nuclei}\\
{MassiveNeutral}    && {All non-charged massive particles}\\
{Nuclei}            && {All nuclei}\\
{Hadrons}           && {All hadrons}\\
{Neutrinos}         && {All neutrinos and anti-neutrinos}\\
{EM}                && $\gamma,\;e^+,\;e^-$\\
{e$+${}$-$}         && $e^+,\;e^-$\\
{mu$+${}$-$}        && $\mu^+,\;\mu^-$\\
{tau$+${}$-$}       && $\tau^+,\;\tau^-$\\
{GPion}             && $\pi^+,\;\pi^-,\;\pi^0$\\
{GChPion}           && $\pi^+,\;\pi^-$\\
{GKaon}             && $K^+,\;K^-,\;K^0_S,\;K^0_L$\\
{GChKaon}           && $K^+,\;K^-$\\
{GRho}              && $\rho^+,\;\rho^-,\;\rho^0$\\
{GChRho}            && $\rho^+,\;\rho^-$\\
{nppbar}            && $n,\;p,\;\bar{p}$\\
{nnbar}             && $n,\;\bar{n}$\\
{Nucnucbr}          && $n,\;\bar{n},\;p,\;\bar{p}$
\end{tabular}
}{AIRES particle groups}{AIRES particle groups.}
%

\subsection{Interactions taken into account in the current version of
AIRES}
\label{S:intlist}

The processes which are most relevant from the probabilistic point of
view are taken into account in AIRES. In the current version
(\currairesversion), the following interactions are
included:
\begin{itemize}
\item {\bf Electrodynamical processes:}
\begin{itemize}
\item Pair production\index{pair production} and $e^+e^-$
annihilation\index{positron annihilation}.
\item Bremsstrahlung\index{bremsstrahlung} (electrons and positrons).
\item Muon bremsstrahlung\index{muon bremsstrahlung} and muonic pair
production \cite{mubrem}.
\item Emission of ``knock-on'' electrons\index{knock-on electrons}
 ($\delta$ rays).
\item Compton\index{Compton effect} and
photoelectric\index{photoelectric effect} effects.
\item LPM effect\index{LPM effect} and dielectric
suppression\index{dielectric suppression}.
\end{itemize}
\item {\bf Hadronic processes:}
\begin{itemize}
\item Inelastic collisions hadron-nucleus.
\item Photonuclear reactions\index{photonuclear reactions}.
\item Nuclear fragmentation, elastic and inelastic.
\end{itemize}
\item {\bf Unstable particle decays.}
\item {\bf Particle propagation:}
\begin{itemize}
\item Medium energy losses (ionization).
\item Coulomb and multiple scattering.
\end{itemize}
\end{itemize}

The hadronic inelastic collisions and photonuclear
reactions\index{photonuclear reactions} are
processed by means of external hadronic interaction
models\index{hadronic models} when their energy is above a certain
threshold\index{threshold energies}; otherwise they are calculated
using an extension of Hillas'\index{Hillas, A. M.} splitting
algorithm\index{splitting algorithm} \cite{splithin,gaisserbk}.

AIRES includes (optionally) links to the well-known external hadronic
interaction packages\index{external packages}\index{hadronic models},
EPOS\index{EPOS} \cite{EPOSLHC}, QGSJET\index{QGSJET}
\cite{QGSJET2R4}, and SIBYLL\index{SIBYLL} \cite{SIBYLL23c}.

\subsection{Processing the interactions}
\label{S:simengine}

We are going to briefly describe how the different interactions are
processed in AIRES. We shall focus in the computational aspects of
these procedures; a more detailed description of the physics involved
in such processes is going to be published elsewhere \cite{papercpc}.

 First of all it is necessary to express that this description is a
general one: The actual algorithms do include a number of technical
details whose complete explanation is beyond the scope of this work,
even if their philosophy is concordant with the scheme here presented.

As mentioned below, in AIRES the particles are stored in arrays
(stacks) and processed sequentially. Each particle entry consists of
different data items containing the different variables that
characterize it: Particle code, energy, position, direction of motion,
etc.

For the simulation engine, the shower starts when the primary particle
is added to the previously empty stack. Then the stack processing loop
begins.

Let $E$, $\vr{r}$, $t$, $\vr{u}$ be respectively the kinetic energy,
position, time and direction of motion of a given particle identified
by its particle code $k_p$. When this particle is going to be
processed it will suffer one of several possible interactions $I_i$,
$i=1,\ldots,n$, $n\ge1$. To fix ideas, let us consider the case of a
positron. The possible interactions, $I_i$, are:
annihilation\index{positron annihilation},
interaction with an atom from the medium and emission of a
``knock-on'' electron\index{knock-on electrons}, and emission of a
bremsstrahlung\index{bremsstrahlung} photon.
\subsubsection*{Evaluating the mean free paths}
\index{mean free path}
Every interaction $I_i$ is characterized by its {\em cross section,\/}
$\sigma_i$, or, equivalently, by its {\em mean free path,\/}
$\lambda_i$. $\lambda_i$ and $\sigma_i$ are connected via:
\begin{equation}                                  \label{eq:mfpsigma}
\lambda_i = {{m_{\mathrm{air}}}\over{\sigma_i}},
\end{equation}
where $m_{\mathrm{air}}$ is the mass of an atom of the medium the
particle propagates trough, that is, an average atom of ``air'' in the
case of air showers. The usual units for $\lambda_i$ are \gcmsq.

The mean free paths do depend on the kind of interaction and on the
particle's instantaneous parameters. They can be calculated
analytically for certain interactions; in other cases they must be
estimated by means of parameterization of experimental data, and this
generally requires extrapolations out of the region corresponding to
the measurements. A typical example of this situation is the case of
the mean free paths for inelastic collisions particle-nucleus, where
``particle'' can be proton, gamma, other nucleus, etc. Such mean free
paths depend on the energy of the projectile particle, and must be
calculated for energies well above the maximum energies attainable in
collider experiments.

Figure \ref{FIG:mfp} contains plots of the mean free paths corresponding
to proton-nucleus, pion-nucleus, kaon-nucleus and Fe-nucleus collisions,
plotted as a function of the projectile energy. All the alternative sets
of mean free paths available in AIRES are displayed\index{hadronic cross
sections}.
\lowfig{[p]}{mfp}{\epsfig{file=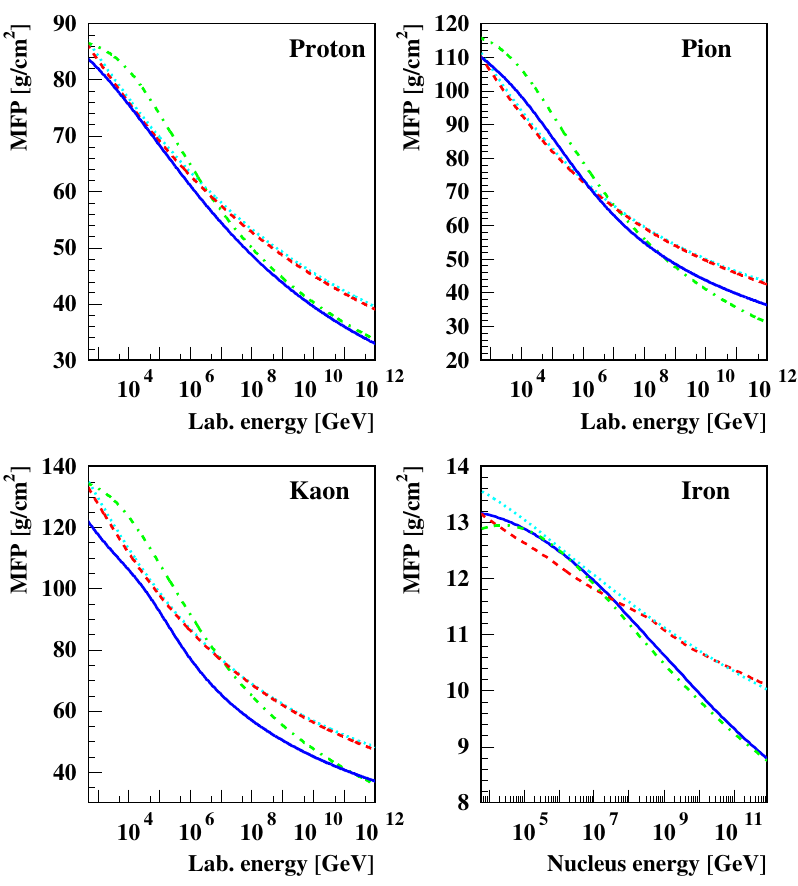}}%
{Hadronic mean free paths}%
{Hadronic mean free
 paths\index{mean free path!hadronic}%
\index{mean free path!nucleus-nucleus collisions}
 versus projectile energy (lab system)\index{hadronic cross sections}.
The solid (blue), and dashed (red) lines represent, respectively the
SIBYLL\index{SIBYLL} 2.1 and QGSJET01\index{QGSJET} models. In the proton,
pion and kaon cases the mean free paths corresponding to the models
SIBYLL 1.6 (dot-dashed, green) and QGSJET99 (dotted, cyan) have been
included for comparison. The iron plot includes also the mean free paths
evaluated using the AIRES built-in algorithm in the SIBYLL
(dot-dashed, green) and QGSJET (dotted, cyan) cases.}
\subsubsection*{Selecting the particle's fate}
For each interaction $i$, $\lambda_i$ represents the mean path
(expressed in ``quantity of matter'', that is, \gcmsq) the particle
should move before actually suffering the interaction. To evaluate the
actual path to a given interaction, it is necessary to sample the
corresponding exponential probability distribution, $P_i(p_i) =
\lambda_i^{-1} \exp(-p_i/\lambda_i)$. Let $p_i$, $i=1,\ldots,n$ the set
of values obtained after sampling the corresponding distributions for
all the possible interactions.

The interaction the particle will actually undergo, also called the
{\em fate\/} of the particle, is then selected: It is the interaction
$j$ corresponding to the minimum of the $p_i$'s, that is, $p_j\le p_i$
for all $i$.
\subsubsection*{Moving the particle and processing the selected interaction}
After the particle's fate has been decided, the corresponding
interaction begins to be processed. First, the particle must be
advanced the path indicated by $p_j$. It is necessary to convert the
path in a geometrical distance, and this depends on the atmospheric
model\index{atmospheric model} and the particle's current position. In
the case of charged particles, the advancing procedure also takes care
of the ionization energy losses, the scattering and the geomagnetic
field deflection. During this step, the particle's coordinates,
direction of motion and energy can be altered.

The final step is to process the interaction itself. This generally
involves the creation of new particles (secondaries) which are added
to the corresponding stacks and remain waiting to be processed, and
eventually the deletion of the current particle, for example in the
case of positron annihilation\index{positron annihilation}.

In some cases, it is necessary to apply corrections to the probability
distributions used to determine the particle's fate. This happens with
processes which have rapidly changing cross sections, or by corrective
processes not taken into account in the original
selection\footnote{The Landau-Pomeranchuk-Migdal (LPM)
effect\index{LPM effect} \cite{LPMigdal,LPM,KleinRev} is an example of
such kind of processes. The LPM effect implies a reduction of the
cross section of $e^\pm,\gamma$ processes at very high energies. In
AIRES it is implemented as a corrective algorithm whose effect is that
of rejecting a fraction of the previously ``approved'' processes. As a
result, the correct cross sections are statistically preserved.}. The
result of the corrective action is that of canceling some
interactions. In such cases the particle is left unchanged and remains
in the stack for further processing.
\subsubsection*{Particles arriving to destination}
The mechanism so far described is capable of generating and propagating
all the secondaries that come after the first interaction of the primary
particle. To let the shower finish it is necessary to determine when a
particle should no more be tracked. In AIRES this corresponds to the case
when one or more of the following conditions hold:
\begin{itemize}
\item The particle's energy is below a given threshold\index{threshold
energies} (low energy particles)\footnote{Unstable particles are
forced to decays.}.
\item The particle's position is out of the interesting region (lost
particles).
\item The particle reached the ground level.
\end{itemize}
It is very simple to show that this is enough to ensure that the
simulation of a shower will end in a finite time.
\subsubsection*{Particle monitoring}
\label{S:monitoring}
The simulation programs include several monitoring routines that
constantly check the status of the particles being propagated and
accumulate data then used to evaluate the different air shower
observables.

The events that are monitored are:
\begin{itemize}
\item Particles that reach ground level.
\item Particles that pass across predetermined observing levels. The
observing levels are constant depth surfaces generally located between
the injection and ground levels, and separated by a constant depth
increment $\Delta X_o$: If $N_o$ is the number of observing levels
($N_o>1$), and $X_o^{(1)}$ ($X_o^{(N_o)}$) is the vertical depth of
the first (last) observing level ($X_o^{(1)} < X_o^{(N_o)}$), then the
vertical depth of the other observing levels is given by
\begin{equation}                                   \label{eq:deltao}
\begin{array}{rcl}
\Delta X_o \!\!&=&\!\! \displaystyle{{X_o^{(N_o)} -X_o^{(1)}}\over
                                    {N_o - 1}} \\*[12pt]
X_o^{(i)}  \!\!&=&\!\! \displaystyle X_o^{(1)} + (i-1) \Delta X_o,\quad
                       i=1,\ldots,N_o  .
\end{array}
\end{equation}
Notice that the first observing level is that of highest altitude.
\item Charged particles that move across the air. For such particles
the continuous energy losses by ionization of the medium are evaluated
and recorded.
\end{itemize}

The data collected by the monitoring routines are used to evaluate
different kind of observables, for example:
\begin{description}
\item[Longitudinal development of the shower.]\index{longitudinal
development} Tabular data giving the number and energy of particles
crossing each defined observing levels.
\item[Shower Maximum.] The data collected for the longitudinal
development of all charged particles are used to estimate the {\em
shower maximum\index{shower maximum},\/}
$X_{\mathrm{max}}$\index{Xmax@$X_{\mathrm{max}}$}, that is, the
vertical depth\index{atmospheric depth!vertical} of the point where
the number of charged particles reaches its maximum (see section
\ref{S:sryfile}).
\item[Lateral distributions.]\index{lateral distributions} Frequency
distributions recording the number of particles reaching ground, as a
function of their distance to the shower core.
\item[Energy distributions.]\index{energy distributions} Energy
spectra of the different particles at ground level.
\item[Arrival time distributions.]\index{time distributions} Mean
ground level arrival time of different particle kinds as a function of
their distance to the shower core.
\end{description}

All the output coming from the monitoring routines is saved in the
form of data tables that can be easily retrieved by the user (see
chapter \ref{C:output1}).

\subsection{Random number generator}
\label{S:randomgen}\index{random number generator}

AIRES contains many procedures that require using random numbers, the
most important example being the propagating procedures that were
described in the preceding paragraphs. Those numbers are adequately
generated by means of a built-in pseudorandom number generator
\cite{MOCCA}, whose source code is included within the AIRES
distribution.

During the early steps of AIRES development, the random number
generator was checked with a series of tests, including uniformity and
correlation tests among others. In particular, this pseudorandom
number generator passed the very stringent ``random walk'' and
``block'' tests described in reference \cite{rwnbtests}.

A more detailed description of the different routines associated with
the generation of random numbers can be found in appendix
\ppref{A:aireslib}.
\section{Statistical sampling of particles: The thinning algorithm}
\label{S:thinning}\index{thinning}

The number of particles that are produced in an air shower grows
significantly when the energy of the primary increases. For ultra high
energy primaries that number can be large enough to make it impossible
to propagate all the secondaries even if the most powerful computers
currently available are used. The total number of particles in a
shower initiated by a $10^{20}$ eV proton primary is approximately
$10^{11}$, being almost impossible even to store the necessary data
for such an amount of particles.

The simulations are made possible thanks to a statistical sampling
mechanism which allows to propagate only a small representative
fraction of the total number of particles. Statistical weights are
assigned to the sampled particles in order to compensate for the
rejected ones. At the beginning of the simulation, the shower primary
is assigned a weight 1.

At the moment of evaluating averages to obtain the physical
observables, each particle entry is weighted with the corresponding
statistical weight. For example, the observables coming from the
monitoring routines, listed in section \ref{S:monitoring}, are
evaluated taking into account those statistical weights. On the other
hand, {\em unweighted distributions\index{unweighted distributions}\/}
are simultaneously calculated in the cases of
longitudinal\index{longitudinal development}, lateral\index{lateral
distributions} and energy distributions\index{energy
distributions}. They are useful to monitor the behavior of the
sampling algorithm.

The sampling algorithm used in AIRES is called {\em thinning algorithm\/}
or simply {\em thinning.\/} It is an extension of the thinning algorithm
originally introduced by A. M. Hillas\index{Hillas, A. M.} \cite{splithin,
MOCCA}, and was implemented modularly as a procedure which is independent
of the units which manage the physical interactions. The original Hillas
algorithm and the AIRES extended thinning algorithm are described in the
following sections.

\def\Eth{E_{\mathrm{th}}}
\subsection{Hillas thinning algorithm}%
\index{thinning!Hillas algorithm|bfpage{ }}
\label{S:Hillasthinning}

Let us consider the process
\begin{equation}                                   \label{eq:th1}
A\to B_1\; B_2\;\ldots\;B_n,\quad n\ge 1
\end{equation}
where a ``primary'' particle $A$ generates a set of $n$ secondaries
$B_1,\ldots,B_n$. Let $E_A$ ($E_{B_i}$) be the energy of $A$ ($B_i$),
and let $\Eth$ be a fixed energy called {\em thinning energy.}

Before incorporating the secondaries to the simulating processes, the
energy $E_A$ is compared with $\Eth$, and then:
\begin{itemize}
\item If $E_A\ge\Eth$, every secondary is analyzed separately, and
  accepted with probability\footnote{The procedure actually used in
  AIRES implements this step in a technically different way, but
  retrieving statistically equivalent results.}
\begin{equation}                                    \label{eq:th2}
P_i = \left\{\begin{array}{ll}
1 & \hbox{if $E_{B_i}\ge\Eth$} \\*[10pt]
{\displaystyle{{E_{B_i}}\over{\Eth}}} & \hbox{if $E_{B_i}<\Eth$}
\end{array}\right.
\end{equation}
\item If $E_A<\Eth$, that necessarily means that the ``primary''
comes from a previous thinning operation. In this case {\bf only one}
of the $n$ secondaries is conserved. It is selected among all the
secondaries with probability
\begin{equation}                                    \label{eq:th3}
P_i= {{E_{B_i}}\over{\sum_{j=1}^n E_{B_j}}}.
\end{equation}
This means that once the thinning energy is reached, the number of
particles is no more increased.
\end{itemize}

In both cases the weight of the accepted secondary particles is equal
to the weight of particle $A$ multiplied by the inverse of $P_i$.

The fact that the statistical weights are set with the inverse of the
acceptance probabilities ensures an unbiased sampling, that is, all
the averages evaluated using the weighted particles will not depend on
the thinning energy, and will be identical to the ``exact'' ones
obtained for $\Eth=0$. Only the fluctuations are affected by the
thinning level: If $\Eth$ is close to the primary energy, then the
thinning process begins early in the shower development, and a low
number of samples is obtained, with relatively large and fluctuating
weights. On the other hand, low thinning energies lead to larger
samples with less statistical fluctuations.

Processing large samples demands more computer time, so lowering the
thinning level makes the simulation more expensive from the
computational point of view.

\subsection{AIRES extended thinning algorithm}
\label{S:AIRESthinning}%
\index{thinning!AIRES extended algorithm|bfpage{ }}

The thinning algorithm of AIRES (\currairesversion) includes an additional
feature which has proved to be helpful to diminish statistical weight
fluctuations in many cases. This extended algorithm was designed to
ensure that all the statistical weights are never larger that a certain
positive number $W_r>1$, specified as an external parameter.

The mechanics of the AIRES extended algorithm can be summarized as
follows: Let $w_A$ be the weight of particle $A$, and $W_y< W_r/2$ be an
additional (internal) positive number. Consider the number of secondaries
in the process (\ref{eq:th1}).
\begin{itemize}
\item If $n\le 3$ then
\begin{itemize}
\item If $w_A > W_y$ or $w_A\,E_{A}/\min(E_{B_1},\ldots,E_{B_n}) > W_r$
then {\em all\/} the secondaries $B_1,\ldots,B_n$ are kept.
\item Otherwise the standard Hillas algorithm\index{thinning!Hillas
algorithm} is used.
\end{itemize}

\item If $n>3$ then {\em the standard Hillas algorithm is always used,\/}
but if the weight of the single selected secondary, $w_B$, happens to be
larger than $W_r$, then $m$ {\em copies\/} of the secondary are kept for
further propagation, each one with weight $w'_B=w_B/m$. The integer $m$ is
adjusted to ensure that $W_y < w'_B <W_r$.
\end{itemize}

In the AIRES algorithm $W_y = W_r/8$ and the limit $W_r$ is defined via
\begin{equation}                              \label{eq:Wfactor}
W_r = A_0\, \Eth\, W_f.
\end{equation}
where $A_0$ is a constant equal to 14 GeV$^{-1}$ and $W_f$ is an external
parameter which can be controlled by the user and that will be referred as
the {\em statistical weight factor\index{statistical weight factor|bfpage{
}}.\/}

In order to optimize the sampling algorithm, it is advantageous to
define different weight limits for different particle types. In AIRES
two weight factors are defined, $W^{(EM)}_f$ and $W^{(H)}_f$,
respectively used when processing electromagnetic or heavy
particles. Parameter $W^{(H)}_f$ is specified indirectly, by means of
the user-controlled ratio
\begin{equation}                                 \label{eq:Wratio}
A_{EH} = \frac{W^{(EM)}_f}{W^{(H)}_f}
\end{equation}
that permits evaluating $W^{(H)}_f$ from $W^{(EM)}_f$.

Notice also that $W_r$ depends on the {\em absolute\/} thinning energy
$\Eth$. The constant $A_0$ was adjusted so that $A_0 \Eth$ gives
approximately the position of the maximum of the all particles weight
distribution (see below).  If $W_f\to\infty$ the extended algorithm
reduces to the standard Hillas procedure.

It is a simple exercise to show that this extended thinning algorithm is
{\em unbiased\/} while ensuring (by construction) that all the particle
weights be smaller than the externally specified maximum value $W_r$ of
equation (\ref{eq:Wfactor}).
\def\putlabel#1{\rput{0}(2.9cm,5.0cm){{\large\bf(#1)}}}%
\lowfig{[tb]}{thinning1}{%
\begin{tabular}{cc}
\hbox{\putlabel{a}\epsfig{file=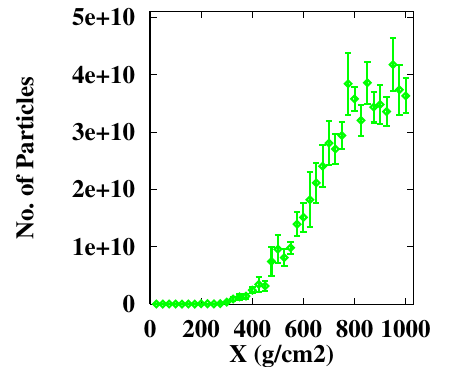,width=7cm}}&
\hbox{\putlabel{b}\epsfig{file=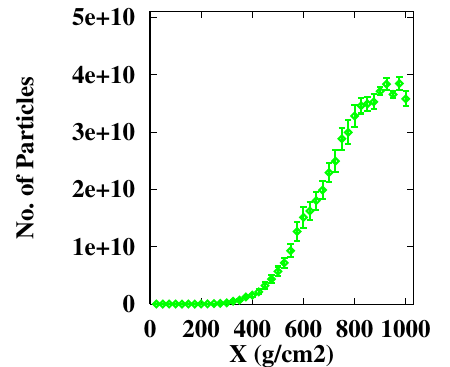,width=7cm}}\\*[7pt]
\hbox{\putlabel{c}\epsfig{file=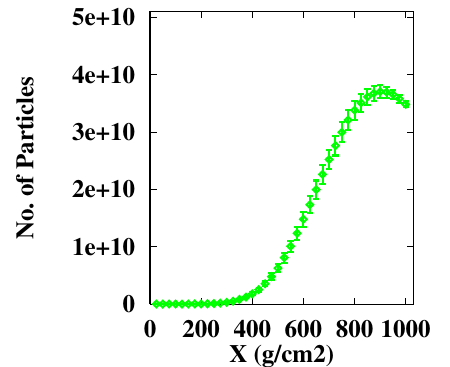,width=7cm}}&
\hbox{\putlabel{d}\epsfig{file=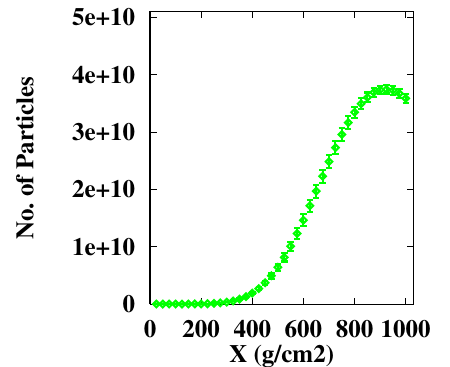,width=7cm}}
\end{tabular}
}%
{Effect of the thinning on the longitudinal development of charged
particles}%
{Effect of the thinning\index{thinning} energy on the fluctuations of
the number of charged particles crossing the different observing
levels during the shower development. Ten $10^{19}$ eV vertical proton
showers were averaged to obtain the data for each thinning level. The
plots labeled (a), (b), (c), (d), correspond to
$\Eth/E_{\mathrm{prim}}= 10^{-3}$, $10^{-4}$, $10^{-6}$ and $10^{-7}$,
respectively.}

It is worthwhile mentioning that this procedure is {\em not\/} equal to
the thinning algorithm of Kobal, Filip\v{c}i\v{c} and Zavrtanik
\cite{slthin}, even if both algorithms do use the concept of keeping
bounded the statistical weights. 
\def\putlabel#1{\rput{0}(2.9cm,1.6cm){{\large\bf(#1)}}}
\lowfig{[tb]}{thinning2}{%
\begin{tabular}{cc}
\hbox{\putlabel{a}\epsfig{file=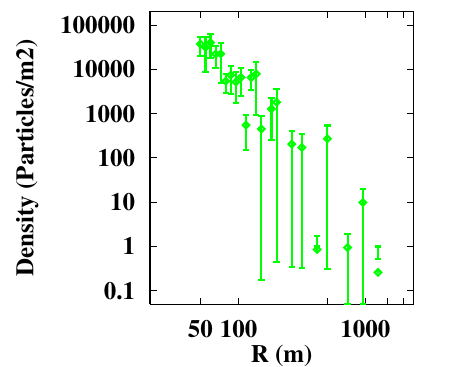,width=7cm}}&
\hbox{\putlabel{b}\epsfig{file=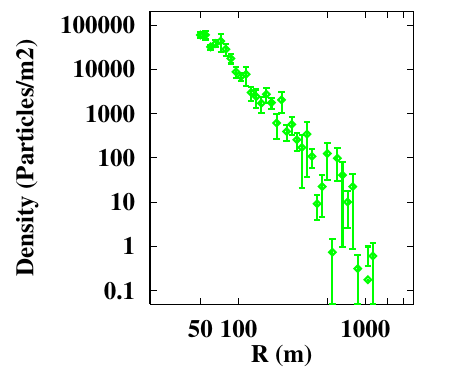,width=7cm}}\\*[7pt]
\hbox{\putlabel{c}\epsfig{file=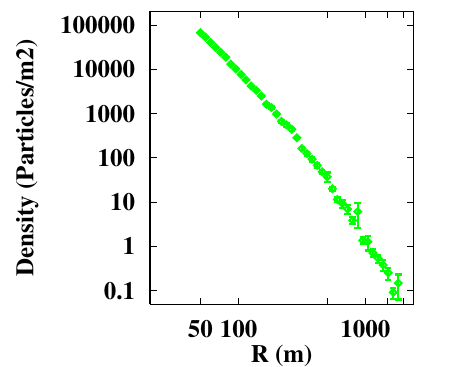,width=7cm}}&
\hbox{\putlabel{d}\epsfig{file=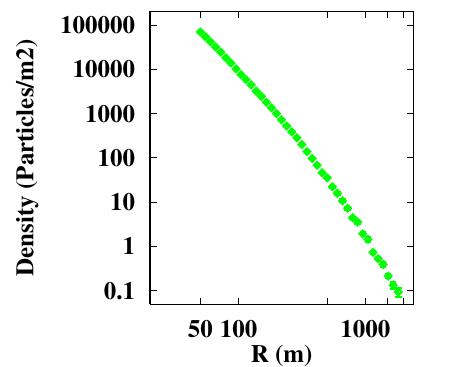,width=7cm}}
\end{tabular}
}%
{Effect of the thinning on the $e^+e^-$ lateral distribution}%
{Effect of the thinning\index{thinning} energy on the fluctuations of
the lateral distribution of electrons and positrons, in the same
conditions as in figure \ref{FIG:thinning1}.}

\subsection{How does the thinning affect the simulations?}
\label{S:thinningsamples}

The effect of the standard thinning\index{thinning} on different
observables evaluated during the simulations can be seen in figures
\ref{FIG:thinning1}-\ref{FIG:thinning3}.  All these simulations were done
using identical initial conditions: $10^{19}$ eV proton showers with
vertical incidence; and considering four different thinning energies,
namely,\\ $\Eth/E_{\mathrm{prim}} = 10^{-3}$, $10^{-4}$, $10^{-6}$ and
$10^{-7}$. In all cases the weight limiting mechanism was disabled.
\def\putlabel#1{\rput{0}(2.9cm,1.6cm){{\large\bf(#1)}}}%
\lowfig{[tb]}{thinning3}{%
\begin{tabular}{cc}
\hbox{\putlabel{a}\epsfig{file=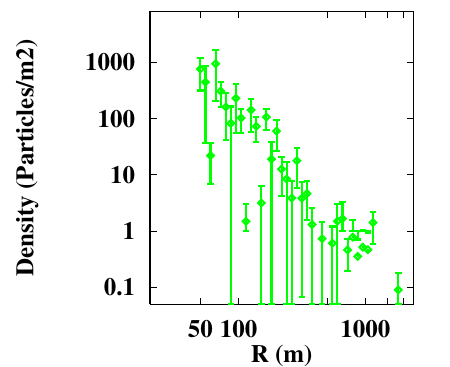,width=7cm}}&
\hbox{\putlabel{b}\epsfig{file=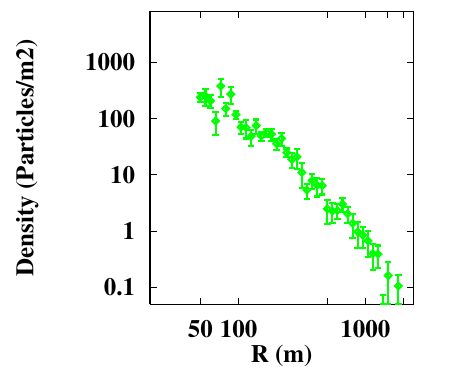,width=7cm}}\\*[7pt]
\hbox{\putlabel{c}\epsfig{file=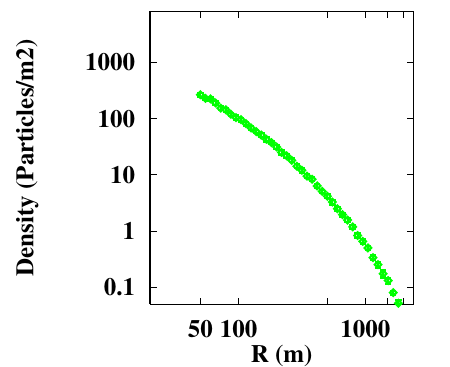,width=7cm}}&
\hbox{\putlabel{d}\epsfig{file=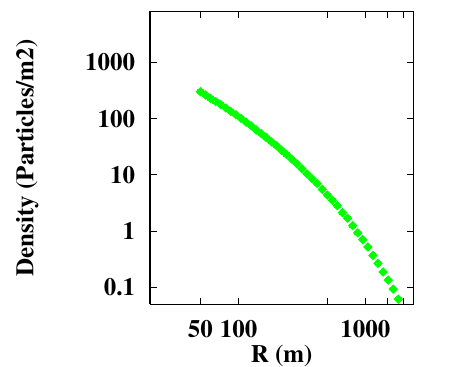,width=7cm}}
\end{tabular}
}%
{Effect of the thinning on the $\mu^+\mu^-$ lateral distribution}%
{Effect of the thinning\index{thinning} energy on the fluctuations of
the lateral distribution of muons, in the same conditions as in figure
\ref{FIG:thinning1}.}

Figure \ppref{FIG:thinning1} corresponds to the longitudinal
development\index{longitudinal development} of all the charged
particles, that is the total number of charged particles crossing the
different observing levels, as a function of the observing levels'
vertical depth.

The plots in this figure show clearly how the statistical fluctuations
diminish systematically as long as the thinning energy is lowered. Compare
the plot for $10^{-3}$ relative thinning with the smooth plots obtained
for the cases $10^{-6}$ and/or $10^{-7}$. As mentioned, the CPU time
required increases each time the thinning energy is lowered. It is
interesting to mention that the simulations done at $10^{-7}$ thinning
level required some 6300 times more CPU time than the ones done with
$10^{-3}$ thinning level.

Notice also that the mean positions of the points corresponding to any
given depth do not present any evident dependence with the thinning
energy, as expected since the Hillas thinning\index{thinning!Hillas
algorithm} algorithm is an unbiased statistical sampling technique. This
observation applies also for the plots of figures \ppref{FIG:thinning2} and
\ppref{FIG:thinning3}.

The degree of reduction of the fluctuation does depend on the
observable considered. In figure \ppref{FIG:thinning2} the lateral
distribution\index{lateral distributions} of ground electrons and
positrons is displayed, again for different thinning levels. It is
noticeable the degree of persistence of the noisy fluctuations, which
are not completely eliminated even in the $10^{-7}$ relative thinning
case.

The lateral distribution of muons displayed in figure
\ppref{FIG:thinning3}\label{P:thinning3} reflects another characteristic of
the thinning algorithm. Even if the fluctuations are very large for
$10^{-3}$ relative thinning level, they reduce immediately when the
thinning is lowered. Compare for example with the plots of figure
\ppref{FIG:thinning1}. To understand the behavior of these distributions it
is necessary to recall that the muons are very penetrating particles, that
is, they undergo a very reduced number of interactions before reaching
ground. Therefore their statistical weights remain small since they are
products of a few factors, and this fact is responsible for the low level
of fluctuations produced. On the other hand, ground electrons and
positrons most likely come out after a long chain of processes involving
many predecessor particles, and in such circumstances very large
statistical weights are unavoidable, and hence the high level of
fluctuations observed in the $e^+e^-$ distribution of figure
\ppref{FIG:thinning2}\index{thinning!Hillas algorithm}.

The AIRES extended thinning algorithm can be useful to reduce such kind of
fluctuations. To illustrate this point let us consider the sample plots
displayed in figure \ppref{FIG:AIRESthin1}.
\lowfig{[tb]}{AIRESthin1}{\epsfig{file=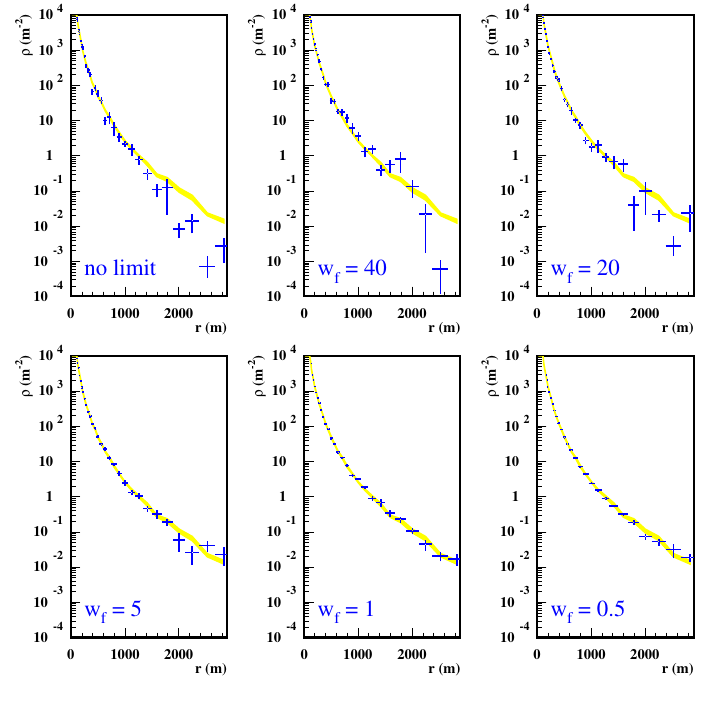}}%
{AIRES thinning algorithm and $e^+e^-$ lateral distribution}%
{Effect of the AIRES extended thinning\index{thinning!AIRES extended
algorithm} on the fluctuations of
the lateral distribution of electrons and positrons. The plots correspond
to $10^{19}$ eV proton showers simulated with $\Eth/E_{\mathrm{prim}}=
10^{-5}$, and different weight factors\index{statistical weight
factor} ($W^{(EM)}_f=W^{(H)}_f=W_f$). The yellow bands
(\/$\hbox{\color{yellow}\vrule height 1ex depth0pt 
width2em}$) correspond to simulations performed in similar conditions, but
using the Hillas algorithm\index{thinning!Hillas algorithm} at $10^{-7}$
relative level. The width of the bands correspond to the average value plus
and minus one RMS error of the mean.}
\lowfig{[p]}{AIRESthinwd}{\epsfig{file=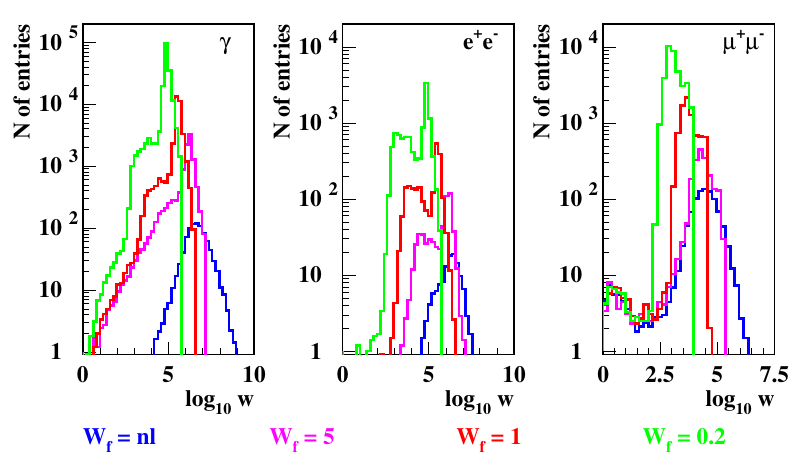}}%
{AIRES thinning algorithm and weight distributions}%
{Effect of the AIRES extended thinning\index{thinning!AIRES extended
algorithm} on the distribution of weights for different particles (gammas,
electrons and positrons, and muons). The plots correspond to $2\times
10^{19}$ eV proton showers simulated with $\Eth/E_{\mathrm{prim}}=
10^{-5}$, and different weight factors\index{statistical weight
factor} ($W^{(EM)}_f=W_f$, $W^{(H)}_f = W^{(EM)}_f/88$).}
\latfig{[p]}{AIRESthincpu}{\epsfig{file=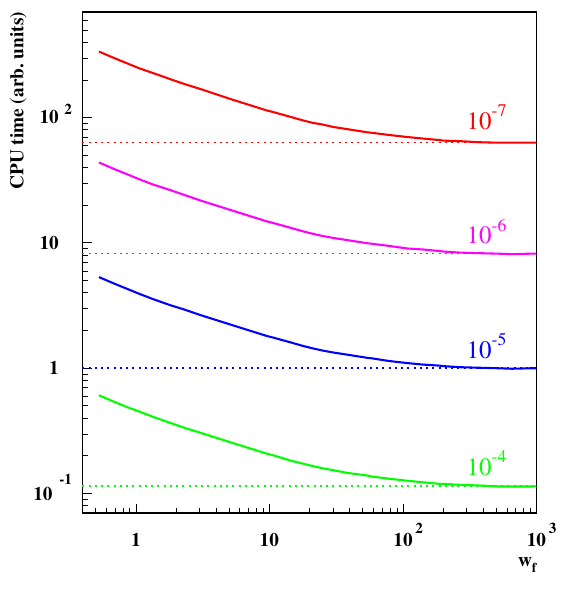,width=8cm}}%
{AIRES thinning algorithm. Processor time requirements}%
{Processor time requirements\index{computer requirements} for the
AIRES extended thinning\index{thinning!AIRES extended algorithm}
algorithm, plotted versus $W_f= W^{(EM)}_f=W^{(H)}_f$ for different
relative thinning levels. All cases correspond to $10^{19}$ eV proton
showers.}

The outstanding characteristic of these plots is the fact that the
density fluctuations diminish when the weight factor\index{statistical
weight factor} ($W^{(EM)}_f=W^{(H)}_f=W_f$) is lowered. In the
particular cases of $W_f=1$ and $W_f=0.5$ the fluctuations
corresponding to the $10^{-5}$ relative thinning are of the order of
the ones corresponding to the $10^{-7}$ (Hillas algorithm) case
(yellow band) which were plotted in all cases for reference.

Looking at the distributions of weights displayed in figure
\ppref{FIG:AIRESthinwd}, it is possible to understand the action of the
weight limiting mechanism. The distributions labeled ``nl'' (blue lines)
correspond to the Hillas algorithm case (no weight limits). Considering
the the distributions of weights for gammas as a typical case, it is
evident that there is a small fraction of particles having weights up to
three orders of magnitude larger than the most probable ones. This rare
cases are generally the cause of many inconvenients that arise when
analyzing the data. The plots for finite $W_f$ show clearly that the
distributions present a sharp end (corresponding to the value of
$W_r$). In the case $W_f=1$ the gamma distribution ends approximately at
the maximum of the ``nl'' case curve, as expected from equation
(\ref{eq:Wfactor}), where the factor $A_0$ is ``tuned'' to give $W_r$ near
the distribution's maximum when $W_f$ is equal to 1.

The muon weights are generally smaller than the electromagnetic
counterparts (see the discussion of figure \ref{FIG:thinning3} in page
\pageref{P:thinning3}). It is therefore necessary to use a smaller
weight limit to modify the corresponding distribution of weights. This
is the case of figure \ref{FIG:AIRESthinwd} that corresponds to the
case $W^{(H)}_f = W^{(EM)}_f/88$.

It is worthwhile mentioning that the weight distributions corresponding to
other thinning energies have the same shape as the ones plotted in figure
\ppref{FIG:AIRESthinwd}, but present a global shift in the abscissas scale
which is proportional to the logarithm of the thinning level (for example,
the weights for the $10^{-6}$ distribution are one order of magnitude
lower than the ones for $10^{-5}$ and so on).

The improvement in the lateral distribution plots is, of course, not free:
The CPU time per shower is increased when $W_f$ decreases. Figure
\ppref{FIG:AIRESthincpu} represents the CPU time
consumption\index{computer requirements} per shower as a function of $W_f$
for various thinning energies. The time unit is the average time required
to complete a shower simulated with $10^{-5}$ relative thinning and
$W_f\to\infty$.

The CPU time per shower increases monotonically when $W_f$ decreases. For
any $\Eth$ and $W_f = 1$, for example the required time is roughly 5 times
larger than the one for $W_f\to \infty$. But it is 1.6 (13) times lower
than the one corresponding to the Hillas algorithm\index{thinning!Hillas
algorithm} for $\Eth/10$ ($\Eth/100$). These figures may represent an
important time saving factor in certain circumstances, for example when
evaluating lateral distributions like the ones of figure
\ppref{FIG:AIRESthin1}.

The use of the AIRES extended thinning algorithm\index{thinning!AIRES
extended algorithm} with finite $W_f$ is always recommended, however. Even
in the least favorable cases, it is possible to get smoother
distributions for every observable setting $W_f$ not larger than 20 and
thus eliminating the particle entries with unacceptably large weights.

%
\chapter{Steering the simulations}
\label{C:steering}
There are many parameters that must be specified before and during an
air shower simulation job. The {\em Input Directive Language (IDL)\/}
is a part of the AIRES system and consists in some 70 human-readable
directives that permit an efficient control of the simulations in a
comfortable environment.

The most common IDL directives are described in this chapter, and many
illustrative examples are discussed; a detailed description of the IDL
language is placed in appendix \ppref{A:idlref}. It is recommended to
properly install the software (see appendix \ref{A:installing}) before
proceeding with the following sections.

\section{Tasks, processes and runs}
\label{S:tpar}\index{tasks, processes and runs}%
\index{ADF or adf|see{internal dump file, portable format}}%
\index{portable dump file|see{internal dump file, portable format}}%
\index{ASCII dump file|see{internal dump file, portable format}}%

The simulation of high energy air showers is a CPU intensive task
which can demand days of processor time to complete.  The AIRES
program was designed taking this fact into account: It includes an
``auto-saving'' mechanism to periodically save into an {\em internal
  dump file\/} (IDF)\index{internal dump file}\index{IDF or
  idf|see{internal dump file}} all relevant simulation data. In case
of a system failure, for example, the simulation process can be
restarted at the point of the last auto-saving operation, thus
avoiding loosing all the previous simulation effort.

The processing block that goes between two consecutive
auto-save\index{task, definition}\index{run,
definition}\index{process, definition} operations is called a {\em
run.\/} With {\em task\/} we mean a specific simulation job, as
defined by the input directives (for example, a task can be ``simulate
ten proton showers''); and with {\em process\/} we identify a system
process, which starts when AIRES is invoked and ends when control is
returned to the operating system.

A task can be completed after one or more processes, and there can be
one or more runs within a process. The limit case consists in having a task
finished in a single run (no auto-save) completed in a single process
(the program invoked just once).

\section{The Input Directive Language (IDL)}%
\index{Input Directive Language}%
\index{IDL|see{Input Directive Language.}}

All the main simulation programs (the default program \kwbf;Aires; and
the executables for the different available external hadronic
models\index{external packages}\index{hadronic models}
\kwbf;Aires;\vkw;Model;), and the summary program
\kwbf;AiresSry;\index{AIRES summary program} read their input
directives from the standard input channel, and use a common language
to receive the user's instructions.  This language is called {\em
  Input Directive Language\/} (IDL)\index{Input Directive Language}.

The IDL directives are written using free format\index{Input Directive
Language!format}, with one directive per line (there are no
``continuation lines'', but each line can contain up to 8,000
characters). Special characters like tab characters, for example, are
treated as blank characters.

All directives are scanned until either an \idlbfx;End; directive or an
end of file is found. Most directives can be placed in any order
within the input stream.

IDL directives can be classified as {\em dynamic\/}\index{Input
Directive Language!dynamic/static directives} or {\em static.\/}
Dynamic directives are processed every time the input data is scanned.
Static ones can be set only at the beginning of a task: Any subsequent
setting will not be taken into account. For instance, the maximum CPU
time per run is controlled by a dynamic directive (it can be changed
at the beginning of every process, and is a parameter that does not
affect the results of the simulation); the ground altitude, instead,
is an example of a static parameter that cannot be changed during the
simulations.

The IDL sentence begins with the directive name. IDL is a case
sensitive language, and in general directive names mix capital and
lowercase letters. The directives can be abbreviated. Consider for
example the following directive name
\akwbf;Primary;Particle;\idlidx;PrimaryParticle;. You must specify the
underlined part, and may or may not use the remaining characters
(\kwbf;Primary;, \kwbf;PrimaryPart;, \kwbf;PrimaryParticle; refer to
the same instruction).

\subsection{A first example}
\label{S:example1}

There are four directives that should be always specified before
starting a simulation task, namely, the ones that control the task
name\footnote{Actually, the \kwbf;TaskName; directive is not mandatory
for a task to start, but its default value
\kw;GIVE\_ME\_A\_NAME\_PLEASE; produces file names which are rather
inconvenient to manage, and so it is strongly recommended to always
set the name of a task before proceeding with the simulations.}, the
statements that provide the primary particle type and energy
specifications and the directive which sets the total number of
showers to be simulated.

Such a minimal set of specifications can be expressed in terms of IDL
directives as follows:
\idlidx;TaskName;\idlidx;PrimaryParticle;\idlidx;PrimaryEnergy;%
\idlidx;TotalShowers;\idlidx;End;%
\kwexample{%
Task \kwt;a\_first\_example;\\
Primary \kwt;proton;\\
PrimaryEnergy \kwt;150 TeV;\\
TotalShowers \kwt;3;\\
End
}

These directives, like most IDL directives, are self-explaining and
posses a simple syntax. They can be placed in any order. Notice that
the particles are specified by their names and physical quantities
like the energy, for example, are entered by means of a number plus a
unit.

\subsection{Errors and input checking}
\label{S:icheck}\index{error messages}\index{input file checking}

Every IDL directive is checked for correct syntax when it is read
in. Additionally, some elemental tests of the values given to the
directive's parameters are also made. When an error is detected, a
message is written to the standard output channel. Directives with
errors are generally ignored. Consider the following directive:
\kwdisplay;PrimaryEnergy  {\tt 100 MeV};
If processed by AIRES, it will give the following error message
\begin{verbatim}
EEEE
EEEE  dd/Mmm/yyyy hh:mm:ss. Error message from commandparse
EEEE  Numeric parameter(s) invalid or out of range.
EEEE  >PrimaryEnergy 100 MeV<
EEEE
\end{verbatim}
which indicates that the energy specification is out of
range.\footnote{The primary energy must be greater than 500 MeV (see
page \pageref{IDL0:PrimaryEnergy}).}

AIRES diagnostic messages always include a brief explanation about the
circumstances that generated the message, together with the name of
the routine that originated it. The messages can be classified in four
categories, accordingly with their severity: \iiset\ii {\em
Informative\/} messages are used to notify the occurrence of certain
events, and are generally associated with successfully concluded
operations. \ii {\em Warning\/} messages. Used to put in evidence
certain not completely ``normal'' situations. In general, processing
continues normally. \ii {\em Error\/} messages indicate abnormal
events like invalid input directives, etc., as illustrated in the
previous example. \ii {\em Fatal\/} messages are issued when a serious
error takes place; in this case the program stops.

The IDL instruction set includes some directives that allow checking
a given input data set. Let us assume that the input directives are
saved into a file named \kwbf;myfile.inp;. Let us consider also that
this file contains the instructions of the first example previously
considered.

The instruction set:
\kwexample{%
Trace\idlidx;Trace;\\
CheckOnly\idlidx;CheckOnly;\\
Input \kwt;myfile.inp;\idlidx;Input;\\
End
}\label{P:inpcheck}
if processed by AIRES will generate an output similar to the
following:
\kwexample{{\tt
. . .\\
\ 0:0002 CheckOnly\\
\ 0:0003 Input myfile.inp\\
\ 1:0001 Task a\_first\_example\\
\ 1:0002 Primary proton\\
\ 1:0003 PrimaryEnergy 150 TeV\\
\ 1:0004 TotalShowers 3\\
\ 1:0005 End\\
\ 0:0004 End\\
. . .
}}
The \idlbfx;CheckOnly; directive instructs AIRES to normally read and
check the input data, and then stop without actually starting any
simulations. The input lines placed after the \idlbfx;Trace; statement
are echoed to the terminal. and the \idlbfx;Input; directives allows
including IDL directives placed in other files. Notice the format used
for directive echoing. It includes the line number as well as the file
nesting level, starting by zero for the standard input
channel. \kwbf;Input; directives can be nested and permit splitting
the input data into separate files. This is most useful for organizing
a set of input files including some common directives in a single
shared file included by every particular file, etc.

In UNIX environments it is possible to use one of the scripts of the
AIRES Runner System\index{AIRES Runner System} to automatically check
a given input file. For details see chapter \ppref{C:ARS}.

\subsection{Obtaining online help}\index{online help}

The AIRES simulation and/or summary programs accept instructions that
permit obtaining information about AIRES IDL instructions. The
information that can be retrieved in this way is not extensive but it
can be useful to the experienced user as a quick guide.

Invoking AIRES interactively and typing ``\idlbfx;?;'' will return a
list of the names of the IDL directives. ``\kwbf;? *;'' will cause the
list to include also the {\em hidden directives\index{Input Directive
Language!hidden directives}\/}. The prompt ``\kwbf;Aires
IDL;$\mathbf>$'' typed at the terminal indicates that AIRES
understands that this session is interactive. The \idlbfx;Help;
command is similar to \kwbf;?; but it will maintain prompting
disabled. During an interactive session it is always possible to
enable or disabling the prompt by means of the directive \idlbfx;Prompt;.

There are two other kinds of help that can be obtained using the
current AIRES version, namely, ``\kwbf;? tables;'' and ``\kwbf;?
sites;'', which display the list of available output data tables (see
section \ref{S:AiresSry} and/or appendix \ref{A:outables}), and the
currently defined geographical sites\index{AIRES site library} (see
section \ref{S:geomagidl}), respectively.

The directive \idlbfx;Exit;, which can be abbreviated as \idlbfx;x;,
will cause AIRES to stop immediately without any further action --not
even completing the IDL instruction scanning phase-- and is useful to end
interactive help sessions.
\lowtable{[ptb]}{idlunits}{%
{\makebox{\begin{tabular}[t]{|l|l|r@{ }l|}
\hline
\thpop{\bfseries\itshape Magnitude}&
{\bfseries\itshape Units} &
\multicolumn{2}{c|}{\bfseries\itshape Conversion factors} \\
\hline
{\bfseries Length}         & \kw;cm;      & $10^{-2\vphantom{9^9}}$&m \\
                           & \akw;m;;     &       $1$&m \\
                           & \kw;km;      & $10^{ 3}$&m \\
                           & \kw;in;      & $0.0254$&m \\
                           & \kw;ft;      & $12$&in \\
                           & \kw;yd;      & $3$&ft \\
$\phantom{\hbox{\bfseries Temperature}}$
                           & \kw;mi;      & $5280$&ft \\
\hline
{\bfseries Time}           & \kw;ns;      & $10^{-9\vphantom{9^9}}$&s \\
                           & \akw;sec;;   &       $1$&s \\
                           & \kw;min;     &      $60$&s \\
                           & \kw;hr;      &    $3600$&s \\
\hline
{\bfseries Energy}         & \kw;eV;      & $10^{-9\vphantom{9^9}}$&GeV \\
                           & \kw;keV;     & $10^{-6}$&GeV \\
                           & \kw;MeV;     & $10^{-3}$&GeV \\
                           & \akw;GeV;;   &       $1$&GeV \\
                           & \kw;TeV;     & $10^{ 3}$&GeV \\
                           & \kw;PeV;     & $10^{ 6}$&GeV \\
                           & \kw;EeV;     & $10^{ 9}$&GeV \\
                           & \kw;ZeV;     & $10^{12}$&GeV \\
                           & \kw;YeV;     & $10^{15}$&GeV \\
                           & \kw;J;       & 
                                            $6.24 \times 10^{9}$&GeV \\
\hline
\end{tabular}}%
~%
\makebox{\begin{tabular}[t]{|l|l|r@{ }l|}
\hline
\thpop{\bfseries\itshape Magnitude}&
{\bfseries\itshape Units} &
\multicolumn{2}{c|}{\bfseries\itshape Conversion factors} \\
\hline
{\bfseries Angle}          & \akw;deg;;   &       $1$&deg
                                            $\vphantom{2^{2^2}_2}$ \\
                           & \kw;rad;     & $180/\pi$&deg \\
\hline
{\bfseries Atmospheric}    & \akw;g/cm2;; & $1$&\gcmsq
                                            $\vphantom{2^{2^2}_2}$ \\
{\bfseries depth}          &              &    &  \\
\hline
{\bfseries Density}        & \akw;g/cm3;; & $1$&${\rm g}/{\rm cm}^3$
                                            $\vphantom{2^{2^2}_2}$\\
                           & \kw;kg/m3;   & $10^{-3}$&${\rm g}/{\rm cm}^3$\\
                           & \kw;kg/dm3;  & $1$&${\rm g}/{\rm cm}^3$\\
\hline
{\bfseries Magnetic}       & \akw;nT;;    &        $1$&nT
                                            $\vphantom{2^{2^2}_2}$\\
{\bfseries field}          & \kw;uT;      &  $10^{ 3}$&nT \\
                           & \kw;mT;      &  $10^{ 6}$&nT \\
                           & \kw;T;       &  $10^{ 9}$&nT \\
                           & \kw;uG;      &  $10^{-1}$&nT \\
                           & \kw;mG;      &  $10^{ 2}$&nT \\
                           & \kw;Gs;      &  $10^{ 5}$&nT \\
                           & \kw;gm;      &        $1$&nT \\
\hline
{\bfseries Temperature}    & \akw;K;;     & $1$ & K
                                            $\vphantom{2^{2^2}_2}$\\
                           & \kw;C;       &
                             \multicolumn{2}{l|}{%
                                $T_{\rm C} + 273.15$}\\
                           & \kw;R;       &
                             \multicolumn{2}{l|}{%
                                $(5/9)T_{\rm R}$}\\
                           & \kw;F; &
                             \multicolumn{2}{l|}{%
                                $(5/9)(T_{\rm F} + 459.67)$}\\
\hline
\vrule height 17.7pt depth 6pt width 0pt\relax & & & \\
\hline
\end{tabular}}}
}%
{Physical units accepted within IDL directives}%
{Physical units accepted within IDL directives\index{Input Directive
    Language!physical units}. The underlined keywords indicate the
  units used internally to store the corresponding quantities. Time
  specifications using \/\kw;hr;, \/\kw;min; and/or \/\kw;sec; may
  consist in the combination of more that one field, like in \/\kw;3
  hr 30 min;, for example. The magnetic field unit \/\kw;T; (\/\kw;Gs;)
  stands for the SI (cgs) unit Tesla (Gauss), while \kw;gm;
  corresponds to $\gamma$ (1 $\gamma= 1$ nT). The formulas for
  temperatures in the Celsius (\kw;C;), Rankine (\kw;R;), and
  Fahrenheit (\kw;F;) scales indicate how to convert a temperature in
  the respective scale to the absolute Kelvin scale.
}

\subsection{Physical units}\index{Input Directive Language!physical units}

There are many IDL directives which include one or more specifications
corresponding to physical quantities. In most cases these
specifications have the format ``number + unit'', like in the
\kwbf;PrimaryEnergy; specification of section \ref{S:example1}, for
instance. ``Number'' and ``unit'' are character strings, the first one
indicates the decimal numerical value for the quantity being
specified, while ``unit'' represents the unit in which ``number'' is
expressed. The characters used for the unit field resemble the name
assigned in the real world to the corresponding unit, e.g. \kw;TeV;
for {\em Tera-electron-volt.\/}

This feature of the IDL language makes the input files more readable,
and diminishes drastically the possibility of errors in the
specifications, especially for those quantities whose validity ranges
may span many orders of magnitude. In such cases a number of commonly used
multiples or sub-multiples of the fundamental unit are surely
available. 

The complete list of units currently implemented is displayed in table
\ref{TAB:idlunits}.

\subsection{Carrying on}
\label{S:carryingon}
In figure \ref{FIG:sampleinp2} a second example of an IDL input data
set is displayed.
Notice first that IDL instructions can be commented: All the
characters following a `\idlbfx;\#;' character are ignored.
%
\lowtable{[p]}{mathopers}{%
{\def\pz{$\phantom{0}$}
\begin{tabular}{|l|c|l|}\hline
\multicolumn{1}{|c}{\bfseries\itshape Function}&
\multicolumn{1}{|c}{\bfseries\itshape Args}&
\multicolumn{1}{|c|}{\bfseries\itshape Operation}\\ \hline
 \kw;ADD;        & 2 & $a_1 + a_2\vphantom{|^{|^|}}$ \\*[1pt]
 \kw;SUBTRACT;   & 2 & $a_1 - a_2$ \\*[1pt]
 \kw;MULTIPLY;   & 2 & $a_1 \times a_2$ \\*[1pt]
 \kw;DIVIDE;     & 2 & $a_1 / a_2 $ \\*[1pt]
 \kw;POWER;      & 2 & $a_1^{a_2}$ \\*[1pt]
 \kw;MIN;        & 2 & $\min(a_1, a_2)$ \\*[1pt]
 \kw;MAX;        & 2 & $\max(a_1, a_2)$ \\*[1pt]
 \kw;MEAN;       & 2 & $(a_1 + a_2)/2$ \\*[1pt]
 \kw;GMEAN;      & 2 & $\sqrt{|a_1 a_2|}$ \\*[1pt]
 \kw;ABS;        & 1 & $|a_1|$ \\*[1pt]
 \kw;NEGATIVE;   & 1 & $- a_1$ \\*[1pt]
 \kw;SQRT;       & 1 & $\sqrt{a_1}$ \\*[1pt]
 \kw;EXP;        & 1 & $e^{a_1}$ \\*[1pt]
 \kw;LOG;        & 1 & $\ln(a_1)$ \\*[1pt]
 \kw;LG10;       & 1 & $\log_{10}(a_1)$ \\*[1pt]
 \kw;SIN;        & 1 & $\sin(a_1)$ \\*[1pt]
 \kw;COS;        & 1 & $\cos(a_1)$ \\*[1pt]
 \kw;SEC;        & 1 & $1/\cos(a_1)$ \\*[1pt]
 \kw;TAN;        & 1 & $\tan(a_1)$ \\*[1pt]
 \kw;ASIN;       & 1 & $\arcsin(a_1)$ \\*[1pt]
 \kw;ACOS;       & 1 & $\arccos(a_1)$ \\*[1pt]
 \kw;ASEC;       & 1 & $\arccos(1 / a_1)$ \\*[1pt]
 \kw;ATAN;       & 1 & $\arctan(a_1)$ \\*[1pt]
 \kw;ATAN2;      & 2 & ${\rm arctan2}(a_1,a_2)$ \\*[1pt]
 \kw;DEG2RAD;    & 1 & $(\pi/180) a_1$ \\*[1pt]
 \kw;RAD2DEG;    & 1 & $(180/\pi) a_1$ \\*[1pt]
 \kw;PI;         & 0 & $\pi$ \\*[1pt]
 \kw;RANDOM;     & 0 & uniform random number in $[0,1)$ \\*[1pt]
 \kw;XVATH;      & 1 & $X_v(a_1)$ atm. depth [g/cm$2$] at alt $a_1$ [m]\\*[1pt]
 \kw;HATXV;      & 1 & $h(a_1)$ alt. [m] at atm. depth [g/cm$2$]\\*[1pt]
 \kw;DENSATH;    & 1 & $\rho(a_1)$ density [g/cm$3$] at alt $a_1$ [m]\\*[1pt]
 \kw;HATDENS;    & 1 & $h(a_1)$ alt. [m] at density [g/cm$3$]\\*[1pt]
\hline
\end{tabular}}
}{%
Available IDL mathematical operations.}%
{Available IDL
  mathematical operations. The function names are NOT case sensitive.
  When there are no ambiguities, they can be abbreviated downto
  the first three characters.}
%
%
\begin{figure}[tp]
\begin{center}
{\baselineskip=9pt\lineskip=1pt\lineskiplimit=1pt\scriptsize
\begin{tabular}{|c|}
\hline\\
\begin{minipage}{14.5cm}
\idlidx;\#;%
\idlidx;Skip;%
\idlidx;Remark;%
\idlidx;TaskName;%
\idlidx;TotalShowers;%
\idlidx;PrimaryParticle;%
\idlidx;PrimaryEnergy;%
\idlidx;PrimaryZenAngle;%
\idlidx;SetGlobal;\index{global variables}%
\idlidx;Import;%
\idlidx;ThinningEnergy;\index{thinning}%
\idlidx;GroundAltitude;%
\idlidx;ObservingLevels;%
\begin{verbatim}
#
# An example of an AIRES IDL input data set.

Skip &next

The directive "Skip" skips all text until the label &label is
found. Notice that it is not equivalent to a "go to" statement
since it is not possible to skip backwards.

As it can easily be seen, most directive names are self-explaining.
&next

Remark    JUST AN EXAMPLE

# It is possible to define variables that can be used later within
# the input file and/or be passed to output files or special
# primary modules.

SetGlobal MyVariable  This string is associated with the variable.
SetGlobal VRem        Another variable.

Remark {VRem} # This expands to: Remark Another variable.

Import HOME   # Importing OS environmental variables.


# The input directives define a "task". Tasks are identified by
# their task name and (eventually) version. If not defined, the
# version is zero.

Task mytask   # Use "Task mytask 5" to explicitly set task
              # version to 5.

# The following three directives are mandatory (have no default
# values)

TotalShowers      2
PrimaryParticle   Proton
PrimaryEnergy     1.5 PeV
#
# The remaining directives allow controlling many parameters of the
# simulations. The respective parameters will take a default value
# whenever the controlling directive is not present.
#

PrimaryZenAngle   15 deg
Thinning          1.e-4 Relative # Relative or absolute
                                 # specifications allowed.

Ground     1000 g/cm2

# You can freely set the number of observing levels to record the
# shower longitudinal development. You can define up to 510
# observing levels and (optionally) altitude of the highest and
# lowest levels.

ObservingLevels 41 100 g/cm2 900 g/cm2
\end{verbatim}
\end{minipage}\\ \\ \hline
\end{tabular}
}\end{center}
\figcaption{sampleinp2}%
{Sample AIRES input}{Sample AIRES input.}
\end{figure}
%
%
\begin{figure}[tp]
\begin{center}
{\baselineskip=9pt\lineskip=1pt\lineskiplimit=1pt\scriptsize
\begin{tabular}{|c|}
\hline\\
\begin{minipage}{14.5cm}
\idlidx;GammaCutEnergy;\index{threshold energies}%
\idlidx;ElectronCutEnergy;%
\idlidx;MuonCutEnergy;%
\idlidx;MesonCutEnergy;%
\idlidx;NuclCutEnergy;%
\idlidx;PrintTables;%
\idlidx;ExportTables;\index{longitudinal development!in energy}%
\idlidx;ADFile;
\idlidx;End;%
\begin{verbatim}
#
# Threshold energies. Particles are not followed below these
# energies.

GammaCutEnergy     200 KeV
ElectronCutEnergy  200 KeV
MuonCutEnergy        1 MeV
MesonCutEnergy     1.5 MeV  # Pions, Kaons.
NuclCutEnergy      150 MeV  # Nucleons and nuclei.

#
# Some output control statements.
#

# Compressed particle data files related directives.

SaveInFile     lgtpcles  e+ e-
SaveNotInFile  grdpcles  gamma

# Saving the ASCII (portable) version of the IDF file (ADF), after
# finishing the simulations.

ADF   On

#
# No tables are printed or exported if no PrintTables ExportTables
# directives are explicitly used.
#
PrintTable 1291         # Longit. devel. of all charged particles.
PrintTable 1707         # Energy longitudinal development of muons.
PrintTable 2207 Opt d   # Setting some options.
PrintTable 3001 Opt M   # Here too.
#
ExportTable 2793 Opt M  # Exported tables are placed in separate,
ExportTable 5501        # plain text files for further processing
                        # (e. g. plotting).

End # End of input data stream.
\end{verbatim}
\end{minipage}\\ \\ \hline
\end{tabular}}
\end{center}
\vskip \abovecaptionskip
\begin{center}
{\itshape{\bfseries Figure \ref{FIG:sampleinp2}.} (continued)}
\end{center}
\end{figure}

The \idlbfx;Skip; statement is also useful to place comments and/or
introduce plain text in the input files (with no need of single line
comment `\kwbf;\#;' characters), as well as to skip a part of the
directives without deleting the lines.

Comments and skipped lines are completely ignored: They just appear in
the input file. Sometimes this is not convenient, and it may be
desirable to save their contents together with the output generated by
the simulating program. The \idlbfx;Remark; directive provides a mean
to do this. The statement
\kwdisplay;Remark \ \ \ {\tt JUST AN EXAMPLE};
placed in the example being discussed, instructs AIRES to place the
comment `JUST AN EXAMPLE' together with the output data. There is no
limit in the number of remark instructions that may appear inside a
given input instruction set. The \kwbf;Remark; directive possesses
another alternative syntax, very useful for multi-line text:
\idlidx;\&;%
\kwexample{%
Rem \&\kwt;eor;\\
\kwt;This is the first line of a multi-line remark.;\\
\kwt;This is the second line;\\
\kwt;S \ p \ a \ c \ e \ s \ \ and TABS \ \ \ \ will be honored.;\\
\kwt;\ \ \ \ \ \ \ \ . . .;\\
\kwt;The label \&eor marks the end of the remark.;\\
\&eor
}

The directives that follow illustrate a very useful feature if the IDL,
which is the possibility of defining {\em global
variables\/}\index{global variables}. Such variables can be used as
replacement text within the IDL input stream, and/or be passed to
output files or external modules called by AIRES. The variables must
be defined before they can be used. This can be done by means of the
\idlbfx;SetGlobal; directive. The \idlbfx;Import; directive permits to
import OS environment variables. Variables can be overwritten, and
deleted using the \idlbfx;DelGlobal; directive.

The global variable replacement mechanism allows also to insert
(simple) calculations when needed. Let us consider the following
example:
\kwexample{%
SetGlobal \kwt;LogEprimeV  18.5;\\
. . .\\
PrimaryEnergy \{= pow \kwt;10;
              \{= subtract \kwt;\{LogEprimeV\} 18\};\} \kwt;EeV;
}
When parsing the \idlbfx;PrimaryEnergy; directive, the expression
between brackets will be processed to return the result of
$10^{18.5-18} = 3.16227766$, and therefore the primary energy will
finally be set to 3.16227766 EeV. The equal (\kwbf;=;) character at the
beginning of the field between backets indicates that what follows is
an expression of the form \kwt;function;~\vkw;args; that will be evaluated
while parsing the directive. The number of arguments can be 0, 1 or 2
according with the function. The available functions are listed in
table \ref{TAB:mathopers}

The input data set of figure \ref{FIG:sampleinp2} continues with the
\kwbf;TaskName; directive and the three mandatory directives already
introduced in section \ref{S:example1}.

The directives that follow set some characteristics of the showers that
are going to be simulated. The \idlbfx;PrimaryZenAngle; directive
gives the shower zenith angle, measured as indicated in figure
\ppref{FIG:coor1}. This directive, and the directive
\idlbfx;PrimaryAzimAngle; permit the user to completely control the
inclination of the shower axis. They can be used to set this
inclination to a fixed value, or to select variable settings selected
at random with adequate probability distributions. In this case the
alternative syntax of the directives should be used. For a more
detailed description see section \ppref{S:showerini} and/or appendix
\ppref{A:idlref}.

The \idlbfx;GroundAltitude; specification indicates the height above sea
level of the ground surface (measured vertically). The specification
can be a length or a vertical atmospheric depth expressed in {\gcmsq}
(see page \pageref{IDL0:GroundAltitude}).
On the other hand the statement
\kwdisplay;ObservingLevels {\tt 41 100 g/cm2 900 g/cm2};
sets the variables $N_o$, $X_o^{(1)}$ and $X_o^{(N_o)}$ of equation
(\ref{eq:deltao}).

The IDL instructions continue with five directives that fix the cut
energies for different particle kinds. Every particle whose kinetic
energy falls below the threshold\index{threshold energies}
corresponding to its kind will be no more propagated by the simulation
program, as explained in section \ppref{S:simengine}.

There are many observables that can be defined and studied to
determine the behavior of air showers with given initial conditions.
Generally only a small fraction of these observables are of interest
for a determined user; and of course, the set of relevant observables
do vary with the particular problem being studied.

These somewhat contradictory facts were taken into account when
designing AIRES output units, together with an analysis of the output
system of existing programs \cite{MOCCA,CORSIKA}. As a result, the
simulation program was provided with two air shower data output units:
The {\em particle data\/} unit and the {\em summary\/} unit.

The particle data unit generates {\em compressed particle data
files\index{compressed output files}\/} containing detailed
information (in a per particle basis) of particles reaching ground or
passing across the different observing levels. The other output unit
processes data stored in a number of internal tables (or histograms)
which were calculated during the simulations and which correspond to
{\em standard\/} observables like lateral distributions, energy
distributions and so on.

The output system will be treated in detail in chapter
\ppref{C:output1}. Nevertheless, it is worthwhile mentioning here that
there are several IDL directives that permit controlling its behavior.

In our example of figure \ppref{FIG:sampleinp2}, the directives
\idlbfx;SaveInFile; and \idlbfx;SaveNotInFile; control the kind of
particles that are saved in the corresponding compressed
files\index{compressed output files}, identified by their extensions
(\kwbf;lgtpcles; and \kwbf;grdpcles;).

The default action for the file containing record for the particles
reaching ground (extension \kwbf;grdpcles;) is that all particle kinds
must be saved. On the other hand, no particles are saved by default in
the longitudinal tracking particle file (extension
\kwbf;lgtpcles;). Therefore, the statements
\kwexample{%
SaveInFile    \kwt;lgtpcles  e+ e-;\\
SaveNotInFile  \kwt;grdpcles  gamma;
}
mean that only electrons and positrons are going to be saved in the
longitudinal file, and that all particles but gamma rays are going to
be recorded in the ground particle file. The particle kind
specifications may include one or more particle or particle group
names (see section \ref{S:acodes}).

There may be more than one of these statements for each file, and
their meaning depends on the order they are placed within the input
data stream. As an example, let us consider the following statements:
\kwexample{%
SaveInFile    \kwt;somefile  None;\\
SaveInFile    \kwt;somefile  Muons;
}
They ensure that only muon records will be saved in
file\footnote{\kwbf;somefile; actually indicates the {\em extension\/}
of the corresponding file, like \kwbf;grdpcles; or \kwbf;lgtpcles; for
example.}  \kwbf;somefile;: The first statement ``clears'', and the
second enables muons. If the order is changed:
\kwexample{%
SaveInFile    \kwt;somefile  Muons;\\
SaveInFile    \kwt;somefile  None;
}
then the result is that \kwbf;somefile; will be considered disabled
because the last \kwbf;None; specification prevents any particle kind
from being saved in the corresponding file.

The logical switch controlled by the instruction \kwbf;ADF
On;\idlidx;ADFile;, enables the {\em\bfseries portable dump
file\index{internal dump file!portable format},\/} the portable version of
the IDF file.

The summary unit manages more than 300 output data tables\index{output
data tables} that can be selectively included within the output
data. Each table is identified by a numerical code, and the directives
\idlbfx;PrintTables; and \idlbfx;ExportTables;\index{exported data
files} permit including a table listing within one of the output
files, or generating a separate plain text file with the corresponding
table, respectively. The complete list of available tables is placed
in appendix \ppref{A:outables}. No tables are exported or ``printed'' if
no \kwbf;Export; or \kwbf;Print; directives are included within the
input data. Notice also that there are several options that modify the
resulting output. Such options control the normalization of
histograms, output format, etc. A more detailed discussion on this
subject is placed in section \ppref{S:AiresSry}.

It is strongly recommended to edit a plain text file containing some
IDL directives, run the simulation program and analyze the obtained
output. In UNIX environments this can be made by means of the command
\kwexample{Aires < \vkw;myfile.inp;}
or, alternatively
\kwexample{Aires\vkw;Model; < \vkw;myfile.inp;}
where \vkw;myfile.inp; is the name of the file containing the IDL
directives, and \vkw;Model; identifies the external hadronic package
linked with the simulation program used (see page \pageref{P:AiresModel}).
\subsubsection*{Input data listing}
The output typed at the terminal by any of the simulation programs
will be similar to the sample displayed in figure
\ref{FIG:samplerlog}. Among other data, AIRES standard output includes
a listing of the most important input parameters. All the parameters
that are not explicitly set will take a default value. When default
values are in effect, it is indicated with a {\bf (D)} symbol placed
before the parameter's description. All The variables included in this
list can be modified by means of IDL instructions.
%
%
\begin{figure}[tp]
{\scriptsize\baselineskip=9pt\lineskip=1pt\lineskiplimit=1pt
\begin{center}
\begin{tabular}{|c|}
\hline\\*[-2pt]
\begin{minipage}{14.5cm}
\begin{verbatim}
>>>>
>>>> This is AIRES version V.V.V (dd/Mmm/yyyy)
>>>> (Compiled by . . . )
>>>> USER: xxxxx, HOST: xxxxx, DATE: dd/Mmm/yyyy
>>>>

> dd/Mmm/yyyy hh:mm:ss. Reading data from standard input unit
> dd/Mmm/yyyy hh:mm:ss. Displaying a summary of the input directives:

>>>>
>>>>          REMARKS.
>>>>

        JUST AN EXAMPLE

>>>>
>>>>          PARAMETERS AND OPTIONS IN EFFECT.
>>>>
>>>> "(D)" indicates that the corresponding default value is being used.
>>>>

                           Task Name: mytask

     RUN CONTROL:
             Total number of showers:          2
     (D)             Showers per run:   Infinite
     (D)            Runs per process:   Infinite
     (D)            CPU time per run:   Infinite

     FILE NAMES:
                            Log file: mytask.lgf
                    Binary dump file: mytask.idf
                     ASCII dump file: mytask.adf
               Compressed data files: mytask.grdpcles
                                      mytask.lgtpcles
                Table export file(s): mytask.tNNNN
                 Output summary file: mytask.sry

     BASIC PARAMETERS:
     (D)                        Site: Site00
                                      (Lat:    .00 deg. Long:     .00 deg.)
     (D)                        Date: dd/Mmm/yyyy

                    Primary particle: Proton
                      Primary energy: 1.5000 PeV
                Primary zenith angle:    15.00 deg
     (D)       Primary azimuth angle:      .00 deg
     (D)      Zero azimuth direction: Local magnetic north
                     Thinning energy: 1.0000E-04 Relative
     (D)          Injection altitude: 100.00 km (1.2829219E-03 g/cm2)
                     Ground altitude: 297.96 m  (1000.000 g/cm2)
           First obs. level altitude: 16.383 km (100.0000 g/cm2)
            Last obs. level altitude: 1.1733 km (900.0000 g/cm2)
          Obs. levels and depth step:         41    20.000 g/cm2
\end{verbatim}
\end{minipage}\\*[-2pt] \\\hline
\end{tabular}
\end{center}}
\figcaption{samplerlog}%
{Sample AIRES terminal output}{Sample AIRES terminal output.}
\end{figure}
\begin{figure}[tp]
\begin{center}
{\baselineskip=9pt\lineskip=1pt\lineskiplimit=1pt\scriptsize
\begin{tabular}{|c|}
\hline\\*[-2pt]
\begin{minipage}{14.5cm}
\begin{verbatim}
     (D)           Geomagnetic field: Off
     (D)         Table energy limits: 10.000 MeV to 1.1250 PeV
     (D)         Table radial limits: 50.000 m  to 2.0000 km
     (D)   Output file radial limits: 100.00 m  to 12.000 km (grdpcles)
     (D)                              100.00 m  to 12.000 km (lgtpcles)

     ADDITIONAL PARAMETERS:
     (D)      Individual shower data: Brief
               Cut energy for gammas: 200.00 KeV
                Cut energy for e+ e-: 200.00 KeV
              Cut energy for mu+ mu-: 1.0000 MeV
               Cut energy for mesons: 1.5000 MeV
             Cut energy for nucleons: 150.00 MeV
     (D)      Bartol inelastic mfp's: On
     (D)        Gamma rough egy. cut: 2.0000 MeV
     (D)         e+e- rough egy. cut: 2.0000 MeV
     (D)    Hadronic Mean Free Paths: SIBYLL
     (D)               SIBYLL switch: On

     MISCELLANEOUS:
     (D)    Seed of random generator: Automatic
     (D)           Atmospheric model: Linsley's standard atmosphere

>>>>
> dd/Mmm/yyyy hh:mm:ss. Beginning new task.
> dd/Mmm/yyyy hh:mm:ss. Initializing SIBYLL 1.6 package.
  Initialization of the SIBYLL event  generator 

  . . . (eventual output from SIBYLL) . . .

> dd/Mmm/yyyy hh:mm:ss. Initialization complete.
> dd/Mmm/yyyy hh:mm:ss. Starting simulation of first shower.
> dd/Mmm/yyyy hh:mm:ss. End of run number 1.
  CPU time for this run: . . . .
> dd/Mmm/yyyy hh:mm:ss. Writing ASCII dump file.
> dd/Mmm/yyyy hh:mm:ss. Task completed.
  Total number of showers: 2
> dd/Mmm/yyyy hh:mm:ss. Writing summary file.
> dd/Mmm/yyyy hh:mm:ss. End of processing.
\end{verbatim}
\end{minipage}\\*[-2pt] \\\hline
\end{tabular}}
\end{center}
\vskip \abovecaptionskip
\begin{center}
{\itshape{\bfseries Figure \ref{FIG:samplerlog}.} (continued)}
\end{center}
\end{figure}

The input parameter listing is divided in sections accordingly with
the different kind of variables that control the computational and
physical environment of the simulations. These sections are
\begin{description}
\item[Run control.] Includes all the parameters controlling the
conditions of the simulations, namely, the total number of showers,
the number of showers per run, the number of runs per process and the
(maximum) CPU time per run. The directives that control these
variables are {\em dynamic\index{Input Directive
Language!dynamic/static directives},\/} and may therefore vary during
the simulations. The quantities displayed in the input parameter listing
correspond thus to instantaneous values of the mentioned parameters.

\item[File names.] A listing with the names of all the files that will
be created during the simulations (excluding, of course, internal
scratch files).
A detailed description of the output files
that can be created by the simulation programs, together with
guidelines on how to manage them can be found in chapter
\ppref{C:output1}; we just give here a brief description of
them:
\begin{description}
\item[Log file] (\vkw;taskname;\kwbf;.lgf;)\index{log file}. This file
contains information about the events that took place during the
simulations. It contains also a summary of the input parameters that
were in effect. Most of the data that goes into the log file is also
written into the standard output channel.
\item[Summary file] (\vkw;taskname;\kwbf;.sry;, also
\vkw;taskname;\kwbf;.tex;)\index{summary file}. Output summary. This
includes general simulation data and all the tables that were printed
using IDL directive \kwbf;PrintTables;.
\item[Exported data files]
(\vkw;taskname;\kwbf;.t;\vkw;nnnn;)\index{exported data files}. Plain
text files containing output tables.
\item[Task summary script file]
(\vkw;taskname;\kwbf;.tss;)\index{task summary script file}. File
containing a summary of input and output data, written in a format
suitable for processing with other programs.
\item[Binary dump file] (\vkw;taskname;\kwbf;.idf;)\index{internal
dump file}. This file contains (in machine-dependent binary format)
all the relevant simulation data. This file is periodically updated
during the task processing. In the case of an interruption, it is
possible to restart the simulations from the last update. The file is
also useful to obtain relevant data after the simulation is completed,
or even during it. This can be done with the help of the summary
program \kwbf;AiresSry;\index{AIRES summary program}.
\item[ASCII dump file] 
(\vkw;taskname;\kwbf;.adf;)\index{internal dump file!portable format}.
Portable version of the IDF file, written at the
end of the task. Like the IDF file, this file can be processed with the
summary program \kwbf;AiresSry;.
\item[Compressed output files] (\vkw;taskname;\kwbf;.grdpcles; and/or
\vkw;taskname;\kwbf;.lgtpcles;)\index{compressed output files}. These
files contain detailed particle data. The ground particle file, for
example, consists of a series of records of all the particles that
reached ground in specified circumstances. Thanks to the compressed
data formatting used, it is possible to save a large number of
particle records using a moderate amount of disk space. The format is
universal, so the files can be written by a given machine and
processed in a different one. The AIRES system includes a
library of subroutines to process such files (see section \ref{S:cio2}).
\end{description}

\item[Basic parameters.] A list of geometrical and physical shower
parameters. These variables define the initial conditions of the
shower simulations (primary particle, axis inclination, etc.), as well
as the settings that are in effect for the parameters of the
monitoring algorithms (number of observing levels, range of radial
distances for output files, etc.).

\item[Additional parameters.] Other shower parameters, generally
depending on the model used. Since the interactions models are
replaceable, the type and number of additional parameters may vary
when changing simulation programs. The variables included in this
section as well as the directives that allow controlling them may also
be changed in future versions of AIRES. By default, only the most
relevant parameters\footnote{This also includes all the variables that
were explicitly set by means of the corresponding IDL instructions.}
are listed: Quantities associated with {\em hidden\index{Input
Directive Language!hidden directives}\/}\label{P:hidden} IDL
directives (see appendix \ref{A:idlref}) are not
included. Nevertheless, AIRES can be instructed to produce a full
listing, by means of the directive: \idlbfx;InputListing; \kwbf;Full;.

\item[Miscellaneous parameters.] Other parameters not included in the
preceding sections.\index{atmospheric model}
\end{description}

\section{More on IDL directives}

\subsection{Run control}
In the example of section \ppref{S:carryingon}, no specifications are
made about the duration of processes and runs. This fact shows up in
the variables listed in the run control section of the listing of
figure \ppref{FIG:samplerlog}, when the default setting, ``Infinite'' is
in effect for the number of showers per run, the number of runs per
process and the CPU time per run. With such settings, the auto-save
mechanism for fault tolerant processing\index{fault tolerant
processing} is disabled: The IDF file will be saved only after
finishing all the simulations specified with the input directives.

This can be acceptable for a short simulation in a reliable computer
system. For heavy tasks it is recommended to split the simulations
into processes and runs.\index{tasks, processes and runs} It is
worthwhile mentioning that {\em the auto-save/restore operations
{\bfseries do not} alter the results of the simulations,\/} which are
bitwise identical independently of the number of such operations
performed.

The IDL directives \idlbfx;ShowersPerRun;, \idlbfx;MaxCpuTimePerRun;
and \idlbfx;RunsPerProcess; provide effective control on the
computational conditions of the simulations. The following examples
illustrate how them can be used.

\kwexample{%
RunsPerProcess \kwt;1;\\
MaxCpuTimePerRun \kwt;2 hr;
}
These two instructions indicate that a new run should begin every two
CPU hours. Since the number of runs per process is 1, a new run will
also imply the beginning of a new process; in other words, the input
file will be scanned every two CPU hours, allowing for eventual
changes in the dynamical parameters of the simulations.

\kwexample{%
RunsPerProcess \kwt;4;\\
ShowersPerRun   \kwt;5;
}
Here the maximum CPU time is not set, indicating that there will be no
time limit for a run to complete. Instead, every run will finish after
concluding the simulations of five showers. The processes will end
when four runs are completed.

The three directives can also be used simultaneously:
\kwexample{%
RunsPerProcess \kwt;2;\\
ShowersPerRun   \kwt;2;\\
MaxCpuTimePerRun \kwt;6 hr;
}
These instructions indicate that a run will finish after six processing
hours or after completing two showers, {\em what happens first.\/}

The run control directives --like any other {\em dynamic\/}\index{Input
Directive Language!dynamic/static directives} directive-- can be
modified during the simulations if needed. The changes will be
effective after a new process is started (see section \ref{S:tpar}).
Let us assume that a certain task is started with the control
parameters of the previous example. After a while it is decided that
the maximum cpu time per run is too high and that there is no need to
limit the number of showers per run. The input file is thus modified:
\iiset\ii The \kwbf;MaxCpuTimePerRun; line is replaced by
\kwexample{MaxCpuTimePerRun \kwt;3 hr;}
\ii The \kwbf;ShowersPerRun; line is deleted. After finishing the
current process (with the old settings this may demand up to 12 CPU
hours), and restarting the simulation program, the input file is
scanned again and the new settings will become effective. The changes
experimented by these dynamic parameters will be recorded in the log
file (extension \kwbf;.lgf;) in the following way
\kwexample{{\baselineskip=9pt\lineskip=1pt\lineskiplimit=1pt
\footnotesize\tt
. . .\\
> dd/Mmm/yyyy hh:mm:ss. Reading data from standard input unit\\
> dd/Mmm/yyyy hh:mm:ss. Changing maximum number of showers per run.\\
\hbox{\ \ }From: 2 to: Infinite\\
> dd/Mmm/yyyy hh:mm:ss. Changing maximum cpu time per run.\\
\hbox{\ \ }From: 6 hr to: 3 hr\\
. . .}}
The dynamic directives can be changed as many times as needed,
including the total number of showers (controlled by directive
\idlbfx;TotalShowers;) which can be modified either during the
simulations or after completing them to append new showers to an
already finished task\footnote{Notice however that it is not possible
to append new showers to any task that was initialized with a {\em
previous\/} version of AIRES.}.

It is important to remark that the mechanism of dividing a task in
several process is possible because all the relevant simulation data
is saved into the internal dump file\index{internal dump file}, and
recovered in successive invocations of AIRES.

In some applications, however, it is necessary to completely disable
this mechanism, and force AIRES to start a new task every time a new
new process starts. This can be done with the help of the
\idlbfx;ForceInit; directive, like in the following example:
\kwexample{%
Task \kwt;myname;\\
ForceInit\\
\kwt;. . .;
}
On the first invocation of AIRES, the task \kwbf;myname; will be
initialized and executed accordingly with the input directives. In
a second call, the AIRES initializing procedures will check for the
existence of the file \kwbf;myname.idf;. After finding it, the task
version will be increased by one, producing a IDF\index{internal dump
file} named \kwbf;myname\_001.idf;, and then the new task will be
executed normally. On successive calls, the version number will be
increased repeatedly, until finding the first non-existent file with
name \kwbf;myname\_vvv.idf;.
\subsection{File directories used by AIRES}
\label{S:filedir}\index{AIRES file directories}%
\index{file directories|see{AIRES file directories}}
The simulation programs read and/or write several files that contain
different kinds of data. By default, all the files generated by AIRES are
located in the {\em\bfseries working directory\index{AIRES file
directories!working directory|bfpage{ }},\/} defined as the current
directory at the moment of invoking AIRES.

There are certain cases, however, where this setting is not adequate. For
that reason, the IDL instruction set contains directives allowing to
control the placement of AIRES files.

 Let us first define the set of directories used by the AIRES system
during the simulations:
\begin{description}
\item[Global\index{AIRES file directories!global directory|bfpage{ }}.]
Containing the log\index{log file}, IDF\index{internal dump file},
ADF\index{internal dump file!portable format} and summary\index{summary
file} files.
\item[Compressed output\index{AIRES file directories!output
directory|bfpage{ }}.] Sometimes referred simply as {\em\bfseries Output
directory,\/} contains the compressed output files\index{compressed output
files}.
\item[Export\index{AIRES file directories!export directory|bfpage{ }}.]
Containing all the exported data files\index{exported data files}.
\item[Scratch\index{AIRES file directories!scratch directory|bfpage{ }}.]
Containing most of the internal files that are generated during the
simulations, including the particle stack scratch files.
\end{description}
The output and scratch directories default to the current working
directory when not specified. On the other hand, the global and export
directory default to the current setting of the output directory.

The IDL directive \idlbfx;FileDirectory; permits complete control on the
listed directories. For example, the sequence of instructions:
\kwexample{%
FileDirectory Scratch \kwt;/mytmpdir;\\
FileDirectory Export \kwt;/myexportdir;
}
sets the scratch\index{AIRES file directories!scratch directory}
(export\index{AIRES file directories!export directory}) to the strings
(must be meaningful to the operating system) \kw;/mytmpdir;
(\kw;/myexportdir;). The directory specifications may be either absolute
or relative. Relative specifications are always with respect to the
working directory\index{AIRES file directories!working directory}. In the
preceding example the remaining directories are not specified, and will
therefore take their respective default settings.

The directive
\kwexample{%
FileDirectory All \kwt;/mydir;
}
simultaneously sets the global, output and export directories.

There is an additional set of directories that can be specified while
scanning the input data. The following instructions, for instance,
\idlidx;Input;\idlidx;InputPath;
\kwexample{%
InputPath Insert \kwt;/dir1:/dir2;\\
InputPath Append \kwt;/dir3;\\
Input \kwt;myinputfile.inp;
}

will cause AIRES to search for file \kwbf;myinputfile.inp; in the
current working directory\index{AIRES file directories!working
  directory} {\em and\/} --if not found there-- in all the directories
specified by means of the \kwbf;InputPath; directives (notice the two
alternative syntaxes). The search path is initially set to a string
that contains the \kwbf;bin; and \kwbf;airesinputs; subdirectories
included within the AIRES distribution. To eliminate those directories
from the file search path (in case you really need it!) you can invoke
the \kwbf;InputPath; directive with no arguments to clear the search
path, or without the \kwbf;Insert; or \kwbf;Append; modifiers to set
the path to {\em exactly\/} the specified directories as in the
following example:
\idlidx;InputPath;
\kwexample{%
InputPath \kwt;/dir1:/dir2 ...;
}
In this last case, input files will be searched for within the current
directory, and the directories specified by the \kwbf;InputPath;
directive.
\subsection{Defining the initial conditions}
\label{S:showerini}
There are two mandatory specifications related to shower parameters
that must always appear within the input data, namely, primary
particle kind and energy.\footnote{When using special primary
  modules\index{special primary particles} the primary energy can be
  defined dynamically from the external module every time it is
  invoked. In this case there in no obligation of specifying it via a
  IDL directive.}

These two specifications, together with other related ones permit a
very wide range of specifications for the shower parameters. Let us
investigate some of the possible alternatives.
\subsubsection*{Mixed composition}
\label{S:mixcomp}\index{mixed composition}%
The primary particle needs not be unique. AIRES allows for simulating
showers with different primary particles each. The following example
illustrates this feature:
\idlidx;PrimaryParticle;%
\kwexample{%
PrimaryParticle \kwt;Proton 0.6;\\
PrimaryParticle \kwt;Iron 0.4;
}
With such settings, the primary will be proton (iron) with 60\% (40\%)
probability. This means that in 100 simulated showers, approximately
60 will be proton showers while the remaining ones will have iron
primaries.

If $n$ alternative primary particles, $p_i$, $i=1,\ldots,n$ were
defined, with weights $w_i$ ($w_i\ne 0$), then the probability for any
shower of being initiated by particle $p_j$, $1\le j\le n$ is given by
\begin{equation}                           \label{eq:mixprim}
P_j = \frac{|w_j|}{\sum_{i=1}^n |w_j|}
\end{equation}
Therefore, the weights entered in the IDL directives need not be
normalized.

Besides this mixed composition feature, AIRES allows also to define
{\em special\/} primary particles processed by external modules. For
details see section \ppref{S:specialprim}\index{special primary
particles}.
\subsubsection*{Varying energy}

The directive
\idlidx;PrimaryEnergy;%
\kwdisplay;PrimaryEnergy $E_{\mathrm{min}}$ $E_{\mathrm{max}}$
$\gamma$;
(see page \pageref{IDL0:PrimaryEnergy}), indicates that the
primary energies will be in the interval $[E_{\mathrm{min}},
E_{\mathrm{max}}]$, selected with probability
\cite{gaisserbk}\index{primary energy spectrum}:
\begin{equation}                             \label{eq:gammadist}
p(E)\>dE = U^{-1}\, E^{-(\gamma+1)}\>dE,\quad
 E_{\mathrm{min}}\le E\le E_{\mathrm{max}},
\end{equation}
where
\begin{equation}
U = \int_{E_{\mathrm{min}}}^{E_{\mathrm{max}}} E^{-(\gamma+1)}\;dE =
\left\{\begin{array}{ll}
{1\over{\gamma}}
\left(E_{\mathrm{min}}^{-\gamma}-E_{\mathrm{max}}^{-\gamma}\right)&
\gamma\ne 0\\*[7mm]
\ln\left(E_{\mathrm{max}}/E_{\mathrm{min}}\right) & \gamma=0
\end{array}\right.
\end{equation}
$\gamma$ can take any value. If not specified it is taken as 1.7.
\subsubsection*{Zenith and azimuth angles}
\label{S:zenazimdist}
The zenith angle directive placed in the example of figure
\ppref{FIG:sampleinp2} corresponds to setting the angle to a fixed
value. In this case the azimuth angle defaults to zero.
On the other hand, the instruction
\idlidx;PrimaryZenAngle;%
\kwexample{PrimaryZenAngle \kwt;0 deg 72 deg S;}
indicates that the zenith angle distributes from $0^\circ$ to
$72^\circ$ with {\em sine\/} distribution\footnote{The sine distribution
is sometimes called cosine distribution, relating it with the
accumulative probability function of the sine distribution:
$F_{\mathrm{sine}}(\Theta)=\int_0^{\Theta} P_{\mathrm{sine}}(u)\>du$.}
\begin{equation}                            \label{eq:sindist}
P_{\mathrm{sine}}(\Theta)\>d\Theta = U^{-1}\,\sin\Theta\>d\Theta,
\quad \Theta_{\mathrm{min}} \le \Theta \le \Theta_{\mathrm{max}} ,
\end{equation}
where
\begin{equation}                            \label{eq:sindistnorm}
U = \int_{\Theta_{\mathrm{min}}}^{\Theta_{\mathrm{max}}}
\sin\Theta\>d\Theta =
 \cos\Theta_{\mathrm{min}} - \cos\Theta_{\mathrm{max}}.
\end{equation}

An alternative to the \kwbf;S; specification is the \kwbf;SC; (or
\kwbf;CS;) specification which corresponds to a {\em sine-cosine\/}
distribution:
\begin{equation}                            \label{eq:sincosdist}
P_{\mathrm{sin\,cos}}(\Theta)\>d\Theta =
U^{-1}\,\sin\Theta\cos\Theta\>d\Theta,
\quad \Theta_{\mathrm{min}} \le \Theta \le \Theta_{\mathrm{max}} ,
\end{equation}
where
\begin{equation}                            \label{eq:sincosdistnorm}
U = \frac{1}{2}\int_{\Theta_{\mathrm{min}}}^{\Theta_{\mathrm{max}}}
\sin(2\Theta)\>d\Theta = \frac{1}{4}
\left[ \cos(2\Theta_{\mathrm{min}}) -
\cos(2\Theta_{\mathrm{max}})\right].
\end{equation}

For varying zenith angles, the default for the azimuth angle is to
uniformly distribute in the interval $[0^\circ,360^\circ]$. In this
case the sine distribution corresponds to showers with directions
having a uniform solid angle distribution.

The azimuth angle can also be set as a varying angle. The directive
\idlidx;PrimaryAzimAngle;%
\kwexample{%
PrimaryAzimAngle  \kwt;37.2 deg 39.5 deg;
}
indicates that the azimuth $\Phi$ will be uniformly distributed in the
interval $[37.2^\circ, 39.5^\circ]$.

Using simultaneously instructions for both the zenith and azimuth
angles, it is possible to simulate showers coming from a determined
direction in the celestial sphere.

 As pointed out in section \ppref{S:coordinates}, the $x$-axis (zero
azimuth axis) corresponds to the local {\em magnetic} north. If
desired, it is possible to specify {\em geographic} azimuths:
\index{geographic azimuth}\index{magnetic azimuth}%
\kwexample{                                    \label{SS:geogazim}%
PrimaryAzimAngle  \kwt;37.2 deg 39.5 deg; Geographic
}
In the preceding directive, the \kwbf;Geographic; keyword indicates
that the origin of the azimuth angles is the direction of the local
geographic north. It is worthwhile mentioning that this {\em does not
alter\/} the axis definitions of section \ref{S:coordinates}; when
geographic azimuths are in effect, the azimuth with respect to the
AIRES coordinate system\index{AIRES coordinate system}, $\Phi$, is
evaluated via
\begin{equation}                                     \label{eq:geophi}
\Phi = \mathrm{D} - \Phi_{\mathrm{geographic}}
\end{equation}
where $\mathrm{D}$ is the geomagnetic declination angle defined in
section \ppref{S:geomag}. Notice that positive geographic azimuths
indicate eastwards directions. For a complete description of this
directive see page
\pageref{IDL0:PrimaryAzimAngle}. 
\subsubsection*{Position of injection, ground and observing levels}
The directives \idlbfx;InjectionAltitude; (or its synonym
\kwbf;InjectionDepth;), \idlbfx;GroundAltitude; (or its synonym
\kwbf;GroundDepth;) and \idlbfx;ObservingLevels; permit controlling
the position of the injection point, the ground surface and the
different observing levels, respectively.

All the altitude specifications refer to {\em vertical altitudes,\/}
noted as $z_v$ in figure \ppref{FIG:coor1}, and can be expressed either
as lengths (above sea level) or vertical atmospheric depths. Whenever
necessary, AIRES transforms lengths into vertical depths and vice-versa
using the current atmospheric model.\index{atmospheric model}

Notice that the vertical altitudes are equal to the corresponding
$z$-coordinates only for points located in the $z$-axis. To illustrate
this point, let us consider the following instructions
\kwexample{%
InjectionAltitude \kwt;100 km;\\
GroundAltitude \kwt;1000 m;\\
PrimaryZenAngle \kwt;60 deg;
}
With such specifications, the primary particles will be injected at an
altitude of 100 km above sea level, measured along the vertical
passing by the injection point. Taking into account that the shower
axis has an inclination of 60 degrees, and applying equation
(\ref{eq:zv}), it is possible to calculate the $z$-coordinate of the
injection point, also referred as {\em central injection altitude.\/}
In this case the result is $z_c=17962$ m.

The positions of the observing levels defined in section
\ppref{S:monitoring} can be set using \kwbf;ObservingLevels;. This
directive has two different formats\index{longitudinal development}:
\begin{enumerate}
\item[(i)] \kwbf;ObservingLevels; $N_o$, with $N_o$ an integer not less
than 4.

In this case the positions of the observing levels are set taking into
account the injection and ground vertical depths. Let $X_i$ ($X_g$) be
the injection (ground) depth, then the spacing between observables and
the positions of the first and last observing levels are set via
\begin{equation}                                 \label{eq:deltao0}
\begin{array}{rcl}
\Delta X_o  \!\!&=&\!\! \displaystyle \frac{X_g -X_i}{N_o + 1} \\*[12pt]
X_o^{(1)}   \!\!&=&\!\! X_i + \Delta X_o \\
X_o^{(N_o)} \!\!&=&\!\! X_g - \Delta X_o
\end{array}
\end{equation}
\item[(ii)] \kwbf;ObservingLevels; $N_o$ $X_a$ $X_b$, with $N_o$ an
integer not less than 4 and $X_a$ and $X_b$ valid vertical depth or
altitude specifications ($X_a \ne X_b$).

In this second case the positions of the first and last observing
levels are set accordingly with $X_a$ and $X_b$, with no dependence on
the positions of the injection and ground levels:
\begin{equation}                               \label{eq:deltaoab}
X_o^{(1)} = \min(X_a, X_b), \quad
X_o^{(N_o)} = \max(X_a, X_b).
\end{equation}
The spacing between consecutive levels is evaluated using equation
\ref{eq:deltao}.
\end{enumerate}
\subsection{Atmosphere}\index{atmosphere}
\label{S:atmosidl}
In AIRES the profile of atmospheric depth and density is conveniently
modeled via multilayer parameterizations like the Linsley described in
section \ref{S:atmosphere}.

The main directive to set the atmospheric profile is
\idlbfx;Atmosphere;. For example,
%
\lowtable{[t]}{addatmosdirs}{%
\def\phigh{$\vphantom{|^{|^|}}$}\def\plow{$\vphantom{|_{|_|}}$}
\begin{tabular}{|l|l|p{8cm}|}\hline
\multicolumn{1}{|c}{\bfseries\itshape Directive}\phigh\plow&
\multicolumn{1}{|c}{\bfseries\itshape Args}&
\multicolumn{1}{|c|}{\bfseries\itshape Action}\\ \hline
 \akw;AddLay;er;\subidlidx;AddAtmosModel;AddLayer;\phigh
               & $h_{\rm beg}$ $h_{\rm end}$ $\rho_{\rm beg}$ $\rho_{\rm end}$ 
               & Adds an exponential layer.\plow
\\ \hline
 \akw;AddLin;Layer;\subidlidx;AddAtmosModel;AddLinLayer;\phigh
               & $h_{\rm beg}$ $h_{\rm end}$ $\rho_{\rm beg}$ $\rho_{\rm lay}$ 
               & Adds a linear layer (for advanced users).\plow
\\ \hline
 \akw;AtmDef;ault;\subidlidx;AddAtmosModel;AtmDefault;\phigh
               & Model id
               & Sets the atmospheric model to use to provide data
                 corresponding to \akw;MatchD;efault; qualifiers or
                  automatically added layers.\plow
\\ \hline
 \akw;AtmIde;nt;\subidlidx;AddAtmosModel;AtmIdent;\phigh
               & Model id string
               & Sets the model identification string
                 (maximum 16 characters).\plow
\\ \hline
 \akw;AtmNam;e;\subidlidx;AddAtmosModel;AtmName;\phigh
               & Model long name
               & Sets the model name (maximum 42 characters).\plow
\\ \hline
 \akw;End; ;\subidlidx;AddAtmosModel;End;\phigh
               & & Ends processing of \kw;AddAtmosModel; instructions.\plow
\\ \hline
\end{tabular}}%
{Instructions recognized by \kwbf;AddAtmosModel;}%
{Instructions recognized by the IDL directive \idlbfx;AddAtmosModel;.}
%
\kwexample{Atmosphere \kwt;SouthPoleAvg;}
sets as current atmospheric model the predefined model identified by
the string ``\kwbf;SouthPoleAvg;''. The available predefined models
are listed in page \pageref{\idlref;Atmosphere;}. The default
atmospheric profile is \kwbf;Linsley;, and corresponds to a
parameterization of the US standard atmosphere\index{US standard
  atmosphere} \cite{usatm76}.

Some of the predefined model accept parameters. Consider the following
example:
\kwexample{Atmosphere \kwt;Isothermic\ ; Temp \kwt;28 C\ \ ;
                       Dens0 \kwt;1.221 kg/m3;}
In this case the \idlbfx;Atmosphere; directive sets a isothermic model
specifying a temperature of 28 C, and a density at sea level of 1.221
kg/m$^3$. The parameters may be specified in any order, and in case of
missing specification a default value is provided.

Besides selecting among the predefined models, it is possible to add
user-defined custom models by means of the \idlbfx;AddAtmosModel;
directive. This directive allows to define a multilayer model by
specifying the beginning and ending altitudes of each layer together
with the densities at both layer ends. Then AIRES uses those data to
calculate the multilayer coefficients, and include the defined model
in the list of available ones. In case the user-specified layers do
not cover the minimum range of altitudes going from sea level up to
100 km.a.s.l., AIRES will automatically add layers (using the default
atmospheric model) to complete the description.

Consider the following example:
\begin{verbatim}
  AddAtmosModel &mylabel
   AtmIdent  MyModelIdStr
   AtmName   My atmospheric model
   AtmDefault  Linsley
   AddLayer     0 m    800 m   MatchDefault     1.2184    kg/m3
   AddLayer   800 m   4000 m   1.2184 kg/m3     0.8422    kg/m3
   AddLayer  4000 m     12 km  0.8422 kg/m3     0.2765    kg/m3
   AddLayer    12 km    35 km  0.2765 kg/m3     6.4846E-6 g/cm3  
   AddLayer    35 km   100 km  6.4846E-6 g/cm3  MatchDefault
  &mylabel
\end{verbatim}
In this case the input data for the \idlbfx;AddAtmosModel; directive
is placed just after the directive as a {\em
  here-document\index{here-document}\/} delimited by the label
\kwbf;\&mylabel;.
The format of the instructions is self-explaining. Each \kwbf;AddLayer; instruction
requires four quantities, namely, layer beginning and ending
altitudes, and the corresponding densities.
The information that the user must provide with these instructions is
redundant, since in the intermediate boundaries a same value must be
specified twice. The advantages of more robustness and human readability
amply compensate the extra effort of duplicating some data. Additionally,
the \kwbf;AddLayer; instructions can be placed in any order.
Notice also that both the altitudes and densities are specified with two
fields in all cases (number + unit). Any of the length or density units
recognized by AIRES can be used (see table \ref{TAB:idlunits}).

Once the model is defined via the \idlbfx;AddAtmosModel; directive, it
can be set as the current model with the \idlbfx;Atmosphere; command
\kwexample{Atmosphere \kwt;MyModelIdStr;}
Table \ref{TAB:addatmosdirs} contains a listing of all the
instructions recognized by directive \idlbfx;AddAtmosModel;. The
demonstration examples that are included in the AIRES distribution
contain several additional examples that explain in more detail how to
define a custom atmospheric model.

\subsection{Geomagnetic field}\index{geomagnetic field}
\label{S:geomagidl}
The components of the Earth's magnetic field used by the simulation
programs can either be set manually or calculated with the help of the
IGRF model
\cite{IGRF}\index{International Geomagnetic Reference Field}%
\index{external packages} (see section \ref{S:geomag}). With the
help of this model it is possible to obtain an accurate estimation of
the geomagnetic field in a given geographic location and for a
determined date.

To activate this mechanism for ``automatic'' evaluation of the magnetic
field, it is necessary to specify both a geographic place and a date.

The directive \idlbfx;Site; tells AIRES the name of the site selected
for the simulations. For example,
\kwexample{Site \kwt;SouthPole;}
indicates that the selected place is ``\kwbf;SouthPole;''. This name
is one of the predefined locations that form the {\em AIRES site
library\index{AIRES site library}.\/} Besides ``SouthPole'', this
library initially contains several other sites related with air shower
experiments. All the predefined sites are listed in table \ref{TAB:sitelib}.
\lowtable{[tb]}{sitelib}{%
\def\pz{$\phantom{9}$}
\begin{tabular}{l@{\kern2.5em}r@{$^\circ$ }lr@{$^\circ$ }lc}
{\bfseries\itshape Site name} &
\multicolumn{2}{c@{\kern1.2em}}{{\bfseries\itshape Latitude}} &
\multicolumn{2}{c}{{\bfseries\itshape Longitude}} &
{\bfseries\itshape Altitude\/} (m.a.s.l)\\*[2pt]
 Site00       &          0.00&    &     0.00&    &  \pz\pz\pz0  \\
 SouthPole    &    \pz  90.00& S  &     0.00&    &  2835  \\
 Malargue     &         35.20& S  &    69.20& W  &  1425  \\
 ElNihuil     &         35.20& S  &    69.20& W  &  1400  \\
 TelescArray  &         39.30& N  &   112.91& W  &  1400  \\
 AGASA        &         35.78& N  &   138.50& E  &  \pz 900  \\
 CASKADE      &         49.09& N  &     8.88& E\ &  \pz 112 \\
 Dugway       &         40.00& N  &   113.00& W  &  1550  \\
 ElBarreal    &         31.50& N  &   107.00& W  &  1200  \\
 FlysEye      &         41.00& N  &   112.00& W  &  \pz 850 \\
 HaverahPark  &         53.97& N  &     1.64& W  &  \pz 220  \\
 Puebla       &         19.50& N  &    98.00& W  &  2200  \\
 SydneyArray  &         30.50& S  &   149.60& W  &  \pz 250 \\
 Yakutsk      &         61.70& N  &   129.40& E  &  \pz 850
\end{tabular}}%
{Predefined sites of the AIRES site library}%
{Predefined sites of the AIRES site library. Site names are case
sensitive. The data for Haverah Park, Sydney Array and Yakutsk sites come
from reference \cite{berezin1}.}

To specify a site that is not included among the predefined ones, it
is first necessary to append it to the site library by means of the
\idlbfx;AddSite; directive. Let us consider, for instance, the
following directive:
\kwexample{AddSite \kwt;cld  -31.5 deg  -64.2 deg  387 m;}
A new site ``\kwbf;cld;'' is defined. The command parameters
represent, respectively, the latitude, longitude and altitude above
sea level that correspond to the defined site. The name string cannot
contain more than 16 characters; names are case sensitive and must be
different to all the previously defined ones.

The \idlbfx;Date; directive defines the date of an event. There are
two alternative syntaxes, as displayed in the following examples:
\kwexample{%
Date \kwt;2018.2;\\
Date \kwt;2018 3 1;
}
In the first statement the date is given as a floating point number
taking the year as the time unit, while in the second the format
``year month day'' is used.

There are no special restrictions on the date specification. However,
the IGRF database\index{external packages} implemented in the current
AIRES version contains data for the years 1955 to 2020. For dates
outside that interval it is necessary to extrapolate the corresponding
data in order to evaluate the geomagnetic field. This may lead to
inaccurate estimations for dates very far from the validity range of
the model (more than ten years away). Nevertheless, extrapolations
near the given boundaries are acceptable, and are of course necessary
for calculations beyond the year 2015.\footnote{The next generation of
IGRF data will be released after the year 2020.}

In case of missing date specification, it is set accordingly with the
system time at the moment of starting the simulations.

Once a site and a date are set, the Earth's magnetic field will be
calculated by means of the IGRF model\index{external packages}, unless
it is explicitly set by means of the \idlbfx;GeomagneticField;
directive. Let us analyze some examples (see also page
\pageref{IDL0:GeomagneticField}):
\kwexample{GeomagneticField Off}
With this instruction the effect of the magnetic field on the motion
of the charged particles will not be taken into account. However, the
field will still be evaluated in order to determine the declination
angle, which is used to transform geographical azimuths into magnetic
ones (see page \pageref{SS:geogazim}).
\kwexample{GeomagneticField  \kwt;32 uT  -60 deg  2 deg;}
The preceding directive instructs AIRES to fully override the IGRF
estimation with the values indicated in the parameters, which
respectively correspond to F, I and D (see section \ref{S:geomag}).
Partial overriding is also supported, like in the following
instruction
\kwexample{GeomagneticField \kwt;32 uT;}
The field strength, F, will be set to the value indicated in the first
 parameter, while I and D will remain as given by the IGRF model.

$xz$-plane Gaussian fluctuations\index{geomagnetic
field!fluctuations}, either absolute or relative, are also supported:
\kwexample{GeomagneticField \kwt;32 uT; Fluctuation \kwt;500 nT;\\
           GeomagneticField \kwt;On; Fluctuation \kwt;10 \%;}
Notice that fluctuations can be introduced with or without overriding
the IGRF field components. It is also possible to specify \kwbf;0.1
Relative; instead of \kwbf;10 \%;.

When magnetic fluctuations are in effect, then the magnetic field used
for each shower will be different. Let $\vr{B}_0$ be the ``central''
value coming from the IGRF model and/or entered manually. Let $\Delta
B$ be the specified fluctuations. Notice that in the case of relative
fluctuations, $\Delta B$ is set using the field strength $B_0$:
$\Delta B= B_0\Delta B_{\mathrm{rel}}$.

Then for each new shower, two independent, Gaussian-distributed random
numbers, $\Delta B_x$ and $\Delta B_z$, having mean zero and standard
deviation $\Delta B/\sqrt{2}$, are generated; and the magnetic field
components are set via
\begin{equation}                                   \label{eq:geobfluc}
\begin{array}{c}
B_x = B_{0x} + \Delta B_x,\\
B_z = B_{0z} + \Delta B_z.
\end{array}
\end{equation}
Notice, however, that the declination angle used for azimuth
transformations will always come from the central value, that is, is
not affected by the fluctuations introduced.
\subsection{Statistical sampling control}\index{thinning}
\label{S:sethin}
The thinning algorithm described in section \ppref{S:thinning} makes use
of several external parameters that can be set by means of IDL
directives.  The thinning energy $E_{\mathrm{th}}$ is the most
important parameter of the thinning algorithm. As illustrated in
figure \ppref{FIG:sampleinp2}, the directive \idlbfx;ThinningEnergy;
permits setting $E_{\mathrm{th}}$, either absolutely or relative to
the primary energy.

The directive \idlbfx;ThinningWFactor; allows controlling the maximum
weight parameter $W^{(EM)}_{\mathrm{max}}$ defined in section
\ppref{S:thinning}. The specification
\kwexample{ThinningWFactor \kwt;2.5;}
sets the {\em weight factor\index{statistical weight factor},\/}
$W_f$, of equation (\ref{eq:Wfactor}) to 2.5, to be used with
electromagnetic particles.

Recommended values for $W_f$ are in the range 0.1 to 50; the default
value is 12. Setting $W_f>100$ is practically equivalent to
$W_f\to\infty$ (see section \ref{S:thinningsamples}).\footnote{{\bf
IMPORTANT:} The statistical weight factor of the AIRES extended
thinning algorithm\index{thinning!AIRES extended algorithm} is not
equivalent to the parameter with the same name defined for AIRES 1.4.2
or earlier. Therefore, the recommended values placed in the AIRES
1.4.2 manual \cite{Aires142} do not apply for the current version.}

The weight factor that is used with non electromagnetic particles,
$W^{(H)}_f$, can also be set by the user: The directive
\idlbfx;EMtoHadronWFRatio; permits setting the ratio $A_{EH}$ defined
in equation (\ref{eq:Wratio}). The default value $A_{EH}=88$ is
normally adequate, but some applications may require performing
simulations with a different relation between electromagnetic and non
electromagnetic weight factors, and in such cases the mentioned
directive is useful to change the ratio as needed.
\subsection{Output table parameters}\index{output data tables}
\label{S:outablepar}
The output tables listed in appendix \ppref{A:outables} are automatically
calculated during the simulations, and the directives to retrieve these
data will be explained in chapter \ppref{C:output1}. Many of these tables
can be customized by means of IDL instructions.

The number of observing levels defined for the longitudinal
tables\index{longitudinal development} (table numbers 1000 to 1999) can be
controlled using the IDL directive \idlbfx;ObservingLevels;, as already
explained in section \ppref{S:carryingon}.

The lateral distribution tables\index{lateral distributions} (table
numbers 2000 to 2499), the energy distribution tables\index{energy
distributions} (table numbers 2500 to 2999), and the mean arrival time
distribution tables\index{arrival time distributions} (table numbers 3000
to 3499) are defined, by default, as histograms with 40 logarithmic bins
(either radial or energy bins depending on the distribution type), plus
two additional ``underflow'' and ``overflow'' bins.

The IDL directives \idlbfx;RLimsTables; and \idlbfx;ELimsTables; allow to
control the radial and energy bins, respectively, as illustrated in the
following examples:
\kwexample{%
RLimsTables \kwt;20 m 2 km;\\
ELimsTables \kwt;2 MeV 1 TeV;}
The first directive sets the range for the standard lateral
distributions. The lowest end of bin 1 (highest end of bin 40) is set to
20 m (2 km). The ``underflow'' bin will thus correspond to all entries
with distances less than 20 m, while the ``overflow'' one to all entries
beyond 2 km.

In a completely similar way, the second directive sets the lower and upper
bounds for the 40 bin energy distributions, and the respective
``underflow'' and ``overflow'' bins.

With the current version of AIRES it is possible to save the tables in a
shower per shower basis, besides the traditional average tables that have
been always available. Since this may generate large IDF\index{internal
dump file} or ADF files in certain cases, the mechanism of individual
shower table saving is disabled by default. The directive
\idlidx;PerShowerData;
\kwexample{PerShowerData \kwt;Full;}
must be used to ensure that the individual shower tables\index{single
shower tables} are being saved.
\subsection{Random number generator}
\index{random number generator}
The AIRES random number generator must be initialized before starting
any set of simulations. The default action is to use a internally
generated seed, generated with an elementary random number
generator\index{random number generator!elementary without seed} that
uses the current clock and CPU usage registers. Therefore, different
invocations of AIRES with the same input directives, will generally
originate different output data because of different initializations
of the random number generator.

The default behavior can be changed if needed. The directive
\idlbfx;RandomSeed; allows the user to set the random seed to a given
number, or to get the seed from an already initialized task. These
features are illustrated in the following examples:
\begin{enumerate}
\item The directive
\kwexample{RandomSeed \kwt;0.1298004637;}
sets the random seed to a fixed constant. The number must be greater
than zero and less than one.
\item The directive
\kwexample{RandomSeed GetFrom \kwt;otheridfile;}
extracts the seed used in the task that created the IDF\index{internal
dump file} file \kwbf;otheridfile;, and uses it to initialize the
generator.
\end{enumerate}
\section{Input parameters for the interaction models}
\label{S:idlmodel}

The expression {\em interaction models\/} identifies a series of
subroutines and functions that contain the actual implementations of
the algorithms that control the propagation of particles. Such
algorithms emulate the physical rules associated with the different
interactions that take place in an air shower.

As it is well-known, there are still many open problems in this area
and therefore the interaction models cannot be considered a
crystallized part of the simulation programs. Furthermore, in the
design of the interaction models and external
packages\index{external packages} units shown in figure
\ppref{FIG:aires_struct}, every effort
was made to make them easily replaceable, in order to be able to
incorporate improved code to be developed in the future.

The IDL directives that are going to be mentioned in this section
allow the user to control different model parameters. Such directives
are defined from within the interaction model section, and for the
reasons explained in the preceding paragraph, they are of a changing
nature: For AIRES versions later than the current version
{\currairesversion} the model related directives may no longer be
supported, be replaced by alternative ones or their syntax be
totally or partially changed.
\subsection{External packages}
\label{S:extp}\index{external packages}
The last versions of EPOS \cite{EPOSLHC}\index{EPOS}, QGSJET
\cite{QGSJET2R4}\index{QGSJET}, and SIBYLL
\cite{SIBYLL23}\index{SIBYLL} hadronic\index{hadronic models}
collisions packages are implemented in AIRES. For technical reasons
they are compile-time implemented, and are available by means of
different executable programs: \kwbf;AiresS23; is a executable
simulation program linked to SIBYLL 2.3, etc.

The current version of AIRES ({\currairesversion}) includes links to
EPOS LHC and 1.99, QGSJET-II 04 and 03, and SIBYLL 2.3, 2.3c, and 2.1.

 All the particle-nucleus and nucleus-nucleus interactions with
projectile kinetic energy above a certain threshold\index{threshold
energies} are processed using the external package, while the low
energy ones are calculated by means of the extended Hillas splitting
algorithm\index{extended Hillas splitting algorithm}
\cite{splithin,gaisserbk}\index{Hillas, A. M.}, or a built-in nuclear
fragmentation model, in the cases of hadron-nucleus or nucleus-nucleus
collisions\index{nucleus-nucleus collisions}.

The IDL directive \idlbfx;ExtCollModel; is an On-Off switch that
allows controlling the use of the external package (EPOS\index{EPOS},
QGSJET\index{QGSJET}, or SIBYLL\index{SIBYLL}\index{hadronic
  models}\index{external packages}, depending on the executable
program being used). The minimum energy required for the external
package to be invoked can be altered using directives
\idlbfx;MinExtCollEnergy; and/or \idlbfx;MinExtNucCollEnergy;, as in
the following example:
\kwexample{%
MinExtCollEnergy \kwt;300 GeV;\\
MinExtNucCollEnergy \kwt;500 GeV;
}
AIRES supports also the directive \idlbfx;ForceModelName; that is
useful to ensure that a given input data set will be processed only
with a determined simulation program. For instance, if an input data
set containing the instruction
\kwexample{ForceModelName \kwt;QGSJET-II-04;}
is processed with other simulation program different from
\kwbf;AiresQIIr04;, the process will immediately be aborted with an error
message. When the directive is not used no check is performed and the
simulations can be started with any program.

The cross sections used to determine the collision mean free
paths\index{mean free path!hadronic} can
also be controlled. In the current version there are several sets of
hadronic cross sections\index{hadronic cross sections} available, namely,
Standard, EPOS, QGSJET and SIBYLL (all versions) cross sections. The

The default mean free paths are the ones corresponding to the external
hadronic package linked to the simulation program. The following
example illustrates how to alter the default settings:
\idlidx;MFPHadronic;\idlidx;MFPThreshold;\index{threshold energies}%
\kwexample{%
MFPHadronic \kwt;SIBYLL23c;\\
MFPThreshold \kwt;120 GeV;
}
These instructions imply that the ``SIBYLL23c'' mean free paths will be
used for collisions with energies over 120 GeV, while the standard
mean free paths will be used for the ones with lower energies.

The mean free path\index{mean free path} sets supported in AIRES
\currairesversion are:\index{hadronic cross sections}\label{P:hmfp}
\kw;Standard;,
\kw;EPOS;\index{EPOS}, \kw;EPOS-LHC3400;, \kw;EPOS1990;,
\kw;QGSJET;\index{QGSJET}, \kw;QGSJET-II-04;, \kw;QGSJET-II-03;,
\kw;SIBYLL;\index{SIBYLL}, \kw;SIBYLL231;, \kw;SIBYLL23c;, \kw;SIBYLL21;.
The generic names
(\kw;EPOS;, \kw;QGSJET;, \kw;SIBYLL;)
refer to the sets associated with the newest installed version of the
respective hadronic models.

The previous directives also indicate that the nucleus-nucleus mean
free paths will be evaluated using special algorithms included within
the external hadronic packages if the projectile's energy per nucleon
falls above the specified threshold; otherwise the mean free path will
be evaluated via a built-in procedure that calculates it by scaling
adequately the proton-nucleus mean free path corresponding to the
model being used. The directive \idlbfx;ExtNucNucMFP; allows to
disable the call to the external routine, and use the built-in
algorithm for all projectile energies.

The hadron-nucleus/nucleus-nucleus and/or the photon-nucleus
collisions can be disabled if desired:
\idlidx;NuclCollisions;\idlidx;PhotoNuclear;%
\kwexample{%
NuclCollisions \kwt;Off;\\
PhotoNuclear   \kwt;Off;
}
These settings are intended to be used only for special purposes: The
results obtained in such conditions may be rather unphysical.
\subsection{Other control parameters}
There are several IDL instructions that allow controlling
different parameters and/or processes of the simulation
algorithms. These IDL directives need not be used for normal
operation. Furthermore, the user should take into account that
improper settings for some of the parameters associated with these
instructions may lead to unphysical results.

\begin{description}

\item[\idlbfx;PropagatePrimary;.] Logical switch to control the
initial propagation of the primary.

\item[\idlbfx;SetTimeAtInjection;.] Logical switch to control whether
or not the shower time is set to zero at the injection point. The
shower clock can be set to zero at the injection point (default) or at
the moment of the first primary interaction.

\item[\idlbfx;GammaRoughCut;, \idlbfx;ElectronRoughCut;.] Threshold
energies\index{threshold energies} for ``normal'' propagation of
gammas and electrons, respectively. Particles with kinetic energies
below those thresholds are ``roughly'' propagated, that is, many
processes are calculated only approximately, or are ignored at all.

\item[\idlbfx;ForceLowEDecays;, \idlbfx;ForceLowEAnnihilation;.] These
directives control the kind of action to be ta\-ken when low energy
particles\index{low energy particles} that can decay or undergo
annihilation reach the low energy threshold.

\item[\idlbfx;LPMEffect;.] IDL switch to enable/disable the LPM
\cite{LPMigdal,LPM} effect\index{LPM effect}. The default is
\kwbf;LPMEffect On;.

\item[\idlbfx;DielectricSuppression;.] IDL switch to enable/disable
the dielectric suppression \cite{LPMigdal,KleinRev}
effect\index{dielectric suppression}. The default is
\kwbf;DielectricSuppression On;.

\item[\idlbfx;MuonBremsstrahlung;.] IDL switch to enable/disable the
muon bremsstrahlung\index{muon bremsstrahlung} and muonic pair
production\index{muonic pair production} processes. The default
is \kwbf;MuonBremsstrahlung On;

\item[\idlbfx;AirZeff;, \idlbfx;AirAvgZ/A;, \idlbfx;AirRadLength;.]
IDL directives associated with internal
parameters. For a detailed explanation see appendix \ppref{A:idlref}.

\end{description}

Since most of these IDL instructions are {\em hidden\/}
directives\index{Input Directive Language!hidden directives} (see page
\pageref{P:hidden}) the respective settings in effect will not be
included in the input data list, unless explicitly indicated by means
of directive \idlbfx;InputListing; (see page
\pageref{IDL0:InputListing}). Additionally, warnings messages will be
issued when using any directive which may lead to simulations with
unphysical results.
\section{Special primary particles}
\label{S:specialprim}\index{special primary particles|bfpage{ }}%
\index{alternative primaries|see{special primary particles}}%
\index{multiple primaries|see{special primary particles}}%
\index{exotic primaries|see{special primary particles}}

In many cases of interest, it is necessary to simulate showers that
cannot be described adequately with the usual scheme of a single primary
particle interacting with a nucleus in the atmosphere and generating a
set of secondaries to be propagated. Instead, one has that a particular
set of interactions that only affect the primary particle, originates
a series of ``normal'' secondary particles that hit the atmosphere and
originate the corresponding cascades. In general, such special interactions
are not modeled adequately by AIRES propagating engine, but it is
possible to overcome this difficulty allowing the simulation program
to start a shower with {\em multiple\/} ``primary'' particles which
are the secondaries coming out from the ``special'' interactions.

The following are examples where the mentioned scheme applies:
\begin{itemize}
\item An {\em exotic\/} cosmic particle (a cosmic
neutrino\index{cosmic neutrinos}, for
instance) interacts and produces a series of particles that can be
normally propagated by AIRES.
\item An electromagnetic particle interacts with the Earth's magnetic
field {\em before\/} reaching the atmosphere, and producing a {\em
pre-shower\index{pre-showers}\/} whose products finally reach the
atmosphere and start interacting with it.
\item A cosmic particle disintegrates (before reaching the Earth) in
two or more fragments that arrive simultaneously in slightly distant
points.
\item Etc.
\end{itemize}

AIRES {\currairesversion} allows the user to simulate showers initiated
in such conditions. An external, user provided, program will be
responsible for generating the particles to be injected at the
beginning of the shower. This process is completely dynamic, and the
sets of generated primary particles may vary from shower to shower.

To implement such an interface is very simple. The user needs to: (i)
Define the special particle within the IDL instructions. (ii) Set
up the external program that will be invoked (via a system call) at
the moment of starting a new showers.
\subsection{Defining special particles}
The AIRES IDL directives allow to specify particles by names
(``proton'', ``gamma'', etc.). The set of known particle names can be
expanded to include those special ``particles'' which need to be
treated separately.

 Consider the following examples:
\idlidx;AddSpecialParticle;%
\kwexample{%
AddSpecialParticle \kwt;myparticX ; \kwt;Xpartsim;\\
AddSpecialParticle \kwt;myparticY ; \kwt;Xpartsim ; \kwt;type Y;
}
The IDL directive \idlbfx;AddSpecialParticle; takes at least two
arguments: (i) A {\em special particle name\/} that uniquely
identifies the added special particle, and (ii) The name of the
executable module that will be invoked when starting the showers
initiated by the respective particles.

In the preceding example, two special particles, namely,
\kwbf;myparticX; and \kwbf;myparticY; are defined and associated to
the same external module, \kwbf;Xpartsim;. In the case of the
definition of \kwbf;myparticY;, some arguments are specified (``\kwbf;
type Y;''). Such arguments are passed (portably) to the module.

Once the special particle(s) are defined\footnote{Up to ten different
special particles can be defined for a given task.}, their names can
be used as argument of the \idlbfx;PrimaryParticle; directive:
\idlidx;PrimaryEnergy;
\kwexample{%
AddSpecialParticle \kwt;myparticX ; \kwt;Xpartsim;\\
\kwt;. . .;\\
\kwt;PrimaryEnergy  20 EeV;\\
PrimaryParticle \kwt;myparticX;
}

Special particles can also be used in the case of mixed
composition\index{mixed composition} (see page \pageref{S:mixcomp}),
like in the following example:
\kwexample{%
AddSpecialParticle \kwt;SSP1 ; \kwt;module\_1;\\
AddSpecialParticle \kwt;SSP2 ; \kwt;module\_2;\\
\kwt;. . .;\\
PrimaryParticle \kwt;SSP1 0.2;\\
PrimaryParticle \kwt;SSP2 0.3;\\
PrimaryParticle \kwt;Proton 0.5;
}
In this case the primary will be \kwbf;SSP1;, \kwbf;SSP2;, or proton
with probabilities 20\%, 30\%, and 50\%, respectively.
\subsection{The external executable modules}
Every time a special primary shower is started, the simulation program
will invoke the executable module associated with the corresponding
primary, defined using the \idlbfx;AddSpecialParticle; directive. Such
an executable program can be a FORTRAN, C or C++ program (or a shell
script running it), and must be capable of providing the calling
module with the list of primary particles that will be added to the
particle stacks before starting the simulation of that shower.

The simulation program and the external module communicate via
internal files in a way that is transparent for the user and
completely portable.

The AIRES object library includes a series of user-friendly routines
(callable from FORTRAN, C or C++) that ease the task of writing such
external modules.

%
%
\begin{figure}[tp]
\begin{center}
{\baselineskip=9pt\lineskip=1pt\lineskiplimit=1pt\scriptsize
\begin{tabular}{|c|}
\hline\\
\begin{minipage}{14.5cm}
\index{special primary particles}%
\libidx;speistart;%
\libidx;urandomt;
\libidx;spaddp0;%
\libidx;speiend;%
\index{AIRES particle codes}%
\begin{verbatim}
c
c     An example of an external module to process "special" primary
c     particles.
c
      program  specialprim0
c
      implicit none
c
c     Declaration of variables retrieved when starting the interface
c     with the calling program.
c
      integer           shower_number
      double precision  primary_energy
      double precision  default_injection_position(3)
      double precision  injection_depth, ground_depth
      double precision  ground_altitude, d_ground_inj
      double precision  shower_axis(3)
c
      integer           rc
      double precision  urandomt
c
c     Some particle codes (AIRES coding system).
c
      integer           pipluscode, piminuscode
      parameter         (pipluscode = 11, piminuscode = -11)
c
c     FIRST EXECUTABLE STATEMENT.
c
c     Starting the AIRES-external module interface.
c
      call speistart(shower_number, primary_energy,
     +               default_injection_position, injection_depth,
     +               ground_altitude, ground_depth,
     +               d_ground_inj, shower_axis)
c
c     Injecting two particles at the initial injection point, and in
c     the direction of the shower axis.
c
      e1 = primary_energy * urandomt(0.05d0)
      e2 = primary_energy - e1
c
      call spaddp0(pipluscode, e1, 1, 0.d0, 0.d0, 1.d0, 1.d0, rc)
      call spaddp0(piminuscode, e2, 1, 0.d0, 0.d0, 1.d0, 1.d0, rc)
c
c     Completing the main program-external module interchange.
c     The integer argument of routine "speiend" is an integer return
c     code passed to the calling program. 0 means normal return.
c
      call speiend(0)
c
      end
\end{verbatim}
\end{minipage}\\ \\ \hline
\end{tabular}
}\end{center}
\figcaption{spextmodule}%
{A module for special primary particles.}{A sample module for
processing special primary particles. The purpose of this example
is to illustrate the basic structure of a program to process the
special primaries; the programmed algorithm is not intended to
have any validity from the physical point of view.}
\end{figure}
Figure \ref{FIG:spextmodule} displays a brief FORTRAN program with the
basic structure needed in every module capable of building a list of
primary particles to start the simulation of a shower.

The program starts with a call to routine \libbfx;speistart; and ends
with a call to \libbfx;speiend;. It is essential to maintain this
structure in any external module: All the calls to any AIRES library
routine {\em must\/} be placed within the mentioned calls.

Once the interface is started, the system is ready to accept primary
particles that will be added to the primary particle list. The basic
routine to add primaries to the list is \libbfx;spaddp0;. For each
invocation of this routine, the corresponding particle is added to the
internal list of particles. There is no limit in the number of
primary particles that can be included in the mentioned list, but the
sum of their energies must not be larger than the primary energy
specified in the input instructions and stored in the variable
\kw;primary\_energy; appearing in figure \ref{FIG:spextmodule}.

\lowfig{[ht]}{shaxsys}{\epsfig{file=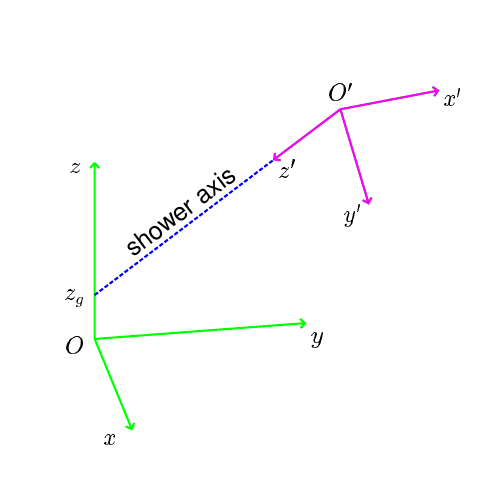}}%
{Shower axis-injection point coordinate system}%
{The shower axis-injection point coordinate system\index{shower
axis-injection point coordinate system}, $x'y'z'$ (magenta),
contrasted with the AIRES coordinate system\index{AIRES coordinate
system}, $xyz$ (green). The origin of the AIRES coordinate system,
$O$, is located at sea level, while $O'$ is located at the original
shower injection point. $z_g$ is the ground altitude. The $z'$-axis is
parallel to the shower axis, the $x'$-axis is always horizontal, and
the $y'z'$-plane contains the $z$-axis.}
Arguments number 3 to 6 of routine \libbfx;spaddp0; define the
direction of motion of the corresponding particle. Argument number 3
is an integer switch selecting the coordinate system to use and the
remaining quantities give the components of a vector, not necessarily
normalized, pointing in the direction of motion of the particle. There
are two options for argument number 3 (variable \kwbf;csys; in the
description of page \pageref{rou0:spaddp0}):
\begin{enumerate}
\item[{\bf 0}] To select the AIRES coordinate system\index{AIRES
coordinate system} defined in section \ppref{S:coordinates}.
\item[{\bf 1}] To select the {\em shower axis-injection point\/}
system\index{shower axis-injection point coordinate system}. This is a
special coordinate system whose $z$-axis is parallel to the shower
axis and its origin is placed at the original injection point (which
remains uniquely determined by the shower zenith and azimuth angles
and by the injection and ground altitudes). In this coordinate system,
illustrated in figure \ref{FIG:shaxsys}, the coordinates $(0,0,z')$ and
the vector $(0,0,1)$ indicate, respectively, the position and
direction of motion of a particle that moves along the shower
axis and towards the ground.
\end{enumerate}

The process is completed with the call to \libbfx;speiend;. This
ensures that all the relevant variables are transmitted back to the
main simulation program, which will recover the control after the
external module ends. Both \libbfx;speistart; and \libbfx;speiend;
must be called only once within the entire external module.

Notice also that one of the AIRES random number\index{random number
generator} routines, namely, \libbfx;urandomt; (see page
\pageref{rou0:urandomt}), is used to evaluate the energy if the $\pi$
mesons being included in the list of primary particles. The random
number generator is {\em not\/} initialized. Instead, its current
status is passed by the main simulation program to the external
module, and read-in within \libbfx;speistart;. As a consequence, the
generated random numbers will be different in different invocations of
the external module. Routine \libbfx;speiend; writes back the final
status of the random number generator, and it is recovered by the main
simulation program, so the numbers used in one and other program are
always independent. If the AIRES random number generator is not used
within the external module, then there are no alteration in the series
of random numbers used by the main simulation program.

An actual external module to process special primaries will surely be
much more complex than the one of the preceding example. The user can
provide special routines with the procedures needed for that purpose,
and use routines from the AIRES object library\index{AIRES object
library} as well. Many of the modules described in appendix
\ppref{A:aireslib} can be used within special primary programs, in
particular the ones directly related with the special particle
interface system, which provide a set of tools covering the needs of
the most common situations, namely:

\def\liblibp;#1;{\libbfx;#1; (page \pageref{rou0:#1})}
\begin{description}
\item[Retrieval of environmental information.] Routine
\libbfx;speistart; starts the AIRES-external module interface and
retrieves some basic variables, namely, shower number, primary energy,
original injection position (three coordinates), vertical atmospheric
depth of the original injection point, ground level altitude and
vertical atmospheric depth, distance between the original injection
point and the ground level (measured along the shower axis), and
unitary vector in the direction of the shower axis. Besides these
variables, it is possible (optionally) to retrieve additional ones
calling other routines included in the AIRES object
library\index{AIRES object library}:
\begin{itemize}
\item \liblibp;speigetpars; returns the parameter string that can be
(optionally) specified in the IDL instruction that defines the
corresponding special particle (see directive
\idlbfx;AddSpecialParticle;, page
\pageref{IDL0:AddSpecialParticle}). The simulation program passes the
argument string directly, without making any special processing on
it.
\item \liblibp;speigetmodname; returns the name of the executable
module specified in the definition of the corresponding special
particle.
\item \liblibp;sprimname; returns the name of the special particle
corresponding to the current invocation of the external module.
\item \liblibp;speitask; returns the current task name.
\item \liblibp;spnshowers; returns three integers that correspond,
respectively, to the total number of showers assigned to the task, and
the numbers of the first and last showers. These quantities are
related to the specifications entered with the directives
\idlbfx;TotalShowers; and \idlbfx;FirstShowerNumber;. The variable
\kwbf;shower\_number; set when calling \libbfx;speistart; (see figure
\ref{FIG:spextmodule}), will always be equal or larger (smaller) than
the first (last) shower number.
\end{itemize}

\item[Adding primary particles to the primary particle list.] Routine
 \libbfx;spaddp0; appends to the particle list the particle defined
with the arguments used in the corresponding call, as illustrated in
the example of figure \ref{FIG:spextmodule}. Additionally, there are
two other related library routines available, namely,
\liblibp;spaddpn; to append with a single call a set of various
primary particles, and \liblibp;spaddnull; to include {\em null\/}
(unphysical) particles. A null particle is not included in the
simulations, but its energy is added to the global null particle
energy counter. Nuclei can be normally appended to the particle list.
 Nuclear codes can be conveniently evaluated using routine
 \liblibp;nuclcode;. 

\item[Changing the injection coordinates and time.] After the initial
call to \libbfx;speistart;, the injection point is set to the original
injection point defined by the global parameters entered within the
input data (zenith and azimuth angles, etc.). The coordinates with
respect to the AIRES coordinate system\index{AIRES coordinate system}
of the original injection point are returned by \libbfx;speistart;
(this corresponds, in the example of figure \ref{FIG:spextmodule}, to
array \kwbf;default\_injection\_position;). The injection coordinates
and time can be changed at any moment using routine
\liblibp;spinjpoint;.

\item[Setting the point of first interaction.] When using normal
primary particles, AIRES evaluates automatically the atmospheric depth
where the first major interaction\index{depth of first interaction}
takes place. This is not possible in the case of a special particle
when a series of primaries are injected before starting the
simulations; and the default action will be to set the first
interaction at the original injection point, regardless whether that
points corresponds or not to an actual point of interaction. As an
alternative to the default action, it is possible to set manually the
coordinates of the point of first interaction using routine
\liblibp;sp1stint;. Of course, this affects only the statistical
analysis of the first interaction depth, and has no effect on the
propagation of the particles.

\item[Version of external module.] The user can assign a version
number to the external module. This version number must be passed to
the main program by means of routine \liblibp;speimv;. The version
number is stored with all the information associated with the current
shower, in particular in the compressed output files\index{compressed
output files}. It is strongly recommended to assign version numbers to
external modules that will be used in production simulations.
\end{description}

We recall here that all the calls to every one of the routines listed
in the previous paragraphs {\em must\/} be placed {\em after\/} the
call to \libbfx;speistart; and {\em before\/} the call to
\libbfx;speiend;.

The IDL directive \idlbfx;SpecialParticLog; allows to print
information about the primary particles that are injected after each
invocation of the external module. Notice that the default is to print
no information in the log (extension lgf) file\index{log
file}.\index{special primary particles!logging}

%
\chapter{Managing AIRES output data}
\label{C:output1}

\section{Using the summary program AiresSry}
\label{S:AiresSry}\index{internal dump file!processing with AIRES
summary program}\index{AIRES summary program}

Every time a task is completed, the simulation programs invoke some
output procedures that create a {\em summary file\index{summary
file}\/} displaying a series of results related with the already
finished simulations; a {\em task summary script
file\index{task summary script file}\/} can also be created. As it
will be discussed
in this section, there are several IDL directives that allow
controlling such AIRES output data.

The summary program, \kwbf;AiresSry;, which is part of the AIRES system
allows the user to process the simulation data contained within the {\em
internal dump file (IDF)\index{internal dump file}\/} or, equivalently,
the {\em portable dump file
(ADF)\index{internal dump file!portable format},\/}
and retrieve any of the available observables, similarly as
the main simulation programs do. It is worthwhile mentioning that
\kwbf;AiresSry; can be used {\em before\/} as well as {\em after\/} the
simulations are finished. In the first case it is possible to monitor the
development of the simulation task while the former alternative is most
convenient for analysis tasks. Backwards compatibility\index{backwards
compatibility} is always ensured: Old IDF's or ADF's generated with any
previous version of AIRES can be processed normally using
\kwbf;AiresSry;.\footnote{Notice, however, that a set of {\em
simulations\/} created using a determined version of AIRES must be
ended using the same version.}

Many observables are of ``tabular'' nature, that is, an array of data
whose elements correspond to a set of values of a determined
variable. For example, the longitudinal development\index{longitudinal
development} of the number of gamma rays is represented by an array
whose elements give the number of gamma rays that have crossed the
different observing levels, as a function of the observing level
altitude.

Most of the tabular observables commonly defined are automatically
calculated by the simulation programs. The corresponding data arrays
are stored in the IDF file and can be retrieved in several ways (see
below). The complete list of currently available output data tables
(more than 300) is placed in appendix \ppref{A:outables}.
\subsection{The summary file}
\index{summary file}
\label{S:sryfile}
The summary file (extension \kwbf;.sry;\/) can be divided into two
parts: \iiset\ii The general section, which includes data on the
evolution of the simulations as well as some basic shower
observables. \ii Tables section, containing data tables accordingly to
user's specifications.

The summary file is generally written as a plain text file (this is
the default). However, the IDL instruction \idlbfx;LaTeX; permits
generating summaries that can be processed using the {\LaTeX}
typesetting system. If the {\LaTeX}\index{LaTeX format for summary
files@{\LaTeX} format for summary files} switch is enabled, then the
AIRES system will generate two files, namely \vkw;taskname;\kwbf;.sry;
and \vkw;taskname;\kwbf;.tex;. The last of these two files can be
normally processed by a standard {\LaTeX} system.

On the other hand, it is also possible to instruct any of the AIRES
programs not to write the summary file. To do this, just include the
directive \idlbfx;Summary; \kwbf;Off; into the input data stream.
\subsubsection*{General section}
The general section of the summary file begins with computer related
information (task and user identification, CPU time usage, etc.). It
also includes information about the input parameters used, and reports
on the number of particle entries processed at each stack. A complete
report on stack usage can be obtained using the IDL directive
\idlidx;StackInformation;%
\kwexample{StackInformation \kwt;On;}

Then general information about the number and energy of particles
reaching ground is displayed. For all the output observables, its
mean\footnote{The statistical analysis is made in a shower-per-shower
basis.}, standard deviation, root mean square error of the mean,
minimum and maximum, are reported.

The IDL directive \idlbfx;OutputListing; \kwbf;Full; will generate an
additional section containing information (generally of computational
nature) on several output quantities defined for different algorithms.

The general section concludes with reports on the vertical depth of the
first interaction\index{depth of first interaction}, and on the location
of the {\em shower maximum\index{shower maximum}\/}.

The data collected for the longitudinal development\index{longitudinal
development} of all charged particles, that is, the number of charged
particles $N_c(i)$ that crossed the observing level $i$, for all
$i=1,\ldots,N_o$, is used to estimate the {\em shower
maximum\index{shower maximum},\/}
$X_{\mathrm{max}}$\index{Xmax@$X_{\mathrm{max}}$}, here defined as
the {\em vertical\/} depth of the point where the number of charged
particles reaches its maximum.  The number of charged particles at the
maximum, $N_{\mathrm{max}}$\index{Nmax@$N_{\mathrm{max}}$} is also
evaluated.

The estimation of the shower maximum is done by means of a {\bf
4-parameter} fit to the Gaisser-Hillas function\index{Gaisser-Hillas
function|bfpage{ }} \cite{gaihfun}\footnote{Our definition of the
Gaisser-Hillas function involves {\em vertical\/} depths. Some
authors, however, use {\em slant\/} depths instead. Both definitions
can be used to parameterize the shower profile. Furthermore, notice
that in the plane Earth approximation both ``vertical'' and ``slant''
forms are equivalent, provided the parameters are adequately
interpreted, that is, taking into account the factor $\cos\Theta$ of
equation (\ref{eq:Xsplane}).
If the Earth's curvature\index{Earth's curvature} is taken into
account, the translations between vertical and slant quantities must
be done numerically (see pages
\pageref{eq:Xsdef} and \pageref{rou0:xslant}).}
\begin{equation}                                  \label{eq:ghf4p}
N_{\mathrm{ch}}(X) = N_{\mathrm{max}} {\left(
\frac{X-X_0}{X_{\mathrm{max}} -X_0} \right)\!}^{\left[\left(
X_{\mathrm{max}} -X_0\right)/\lambda\right]}
\exp\!\left(\frac{X_{\mathrm{max}}-X}{\lambda}\right),
\quad X \ge X_0.
\end{equation}
$X_{\mathrm{max}}$, $N_{\mathrm{max}}$, $X_0$ and $\lambda$ are the
free parameters to be adjusted. Notice that $N_{\mathrm{ch}}(X_0) =
0$,\footnote{The depth $X_0$ refers to the point where the
  Gaisser-Hillas function is zero, and is not equal and not even
  necessarily related to the {\em depth of the first
    interaction\index{depth of first interaction},\/} noted $X_1$ in
  this manual.} and that $N_{\mathrm{ch}}(X)$ is taken as 0 for $X<
X_0$.

A weighted nonlinear least squares fit performed with the aid of the
very robust Levenberg-Mardquardt algorithm --as implemented in the
public domain software library
Netlib\index{external packages}\index{Netlib} \cite{netlib}--
is done after the simulation
of every individual shower is completed. The values reported in the
summary file correspond to the plain average of all the fits with
``reasonable'' results (converged fits). The number of such converged
fits is also reported.
\subsubsection*{Tables section}
The output data tables listed in appendix \ppref{A:outables} that are
automatically calculated during the simulations can be totally or
partially included within the output summary file. An index of such
tables can also be printed using the directive \idlbfx;TableIndex;

The \idlbfx;PrintTables; directive must be used to include one or more
tables within the output summary file. Its syntax is shown in the
following example:
\kwexample{PrintTables \kwt;1291; Options \kwt;RM;}
This instruction orders AIRES to ``print'' table 1291 (longitudinal
development of all charged particles) into the summary file. The
options used are: \kwbf;R;, to list RMS errors of the means, and
\kwbf;M; to include maximum and minimum data as numerical entries.
For a detailed explanation of the directive \kwbf;PrintTables; see
page \pageref{IDL0:PrintTables}.
\subsection{Exporting data}\index{exported data files}
\label{S:exporting}
An interesting feature of both the summary and the main simulation
programs\index{AIRES summary program}, is that they are able to generate
output files containing any one of the tables listed in appendix
\ppref{A:outables}.
Let us consider, for instance, the following IDL instructions:
\idlidx;TaskName;\idlidx;Summary;\idlidx;ExportTables;%
\kwexample{%
Task \kwt;mytask;\\
Summary \kwt;Off;\\
Export  \kwt;1205 1211;\\
Export  \kwt;1293; Option \kwt;a;\\
Export  \kwt;2001; Options \kwt;ds;\\
Export  \kwt;2791 2793; Options \kwt;ML;\\
End
}
Here \vkw;mytask\/; is a string that represents an already finished or
currently running task. The \kwbf;Summary Off; directives disables
the summary file. This is, of course, optional, but might be useful
when the user is just interested in creating table files.

The first \kwbf;ExportTables; directive (the abbreviated name will be
correctly interpreted by any of the AIRES main programs) indicates
that all the tables whose numbers are in the range
$[\hbox{1205},\hbox{1211}]$ must be exported with the default
options. Looking at the listing in appendix \ppref{A:outables}, it comes
out that the involved tables are tables number 1205, 1207 and 1211.

The second export directive instruct AIRES to export table 1211 with the
option of listing the {\em slant\/} depths of the observing levels, that
is, measured along the shower $\underline{\hbox{\bf a}}$xis (equation
(\ref{eq:Xsdef})). By default (Option \kwbf;r;) all atmospheric depths
listed within exported tables are vertical depths.

In the second export directive, the option string \kwbf;ds; modifies
the default settings. \kwbf;d; indicates that the particle numbers
must be normalized to particle densities, expressed in
particles/m$^2$; and \kwbf;s; suppresses the file header (only the
tables will be written). This last option may be useful when the
exported files are read by other applications (piped).  Suppressing
the file header, however, may lead to not understandable files,
especially if they are not processed at the moment they are
produced. It is therefore recommended to always keep such information;
and it must be also taken into account that all the header lines are
``commented out'' by means of a leading comment character which
defaults to ``\idlbfx;\#;'', but can be changed\index{comment
characters in output files, changing} by means of the directive
\idlbfx;CommentCharacter; (see page \pageref{IDL0:CommentCharacter}).

In the last example, the energy distributions\index{energy
distributions} 2791, 2792 and 2793 are exported. The option \kwbf;M;
indicates that energies must be expressed in MeV (the default is GeV);
while \kwbf;L; indicates that the corresponding data are normalized to
$dN/d\log_{10}{E}$ distributions. The alternative option \kwbf;l;
corresponds to $dN/d\ln{E}$ normalization.

To process the preceding code, it might be useful to edit a small text
file containing them and then use --for instance-- the summary program to
process it:
\kwexample{AiresSry < \kwt;myfile.inp;}
The files \vkw;mytask;\kwbf;.t1205;, \vkw;mytask;\kwbf;.t1207;,
\ldots, etc., will be created. If such files already exist, they will
be overwritten.

If the simulations that generated the data being processes were run with
the \idlbfx;PerShowerData; option \kwbf;Full; (see section
\ref{S:outablepar}), then it is possible to export {\em single shower
tables\index{single shower tables}\/} by placing the directive
\idlidx;ExportPerShower;%
\kwexample{ExportPerShower}
together with the \idlbfx;ExportTables; one(s). Returning to our previous
example, if such directive is placed inside the file \kwbf;myfile.inp;,
then for each one of the exported tables, the files
\[
\begin{array}{l}
\hbox{\vkw;mytask;\kwbf;.t;\vkw;nnnn;}\\
\hbox{\vkw;mytask;\kw;\_;\kwbf;s0001.t;\vkw;nnnn;}\\
\hbox{\vkw;mytask;\kw;\_;\kwbf;s0002.t;\vkw;nnnn;}\\
\ldots
\end{array}
\]
will be created, corresponding respectively to the usual average table,
and the tables for shower 1, 2, etc.
\subsection{The task summary script file}
\index{task summary script file}
\index{TSS or tss|see{task summary script file}}
\label{S:tssfile}
The task summary script file file (extension \kwbf;.tss;\/) is a text
file containing information about the main input and output parameters
of the simulation, in a format suitable for further processing by
other programs.
%
%
\begin{figure}[p]
{\scriptsize\baselineskip=9pt\lineskip=1pt\lineskiplimit=1pt
\begin{center}
\begin{tabular}{|c|}
\hline\\*[-2pt]
\begin{minipage}{14.5cm}
\begin{verbatim}
#### AIRES TSS --- version V.V.V
#>>>
#>>> This is AIRES version V.V.V (dd/Mmm/yyyy)
#>>> (Compiled by . . .)
#>>> USER: xxxxx, HOST: xxxxx, DATE: dd/Mmm/yyyy
#>>>
#
# TSS file for task mytask
#
#
# Program and compilation parameters.
#
AiresVersion = V.V.V
#
# Units
#
LengthUnit = m
TimeUnit = sec
EnergyUnit = GeV
DepthUnit = g/cm2
AngleUnit = deg
MagneticFieldUnit = nT
#
# General data
#
TaskName = mytask
TaskVersion = 0
#
TotalShowers = 3
CompletedShowers = 3
. . .
. . . (Other general data)
. . .
#
# Basic Input Parameters.
#
Site = Site00
SiteLatitude = 0.000000
SiteLongitude = 0.000000
EventDate = dd/Mmm/yyyy
#
NumberOfDifferentPrimaries = 1
PrimaryParticle = Proton
PrimaryParticleCode = 31
. . .
. . . (Other task input parameters)
. . .
#
# Parameters relative to each shower
#
ShowerPerShowerKey = PrCode PrEgy Zenith Azim . . .
#
DataSh000001 =     31   350000.       0.00000       0.00000 . . .
DataSh000002 =     31   350000.       0.00000       0.00000 . . .
DataSh000003 =     31   350000.       0.00000       0.00000 . . .
#
# End of tss file
#
\end{verbatim}
\end{minipage}\\*[-2pt] \\\hline
\end{tabular}
\end{center}}
\figcaption{sampletss}%
{Sample AIRES TSS file}{Sample AIRES task summary script (TSS)
\index{task summary script file} file.}
\end{figure}

The format of this file is very simple: Each data item is written
using a single line, in the format
\kwexample{Keyword = \vkw;value;}
Some comments are included to make the file more human readable. The
comment lines begin with a comment character (`\#' or the character
set with the \idlbfx;CommentCharacter;\index{comment characters in
output files, changing} directive).

In figure \ref{FIG:sampletss} a typical TSS file is displayed. Some of
the records were not displayed for brevity. The
first data items correspond to the version of AIRES used for the
simulations, and the units corresponding to different quantities
written in the file. The general data, basic and additional input
parameters, and miscellaneous sections contain the current values of
all the input parameters. The last section contains a summary of
shower observables. Using one line per shower, a series of shower
output data is displayed. The \kwbf;ShowerPerShowerKey; line gives the
key with the meaning of each one of the columns of the shower data
lines, starting with the primary code (key \kwbf;PrCode;) for the
first item after the equal sign, and continuing until completing all
the following items:
\begin{center}
\def\phb{$\phantom{0}$}
\begin{tabular}{cll}
 \phb1 & \kwbf;PrCode; & Primary particle code \\
 \phb2 & \kwbf;PrEgy;  & Primary energy \\
 \phb3 & \kwbf;Zenith; & Shower zenith angle\\
 \phb4 & \kwbf;Azim;   & Shower azimuth angle\\
 \phb5 & \kwbf;X1v;    & Vertical depth of first interaction\\
 \phb6 & \kwbf;Xmaxv;  & Vertical depth of shower maximum\\
 \phb7 & \kwbf;Nmax;   & Number of particles at shower maximum\\
 \phb8 &  \kwbf;X0v;   & Fitted parameter $X_0$ (vertical) of Gaisser-Hillas
                         function\index{Gaisser-Hillas function} (equation
                         (\ref{eq:ghf4p}))  \\
 \phb9 & \kwbf;Lambda; & Fitted parameter $\lambda$ (vertical) of
                         Gaisser-Hillas function (equation (\ref{eq:ghf4p}))\\
    10 & \kwbf;SofSqr; & Normalized sum of squares (equation
                         (\ref{eq:nsofsq})) from the longitudinal
                         profile fit\\
    11 & \kwbf;FitRc;  & Return code of the longitudinal profile fit
\end{tabular}
\end{center}

By default, AIRES does not create any task summary script file. The
directive \idlbfx;TSSFile; must be used to enable this feature.

\section{Processing compressed particle data files}
\label{S:cio2}\index{compressed output files}

Like other simulation systems \cite{MOCCA,CORSIKA}, AIRES can produce
output files containing detailed information about the particles
generated during the simulations. The well-known fact that that
detailed information generates huge amounts of data has been
especially taken into account in the design of AIRES, which includes
an {\em ad hoc\/} data compressing algorithm to save file space.

A detailed explanation of the compressing algorithm --a rather
technical matter-- is beyond the scope of this manual. We shall limit
ourselves to briefly list its main characteristics:
\begin{description}
\item[Format.] The compressed files are plain text files that
can be generated in any computer and copied and processed in any other
one. This is valid even if the machines do not have the same operating
system and/or do not use the same character codes (for example
non-ASCII machines).
\item[Organization.] The files contain a header with data related to
its structure and the conditions of the simulation. The particle data
section represents the bulk of the file and, in general, the records
corresponding to any one of the simulated showers are delimited by
``beginning of shower'' and ``end of shower'' records. There is
practically no limit in the number of showers that can be included in
a single file.\footnote{It is possible to store up to 759375 showers
in a single compressed file.}  On the other hand, groups of showers
can be saved into separate files, up to the limit of storing each
shower in a different file (see page
\pageref{IDL0:SeparateShowers}).
\item[Compression rate.] The data compression algorithms were designed
to take profit of the physical properties of the quantities being
stored. This involves information about lower and upper bounds for a
variable, possibility of subtracting a given fixed value\footnote{This
refers to internal operations which do not alter any user-level
results.}, etc. Precision requirements were also taken into account,
imposing a minimum of five significant figures in most cases. To give
an idea of the size of compressed records, let us consider the default
ground particle record (see below) whose fields are: Particle code,
logarithm of the energy, logarithm of the distance to the core, polar
angle in the ground plane, arrival time, $x$ and $y$ components of the
direction of motion, and statistical weight. This record thus has one
integer field and six real ones, and its length is 18 characters
(bytes). This figure should be contrasted with a standard FORTRAN
internal write statement with single precision for real variables,
which generates 28 data bytes when writing the same fields. Taking
into account that such records usually include additional formatting
fields, the compression rate of AIRES algorithm compared with standard
unformatted FORTRAN i/o should be larger than 36\%.
\end{description}

It is worthwhile mentioning the the AIRES package includes a library
of subroutines, namely, the {\em AIRES object library\index{AIRES
object library},\/} which contains many routines to read and process
the compressed output files. Backwards compatibility\index{backwards
compatibility} is always ensured: Old compressed files generated with any
previous version of AIRES can be read normally using
the library routines.
\subsection{Customizing the compressed files}
\label{S:ciocustom}
Two kinds of compressed files are implemented in the current version
of AIRES (\currairesversion):
\begin{description}
\item[Ground particle file.] (Extension \kwbf;.grdpcles;) This file
contains records with data of particles that reached the ground
surface.
\item[Longitudinal tracking particle file.] (Extension
\kwbf;.lgtpcles;) Compressed file containing detailed data related
with particles crossing the predefined observing levels (see section
\ref{S:monitoring}).
\end{description}
\lowtable{[tb]}{bosr}{%
\begin{tabular}{rrp{5.5cm}p{6.2cm}}
&\multicolumn{1}{c}{\itbf Field}&
\makebox[5.5cm][c]{\itbf Name}&
\makebox[6.2cm][c]{\itbf Description}\\*[4pt]
{\em Integer\/}&
    1 & Primary particle code & Stores the code of the
                                primary particle.  \\*[2pt]
&   2 & Shower number & Shower number. By default, the first shower is
                        labeled with the number 1, but the user can
                        manually set the first shower number\index{first
                        shower number} by means of
                        the IDL directive \idlbfx;FirstShowerNumber;
                        (see page \pageref{IDL0:FirstShowerNumber}). \\*[2pt]
&3--8 & Starting date and time & Six fields containing,
                               res\-pec\-ti\-ve\-ly, the
                               year, month, day, 
                               hours, minutes and seconds
                               corresponding to the beginning of the
                               simulation of the corresponding
                               shower. \\*[4pt]
{\em Real\/}&
   1 & Primary energy (GeV) (log) & The logarithm of the
                                  primary particle's energy. \\*[2pt]
&  2 & Primary zenith angle (deg) & The zenith angle of
                                    the primary particle. \\*[2pt]
&  3 & Primary azimuth angle (deg) & The azimuth angle of
                                     the pri\-ma\-ry par\-ti\-cle. \\*[2pt]
&  4 & Thinning energy (GeV)       & The absolute thinning\index{thinning}
                                     energy used for the respective
                                     shower. \\*[2pt]
&  5 & First interaction depth (g/cm2) & The vertical
                                        depth of the point where the
                                        first interaction took place,
                                        $X_1$\index{depth of first
                                        interaction}. \\*[2pt]
&  6 & Central injection altitude (m) & The $z$-coordinate of the
                                       primary's injection point (see
                                       figure \ref{FIG:coor1}). \\*[2pt]
&  7 & Global time shift (sec)        & The time $t_0$ required for a
                                       particle moving along the
                                       shower axis at the speed of
                                       light, to go from the injection
                                       point to the ground level. 
\end{tabular}}%
{Fields contained in the ``\/{\em beginning of shower\/}'' record of
compressed particle files}%
{Fields contained in the ``\/{\em beginning of shower\/}'' record of
compressed particle files. The structure of this record does not
depend on the compile-time option selected for the particle record.}
\subsubsection*{Ground particle file}
There are three basic types of data records in this file: ``Beginning
of shower'' record, ``end of shower'' record and particle record (also
referred as default record). The ``external primary particle'' and
``special primary trailer record'' are also defined. These last two
records are used only in connection with the special primary
particles\index{special primary particles} described in section
\ppref{S:specialprim}.

All the particle records written out during the simulation of a single
shower will appear in the file preceded by a beginning of shower
record, and followed by the corresponding ``end of shower'' one.
\lowtable{[p]}{eosr}{%
\begin{tabular}{rrp{5.5cm}p{6.2cm}}
&\multicolumn{1}{c}{\itbf Field}&
\makebox[5.5cm][c]{\itbf Name}&
\makebox[6.2cm][c]{\itbf Description}\\*[3.5pt]
{\em Integer\/}&
    1 & Shower number &
Shower number (matching the shower number of the
    corresponding ``beginning of shower'' record). \\*[1.4pt]
&   2 & Xmax fit return code &
Integer code returned by the $X_{\mathrm{max}}$ and $N_{\mathrm{max}}$
    fitting routine described in section \ppref{S:AiresSry}. \\*[1.4pt]
&3--8 & Ending date and time &Six fields containing,
                               res\-pec\-ti\-ve\-ly, the
                               year, month, day, 
                               hours, minutes and seconds
                               corresponding to the end of the
                               simulation of the corresponding
                               shower. \\*[3.5pt]
{\em Real\/}&
   1 & Total number of shower particles &
Total number of particles processed during the simulation of
   the corresponding shower. \\*[1.4pt]
&  2 & Total number of lost particles &
Total number particles that went outside the region of interest for
   the simulations. \\*[1.4pt]
&  3 & Number of low energy particles &
Total number of particles whose kinetic energies fell below the
   corresponding thresholds\index{threshold energies}. \\*[1.4pt]
&  4 & Number of particles reaching\hfil\break ground &
Total number of particles that reached the ground level, including
   also those 
   particles not saved in the compressed file. \\*[1.4pt]
&  5 & Total number of unphysical\hfil\break particles &
Number of ``particles'' generated by special procedures --like the
   splitting algorithm\index{splitting algorithm}, for example-- which
   cannot be associated with
   physical particles. This number is generally very small. \\*[1.4pt]
&  6 & Total number of neutrinos &
Total number of neutrinos ($\nu_e$, $\bar\nu_e$, $\nu_\mu$,
   $\bar\nu_\mu$) generated during the simulation of the current
   shower. \\*[1.4pt]
&  7 & Particles too near to the shower\hfil\break core &
Number of particles that were {\em not\/} saved in the compressed file
   because they were too near to the shower axis (see text). \\*[1.4pt]
&  8 & Particles in the resampling region&
Number of particles that were processed with the resampling
  algorithm\index{resampling algorithm} (see text). 
\end{tabular}}%
{Fields contained in the ``\/{\em end of shower\/}'' record of compressed
particle files}%
{Fields contained in the ``\/{\em end of shower\/}'' record of compressed
particle files. The structure of this record does not
depend on the compile-time option selected for the particle record.}
%
%
\begin{table}[p]
\begin{center}
\begin{tabular}{rrp{5.5cm}p{6.2cm}}
&\multicolumn{1}{c}{\itbf Field}&
\makebox[5.5cm][c]{\itbf Name}&
\makebox[6.2cm][c]{\itbf Description}\\*[3.5pt]
{\em Real\/}&
  $\phantom{\hbox{3--}}$9 
& Particles too far from the shower\hfil\break core &
Number of particles that were {\em not\/} saved in the compressed file
   because they were too far from the shower axis (see text). \\*[1.4pt]
& 10 & Shower maximum depth (Xmax)\hfil\break (g/cm2) &
Vertical depth\index{atmospheric depth!vertical} of the point where
  the number of charged particles is maximum, $X_{\mathrm{max}}$,
  obtained  from a fit to the simulation
   data (see section \ref{S:AiresSry}). \\*[1.4pt]
& 11 & Total charged particles at shower\hfil\break maximum &
Number of charged particles at $X_{\mathrm{max}}$, $N_{\mathrm{max}}$,
   calculated as explained in section \ppref{S:AiresSry}. \\*[1.4pt]
& 12 & Energy of lost particles (GeV) &
Total energy of the particles of real field number 2. \\*[1.4pt]
& 13 & Energy of low-energy particles &
Total energy of the particles of real field number 3. \\*[1.4pt]
& 14 & Energy of ground particles (GeV) &
Total energy of the particles of real field number 4. \\*[1.4pt]
& 15 & Energy of unphysical particles\hfil\break (GeV) &
Total energy of the particles of real field number 5. \\*[1.4pt]
& 16 & Energy of neutrinos (GeV) &
Total energy of the particles of real field number 6. \\*[1.4pt]
& 17 & Energy lost in the air (GeV) &
Total amount of energy lost by continuous medium losses (ionization
   losses) due to charged particles moving through the air. \\*[1.4pt]
& 18 & Energy of particles too near to the\hfil\break core &
Total energy (in GeV) of the particles of real field number 7. This
   field is not defined for the longitudinal tracking particle
   file. \\*[1.4pt]
& 19 & Energy of resampled particles &
Total energy (in GeV) of the particles of real field number 8. This
   field is not defined for the longitudinal tracking particle
   file. \\*[1.4pt]
& 20 & Energy of particles too far from the\hfil\break core &
Total energy (in GeV) of the particles of real field number 9. This
   field is not defined for the longitudinal tracking particle
   file. \\*[1.4pt]
& 21 & CPU time (sec) &
Amount of processor time required for the simulation of the current
   shower. 
\end{tabular}
\end{center}
\vskip \abovecaptionskip
\begin{center}
{\itshape{\bfseries Table \ref{TAB:eosr}.} (continued)}
\end{center}
\end{table}
\lowtable{[p]}{eppr}{%
\begin{tabular}{rrp{5.5cm}p{6.2cm}}
&\multicolumn{1}{c}{\itbf Field}&
\makebox[5.5cm][c]{\itbf Name}&
\makebox[6.2cm][c]{\itbf Comment}\\*[4pt]
{\em Integer\/}&
    1 & Particle code & Stores the code of the corresponding
                                primary particle.  \\*[4pt]
{\em Real\/}&
   1 & Energy (GeV) (log) & The logarithm of the corresponding
                                  primary particle's energy. \\*[2pt]
&  2 & Direction of motion (x component) & With respect to
   the AIRES 
   coordinate system (see page \pageref{S:coordinates}). \\*[2pt]
&  3 & Direction of motion (y component) \\*[2pt]
&  4 & Direction of motion (z component) \\*[2pt]
&  5 & X coordinate (m) & Particle injection coordinate,
   with respect to the AIRES 
   coordinate system (see page \pageref{S:coordinates}). \\*[2pt]
&  6 & Y coordinate (m) \\*[2pt]
&  7 & Z coordinate (m) \\*[2pt]
&  8 & Injection depth (g/cm2) \\*[2pt]
&  9 & Injection time (ns) \\*[2pt]
& 10 & Particle weight & Initial statistical weight of the
   corresponding particle. 
\end{tabular}}%
{Fields contained in the ``\/{\em external primary particle\/}'' record of
compressed particle files}%
{Fields contained in the ``\/{\em external primary particle\/}''
record of compressed particle files\index{special primary
particles}. The structure of this record does not depend on the
compile-time option selected for the particle record.}
\lowtable{[p]}{sptr}{%
\begin{tabular}{rrp{5.5cm}p{6.2cm}}
&\multicolumn{1}{c}{\itbf Field}&
\makebox[5.5cm][c]{\itbf Name}&
\makebox[6.2cm][c]{\itbf Description}\\*[4pt]
{\em Integer\/}&
    1 & Version of external module & User-settable integer in the
    range\hfil\break $[0, 759375]$.                     \\*[4pt]
{\em Real\/}&
   1 & Total number of primaries & Total number of primary
   particles.\\*[2pt]
&  2 & Unweighted primary entries & Unweighted number of primary
   entries.\\*[2pt]
&  3 & Total energy of primary particles (GeV) & Total energy of primary
   particles\hfil\break (weighted). 
\end{tabular}}%
{Fields contained in the ``\/{\em special primary trailer record\/}'' of
compressed particle files}%
{Fields contained in the ``\/{\em special primary trailer record\/}''
of compressed particle files\index{special primary particles}. The
structure of this record does not depend on the compile-time option
selected for the particle record.}

The fields that make the beginning (end) of shower record are listed
in table \ref{TAB:bosr} (\ref{TAB:eosr}). Tables \ref{TAB:eppr} and
\ref{TAB:sptr} describe the fields of the special primary related
records. In these and in any other records, the data fields can be
classified in {\em integer\/} and {\em real\/} fields.

The fields contained in such delimiting records account for general
air shower parameters or observables and were included for special
analysis tasks.

In the case of showers initiated by special primary
particles\index{special primary particles} (see section
\ref{S:specialprim}), the ``Primary particle'' code of the
corresponding ``beginning of shower'' record will not correspond to a
standard particle code. Instead, the returned code will be a negative
integer with an absolute value slightly smaller than
100000.\footnote{The library routine \libbfx;crospcode; is the
adequate one to manage such special particle codes.}

In those cases, the ``beginning of shower'' record will be followed by
a series of ``external primary particle'' records (one for each
injected primary particle). This series ends with a ``special primary
trailer record'' which will precede the default particle records
written for that shower.

The fields included in the default records, associated with particle
data, can be selected at compile time among the various available
alternatives. The installation configuration file (see appendix
\ref{A:installing}) contains detailed instructions on how to select
the particle record options.

The most relevant physical properties of the ground particles can be
saved in the ground compressed file, namely,
\begin{description}
\item[Particle code.] An integer code that identifies the particle.
\item[Energy.] The logarithm of the kinetic energy of the particle.
\item[Coordinates.] The polar coordinates $(R,\varphi)$ of the particle at
ground, measured from the intersection of the shower axis with the ground
surface. $R$ is the distance to the core and $\varphi$ is the angle with
respect to the $x$-axis.
\item[Direction of motion.] The $x$ and $y$ components, $u_x$, $u_y$,
of the unitary vector $\vr{u}$ which indicates the particle's
direction of motion. The $u_z$ component must be negative for the
particles reaching ground because such particles move
downwards. It can be calculated via:
\begin{equation}                                   \label{eq:uzg}
u_z = -\sqrt{1- u_x^2 -u_y^2} .
\end{equation}
\item[Arrival time.] The saved quantity is the arrival time delay
$t-t_0$, where $t$ is the absolute time (measured from the beginning
of the shower), and $t_0$ is the global time shift described in table
\ppref{TAB:bosr}.
\item[Particle weight.] The statistical weight of the particle (see
section \ref{S:thinning}).
\item[Creation depth.] The vertical atmospheric\index{atmospheric
depth!vertical} depth of the point where the particle was inserted
into the simulating program's stacks.
\item[Parent particle code.] The code of the parent particle that
  generated (after a decay for example) the corrent one.
\item[Parent particle energy.] The logarithm of the kinetic energy of
  the parent particle.
\item[Last hadronic depth.] The vertical atmospheric
depth\index{atmospheric depth!vertical} corresponding to the last
hadronic interaction suffered by the particle or by one of its
ancestors.
\item[Last hadronic projectile code.] The particle code corresponding
  to the projectile of the last hadronic interaction suffered by the
  particle or by one of its ancestors.
\item[Last hadronic projectile energy.] The logarithm of the kinetic
  energy of the already mentioned projectile particle.
\end{description}
\lowtable{[tb]}{gpr}{%
\begin{tabular}{rccccl}
&\multicolumn{4}{c}{\itbf Field}&
\makebox[5.5cm][c]{\itbf Name}\\
&{\bf short} & {\bf normal} & {\bf long} & {\bf xlong} \\*[4pt]
{\em Integer\/}&
   1 &  1 &  1 &  1 & Particle code\\
& -- & -- & -- &  2 & Parent particle code\\
& -- & -- & -- &  3 & Last hadronic projectile code\\*[4pt]
{\em Real\/}&
  -- & -- & -- &  1 & Parent particle energy (GeV) (log) \\
& -- & -- & -- &  2 & Last hadronic proj. energy (GeV) (log) \\
&  1 &  1 &  1 &  3 & Energy (GeV) (log) \\
&  2 &  2 &  2 &  4 & Distance from the core (m) (log)\\
&  3 &  3 &  3 &  5 & Ground plane polar angle (radians)\\
& -- &  4 &  4 &  6 & Direction of motion (x component)\\
& -- &  5 &  5 &  7 & Direction of motion (y component)\\
&  4 &  6 &  6 &  8 & Arrival time delay (ns)\\
&  5 &  7 &  7 &  9 & Particle weight\\
& -- & -- &  8 & 10 & Particle creation depth (g/cm2)\\
& -- & -- &  9 & 11 & Last hadronic interaction depth (g/cm2)
\end{tabular}}%
{Fields contained in the particle records of compressed ground
particle files}%
{Fields contained in the particle records of compressed ground
  particle files. The field numbers for the different particle records
  selectable at compilation time (see text), named {\bfseries short},
  {\bfseries normal}, {\bfseries long}, and {\bfseries xlong} records,
  are tabulated. Notice that a given field can have different field
  numbers.}%

For each one of these quantities, a corresponding record field is
defined. The complete list of fields is placed in table \ppref{TAB:gpr}.
As mentioned previously, there are several record formats each one
including a different subset of all the available fields.

In contrast with the beginning of shower and end of shower records,
a given field of the particle record can be assigned different field
numbers. As it will be seen below in this chapter, this does not
affect the user's processing of compressed files, which can be done
independently of the field number assignments.

There are specific IDL directives that can control the particles that
are actually saved in the ground particle file.

To start with, let us consider the directives \idlbfx;RLimsFile; and
\idlbfx;ResamplingRatio;, whose syntax is
\kwexample{%
RLimsFile \kwt;grdpcles; $r_{\mathrm{min}}$ $r_{\mathrm{max}}$ \\
ResamplingRatio          $s_r$
}
\kwbf;grdpcles; identifies the file the directive refers to and
$r_{\mathrm{min}}$ and $r_{\mathrm{max}}$ represent length
specifications ($0<r_{\mathrm{min}} < r_{\mathrm{max}}$). $s_r$ is a
real number ($s_r\ge 1$).

Such directives instruct AIRES to save {\em unconditionally\/} those
particles whose distances to the shower axis lie within the interval
$[r_{\mathrm{min}}, r_{\mathrm{max}}]$.

 On the other hand, all the particles whose distances to the shower
axis are smaller than
\begin{equation}                                     \label{eq:sr0}
r_0 = \frac{r_{\mathrm{min}}}{s_r}
\end{equation}
($r_0$ is, by definition, not larger than $r_{\mathrm{min}}$) are {\em
not\/} included in the ground file, but their number and energy are
recorded and the totals are included in the ``end of shower record''
(fields 7, 9, 18, and 20).

Finally, all the particles whose distances $r$ to the shower axis lie
in the interval $[r_0,r_{\mathrm{min}}]$ are processed by a {\em
resampling algorithm\index{resampling algorithm}\/}\label{def:rsa}
which {\em conditionally\/} keeps the particles accordingly with the
following rules: (i) ``Nonnumerous'' particles --like pions, nucleons,
etc.-- are always saved. (ii) For every ``numerous'' particle --i.e.,
gammas, electrons, positrons and muons-- in the mentioned region, the
acceptance probability is\footnote{The expression of the acceptance
probability is inspired in a suggestion by P. Billoir
\cite{pierresample}.}
\begin{equation}                                     \label{eq:srp}
p_s = \left(\frac{r}{r_\mathrm{min}}\right)^2 .
\end{equation}
(iii) The statistical weights of the accepted particles are increased
via
\begin{equation}                                     \label{eq:srw}
w' = \frac{w}{p_s},
\end{equation}
in order to keep unbiased the sampling algorithm.

The total number and energy of particles that fall within the
resampling area, are recorded in the ``end of shower record'' (fields
8 and 19).

The \idlbfx;SaveInFile; (\idlbfx;SaveNotInFile;) directive permits
including (excluding) one or more particle kinds into (from) the
compressed file. Section \ppref{S:carryingon} contains several
illustrative examples on how to use them.

Notice that by default, {\em all\/} particle kinds are enabled to be
saved into the ground particle file.
\subsubsection*{Longitudinal tracking particle file}
The structure of the longitudinal tracking particle file is very
similar to the already described ground particle file: Both files have
virtually the same ``beginning of shower'', ``end of shower'',
``external primary particle'', and ``special primary
trailer''\index{special primary particles} records; and there are
alternative formats for the particle records.
\lowtable{[tb]}{ltr}{%
\begin{tabular}{rccccccl}
&\multicolumn{6}{c}{\itbf Field}&
\makebox[5.5cm][c]{\itbf Name}\\
&{\bf short} & {\bf norm}& {\bf norm} & {\bf long}
&{\bf x-} &{\bf x-x-} \\
&            & {\bf (a)}   & {\bf (b)}    &
& {\bf long} & {\bf long}\\*[4pt]
{\em Integer\/}&
   1 &  1 &  1 &  1 &  1 &  1 & Particle code\\
&  2 &  2 &  2 &  2 &  2 &  2 & Observing levels crossed\\
& -- & -- & -- & -- & -- &  3 & Parent particle code\\
& -- & -- & -- & -- & -- &  4 & Last hadronic projectile code\\*[4pt]
{\em Real\/}&
  -- & -- & -- & -- & -- &  1 & Parent particle energy (GeV) (log) \\
& -- & -- & -- & -- & -- &  2 & Last hadronic proj. energy (GeV) (log) \\
& -- &  1 & -- &  1 &  1 &  3 & Energy (GeV) (log) \\
& -- & -- &  1 &  2 &  2 &  4 & Direction of motion (x component)\\
& -- & -- &  2 &  3 &  3 &  5 & Direction of motion (y component)\\
&  1 &  2 &  3 &  4 &  4 &  6 & Particle weight\\
&  2 &  3 &  4 &  5 &  5 &  7 & Crossing time delay (ns)\\
&  3 &  4 &  5 &  6 &  6 &  8 & X coordinate (m)\\
&  4 &  5 &  6 &  7 &  7 &  9 & Y coordinate (m)\\
& -- & -- & -- & -- &  8 & 10 & Particle creation depth (g/cm2)\\
& -- & -- & -- & -- &  9 & 11 & Last hadronic interaction depth (g/cm2)
\end{tabular}}%
{Fields contained in the particle records of compressed longitudinal
tracking particle files}%
{Fields contained in the particle records of compressed longitudinal
  tracking particle files. The field numbers for the different
  particle records selectable at compilation time (see text), named
  {\bfseries short}, {\bfseries normal (a)}, {\bfseries normal (b)},
  {\bfseries long}, {\bfseries extra-long}, and {\bfseries
    extra-extra-long} records, are tabulated. Notice that a given
  field can have different field numbers.}%

For that reason, it is highly recommended to the reader be familiar
with the contents of the previous section describing the ground
particle file before proceeding to read the present section. We shall
limit here to briefly describe only those aspects that are somehow
different in both files.

The longitudinal tracking particle file contains records storing
detailed information about the particles that cross the defined
observing levels. Since the observing levels are generally located at
altitudes that include the shower maximum\index{shower maximum}, and
due to the fact that a single particle can cross more than one
observing level during its life, it is clear that the longitudinal
files can potentially be much larger than the average ground particle
files.

For that reason, a special effort was made to save as much space as
possible, and various record formats were defined to allow the user to
select just the necessary fields. The record format selection can be
done during installation, following the instructions placed in
appendix \ppref{A:installing}. Table \ppref{TAB:ltr} lists all the defined
data fields for the different default records.

The second integer field, named ``Observing levels crossed'' contains
information about the observing levels the particle has crossed, and
simultaneously about its direction of motion.

Let $N_o$ ($N_o\le 510$) be the number of defined observing levels. At
a certain monitoring operation, a given particle crosses several
observing levels from level $i_f$ to level $i_l$ ($i_l$ may be equal
to $i_f$). Let $u_z$ be the $z$-component of the particle's direction
of motion. If $u_z>0$ ($u_z<0$) the particle goes upwards (downwards),
and therefore $i_f\ge i_l$ ($i_f\le i_l$).

All this information is encoded in a single integer number called the
{\em crossed observing levels key\index{crossed observing levels
key|bfpage{ }},\/} $L$, defined by the following equation:
\begin{equation}                            \label{eq:olcrossed}
L= i_f + 512\,i_l + 512^2\, s_{\mathrm{ud}}
\end{equation}
where
\begin{equation}                            \label{eq:olsud}
s_{\mathrm{ud}} = \left\{
\begin{array}{ll}
1 & \hbox{if $u_z >0$} \\
0 & \hbox{if $u_z \le 0$}
\end{array}
\right.
\end{equation}

The three variables that appear in the right hand side of equation
(\ref{eq:olcrossed}) can be easily reconstructed when $L$ is known (see
page \pageref{P:olutil}).

The real fields listed in table \ppref{TAB:ltr} are defined similarly to
the corresponding ground particle record fields, with the exception of
the $x$ and $y$ coordinates which are defined as follows:
\begin{description}
\item[Coordinates.] The $(x, y)$ coordinates are the Cartesian
coordinates of the point where the particle crossed the level $i_f$,
{\bf measured from the intersection between the shower axis and the
corresponding observing level's surface.}
\item[Time delay.] Defined as the difference $t-t_f$ where $t$ is the
particle's absolute time and $t_f$ is the time required for a particle
moving along the shower axis at the speed of light to go from the
injection point to observing level $i_f$.
\end{description}

The IDL directives \idlbfx;RLimsFile;, \idlbfx;ResamplingRatio;,
\idlbfx;SaveInFile; and \idlbfx;SaveNotInFile; can be used with
longitudinal files to control when a particle must be saved or
not. The last two directives do not present special difficulties, and
work as explained in section \ppref{S:carryingon}.

On the other hand, the directives\footnote{Notice that the parameter
controlled by directive \idlbfx;ResamplingRatio; is {\em global,\/}
that is, its last setting applies to every one of the compressed files
in use.}
\kwexample{%
RLimsFile \kwt;lgtpcles; $r_{\mathrm{min}}$ $r_{\mathrm{max}}$\\
ResamplingRatio          $s_r$
}
define three parameters, $r_{\mathrm{min}}$, $r_{\mathrm{max}}$ and $s_r$,
that are used to determine whether a particle record must be saved or
not. The rules are the following:
\begin{enumerate}
\item Let $X_c = 0.8\,X_i + 0.2\,X_g$, where $X_i$ and $X_g$ are the
vertical injection and ground depths, respectively.
\item For each observing level $i$, $i=1,\ldots,N_o$, let
\begin{equation}                             \label{eq:rmini}
r_i =\left\{\begin{array}{ll}
0 & \hbox{if $\;X_o^{(i)} \le X_c$} \\*[8pt]
\displaystyle \left(\frac{X_o^{(i)} - X_c}{X_g-X_c}\right)
 r_{\mathrm{min}} &
\hbox{if $\;X_c < X_o^{(i)} < X_g$} \\*[10pt]
r_{\mathrm{min}} & \hbox{if $\;X_o^{(i)}\ge X_g$}
\end{array}
\right.
\end{equation}
where $X_o^{(i)}$ is the vertical depth of observing level $i$; and
let $r_{0i} = r_i/s_r$.

\item Any particle crossing observing levels will {\em not\/} be saved
in the longitudinal file if one of the following conditions is true:
\begin{enumerate}
\item $|x| < r_{0i}$ {\bf and} $|y| < r_{0i}$.
\item $|x| > r_{\mathrm{max}}$ {\bf or} $|y| > r_{\mathrm{max}}$.
\end{enumerate}
$x$ and $y$ are the Cartesian coordinates of the particle at observing
level $i_f$, measured from the intersection between the shower axis
and the corresponding observing level.

\item Gammas, electrons, positrons and muons crossing observing levels
and verifying the two following conditions
\begin{enumerate}
\item $|x| < r_{\mathrm{min}}$ {\bf and} $|y| < r_{\mathrm{min}}$.
\item $|x| > r_{0i}$ {\bf or} $|y| > r_{0i}$.
\end{enumerate}
(in the same notation of the previous point), will be {\em
conditionally\/} kept, with probability and reweighting factor given
by equations (\ref{eq:srp}) and (\ref{eq:srw}), respectively.

\item All the particles not fulfilling the conditions of the preceding
points will be {\em unconditionally\/} saved in the file.
\end{enumerate}

These rules set varying limits for the zone of excluded particles. In
the zone near the shower axis, all particles crossing observing levels
placed above $X_c$ will be saved, then the exclusion zone enlarges
proportionally to the depth of the observing level, reaching the
value indicated in the \kwbf;RLimsFile; directive at the ground
depth. Notice that $X_c$ divides the complete shower path (as measured
in atmospheric depth) into two, upper-lower, 20\%-80\%,
zones.

The number of defined observing levels affects the degree of detail of the
monitoring of the longitudinal shower development, and some applications
usually require that this number be as large as possible. On the other
hand, such setting may lead to the generation of very big longitudinal
particle files since a large number of data records are generated as long
as every particle crosses the observing levels. To overcome this
difficulty, AIRES includes a selection mechanism to avoid including in the
compressed file the information related with all the defined observing
levels. Consider the following illustrative example:
\idlidx;ObservingLevels;\idlidx;SaveInFile;\idlidx;RecordObsLevels;%
\kwexample{\label{P:recobslev}%
\kwt;ObservingLevels 100;\\
\kwt;SaveInFile lgtpcles e+ e-;\\
RecordObsLevels \kwt;None;\\
RecordObsLevels \kwt;1;\\
RecordObsLevels \kwt;4;\\
RecordObsLevels \kwt;10 90 10;\\
RecordObsLevels Not \kwt;20;
}%
The first directive sets the number of observing levels to 100, and
the second one enables particle saving in the longitudinal particle
tracking file. In this case only electrons and positrons will be
recorded (Notice that the longitudinal file is disabled by default,
and therefore it is necessary to use unless one \idlbfx;SaveInFile;
instruction to enable it). The default action is to record particles
crossing {\em any\/} of the defined observing levels, and the
remaining instructions are placed to override this default
setting. The directive \kwbf;RecordObsLevels None; eliminates all the
defined observing levels from the set of levels to be taken into
account to save particle records into the compressed file. The actions
of the instructions that follow are, respectively: Mark level 1 for
recording particles crossing it; idem level 4; idem all levels from 10
to 90 in steps of 10 levels; unmark level 20. The resulting set of
marked levels is $\{ 1, 4, 10, 30, 40, 50, 60, 70, 80, 90 \}$.

\subsection{Using the AIRES object library}
\label{S:usinglib}\index{AIRES object library}

The {\em AIRES object library\/} is a set of routines designed with
the main purpose of providing adequate tools to analyze the data
saved in the compressed output files.

Appendix \ppref{A:aireslib} explains in detail the contents of the
library and how to use it. In this section some illustrative examples
are presented.

From now on we are going to assume that the AIRES file is being
processed by a program, provided by the user and similar to
the demonstration programs that are included with the AIRES software
distributions.

We are going to use FORTRAN in our examples, but this is not a
restriction since the AIRES library includes routines for a {\em C
interface\index{AIRES object library!C/C++ interface},\/} which allow the
C user to fully exploit the library's resources.
\subsubsection*{Output particle codes}
Every analysis program must begin with a call to routine
\libbfx;ciorinit;. This routines sets up the environment where the
library routines can work adequately.
\lowtable{[p]}{codsys}{%
\vbox{
\begin{displaymath}
\begin{array}{ccrrrrrr}
\hbox{\itbf Particle}&\ \ &\multicolumn{6}{c}{\hbox{\itbf Codes}}\\
&&
\hbox{\itbf AIRES}&
\hbox{\itbf PDG~}&
\hbox{\itbf CORSIKA}&
\hbox{\itbf GEANT}&
\hbox{\itbf SIBYLL}&
\hbox{\itbf MOCCA}\\*[6pt]
\gamma       &&   1\;\;&   22 &  1 \;\;\;\;&  1 \;\;\;&  1 \;\;\;&  1
             \;\;\;\\*[3pt]
e^+          &&   2\;\;&  -11 &  2 \;\;\;\;&  2 \;\;\;&  2 \;\;\;&  2
             \;\;\;\\*[3pt]
e^-          &&  -2\;\;&   11 &  3 \;\;\;\;&  3 \;\;\;&  3 \;\;\;& -2
             \;\;\;\\*[3pt]
\mu^+        &&   3\;\;&  -13 &  5 \;\;\;\;&  5 \;\;\;&  4 \;\;\;&  3
             \;\;\;\\*[3pt]
\mu^-        &&  -3\;\;&   13 &  6 \;\;\;\;&  6 \;\;\;&  5 \;\;\;& -3
             \;\;\;\\*[3pt]
\tau^+       &&   4\;\;&  -15 & 86 \;\;\;\;& 86 \;\;\;& 20 \;\;\;& 10
             \;\;\;\\*[3pt]
\tau^-       &&  -4\;\;&   15 & 87 \;\;\;\;& 87 \;\;\;&-20 \;\;\;&-10
             \;\;\;\\*[3pt]
    \nu_e    &&   6\;\;&   12 & 66 \;\;\;\;&  4 \;\;\;& 15 \;\;\;&  0
             \;\;\;\\*[3pt]
\bar\nu_e    &&  -6\;\;&  -12 & 67 \;\;\;\;&  - \;\;\;& 16 \;\;\;&  0
             \;\;\;\\*[3pt]
    \nu_\mu  &&   7\;\;&   14 & 68 \;\;\;\;&  4 \;\;\;& 17 \;\;\;&  0
             \;\;\;\\*[3pt]
\bar\nu_\mu  &&  -7\;\;&  -14 & 69 \;\;\;\;&  - \;\;\;& 18 \;\;\;&  0
             \;\;\;\\*[3pt]
    \nu_\tau &&   8\;\;&   16 &  4 \;\;\;\;&  4 \;\;\;&  - \;\;\;&  0
             \;\;\;\\*[3pt]
\bar\nu_\tau &&  -8\;\;&  -16 &  - \;\;\;\;&  - \;\;\;&  - \;\;\;&  0
             \;\;\;\\*[3pt]
\pi^0        &&  10\;\;&  111 &  7 \;\;\;\;&  7 \;\;\;&  6 \;\;\;&  5
             \;\;\;\\*[3pt]
\pi^+        &&  11\;\;&  211 &  8 \;\;\;\;&  8 \;\;\;&  7 \;\;\;&  4
             \;\;\;\\*[3pt]
\pi^-        && -11\;\;& -211 &  9 \;\;\;\;&  9 \;\;\;&  8 \;\;\;& -4
             \;\;\;\\*[3pt]
K^0_S        &&  12\;\;&  310 & 16 \;\;\;\;& 16 \;\;\;& 12 \;\;\;& 12
             \;\;\;\\*[3pt]
K^0_L        &&  13\;\;&  130 & 10 \;\;\;\;& 10 \;\;\;& 11 \;\;\;& 13
             \;\;\;\\*[3pt]
K^+          &&  14\;\;&  321 & 11 \;\;\;\;& 11 \;\;\;&  9 \;\;\;& 11
             \;\;\;\\*[3pt]
K^-          && -14\;\;& -321 & 12 \;\;\;\;& 12 \;\;\;& 10 \;\;\;&-11
             \;\;\;\\*[3pt]
\eta         &&  15\;\;&  221 & 17 \;\;\;\;& 17 \;\;\;& 22 \;\;\;& 14
             \;\;\;\\*[3pt]
n            &&  30\;\;& 2112 & 13 \;\;\;\;& 13 \;\;\;& 14 \;\;\;&  6
             \;\;\;\\*[3pt]
\bar{n}      && -30\;\;&-2112 & 25 \;\;\;\;& 25 \;\;\;&-14 \;\;\;& -6
             \;\;\;\\*[3pt]
p            &&  31\;\;& 2212 & 14 \;\;\;\;& 14 \;\;\;& 13 \;\;\;&  7
             \;\;\;\\*[3pt]
\bar{p}      && -31\;\;&-2212 & 15 \;\;\;\;& 15 \;\;\;&-13 \;\;\;& -7
             \;\;\;
\end{array}
\end{displaymath}}}%
{Particle coding systems supported by AIRES library routines}%
{Elementary particle codes\index{particle codes} corresponding
  to several commonly used coding systems, for the most relevant
  particles.
 The routines that process
  AIRES compressed output files allow the user to select any one of
  these coding schemes.
}

This routine permits setting the particle coding system that the user
wants to work with. 
It is possible to select either the AIRES coding system already
described in section \ppref{S:acodes}, or other usual coding systems. The
coding systems known by AIRES {\currairesversion} are the following:
\begin{enumerate}
\item Aires internal coding\index{AIRES particle codes}.
\item Aires coding for elementary particles and decimal nuclear codes ($A
+ 100\, Z$).
\item Particle Data Group coding system \cite{PDG}, extended with decimal
nuclear codes ($A+ 10^5\,Z$).
\item CORSIKA\index{CORSIKA} simulation program particle coding system
\cite{CORSIKA}.
\item GEANT particle coding system \cite{GEANT}.
\item SIBYLL\index{SIBYLL} \cite{SIBYLL23c} particle coding system,
extended with decimal nuclear codes ($A+ 100\,Z$).
\item MOCCA-style particle codes\index{MOCCA} \cite{MOCCA}, extended
to match all AIRES particles.
\end{enumerate}
The codes corresponding to elementary particles are listed in table
\ref{TAB:codsys}\index{CORSIKA!particle codes}\index{GEANT!particle
codes}\index{SIBYLL!particle codes}\index{MOCCA!particle
codes}\index{AIRES particle codes}\index{particle codes}.
\subsubsection{Opening existing files}
Once the proper environment is set up by means of the initializing
routine, the system is ready to open any existing compressed file. The
open routine \libbfx;opencrofile; will use the header information to
initialize the internal variables that permit processing the different
fields defined for the file. 
The following example illustrates how to open a file:
\kwexample{%
\kwt;program sample;\\
\kwt;character*80 mydir, myfile;\\
\kwt;integer channel, irc;\\
\kwt;. . .; \\
\kwt;call; ciorinit\kwt;(0, 1, 0, irc);\\
\kwt;. . .; \\
\kwt;call; opencrofile\kwt;(mydir, myfile, 0, 10, 4, channel, irc);\\
\kwt;. . .;}
\kwbf;myfile; and \kwbf;mydir; are character strings containing
respectively the file name and the directory where it is placed. The
integer argument ``10'' indicates that the logarithmic fields are
going to be transformed into decimal logarithms. \kwbf;channel; is an
output parameter identifying the opened file; it should not be set by
the calling program.

 It is important to remark that this call will transparently open {\bf
any} compressed file, regardless of its type or format (ground
particle as well as longitudinal tracking particle files in all their
variants), the AIRES version used to write it and/or the machine used
when writing it.
\subsubsection{Getting information about the file}
The headers of the compressed files are divided into two parts: One
part containing the definitions of the file's data records and another
section with information about the simulations that originated the
file.

The file definitions are specific to each opened file, and therefore
the system must store them separately for each one of the files that
are simultaneously open.

The other information, however, is of global character, and so the
available data always corresponds to the last opened file. These data
are superseded each time \kwbf;opencrofile; is called.

Routine \libbfx;croheaderinfo; prints a summary of this global
 information while \libbfx;croinputdata0; copies some of those data
 into arrays to make them available to the user (see page
 \pageref{rou0:croinputdata0}) and \libbfx;crotaskid; returns task
 name information. Functions \libbfx;getinpint;, \libbfx;getinpreal;,
 \libbfx;getinpstring; and \libbfx;getinpswitch; (see pages
 \pageref{rou0:getinpint}--\pageref{rou0:getinpswitch}) allow to
 obtain other input data items not returned by
 \kwbf;croinputdata0;. \libbfx;getglobal; can be called to retrieve
 information about global variables\index{global variables} that were
 defined during the simulations. \libbfx;idlcheck; returns information
 about the IDL instructions that were valid when the file was
 generated, and \libbfx;crofileversion; and \libbfx;thisairesversion;
 return version information that might be useful when reading
 compressed files written with old AIRES versions.

In some special applications it is necessary to access
information that can be stored only in the internal dump
file\index{internal dump file!accessing}. In these cases, it can be
helpful to invoke the routine \libbfx;loadumpfile; right after opening
the corresponding compressed file, and then use some of the routines
\libbfx;dumpinputdata0;, \libbfx;dumpfileversion;, etc., to access
the mentioned data.

The structure of any already opened file can be printed calling
routine \libbfx;crofileinfo; which prints a list of the different
records defined for the corresponding file and the names of the fields
within records. It is also possible to load into arrays such
information by means of routine \libbfx;crorecstrut; in order to make
it available to the analysis program.
\subsubsection*{Reading the data records}
Once a file is open, it remains positioned at the beginning of the
compressed data section. From then on, the file can be sequentially
read using routine \libbfx;getcrorecord;:
\kwexample{%
\begin{tabbing}
\kwt;okflag =; getcrorecord\kwt;(;\=
                             \kwt;channel, indata, fldata, altrec,;\\
                                  \>\kwt;0, irc);
\end{tabbing}
}%
\kwbf;getcrorecord; returns logical data, which in this case are
stored in the logical variable \kwt;okflag;. The returned value is
``\kwbf;true;'' if the reading operation was completed successfully,
``\kwbf;false;'' otherwise (end of file, I/O error, etc.).

\kwbf;ciochann; should be the same integer variable used when opening
the file; it identifies the file to be processed.

 \kwbf;irc; is an integer return code. If \kwbf;okflag; is ``false'',
then the return code contains information about the error that
generated the abnormal return, as explained in page
\pageref{rou0:getcrorecord}. For successful read operations,
\kwbf;irc; indicates the record type that has been just read in: 0 for
the default particle record, 1 (2) for the ``beginning (end) of
shower'' record, etc. At the same time, the logical variable
\kwbf;altrec; distinguishes between ``alternative'' (non default)
records (\kwbf;true;), from default ones (\kwbf;false;).

The data stored in the different fields of the record is retrieved by
means of the arrays \kwbf;indata; and \kwbf;fldata;. Both are single
index arrays, containing integer and double precision data,
respectively. The data items stored in these arrays does vary with the
kind of file being processed and the type of record that was
scanned. In all cases, the routine will automatically set the relevant
elements of these arrays accordingly with the logical definition of
the record, regardless of the physical structure of it which remains
absolutely hidden at the user's level.

To fix ideas, let us suppose that a ground particle file with {\em
normal\/} particle records is being processed. Every time \kwbf;irc;
is zero (default record), the integer and real data arrays will
contain the elements listed in table \ppref{TAB:gpr}, that is
\begin{displaymath}
\hbox{\begin{tabular}{lcl}
\kw;indata;\kwt;(1); & $\leftarrow$ & Particle code\\*[4pt]
\kw;fldata;\kwt;(1); & $\leftarrow$ & Energy (GeV) (log)\\
\kw;fldata;\kwt;(2); & $\leftarrow$ & Distance from the core (m) (log)\\
\kw;fldata;\kwt;(3); & $\leftarrow$ & Ground plane polar angle (radians)\\
\kwt;. . .;        \\
\kw;fldata;\kwt;(7); & $\leftarrow$ & Particle weight
\end{tabular}}
\end{displaymath}

For different return codes, the number of assigned array elements may
be different, as well as their meanings; but in all cases such data
items will be set accordingly with the corresponding record sequence
(tables \ref{TAB:bosr}, \ref{TAB:eosr}, etc.).

In order to make the analysis programs simpler and more robust, a
special routine has been included in the AIRES library to {\em
automatically\/} set the adequate field indices corresponding to a
given record of a certain compressed file, as illustrated in the
example of figure \ref{FIG:autoidx}.
\lowfig{[tp]}{autoidx}{%
{\openup-1pt\small
\begin{tabular}{|c|}
\hline\\
\begin{minipage}{14.5cm}
\libidx;ciorinit;%
\libidx;opencrofile;%
\libidx;crofieldindex;%
\libidx;getcrorecord;%
\kwexample{%
\kwt;program sample;\\
\kwt;. . .;\strut\\
\kwt;integer datype, irc, icode, idist, inear;\\
\kwt;integer indata(30);\\
\kwt;double precision fldata(30);\\
\kwt;integer particlecode;\\
\kwt;double precision logdistance, numberofnear;\\
\kwt;. . .;\strut\\
\kwt;call ciorinit(0, 1, 0, irc);\\
\kwt;call opencrofile(mydir, myfile, 0, 10, 4, channel, irc);\\
\kwt;. . .;\strut\\
\vbox{\begin{tabbing}
\kwt;icode =; crofieldindex\kwt;(;\=\kwt;channel,;
                                 0\kwt;,; 'Particle code'\kwt;,;\\
                                \>\kwt;4, datype, irc);\\
\kwt;idist =; crofieldindex\kwt;(;\>
                             \kwt;channel,;
                                 0\kwt;,;\\
                                 \> 'Distance from the core'\kwt;,;\\
                                 \>\kwt;4, datype, irc);\\
\kwt;inear =; crofieldindex\kwt;(;\>
                              \kwt;channel,;
                                 2\kwt;,; 'Particles too near'\kwt;,;\\
                                 \>\kwt;4, datype, irc);
\end{tabbing}}\\
\kwt;. . .;\strut\\
\vbox{\begin{tabbing}
\kwt;okflag = getcrorecord(;\=
                             \kwt;channel, indata, fldata, altrec, 0,;\\
                                  \>\kwt;irc);
\end{tabbing}}\\
\kwt;if (irc .eq.; 0\kwt;) then;\\
\kwt;\hbox{}\kern1.2em particlecode  = indata(;icode\kwt;);\\
\kwt;\hbox{}\kern1.2em logdistance \ = fldata(;idist\kwt;);\\
\kwt;\hbox{}\kern1.2em . . .;\strut\\
\kwt;else if (irc .eq.; 2\kwt;) then;\\
\kwt;\hbox{}\kern1.2em numberofnear  = fldata(;inear\kwt;);\\
\kwt;\hbox{}\kern1.2em . . .;\strut\\
\kwt;end if;\\
\kwt;. . .;\strut\\
\strut}
\end{minipage} \\ \hline
\end{tabular}}
}%
{Processing compressed data files, an example}%
{Processing compressed data files, an example illustrating how to set
field indices automatically.}

The outstanding characteristic of this piece of code is that the
elements of arrays \kwbf;indata; and \kwbf;fldata; are not referenced
directly using numeric indices, but by means of integer variables like
\kwbf;icode; for instance (see figure \ref{FIG:autoidx}).

Those index variables are set by means of routine
\kwbf;crofieldindex;. The arguments required by this routine include:
\iiset\ii The identification of the file (\kwbf;channel;). \ii The
record type, coincident with the return codes of \kwbf;getcrorecord;
already mentioned. In this example 0 for the default record and 2 for
the ``end of shower'' record. \ii The first characters of the field
name. Fields are identified by their {\em names,\/} providing
therefore absolute transparency to the fact that the order and number
of fields may change with the file being processed. The next argument
of \kwbf;crofieldindex; is set to 4 to force the program to stop in
case of ambiguous or erroneous field specification, thus providing a
very safe processing environment. \ii The output argument
\kwbf;datype; returns information about the data type corresponding to
the specified field, as explained in page
\pageref{rou0:crofieldindex}.
In the particular case of longitudinal particle tracking files, it is
generally convenient to use the routine \libbfx;getlgtrecord; in place
of \libbfx;getcrorecord;. A complete description of
\kwbf;getlgtrecord; can be found in page \pageref{rou0:getlgtrecord};
this routine must be used jointly with \libbfx;getlgtinit;.
\subsubsection*{Closing files and ending a processing session}
The AIRES library routines support simultaneous processing of more
than one compressed file. Several compressed files can be
opened at the same time, each one identified by the corresponding
channel integer variable.

The opened files can be closed using two alternative procedures (see page
\pageref{rou0:cioclose1}): \iiset\ii Routine \libbfx;cioclose1; closes
individual files. \iiset \libbfx;cioclose; closes {\em all} the
currently opened files.

Routine \kwbf;cioclose; should be used only if the processing session
will continue after closing all files. To finish an analysis program in
an ordered fashion use the routine \libbfx;ciorshutdown;. This
procedure performs all the required tasks to properly set down the
processing system, including a call to \kwbf;cioclose;.
\subsubsection*{Other operations}
There are many other routines included in the AIRES library that
provide useful tools for special analysis tasks. Such routines are
explained in detail in appendix \ppref{A:aireslib}, we shall limit here
to a brief presentation of the most relevant ones:
\begin{description}
\item[Counting records.] Routines \libbfx;crorecinfo; and
  \libbfx;croreccount; count the data records contained within a
  compressed file.
\item[Repositioning.] Routine \libbfx;crorewind;\index{rewinding
compressed files} repositions an already opened file at the beginning of
the data section. Routines \libbfx;crorecnumber; and \libbfx;crogotorec;,
used jointly, permit accessing the data records in arbitrary order.
\item[Fast scanning of a file.] Routine \libbfx;crorecfind; finds the
next appearance of a record of a given type (to locate shower headers,
for example). \libbfx;getcrorectype; returns the type of the next
record, and \libbfx;regetcrorecord; re-reads the current record.
\item[Longitudinal tracking file utilities.]\label{P:olutil} Routine
\libbfx;crooldata; returns basic information about the positions of the
observing levels defined for the simulations; while \libbfx;olcoord;
returns the coordinates of the intersections between the observing levels
and the shower axis and \libbfx;olv2slant; evaluates the slant depths
corresponding to each observing level. Routines \libbfx;olcrossed; and
\libbfx;olcrossedu; decode the crossed observing levels key\index{crossed
observing levels key} defined in equation (\ref{eq:olcrossed}), returning
the variables $i_f$, $i_l$ and $s_{\mathrm{ud}}$ (see section
\ref{S:ciocustom}). The logical function \libbfx;olsavemarked; permits
determining whether or not a given observing level is recorded into a
compressed file.
\item[Special primary utilities.] Besides the specific routines
designed to process special primary particles, described in detail in
section \ref{S:specialprim}\index{special primary particles}, the
AIRES library includes also some auxiliary routines that are useful to
obtain data about the special primaries that were defined at the
moment of creating the compressed file that is being analyzed.
\libbfx;crospcode; and \libbfx;crospmodinfo; are examples of such
procedures.
\item[Miscellaneous routines.] The library contains some other routines
than may be useful in certain applications, for example the pseudo-random
number\index{random number generator} utilities \libbfx;raninit;,
\libbfx;urandom;,\libbfx;urandomt;, and \libbfx;grandom;;
Gaisser-Hillas function\index{Gaisser-Hillas function} related
routines: \libbfx;fitghf;, \libbfx;ghfpars;, \libbfx;ghfx;,
\libbfx;ghfin;; atmospheric depth\index{atmospheric depth!slant}
utility routines line \libbfx;xslant;; etc.
\end{description}

The AIRES object library is continously evolving, so
additional procedures will be surely included in future AIRES
versions.

%
\chapter{The AIRES Runner System}\index{AIRES Runner System}
\label{C:ARS}\index{ARS o ars|see{AIRES Runner System}}

Production simulation tasks usually require large amounts of computer
time to complete, and in such cases the user risks loosing all the
simulation run if the system goes down before the task is finished.
To avoid this inconvenient situation, the AIRES simulation system
provides a special auto-saving mechanism that permits splitting the
simulation job into small runs. In case of abnormal interruption, the
simulations can be restarted at the point they were when the last
auto-saving was performed\index{fault tolerant processing}.

As explained in chapter \ppref{C:steering}, a simulation task may
require several invocations of the simulation program if the auto-save
mechanism is enabled. If this is done manually, the user must control
the sequence of instructions needed to complete the simulations.  To
ease the management of such sequential series of processes, a set of
scripts were developed with the capability of automatically launching
the corresponding jobs. These scripts are part of the {\em AIRES Runner
System (ARS)\/}, designed as a set of interactive procedures to manage
complex simulations tasks.

The AIRES Runner System works only on UNIX platforms, and provides
tools for input file checking, sequential and concurrent task
processing, event logging, etc. This chapter is devoted to present
some examples that will help the user to get familiar with the Runner
System.

There are many parameters that modify the behavior of the AIRES
Runner System. Most of them are user-settable and their definition
statements are placed within the ARS initializing file
\kwbf;.airesrc;\index{airesrc@{\kw;.airesrc; initialization file}}. In
standard AIRES installations, this file is placed in the user's home
directory.
\section{Checking input files}\index{input file checking}

In section \ppref{S:icheck}, the IDL directives \idlbfx;CheckOnly; and
\idlbfx;Trace; were used to instruct AIRES simulation programs to scan
a given input file, report on its contents and stop without actually
starting the simulations.

The ARS command
\arsidx;airescheck;%
\kwexample{airescheck \kwt;-t myfile.inp;}
will invoke \kwbf;Aires; with the same input as displayed in page
\pageref{P:inpcheck}. The \kwbf;-t; qualifier is placed to enable
typing the input lines as long as they are scanned.

There are additional qualifiers accepted by this command, for example:
\kwexample{airescheck \kwt;-tP -p AiresQIIr04  myfile.inp;}
The \kwbf;-p; qualifier overrides the default simulation program used to
process the input file, and the \kwbf;P; switch indicates that the output
must be printed instead of being typed at the terminal. The print command
to use can be set modifying the
\kwbf;.airesrc;\index{airesrc@{\kw;.airesrc; initialization file}}
initializing file.
\section{Managing simulation tasks}
Once the input file has been checked, the simulations can be
started. The command
\arsidx;airestask;%
\kwexample{airestask \kwt;myfile;}
will first check that file \kwbf;myfile.inp;\footnote{\kwbf;airestask;
first assumes a default extension \kwbf;.inp; for the input file name,
and as a second alternative, tries to find the file whose {\em
complete\/} name is as specified in the input parameter.} exists, and
then will create an entry in the corresponding {\em ARS spool.\/}
Finally, \arsbfx;aireslaunch; will be executed.

The \kwbf;aireslaunch; script will detect that there is a task pending
completion and so will prompt the user to start the simulations. In
case of positive answer, the simulation program will be started with
the corresponding input, and will be repeatedly invoked if necessary
until the task is completed\footnote{The simulation program
communicates with the script via a file that contain information about
the status of the simulations.}. All those operations are completely
automatic, no further user intervention is normally required.

If there are more than one task to be processed, they can be spooled
at any moment after launching the first simulations. The command
\kwexample{airestask \kwt;my\_other\_file;}
will make a new spool entry which will be queued after the first
one. Execution of this task will start as soon as the previous one is
finished. There is no limit in the number of tasks that can be queued
in the ARS spools.

At any moment during the simulations, it is possible to inspect the
evolution of the spooled tasks by means of the ARS command
\arsbfx;airesstatus;.

In the preceding examples, the default simulation program (which normally
is the \kwbf;Aires; program) will be used. There are two alternatives to
override the default specification: \iiset\ii Modify the default program
setting of the initialization file
\kwbf;.airesrc;\index{airesrc@{\kw;.airesrc; initialization file}}. \ii
Use the \kwbf;-p; qualifier of the \kwbf;airestask; command:
\kwexample{airestask -p \kwt;AiresEPLHC; \kwt;yet\_another\_file;}
\kwbf;AiresEPLHC; is the name of a variable defined within the
initialization file, which indicates the executable program that
contains a link to the EPOS LHC\index{EPOS} hadronic
package.
\subsection{Canceling tasks and/or stopping the simulations}
Every spooled task can be canceled by means of the command
\arsbfx;airesuntask;, for example:
\kwexample{airesuntask \kwt;my\_other\_file;}
will erase the second spooled task of the preceding section.
If the \kwbf;airesuntask; command is invoked with no parameters, then
it will prompt the user to cancel each one of the spooled tasks.

It is not recommendable to remove the spool entries corresponding to
tasks that are currently running. In such cases it is better to first
stop the simulation program, and wait until the AIRES Runner System
shuts down.

The simulation program can be stopped with the ARS command
\arsbfx;airesstop;, which generally is invoked with no arguments. This
script originates an ordered shutdown of the simulations, which
includes an update of the internal dump file\index{internal dump
file}, and may take up to several minutes to effectively interrupt the
simulations.  The command \kwbf;airesstatus; can be used to monitor
the status of the system during this process.

On the other hand, a currently running simulation can be immediately
aborted by means of command \arsbfx;aireskill;. In this case the
corresponding processes are killed without any previous auto-saving
operation.

Stopped simulations can always be restarted using \kwbf;aireslaunch;.
\subsection{Performing custom operations between processes}
\label{S:AfterProcess}
Every time a process\footnote{See section \ppref{S:tpar}.}\index{tasks,
processes and runs} ends, the ARS checks for the existence of a executable
script named \kwbf;AfterProcess; (case sensitive!), first in the current
working directory\index{AIRES file directories!working directory}, and
then --if not found-- in the default directory of the user's account (HOME
directory). If the file is found, it is executed.

The complete command line used when invoking the \kwbf;AfterProcess; macro
is the following:
\kwexample{\kwt;AfterProcess\ ; \vkw;spool; \vkw;tn; \vkw;msg; \vkw;rc;
\vkw;trial; \vkw;totsh; \vkw;lastsh; \vkw;prog;
}
where
\begin{description}
\item[\vkw;spool;] is the spool identification.
\item[\vkw;tn;] is the task name.
\item[\vkw;msg;] is a message string coming from the simulation
program. Normally it takes the values \kwbf;EndOfTask; or
\kwbf;EndOfRun;, indicating if the current task was or not finished,
respectively. Other values are also possible and correspond to abnormal
situations.
\item[\vkw;rc;] is a numeric parameter, taking the value 2 if the run has
been stopped using an AIRES.STOP file (command \arsbfx;airesstop;).
\item[\vkw;trial;] is a numeric variable counting the number of trials for
the current run. Generally takes the value 1, but in certain
circumstances, for example when relaunching AIRES after a system crash, it
can take larger values.
\item[\vkw;totsh;] is the total number of showers for the current task.
\item[\vkw;lastsh;] is the last completed shower.
\item[\vkw;prog;] is the instruction used to invoke AIRES, which includes
the full name of the simulation program used in the last run.
\end{description}

This powerful ARS option makes it possible for the user to perform
operations of almost every kind after ending the processes. Of course, a
certain degree of expertise with UNIX systems may be required in certain
cases. Typical examples of operations that can be done using this facility
are: File movement after completion of tasks (for example to massive
storage systems), alerts of any type about conditions of the system, like
full disks, etc.

On return, the \kwbf;AfterProcess; script can communicate with the ARS via
the exit code. If it is zero then processing will continue normally,
otherwise the ARS will send a mail notifying the abnormal return code
and then will stop. If it is necessary to restart the simulations, it can
be done using the ARS command \arsbfx;aireslaunch;.

The following shell script is a very simple example of an ``after
process'' macro:
\kwexample{\tt
\begin{tabbing}
\#!/bin/sh\\
\#\\
if\= [ \$3 = EndOfTask ]\\
then\\
\#\\
\# This code will be executed only after ending a task.\\
\#\\
\>mv \$\{2\}.grdpcles /mysafeplace\\
fi\\
exit 0\\
\end{tabbing}}
Notice that no action will be taken up to the end of a task. Whenever this
happens, the corresponding ground particle file is moved to another
directory. The command \kwbf;exit 0; ensures normal return code;
\kwbf;exit; \vkw;n; with $\hbox{\vkw;n;} \neq 0$ means an abnormal exit
and in this case the simulations will be stopped.

Similarly as in the case of the \kwbf;AfterProcess; macro, the ARS
system will search for a \kwbf;BeforeProcess; macro, right before
invoking the simulation program. The existence of the
\kwbf;BeforeProcess; macro is checked in the working
directory\index{AIRES file directories!working directory}, and in the
user's account (HOME) directory (in that order). If the file is found,
it is executed.

The complete command line used when invoking the \kwbf;BeforeProcess; macro
is the following:
\kwexample{\kwt;BeforeProcess\ ; \vkw;spool; \vkw;tn;
            \vkw;trial; \vkw;ifile; \vkw;prog;
}
where
\begin{description}
\item[\vkw;spool;] is the spool identification.
\item[\vkw;tn;] is the task name, or \kwbf;UNKNOWN; if the task is not
initialized yet.
\item[\vkw;trial;] is a numeric variable counting the number of trials for
the current run. Generally takes the value 1, but in certain
circumstances, for example when relaunching AIRES after a system crash, it
can take larger values.
\item[\vkw;ifile;] is the name of the input file to be used when
running the simulation program.
\item[\vkw;prog;] is the instruction used to invoke AIRES, which includes
the full name of the simulation program used in the next run.
\end{description}

After completing execution of the \kwbf;BeforeProcess; macro, the ARS
checks the corresponding return code, continuing with the next step
only if it is zero.
\section{Concurrent tasks}

In many cases it is necessary to simultaneously process more than one
task. Systems having more than one CPU and/or clusters of machines
sharing the same file system, are examples of such situation.

The AIRES Runner System provides certain tools designed to work under
such circumstances. The key idea is to define more than one spool, and
assign one spool to each processing unit, either a CPU or a machine
inside the cluster.

In the preceding examples, the \arsbfx;airestask; command was invoked
without spool specification. The default spool is used in case of
missing specification, and that is what was actually done in those
examples.

In the standard configuration there are 9 predefined spools, named
respectively ``1'', ``2'', \ldots, etc. Spool ``1'' is the default
spool\footnote{The ARS includes also the commands
\arsbfx;mkairesspool; and \arsbfx;rmairesspool; which allow the user
to respectively create and delete spool directories.}. The command
\kwexample{airestask -s \kwt;2 myfile;}
will create a spool entry placed in spool ``2''. The user will be
prompted to start the simulations if there is currently no activity
related with that spool. The command
\kwexample{airesstatus \kwt;2;}
will report on the simulations that are running at spool ``2''. Similarly,
\kwexample{airesstatus \kwt;all;}
will report on the simulations that are running at every active spool.

In the following interactive session, it is illustrated how to launch
three simultaneous tasks (it is assumed that the machine possesses
various CPU's which can be automatically assigned to the launched
processes):
\kwexample{%
\kwt;cd directory1;\\
aireslaunch -s 1 \kwt;task1;
}\kwexample{\kwt;. . .;}\kwexample{%
\kwt;cd directory2;\\
aireslaunch -s 2 \kwt;task2;
}\kwexample{\kwt;. . .;}\kwexample{%
\kwt;cd directory3;\\
aireslaunch -s 3 \kwt;-p AiresQIIr4 task3;\\
\kwt;. . .;
}

It is most important that the working directories of different tasks
be also different: Concurrent simulation programs running with the
same working directory may generate conflicts when communicating with
the ARS scripts. This fact is stressed by means of the \kwbf;cd;
commands of the example, where \kwbf;directory1;, \kwbf;directory2;
and \kwbf;directory3; {\em must\/} be different directory
specifications.

Notice also that the third spooling command makes use of an alternative
simulation program in order to perform a different kind of
simulation. Alternative programs may also be necessary when running
simulations on clusters sharing the same file system but made with non
compatible platforms. In those cases it is necessary to have different
executable modules for each platform. Once such modules are available, it
is possible to change the default programs corresponding to the different
spools by means of suitable modifications to the
\kwbf;.airesrc;\index{airesrc@{\kw;.airesrc; initialization file}}
initialization.

The details about how to make the AIRES Runner System work in complex
operating environments are rather technical and go beyond the scope of
this manual. Such a job requires normally a good degree of expertise on
UNIX systems.
\section{Some commands to manage dump file data}
\label{S:ARSexport}

Chapter \ppref{C:output1} explains in detail the operations needed to
retrieve data stored within the internal dump file\index{internal dump
file} in either its binary or ASCII versions. Some of them are frequently
used and generally involve very similar sequences of instructions. A
typical example is to export one or more tables corresponding to an
already finished task\index{exported data files}\index{output data
tables}.

The ARS includes a shell script that can be helpful in those
cases. Consider for example the command (under UNIX)
\arsidx;airesexport;
\kwexample{airesexport \kwt;mytask 1001 1205 to 1213;}
Its action is to invoke the AIRES summary program\index{AIRES summary
program} with the following input
\idlidx;Summary;\idlidx;TaskName;\idlidx;ExportTables;\idlidx;End;
\kwexample{\tt
Summary Off\\
TaskName mytask\\
ExportTable 1001\\
ExportTables 1205 1213\\
End
}
generating text files for tables 1001, 1205, 1207, 1211 and 1213 (see
appendix \ref{A:outables}).

In some cases it may be necessary to specify other parameters, like in the
following example
\kwexample{%
airesexport -w \kwt;idfdir; -O \kwt;LM; -s \kwt;mytask 2501;
}
This command will generate single shower tables\index{single shower
tables} (enabled by the \kwbf;-s; qualifier) as well as average ones. The
options \kwbf;LM; correspond to $dN/d\log_{10}E$ distributions with
energies expressed in MeV (see section \ref{S:exporting}), and the string
following the \kwbf;-w; qualifier (\kwbf;idfdir;) indicates the directory
where the IDF and/or ADF\index{internal dump file!portable format} files
are located (The global\index{AIRES file directories!global directory}
directory accordingly with the definitions of section \ref{S:filedir}).
\subsection{Converting IDF binary files to ADF portable format.}
\label{S:IDF2ADF}\index{internal dump file!portable format}%
\index{converting IDF files to ADF portable format}

ADF files were implemented for AIRES version 2.0.0, and to have them
written by the simulation programs after a task is completed, it is
necessary to explicitly enable them by means of the IDL directive
\idlbfx;ADFile;. The (binary) IDF is {\em always\/} generated,
regardless of the input settings and/or the version of AIRES used.

 Of course, the IDF stores all the data associated with both input
parameters and output observables, and is enough for any kind of analysis
provided the user always works with compatible computers. But this may not
be the case when a person or group is working at different locations. For
such cases, a {\em portable\/} file format is needed and the ADF
becomes essential to enable data analysis in non-compatible workstations.

If the ADF was not generated during the simulations, or if the
simulations were performed using a version of AIRES previous to version
2.0.0, it must be created manually. The current AIRES distribution
includes an IDF to ADF converting program\index{AIRES IDF to ADF converting
program}, whose default name is \kwbf;AiresIDF2ADF;.

This program can be used directly. It is just necessary to invoke it (no
arguments needed) and answer to the prompts that will be appearing.

On the other hand, the ARS includes a special shell script that permits
converting files without calling \kwbf;AiresIDF2ADF; manually. Let us
illustrate how to use this command with an example. Suppose in a certain
place there are some IDF files that need to be converted to ADF
format. The UNIX command
\kwexample{%
idf2adf \kwt;taskname1 taskname2 taskname3;
}
will search for the files \kwbf;taskname1.idf;, \kwbf;taskname2.idf;,
etc., and will call \kwbf;AiresIDF2ADF; as many times as necessary, to
create the portable files \kwbf;taskname1.adf;, \kwbf;taskname2.adf;,
etc. Of course, the old IDF files will remain unchanged.

This script will work well in most cases. However, there might be special
situations where it is necessary to use \kwbf;AiresIDF2ADF;\index{AIRES
IDF to ADF converting program} manually, for example when the IDF file is
renamed with a new name not ending with ``.idf''.

%
%
%
%
\cleardoublepage
%
%
\appendix

%
\chapter{Installing AIRES and maintaining existing installations}%
\index{AIRES!installation}\index{installing AIRES}%
\label{A:installing}

As mentioned in section \ppref{S:downloading}, every AIRES distribution
is currently packed in a single compressed UNIX tar file. In this
appendix it is assumed that the software distribution was successfully
decompressed and tar expanded.

\section{Installing AIRES \currairesversion}
\label{AS:installing}

In UNIX platforms,
the installing procedure is quite simple: Almost everything is done
automatically. The key points to take into
account are:
\begin{itemize}
\item[{\bf(a)}] A Unix shell script \kwbf;doinstall; is provided. This
        script will install the software automatically.

\item[{\bf(b)}] The file \kwbf;config; contains all the customizable
        variables. You should check its contents and edit it if needed
        before invoking \kwbf;doinstall;.

\item[{\bf(c)}] There will be two main directories:
  \begin{enumerate}
   \item {\em Aires root directory\/}
     (hereinafter named \kwbf;Aroot;), which is the highest level
     directory for the installed files. Normally the AIRES
     distribution compressed tar file and the AIRES External
     Input data compressed tar files are located in this directory before
     starting the installation process. You might need to specify
     \kwbf;Aroot; by editing the \kwbf;config; file (located within
     the \kwbf;Iroot; directory). For standard, personal installation,
     the default (creating a directory named \kwbf;aires; in your home
     directory) will be OK.
    \item {\em Installation root directory\/} (hereinafter named
     \kwbf;Iroot;), which is the directory where the distribution file
     was downloaded (that is, the directory containing the
     \kwbf;doinstall; script).
  \end{enumerate}
   Notice that the \kwbf;Iroot; and \kwbf;Aroot;
   directories may or may not be the same directory (Do not worry
   about this: The installation program will manage every case properly.).
\item[{\bf(d)}] Your account must have access to a FORTRAN compiler
        (normally, command \kwbf;gfortran;), and in some
        cases to a C compiler (commands \kwbf;gcc;, etc.);
        and these compilers must be placed in one of the PATH
        directories (in other words, if you type at your terminal,
        say, \kw;gfortran;, the machine will take \kw;gfortran; as a known
        command).  If the compilers are not in the PATH you will have
        to enter their absolute location manually in the \kwbf;config;
        file (Our recommendation, however, is to ensure that the
        compilers are in the PATH. It is something not difficult to
        achieve. If you do not know how to proceed or what we are
        speaking about, then ask your local UNIX expert).
\end{itemize}

\subsection{Installation procedure step by step}

\begin{enumerate}
\item Ensure that you have write permission on both \kwbf;Iroot; and
   \kwbf;Aroot; directories, and in all their sub-directories.

\item \kwbf;cd; to \kwbf;Iroot;, and edit the file \kwbf;config; if
  necessary. The current AIRES installation program will automatically
  set most of the required installation parameters, including
  automatic detection of the operating system (Linux and Mac OSX are
  the ones currently supported).

\item If you need a full installation that will allow you to perform
  simulations using the hadronic packages QGSJET and/or EPOS, you must
  download and place inside the \kwbf;Aroot; directory the {\em\bfseries
    External Input data file\/}\index{External input data file}
  (available from the AIRES repository). The size of this file is
  about 300 MB.

\item Enter the command
       \kwdisplay;doinstall 0;
   if you are installing AIRES for the first time, or
       \kwdisplay;doinstall 1;
   if you are upgrading your current installation (This is the case
   for those users that are already employing a previous version of
   AIRES. Note that you {\bf should not erase} any existing installation
   of AIRES before completing the upgrade.).

   This procedure will install the software using the data you set in step 2.
   This may take some minutes to complete. A message will be typed at your
   terminal indicating whether the installation was successful or not. If you
   get any error message(s), you should check all the requirements described
   previously, in particular points (d) and (1). Try also modifying the
   \kwbf;config; file.

\item Type the command (case sensitive)\footnote{The name \kwbf;Aires;
      can be changed modifying adequately the \kwbf;config; file. If
      this name was changed, then the user supplied name must be typed
      in place of the default one.}
      \kwdisplay;Aires;
   to see if the program is running and is in your search path. You
   should see typed at your terminal something like the following
   text\footnote{You should also obtain a similar output if you invoke
     the
     AIRES/EPOS/QGSJET/SIBYLL\index{EPOS}\index{QGSJET}\index{SIBYLL}
     simulation program instead of the default \kwbf;Aires;.}:
\medskip
\begin{verbatim}
 >>>>
 >>>> This is AIRES version V.V.V (dd/Mmm/yyyy)
 >>>> (Compiled by . . . . .
 >>>> USER: uuuuu, HOST: hhhhhhh, DATE: dd/Mmm/yyyy
 >>>>

 > dd/Mmm/yyyy hh:mm:ss. Reading data from standard input unit
\end{verbatim}
\medskip
   where \kwbf;V.V.V; indicates the current version of AIRES
   (\currairesversion) and goes together with the release date.
   Type \kw;x; and press $\langle$ENTER$\rangle$ to leave the program.

   If step 4 ended successfully and you fail to run the program, it is
   likely that the AIRES \kwbf;bin; directory is not in your
   environment search path (Unix environment variable PATH). In some
   systems you need to log out and log in again to make effective any
   PATH change. If you cannot place the AIRES \kwbf;bin; directory
   into your account's PATH, then ask a Unix expert to do that for
   you. Once you are sure that the directory is in the search path,
   and if the problem still persists, check if the executable file
   \kwbf;Aires; exists. If it does not exist that means that step 4
   was not successfully completed. Do not continue with the next step
   until you succeed with this one.

\item {\bf Installation of extensions.} The current version of AIRES
  permits the installation of extensions that add new capabilities to
  the original system. At present, there is one of such extensions
  available: ZHAireS\index{ZHAireS}, that allows to simulate the radio
  wave emission that takes place during shower development (see
  reference \cite{ZHAireS} and the corresponding manual for details).
  Installation of extensions is not mandatory; if you do not need to
  do so you can skip this point and go directly to
  (\ref{itlab:goonafterinst}).

   To install the extension(s):
   \begin{enumerate}
   \item Download the extensions available (compressed tar files of
     the form \vkw;Name;\kwbf;-Exten-;\vkw;v-v-v;\kwbf;.tar.gz;), and
     place then in the same directory where the current AIRES
     distribution tar file is.

   \item \kwbf;cd; again to the \kwbf;Iroot; directory. Together with the
     \kwbf;doinstall; script that you have already used to install
     AIRES, you will find another script: \kwbf;addextensions;.

   \item If you want to perform a quick installation of the
     extensions, using the same installation parameters that are in
     the \kwbf;config; file, just execute
          \kwdisplay;addextensions;
        This command will do all the necessary to install the extensions,
        including compilation of sources and building executable programs.
        Then go directly to (\ref{itlab:goonafterext0}).

   \item Instead, if you need to customize the configuration parameters that
        are relevant to the extension, then execute   
          \kwdisplay;addextensions noinstall;
        This command will just expand the corresponding tar file and
        perform some checks, without compiling and/or building any
        program or library.  After this is complete, for each
        installed extension you will see the corresponding
        \kwbf;config.;\vkw;extension\_name\_in\_lowercase; file. You
        can edit this file manually, perform all the needed changes,
        and complete the installation of the extension using the
        command \kwdisplay;doinstall 3
        \hbox{\vkw;extension\_name\_in\_lowercase;};

   \item\label{itlab:goonafterext0} Follow the instructions placed in
     the specific \kwbf;Install.;\vkw;Extension;\kwbf;.HowTo; file to
     ensure that the corresponding program is working properly.
   \end{enumerate}

\item\label{itlab:goonafterinst} \kwbf;cd; to your HOME directory and
  verify the presence of a file named
  \kwbf;.airesrc;\index{airesrc@{\kw;.airesrc; initialization file}}.

   Normally it is not necessary to change anything in this file, but the
   need may appear in the future, specially if you decide to use the
   UNIX scripts that are provided to help running AIRES (see chapter
   \ref{C:ARS}). 

\item If you completed successfully these steps, the software should
   be properly installed.

\item After successfully completing these steps you can optionally
  delete the files corresponding to old versions of AIRES. Such files
  are placed within the Aroot directory. For example, directory 18-09-00
  contains AIRES 18.09.00 files, etc.

\item If you completed successfully these steps, the software should
  be properly installed. In that case, and if you do not have previous
  experience using AIRES, we strongly recommend you to go to the Iroot
  directory again, enter the doc sub-directory, print the file
  LearnByExamples.txt, and follow the instructions that are in this
  file to learn how to use AIRES.

\end{enumerate}
\section{Recompiling the simulation programs}
\index{recompiling simulation programs}

In many cases it may be necessary to recompile the simulation programs
after having successfully installed the AIRES system. Some examples of
such situations are:
\begin{itemize}
\item Some compilation parameters were not set accordingly with the
user needs; or the required configuration is no more the one set up at
the moment of installing the software.

\item It is necessary to install AIRES in different (not compatible)
platforms sharing the same directory tree.

\item It is necessary to create more than one executable program, each
one compiled with different compilation parameters. As an example of
this case, consider that the number and kind of records that are
written in the compressed particle files\index{compressed output
files} can be controlled by means of compilation parameters (see
section \ref{S:ciocustom}), and that it is required to have the
executables for different file formats.
\end{itemize}

The arguments recognized by the \kwbf;doinstall; executable script
allow the user to easily perform the different operation required in
cases like the ones previously enumerated.

The general syntax of \kwbf;doinstall; is:
\kwexample{doinstall \vkw;ilev; [ \vkw;cfext; ]}
\vkw;ilev; is an integer ranging from 0 to 4 indicating the ``level''
of installation:
\begin{description}
\item[0] Complete installation of the AIRES system. Necessary only
when installing AIRES for the first time.
\item[1] Upgrade of an existing installation, making the installed
version the new current version.
\item[2] {\em Recompiling.\/} All the simulation programs and the
summary program are compiled and linked. The AIRES object
library\index{AIRES object library} is rebuilt.
\item[3] {\em Relinking.\/} New executables for all the simulation
programs and the summary program are created using the existing object
files.
\item[4] {\em Rebuilding the library.\/} The AIRES object library is
rebuilt using the existing object files.
\end{description}
\vkw;cfext; is an optional argument. It is a character string
indicating that an alternative configuration file must be used to set
the installation parameters. If \vkw;cfext; is no null, then the file
\kwbf;config.;\vkw;\/cfext; is used instead of the default \vkw;config;
file used when \vkw;cfext; is not specified.

To perform different compilation/installation jobs, it might be useful
to have several configuration files. For example, the \vkw;config;
file is first copied to a new \kwbf;config.short; file. Then
\kwbf;config.short; is edited changing the following parameters:
\iiset\ii The format for both ground and longitudinal tracking
compressed files is set to ``short''. \ii The name of the executable
program \kwbf;Aires; is changed into \kwbf;Aires\_sht;. Finally the
command
\kwexample{doinstall 2 short}
is executed. This will generate several new executable programs,
namely, \kwbf;Aires\_sht;, \kwbf;Aires\_shtS23;, etc., which will be
capable of producing compressed files with short format particle
records.

%
\chapter{IDL reference manual}%
\index{Input Directive Language!reference manual}
\label{A:idlref}

All the main simulation programs and the summary program
\kwbf;AiresSry;\index{AIRES summary program} use a common language to
receive the user's instructions.  This language is called {\em Input
  Directive Language\/} (IDL), and currently consists of some 70
different instructions to set simulation parameters, control the
output data, etc. In this section we list, alphabetically ordered, all
AIRES {\currairesversion} IDL directives.

The IDL directives can be written using no special format\index{Input
Directive Language!format}, with one directive per line (there are no
``continuation lines'', but each line can contain up to 176
characters). The directives start with the directive name followed by
the corresponding parameters. All the ``words'' that form a sentence
must be separated by blanks and/or tab characters.

All directives are scanned until either an \idlbfx;End; directive or an
end of file is found. Most directives can be placed in any order
within the input stream. The \idlbfx;Input; directive permits
inserting instructions placed in separate files letting the user to
conveniently organize complex input data sets. \kwbf;Input; directives
can be nested.

{\em Dynamic\/}\index{Input Directive Language!dynamic/static directives}
(can be set every time the input file is scanned),
{\em static\/} (can be set only at task initialization time) and {\em
hidden\/}\footnote{Hidden
directives are connected to parameters that seldom need to be
modified. They are not printed in the input data summary, unless were
explicitly set or a full listing mode was enabled. Notice that this
only affects output data printing: All other directive properties
remain unchanged.}\index{Input Directive Language!hidden directives}
 (associated with rarely changing parameters) directives are
respectively marked as {\bf d}, {\bf s}, {\bf h}. Names in
\kw;typewriter; or \kwbf;boldface; font refer to keywords, while names in
\vkw;italics\/; refer to variable parameters. \akwbf;Underlined;;
parts of keywords refers to shortest abbreviations: Not underlined
characters are optional. Expressions between
square brackets ([ {\it expression\/} ]) are optional, while alternatives
are written in the following way: \{ {\it alt\_1\/} $|$ {\it alt\_2\/} \}.
To specify angles, lengths, times, energies, atmospheric depths,
magnetic fields, etc., it is required to give two fields separated by
blank space:
\kwexample{\vkw;number; \vkw;unit;}
\vkw;number; is a decimal number and \vkw;unit; is a character string
representing the physical unit used in the specification.  All the
valid units are listed in table \ppref{TAB:idlunits}.  Additionally,
time specifications may be of the form: [ \vkw;number; \kwbf;hr; ] [
\vkw;number; \kwbf;min; ] [ \vkw;number; \kwbf;sec; ], where
\vkw;number; represents a floating point number.

%
%
%
\def\bexprule{\nopagebreak\vrule height4ex width0pt}
\def\sdir{\bexprule {\bf (s) }}
\def\ddir{\bexprule {\bf (d) }}
\def\shdir{\bexprule {\bf (s,h) }}
\def\stx#1{\nopagebreak\makebox[4em][l]{\em Syntax:}
\parbox[t]{12.8cm}{\raggedright\strut#1\strut}\hfil\break}
\def\stxp#1{\nopagebreak\makebox[4em][l]{\ }
\parbox[t]{12.8cm}{\raggedright\strut#1\strut}\hfil\break}
\def\aop#1{\{ #1 \}}\def\asep{ $|$ }
\def\dft#1{\nopagebreak\makebox[4em][l]{\em Default:}
\parbox[t]{12.8cm}{\raggedright\strut#1\strut}\hfil\break}
\def\dftp#1{\nopagebreak\makebox[4em][l]{\ }
\parbox[t]{12.8cm}{\raggedright\strut#1\strut}\hfil\break}
\def\sdswitch;#1;#2; #3 #4
{\stx{\akw;#1;#2\ ; [ \aop{\akw;On;;\asep\akw;Off;;} ]}
\dft{\kw;#1#2\ ; is equivalent to \kw;#1#2 #3;}
\dftp{\kw;#1#2 #4\ ; is assumed in case of missing specification.}}
\def\sdbf;#1;#2; #3 #4
{\stx{\nopagebreak\akw;#1;#2\ ; [ \aop{\kw;Brief;\asep\kw;Full;} ]}
\dft{\kw;#1#2\ ; is equivalent to \kw;#1#2 #3;}
\dftp{\kw;#1#2 #4\ ; is assumed in case of missing specification.}}
{\raggedbottom
\section{List of IDL directives.}
\begin{description}
\def\idl #1
{\pagebreak[3]\item[{\color{airesblue}\large\bf
#1}]\ifcat\##1\else\label{IDL0:#1}\fi\idlbfpx;#1;\hfil\break\nopagebreak}
%
%
\idl \#
\bexprule Comment character. For every scanned input line,
all characters placed after the comment character `\#' are ignored.
%
\idl \&
\stx{\akw;\&;;\vkw;label;}
\bexprule IDL label. Labels are used by several directives, for
example \kwbf;Remark; and
\kwbf;Skip;. The \kwbf;\&; must be the first non-blank character in the
line, and all characters after \vkw;label; are treated as a
comment. \vkw;label; is a non null string which can contain any
character excluding blanks and the comment character \kwbf;\#;.
%
\idl AddAtmosModel
\stx{\akw;AddAtm;osModel\ ;  [ \vkw;modidstr; ]\ 
                             [ \kw;<<; ]\  \vkw;input\_file;}
\stxp{\akw;AddAtm;osModel\ ; [ \vkw;modidstr; ]
                             [ \kw;<<; ]\  \kw;\&;\vkw;label;}
\stxp{{\it\ . . . \ \ (instructions for model definition)}}
\stxp{\kw;\&;\vkw;label;}
\ddir Adding a custom atmospheric model. \vkw;modidstr; is a string
having no more that 16 characters that uniquely identifies the model
being defined. The instructions that define the model can either be
read-in from a separate external file \vkw;input\_file;, or from a
here-document\index{here-document} delimited by a label
\kw;\&;\vkw;label;. \vkw;modidstr; can be specified either in the
directive line, or within the input data file or here-document. For a
more detailed description of the directives to define an atmospheric
model, see section \ppref{S:atmosidl}.
%
\idl AddSite
\stx{\akw;AddSite;\ ; \vkw;name\ ; \vkw;lat\ ; \vkw;long\ ; \vkw;height;}
\ddir Appending a new site to the {\em AIRES site library\index{AIRES site
library}.\/} \vkw;name; is a string having no more than 16 characters, and
must be different to all the previously defined sites including the
predefined entries listed in table \ppref{TAB:sitelib}. Site names are case
sensitive. \vkw;lat; and \vkw;long; are angle specifications defining
respectively the geographic latitude and longitude of the site. \vkw;lat;
(\vkw;long;) must be in the range $[-90^\circ,90^\circ]$
($[-180^\circ,180^\circ]$). \vkw;height; is a length specification
defining the site's altitude above sea level.  The directive \idlbfx;Site;
permits to select already defined locations.
\clearpage
\idl AddSpecialParticle
\stx{\akw;AddSpec;ialParticle\ ; \vkw;pname\ ; \vkw;module\ ;
                                 [ \vkw;parstring; ]}
\stxp{\akw;AddSpec;ialParticle\ ; \vkw;pname\ ; \vkw;module\ ;
           [ \vkw;parstring; ] \kw;<<; \vkw;input\_file;}
\stxp{\akw;AddSpec;ialParticle\ ; \vkw;pname\ ; \vkw;module\ ;
           [ \vkw;parstring; ] \kw;<<; \kw;\&;\vkw;label;}
\stxp{\vkw;\ \  . . .\ ; \vkw;(input data for the external module);}
\stxp{\kw;\&;\vkw;label;}
\ddir Adding a new definition to the list of special
particles\index{special primary particles}. \vkw;pname; is a string
having no more than 16 characters that uniquely identifies the special
particle being defined. \vkw;module; is the name of the executable
module associated to the special particle. The file \vkw;module; must
exist in the current ``working directory''\index{AIRES file
  directories!working directory} or in one of the directories
currently included in the file search path. Every time a new shower
with ``primary'' \vkw;pname; starts, the module \vkw;module; will be
executed by the main simulation program to generate a list of
(standard) primary particles that will be the actual shower
primaries. Section \ppref{S:specialprim} contains a detailed
description about how to build and use such kind of
modules. \vkw;parstring; is an optional parameter string (can contain
embedded blanks) that is (portably) passed to the external
module. Additional input data can be passed to the external module via
a \vkw;input\_file; or by a here-document\index{here-document}
delimited by a label \kw;\&;\vkw;label;.
%
\idl ADFile
\sdswitch;ADF;ile; On Off
\ddir If \kwbf;ADFile On; is specified, then an ASCII dump
file\index{internal dump file!portable format} will be generated upon task
completion. The ASCII dump file (ADF) is a portable version of the
internal dump file (IDF) that can be transferred among different
platforms.
%
\idl AirAvgZ/A
\stx{\akw;AirAvgZ/A;\ ; \vkw;number;}
\dft{\kw;AirAvgZ/A 0.5;}
\shdir Sets the value of the average ratio $Z/A$ for air.

 This directive belongs to the model-dependent IDL instruction set and
 may be changed or not implemented in future versions of AIRES.
\clearpage
\idl AirRadLength
\stx{\akw;AirRadL;ength\ ; \vkw;number;}
\dft{\kw;AirRadLength 36.62;}
\shdir Sets the value of the radiation length for air, expressed in
\gcmsq.

 This directive belongs to the model-dependent IDL instruction set and
 may be changed or not implemented in future versions of AIRES.
%
\idl AirZeff
\stx{\akw;AirZ;eff\ ; \vkw;number;}
\dft{\kw;AirZeff 7.3;}
\shdir Sets the value of the effective atomic number $Z$ for air.

 This directive belongs to the model-dependent IDL instruction set and
 may be changed or not implemented in future versions of AIRES.
%
\idl Atmosphere
\stx{\akw;Atmos;phere\ ; \vkw;modname; [ \vkw;modpars; ]}
\dft{\kw;Atmosphere Linsley;}
\sdir Switches among different atmospheric models\index{atmospheric
model}. \vkw;modname; is a string that uniquely identifies the atmospheric
model to use. The optional argument(s) represented by \vkw;modpars;
correspond to parameters that some of the supported models may accept.
The current version of AIRES includes the following predefined
atmospheric models:
\begin{enumerate}
\item \kwbf;Linsley;. Linsley's standard
atmosphere\index{US standard atmosphere} model. This is the default
model and has no parameters (see section \ref{S:atmosphere}).
\item \kwbf;SouthPoleAvg;. South Pole average atmosphere. This model
  has no parameters, and gives a profile obtained from
  the average of four atmospheric profiles corresponding to typical
  profiles for the months of March, July, October, and December. This
  option is recommended for simulations at the South Pole site. 
\item \kwbf;LSouthPole;.  Linsley's model for the South Pole. No
  user-settable model parameters. This model should be used only for
  simulations with ground level not less than 2000
  m.a.s.l.
\item \kwbf;MalargueAvg;. Malargue site annual average atmosphere.
\item \kwbf;GAMMA;. GAMMA atmosphere model developed by the La Plata
(Argentina) group \cite{JuanCruzAtm}. Parameter available:
  \kwbf;GrdTemp;, temperature at ground.
\item \kwbf;Isothermic;. Isothermic atmosphere. Parameters available:
  \kwbf;Temp;, temperature; \kwbf;Dens0;, density at sea level.
\item \kwbf;Homogeneous;. Constant density atmosphere. Parameter
  available: \kwbf;Density;\hfil\break
  (e.g., \kw;Density; \kw;1.22 kg/m3;); if
  not specified it defaults to 1.2041 kg/m$^3$
\end{enumerate}
Not specified ground temperatures default to 295 K.

In addition to the predefined atmospheric models, the user may add
custom models using the directive \idlidl;AddAtmosModel;.
%
\idl Brackets
\stx{\akw;Bra;ckets\ ; \aop{\akw;On;;\asep\akw;Off;;}}
\stxp{\akw;Bra;ckets\ ; [ \akw;On;; ] \vkw;ob; \vkw;cb; [ \vkw;ec; ]}
\dft{\kw;Brackets On $\{$ $\}$ \&;}
\ddir Controls the behavior of the variable replacement algorithm
used while scanning the input file. When the feature is disabled
(\kwbf;Brackets Off;) the input lines are not scanned to search for
defined variables to be replaced. When \kwbf;Brackets On; is in effect
and there are defined variables, then variable substitution is
performed when it corresponds. The active variable names must be
enclosed using the current brackets, which can be changed using this
directive. The arguments \vkw;ob;, \vkw;cb;, and \vkw;ec; correspond,
respectively, to the opening and closing brackets, and the bracket
escape character. These single character variables must be different,
and can be specified with the same rules that apply for the argument
of the \kwbf;CommentCharacter; directive.
%
\idl CheckOnly
\sdswitch;Check;Only; On Off
\ddir When \kwbf;CheckOnly; is enabled, the simulation program reads
and process all the input data normally, performs the internal
consistency checks and then exits without starting the
simulations. This directive is useful for input file debugging.
%
\idl CommentCharacter
\stx{\akw;CommentChar;acter\ ; \aop{\vkw;char;\asep\vkw;nnn;}}
\dft{The default comment character is `\#'}
\ddir The plain text files produced with the
\kwbf;ExportTables;\index{exported data files} directive can have
heading and trailing lines. All these lines start with a comment
character\index{comment characters in output files, changing} in
their first column. The default comment character (`\#') is normally
OK, but if the \kwbf;Export;'ed files could be used as input of
another program (a plotting utility, for example) which recognizes a
different comment character; in such cases the \kwbf;CommentCharacter;
directive permits setting this mentioned character. \vkw;char; can be
any {\em single\/} character (with no quotes). Alternatively, the
comment character can be specified by means of its ASCII decimal code,
expressed in the form of a {\em three-figure\/} number \vkw;nnn; (This
permits using non-printable comment characters as well as resetting
the comment character to `\#').
\clearpage
\idl Date
\stx{\akw;Date;\ ; \vkw;fpyear;}
\stxp{\akw;Date;\ ; \vkw;year\ ; \vkw;month\ ; \vkw;day;}
\dft{The current date at the moment of invoking the program.}
\sdir This directive sets the date assumed for the simulations. The
date is used at the moment of evaluating the geomagnetic
field\index{geomagnetic field} by means of
the\index{external packages}
IGRF\index{International Geomagnetic Reference Field}
model (see sections \ref{S:geomag} and
\ref{S:geomagidl}). Setting the date may be necessary when performing
simulations with the purpose of analyzing a certain air shower event
reported by an experiment. The date can be specified either as three
integers (\vkw;year; \vkw;month; \vkw;day;) or a floating point number
with the format ``year.part\_of\_the\_year''.
%
\idl DelGlobal
\stx{\akw;DelGl;obal\ ; \vkw;var;}
\ddir Deletes an already defined global variable\index{global
variables}. See also directives
\kwbf;Import; and \kwbf;SetGlobal;.
%
\idl DielectricSuppression
\sdswitch;DielectricSup;pression; On On
\shdir Switch to include/exclude the dielectric
 suppression\index{dielectric suppression} effect from the LPM
 algorithms\index{LPM effect} \cite{LPMigdal,KleinRev} for the case of
 elec\-tron or positron bremsstrahlung\index{bremsstrahlung}. The
 effect is {\bf enabled} by default. Disabling it may lead to non
 realistic air shower simulations. If \kwbf;LPMEffect Off; is in
 effect (see page \pageref{IDL0:LPMEffect}), then the dielectric
 suppression is always disabled.

 This directive belongs to the model-dependent IDL instruction set and
 may be changed or not implemented in future versions of AIRES.
%
\idl DumpFile
\stx{\akw;Dump;File;}
Reserved for future use.
%
\idl Echo
\stx{\akw;Echo;\ ; \vkw;string;}
\ddir The action of this directive is to write \vkw;string; to the
standard output channel. Useful to have messages typed while the
AIRES input file(s) are being processed. Notice that the message is
written {\em only\/} to standard output: use the \idlidl;Remark;
directive if it is necessary to save it within the AIRES output files.
\clearpage
\idl ElectronCutEnergy
\stx{\akw;ElectronCut;Energy\ ; \vkw;energy;}
\dft{\kw;ElectronCutEnergy 80 KeV;}
\sdir Minimum kinetic energy for electrons and positrons. Every
electron having a kinetic energy below this threshold\index{threshold
energies} is not taken into account in the simulation; positrons are
forced to annihilation\index{positron annihilation}.
 \vkw;energy; must be greater than or equal to 80 keV.
%
\idl ElectronRoughCut
\stx{\akw;ElectronR;oughCut\ ; \vkw;energy;}
\dft{\kw;ElectronRoughCut 900 KeV;}
\sdir Electrons and positrons are not followed using detailed
calculations when their energy\index{threshold energies} is below the
one specified by means of this directive. This means that several
processes are not taken into account, for example Coulomb scattering.
\vkw;energy; must be greater than or equal to 45 keV.

 This directive belongs to the model-dependent IDL instruction set and
may be changed or not implemented in future versions of AIRES.
%
\idl ELimsTables
\stx{\akw;ELimsT;ables\ ; \vkw;minenergy\ ; \vkw;maxenergy;}
\dft{\kw;ELimsTables 10 MeV\ ; \vkw;emax;}
\dftp{\vkw;emax; is the maximum between 10 TeV and $0.75\,
E_{\mathrm{primary}}$.}
\sdir This directive defines the energy interval to use in the energy
distribution\index{energy distributions} tables (histograms). Each energy
distribution histogram consists of 40 logarithmic bins starting with
\vkw;minenergy; (lower energy of bin 1) and ending with \vkw;maxenergy;
(upper energy of bin 40).
%
\idl EMtoHadronWFRatio
\stx{\akw;EMtoHadronWF;Ratio\ ; \vkw;ratio;}
\dft{\kw;EMtoHadronWFRatio 88;}
\shdir Ratio between the electromagnetic and hadronic
thinning\index{thinning} weight factors\index{statistical weight
factor}. This instruction permits setting the ratio
$A_{EM}$ of equation (\ref{eq:Wfactor}). \vkw;ratio;
must be equal or greater than 1. The default value of 88, adjusted
taking into account the results of representative simulations, is
normally adequate.
%
\idl End
\stx{\akw;End;;}
\ddir End of directive stream for the current input file. The file is
no more scanned when this directive is found. If \kwbf;End; is not
present, the file is entirely scanned.
%
\idl Exit
\stx{\akw;Exi;t;}
\stxp{\akw;x;;}\idlbfpx;x;
\ddir The program is stopped without taking any further action. This
directive is useful to end an interactive session.
%
\idl ExportPerShower
\sdswitch;ExportP;erShower; On Off
\ddir This directive affects only those tasks simulated with the
\idlbfx;PerShowerData; \kwbf;Full; option (see page
\pageref{IDL0:PerShowerData}). If \kwbf;ExportPerShower On; is
specified, then a set of plain text files\index{exported data files!for
single showers} (one file per simulated shower) will be written for all
the tables selected for exporting (see directive
\kwbf;ExportTables;). Each one of these ``single shower'' tables contains
the values adopted by the corresponding observable in the respective
shower. The normal table containing the average over showers is also
exported, and is not affected by this directive.
%
\idl ExportTables
\stx{\akw;Export;Tables\ ; \vkw;mincode; [ \vkw;maxcode; ] [
\akw;O;ptions\ ; \vkw;optstring; ]}
\stxp{\akw;Export;Tables\ ; \akw;Cl;ear;}
\dft{No tables are exported by default.}
\ddir Tables whose codes range from \vkw;mincode; to \vkw;maxcode; are
selected for exporting as plain text files\index{exported data
files}. If \vkw;maxcode; is not specified, it is taken equal to
\vkw;mincode;. The table codes are integers. A complete list of
available tables (more than 180) is placed in appendix B, or can be
obtained with directives \kwbf;Help tables; and/or
\kwbf;TableIndex;. The \kwbf;Clear; option permits clearing the list
of exported tables, thus overriding all the previous
\kwbf;ExportTables; directives. \vkw;opstring; is a string of
characters to set available options: \kwbf;s; (\kwbf;h;) suppress
(include) file header; \kwbf;x; (\kwbf;X;) include ``border'' bins as
comments (within the data); \kwbf;U; do not include ``border'' bins;
\kwbf;r; (\kwbf;d;) normal (density) lateral distributions; \kwbf;L;
(\kwbf;l;) distributions normalized as $d/d\log_{10}$ ($d/d\ln$); \kwbf;r;
(\kwbf;a;) express atmospheric depth as vertical (slant) depths;
\kwbf;K;, \kwbf;M;, \kwbf;G;, \kwbf;T;, \kwbf;P;, \kwbf;E;, express
energies in keV, MeV, \ldots, EeV. The default options are:
\kwbf;hxrG;.
%
\idl ExtCollModel
\sdswitch;ExtColl;Model; On On
\sdir Switch to enable/disable the external hadronic interactions
model\index{hadronic models}\index{external packages}.

 This directive belongs to the model-dependent IDL instruction set and
may be changed or not implemented in future versions of AIRES.
%
\idl ExtNucNucMFP
\sdswitch;ExtNucNuc;MFP; On On
\shdir Switch to enable/disable calculation of mean free
 paths\index{mean free path!nucleus-nucleus collisions} for
 nucleus-nucleus collisions\index{nucleus-nucleus collisions} via the
 corresponding external hadronic interactions model\index{hadronic
   models}\index{external packages}.

 If the switch is set to \kwbf;Off; then the nucleus-nucleus mean free
 paths are evaluated using an AIRES built-in procedure. In this case
 the mean free paths are obtained by scaling properly the
 corresponding proton-nucleus mean free path.

 This directive belongs to the model-dependent IDL instruction set and
may be changed or not implemented in future versions of AIRES.
%
\idl FileDirectory
\stx{\akw;FileDir;ectory\ ; \vkw;dopt\ ; \vkw;directory;}
\dft{The output and scratch directories default to the current
(working) directory. The global and export directories default to
the current value of the output directory.}
\ddir This directive sets the output file directories. \vkw;dopt; is a
character string that can take any one of the following values:
\akw;A;ll;, \akw;O;utput;\index{AIRES file directories!output directory},
\akw;G;lobal\index{AIRES file directories!global directory};,
\akw;E;xport\index{AIRES file directories!export directory};, or
\akw;S;cratch;\index{AIRES file directories!scratch directory}. These
alternatives permit setting all the AIRES directories defined in section
\ppref{S:filedir}. The option \kwbf;All; can be used to simultaneously set
the ``output'' (compressed file), ``global'' and ``export''
directories. \vkw;directory; is a character string not longer than 94
characters that must be recognized by the operating system as a valid
directory.
%
\idl FirstShowerNumber
\stx{\akw;FirstSh;owerNumber\ ; \vkw;fshowerno;}
\dft{\kw;FirstShowerNumber 1;}
\sdir A positive integer in the range $[1, 759375]$ indicating the number
to be assigned to the first simulated shower\index{first shower
number}. The shower number is used in tables 5000 to 5513, and in the
``beginning of shower'' and ``end of shower'' compressed file records (for
details see chapter \ref{C:output1}).
%
\idl ForceInit
\sdswitch;ForceInit;; On Off
\ddir If \kwbf;ForceInit; is enabled, then a new task is started at the
beginning of every process. If the corresponding IDF file exists,
then the task version is increased until an unused version is found. This
directive is useful for debugging purposes.
%
\idl ForceLowEAnnihilation
\stx{\akw;ForceLowEA;nnihilation\ ; \vkw;opt;}
\dft{\kw;ForceLowEAnnihilation Normal;}
\dftp{\kw;ForceLowEAnnihilation; with no specification is equivalent
to \kw;ForceLowEAnnihilation Always;.}
\shdir Directive to control the action to take when processing a low
energy particle\index{low energy particles!annihilation} that can
annihilate with its respective anti-particle. The variable \vkw;opt;
can take the values \akw;Al;ways;, \akw;Ne;ver;, or \akw;No;rmal;. The
first two alternatives correspond, respectively, to the cases where
the low energy particles will always be forced to annihilation or be
discarded without producing any secondary particle. In the (default)
\kwbf;Normal; option the action to take for annihilating low energy
particles depends on the particle cut energy and mass: If the cut
energy is less (greater) than the rest mass then the particle is (is
not) forced to annihilation.
%
\idl ForceLowEDecays
\stx{\akw;ForceLowED;ecays\ ; \vkw;opt;}
\dft{\kw;ForceLowEDecays Normal;}
\dftp{\kw;ForceLowEDecays; with no specification is equivalent
to \kw;ForceLowEDecays Always;.}
\shdir Directive to control the action to take when processing a low
energy unstable particle\index{low energy particles!decay} that
can decay into other particles.  The variable \vkw;opt; can take the
values \akw;Al;ways;, \akw;Ne;ver;, or \akw;No;rmal;. The first two
alternatives correspond, respectively, to the cases where the low
energy particles will always be forced to decays or be dicarded
without producing any secondary particle. In the (default)
\kwbf;Normal; option the action to take for decaying low energy
particles depends on the particle cut energy and mass: If the cut
energy is less (greater) than the rest mass then the particle is (is
not) forced to decays.
%
\idl ForceModelName
\stx{\akw;ForceMod;elName\ ; \vkw;modsel;}
\dft{No model name check is performed when this directive is not used.}
\sdir This directive allows the user to force that a given input data
set will be processed with the simulation program linked with the
external collision package\index{hadronic models}\index{external
  packages} specified with \vkw;modsel;. Currently \vkw;modsel; can be
one of (case dependent!)  EPOS-LHC3400\index{EPOS}, EPOS1990,
QGSJET-II-04\index{QGSJET}, QGSJET-II-03, SIBYLL23c\index{SIBYLL},
SIBYLL231, or SIBYLL21. This directive is useful as a security tool to
allow execution of simulations only if the executable being used is
the one that corresponds to the selected hadronic model. Consider also
using it together with \idlbfx;StopOnError;.

 This directive belongs to the model-dependent IDL instruction set and
 may be changed or not implemented in future versions of AIRES.
\clearpage
\idl GammaCutEnergy  
\stx{\akw;GammaCut;Energy\ ; \vkw;energy;}
\dft{\kw;GammaCutEnergy 80 KeV;}
\sdir Minimum energy for gammas. Every gamma ray having an energy
below this threshold\index{threshold energies} is not taken into
account in the simulation. \vkw;energy; must be greater than or equal
to 80 keV.
%
\idl GammaRoughCut
\stx{\akw;GammaR;oughCut\ ; \vkw;energy;}
\dft{\kw;GammaRoughCut 750 KeV;}
\sdir Gamma rays are not followed using detailed
calculations when their energy is below the one specified by means of
this directive. This means that several processes are not taken into
account, for example pair production\index{pair production}.
\vkw;energy; must be greater than or equal to 45 keV.

 This directive belongs to the model-dependent IDL instruction set and
may be changed or not implemented in future versions of AIRES.
%
\idl GeomagneticField
\stx{\akw;Geomag;neticField\ ; [ \aop{\akw;On;;\asep\akw;Off;;} ]}
\stxp{\akw;Geomag;neticField\ ; \vkw;stg; [ \vkw;inc; [
\vkw;dec; ] ] [ \akw;F;luctuations\ ; \vkw;fluc; ]}
\stxp{\akw;Geomag;neticField\ ; [ \akw;On;; ]
 \ \ \akw;F;luctuations\ ; \vkw;fluc;}
\dft{\kw;GeomagneticField Off\ ; when there is no \kw;Site;\idlidx;Site;
specification; \kw;GeomagneticField On\ ;  otherwise.}
\sdir Setting the geomagnetic field\index{geomagnetic field} manually
and/or enabling magnetic fluctuations. \vkw;stg; must be a valid
magnetic field strength specification, and \vkw;inc; and \vkw;dec; are
angle specifications. Such fields correspond respectively to the
geomagnetic field strength, F, and to the inclination, I, and
declination, D, angles defined in section \ppref{S:geomag}. When one or
more of such parameters are entered by means of the
\kwbf;GeomagneticField; directive, they override the respective values
that are calculated automatically using the IGRF model
\cite{IGRF}\index{International Geomagnetic Reference Field}%
\index{external packages}, as explained in section
\ppref{S:geomagidl}. The fluctuation\index{geomagnetic
field!fluctuations} specification \vkw;fluc; adopts three different
formats: \iiset\ii {\em Absolute:\/} In this case \vkw;fluc;
represents a (positive) magnetic field strength. \ii {\em Relative:\/}
\vkw;fluc; adopts the format \vkw;number; \akwbf;R;elative;, and
refers to the ratio between the actual fluctuation strength and the
average value of the magnetic field. \ii {\em In percent:\/}
\vkw;fluc; adopts the format \vkw;number; \akwbf;\%;;. \vkw;number;
corresponds to a relative specification multiplied by 100. The effect
of magnetic field fluctuations is explained in section
\ppref{S:geomagidl}.
\clearpage
\idl GroundAltitude
\stx{\akw;Ground;Altitude\ ; \vkw;altdepth;}
\stxp{\akw;GroundD;epth\ ; \vkw;altdepth;}%
\index{Input Directive
Language!directives!GroundDepth@{\kw;GroundDepth; (synonym of
\kw;GroundAltitude;)}}
\dft{The altitude of the site currently in effect.}
\sdir Ground level altitude. \vkw;altdepth; can be either a length
specification (ranging from 0 to 112 km) or an atmospheric depth
specification (ranging from 0 to 1033 g/cm$^2$).
%
\idl Help
\stx{\akw;He;lp\ ; [ \aop{\kw;*;\asep\akw;table;s;\asep\akw;site;s;} ]}
\stxp{\akw;he;lp\ ; [ \aop{\kw;*;\asep\akw;table;s;\asep\akw;site;s;} ]}
\stxp{\akw;?;\ ; [ \aop{\kw;*;\asep\akw;table;s;\asep\akw;site;s;}
                 ]}\idlbfpx;?;
\ddir The action of the \kwbf;Help; directive is to type a brief
summary of IDL directives, output data tables (histograms) or sites
defined in the AIRES site library\index{AIRES site
library}. \kwbf;Help *; gives a full IDL directive list, including all
``hidden'' directives. The \kwbf;?; form is equivalent to the combined
action of \kwbf;Help; and \kwbf;Prompt On;
%
\idl Import
\stx{\akw;Imp;ort\ ; [ \aop{\akw;Dyn;amic;\asep\akw;Sta;tic;} ] 
     \vkw;varname;}
\dft{No environmental variables are imported by default.}
\ddir Importing environment variables. The operating system
environment variable \vkw;varname; is imported and stored as an active
variable that can either be used within the IDL input stream or passed
to the compressed output files or special primary
modules\index{special primary particles}. The
\kwbf;Dynamic; qualifier (default) indicates the {\em dynamic\/}
character of the corresponding variable. This means that the value
currently passed to the external modules is modified each time AIRES
is invoked for a given task. On the other hand, \kwbf;Static;
variables are set at the first invocation of AIRES; and further
settings have no effect.
%
\idl ImportShell
\stx{\akw;ImportSh;ell\ ; [ \aop{\akw;Dyn;amic;\asep\akw;Sta;tic;} ]
     \vkw;pname\ ; \vkw;shell\_instruction;}
\dft{Nothing imported by default.}
\ddir Importing the output of shell commands. The data written to
standard output when the operating system executes the instruction
\vkw;shell\_instruction; is imported and stored as the active variable
\vkw;pname; that can either be used within the IDL input stream or
passed to the compressed output files or special primary
modules\index{special primary particles}. The
\kwbf;Dynamic; qualifier (default) indicates the {\em dynamic\/}
character of the corresponding variable. This means that the value
currently passed to the external modules can be modified each time
AIRES is invoked for a given task. On the other hand, \kwbf;Static;
variables are set at the first invocation of AIRES; and further
settings have no effect.
%
\idl InjectionAltitude
\stx{\akw;Inject;ionAltitude\ ; \vkw;altdepth;}
\stxp{\akw;InjectionD;epth\ ; \vkw;altdepth;}%
\index{Input Directive
Language!directives!InjectionDepth@{\kw;InjectionDepth; (synonym of
\kw;InjectionAltitude;)}}
\dft{\kw;InjectionAltitude 100 km;}
\sdir Primary injection altitude. \vkw;altdepth; can be either a length
specification (ranging from 0 to 112 km) or an atmospheric depth
specification (ranging from 0 to 1033 g/cm$^2$).
%
\idl Input
\stx{\akw;Inp;ut\ ; \vkw;file\ ; 
          [ \vkw;arg1; [ \vkw;arg2; [ \vkw;. . .; ] ] ]}
\ddir File \vkw;file; is inserted in the input data
stream. \kwbf;Input; directives can be nested. The search path for
locating input files include the ``working directory''\index{AIRES
  file directories!working directory} (see section \ref{S:filedir}),
the directories that are included by default, and all the directories
that were specified with directive \idlbfx;InputPath;. Optionally, it
is possible to pass arguments to the included file. They are assigned
to global variables\index{global variables} named ``\kwbf;1;'',
``\kwbf;2;'', ..., and will be visible for the directives placed
within \kwbf;file;.
%
\idl InputListing
\sdbf;InputLis;ting; Brief Brief
\ddir Data related to hidden input directives are not printed in the
output summary file unless the corresponding variables were explicitly
set or \kwbf;InputListing Full; was specified.
%
\idl InputPath
\stx{\akw;InputP;ath\ ; [ \aop{\akw;Ins;ert;\asep\akw;App;end;} ] %
       [\vkw;dir1;[:\vkw;dir2;[: . . . ] ] ]}
\dft{The file search path contains by default the ``working
  directory''\index{AIRES file directories!working directory}, and the
  directories coming with the AIRES distribution that contain input
  files and/or executable modules that could be invoked from within a
  set of IDL directives.}
\ddir Modifying the directory search path for the files included with
the \idlbfx;Input; directive and/or other directives that request
external files. This directive can be used multiple times if
required. Different search directories can be specified in a single
invocation separating them with colons (:) with no embedded
blanks. The keyword \kwbf;Append; indicates that the specified
directory(ies) must be appended to the ones already inserted, while
\kwbf;Insert; indicates that such directory(ies) must be inserted at
the beginning of the current path string. If \kwbf;InputPath; is
invoked with no arguments, then the search path is cleared.
\clearpage
\idl LaTeX
\sdswitch;LaT;eX; On Off
\ddir If \kwbf;LaTeX On; is specified, then the output summary file is
written using the {\LaTeX}\index{LaTeX format for summary
files@{\LaTeX} format for summary files} word processor
format. Otherwise it is written as a plain text file. When this option
is enabled, a \TeX\ file \vkw;taskname;\kwbf;.tex; is created
simultaneously with the summary file.
%
\idl LPMEffect
\sdswitch;LPM;Effect; On On
\shdir Switch to include/exclude the Landau-Pomeranchuk-Migdal
 effect\index{LPM effect} \cite{LPM,LPMigdal} from the
 elec\-tron-positron and gamma propagating algorithms. The effect is
 {\bf enabled} by default. Disabling it may lead to non realistic air
 shower simulations. If \kwbf;LPMEffect Off; is in effect, then the
 dielectric suppression\index{dielectric suppression} is also disabled
 (see page \pageref{IDL0:DielectricSuppression}).

 This directive belongs to the model-dependent IDL instruction set and
may be changed or not implemented in future versions of AIRES.
%
\idl MaxCpuTimePerRun
\stx{\akw;MaxCpu;TimePerRun\ ; \aop{\vkw;time;\asep\akw;Inf;inite;}}
\dft{\kw;MaxCpuTimePerRun Infinite;}
\ddir This directive sets the maximum CPU time for individual runs,
being a {\em run\/} the processing chunk that goes between two
consecutive updates of the internal dump file. This parameter does not
impose any restriction on the CPU time available for the simulation of
a single shower (or a group of them), which is always
infinite. \vkw;time; is any valid time specification. See also
directives \kwbf;RunsPerProcess; and \kwbf;ShowersPerRun;.
%
\idl MesonCutEnergy
\stx{\akw;MesonCut;Energy\ ; \vkw;energy;}
\dft{\kw;MesonCutEnergy 60 MeV;}
\sdir Minimum kinetic energy for mesons (pions, kaons, etc.). Every
meson having a kinetic energy below this threshold\index{threshold
energies} is not taken into account in the simulation; unstable
particles are forced to decays. \vkw;energy; must be greater than or
equal to 500 keV.
%
\idl MFPHadronic
\stx{\akw;MFP;Hadronic\ ; \vkw;mfpsel;}
\sdir Directive to select among different sets of mean free
paths\index{hadronic cross sections}%
\index{mean free path!hadronic}%
\index{mean free path!nucleus-nucleus collisions}
parameterizations. \vkw;mfpsel; is a character string that identifies
the set to be used (see page \pageref{P:hmfp}).

 This directive belongs to the model-dependent IDL instruction set and
 may be changed or not implemented in future versions of AIRES.
%
\idl MFPThreshold
\stx{\akw;MFPThre;shold\ ; \vkw;energy;}
\dft{\kw;MFPThreshold 50 GeV;}
\shdir Threshold energy\index{threshold energies} for the currently
effective mean free paths. All hadronic collisions
with energy greater than or equal to this threshold will be processed
using the current mfp\index{hadronic cross sections} parameterization
(that can be set using directive \kwbf;MFPHadronic;); otherwise
standard MFP's will be used\index{hadronic cross sections!low energy}.
\vkw;energy; must be greater than or
equal to 200 MeV.

 This directive belongs to the model-dependent IDL instruction set and
 may be changed or not implemented in future versions of AIRES.
%
\idl MinExtCollEnergy
\stx{\akw;MinExtCollE;nergy\ ; \vkw;energy;}
\dft{\kw;MinExtCollEnergy 200 GeV; for the SIBYLL\index{SIBYLL} model;
     \kw;MinExtCollEnergy 80 GeV; for the QGSJET\index{QGSJET} model.}
\shdir Threshold energy\index{threshold energies} for invoking the
external hadronic
collision\index{hadronic models}\index{external packages} routine (if
enabled). \vkw;energy; must be greater than or
equal to 25 GeV.

 This directive belongs to the model-dependent IDL instruction set and
may be changed or not implemented in future versions of AIRES.
%
\idl MinExtNucCollEnergy
\stx{\akw;MinExtNucCollE;nergy\ ; \vkw;energypernucleon;}
\dft{\kw;MinExtCollEnergy 200 GeV; for the SIBYLL\index{SIBYLL} model;
     \kw;MinExtCollEnergy 80 GeV; for the QGSJET\index{QGSJET} model.}
\shdir Threshold energy\index{threshold energies} per nucleon for
invoking the external nucleus-nucleus
collision\index{nucleus-nucleus collisions}\index{external packages}
routine (if enabled). \vkw;energypernucleon; must be greater than or
equal to 25 GeV.

 This directive belongs to the model-dependent IDL instruction set and
may be changed or not implemented in future versions of AIRES.
%
\idl MuonBremsstrahlung
\sdswitch;Muonbrem;sstrahlung; On On
\shdir Switch to include/exclude the muon
 bremsstrahlung\index{muon bremsstrahlung} \cite{mubrem} and
 muonic pair production\index{muonic pair production} processes from
 the muon propagating algorithms. These
 interactions are {\bf enabled} by default. Disabling them may lead to
 non realistic air shower simulations.

 This directive belongs to the model-dependent IDL instruction set and
 may be changed or not implemented in future versions of AIRES.
%
\idl MuonCutEnergy
\stx{\akw;MuonCut;Energy\ ; \vkw;energy;}
\dft{\kw;MuonCutEnergy 10 MeV;}
\sdir Minimum kinetic energy for muons\index{threshold
energies}. Every muon having a kinetic energy below this threshold is
not taken into account in the simulation; it is forced to a decay.
\vkw;energy; must be greater than or equal to 500 keV.
%
\idl NuclCollisions
\sdswitch;NuclColl;isions; On On
\shdir Switch to include/exclude the hadronic inelastic collisions
 with air nucleus from the heavy particles propagating algorithms. The
 collisions are {\bf enabled} by default. Disabling them may lead to
 non realistic air shower simulations.

 This directive belongs to the model-dependent IDL instruction set and
may be changed or not implemented in future versions of AIRES.
%
\idl NuclCutEnergy
\stx{\akw;NuclCut;Energy\ ; \vkw;energy;}
\dft{\kw;NuclCutEnergy 120 MeV;}
\sdir Minimum kinetic energy for nucleons and nuclei. Every such
particle having a kinetic energy below this threshold\index{threshold
energies} is not taken into account in the simulation. \vkw;energy;
must be greater than or equal to 500 keV.
%
\idl ObservingLevels
\stx{\akw;Observing;Levels\ ; \vkw;nofol\ ; [ \vkw;altdepth1\ ;
     \vkw;altdepth2; ]}
\dft{\kw;ObservingLevels 19;}
\sdir This directive defines the number and position of the observing
levels used for longitudinal development\index{longitudinal
development} recording (see page \pageref{S:monitoring})
. \vkw;altdepth1; and \vkw;altdepth2; are altitude (or atmospheric
depth) specifications that define the positions of the first and last
observing levels. \vkw;nofol; is an integer that sets the number of
observing levels. It must lie in the range $[4, 510]$. The observing
levels are equally spaced in atmospheric depth units. The first (last)
level corresponds to the highest (lowest) altitude.

If \vkw;altdepth1; and \vkw;altdepth2; are not specified, then the
observing levels are placed between the injection and ground planes,
but spacing them differently (see section \ref{S:showerini}): The
injection level corresponds to observing level ``0'' while the ground
level corresponds to observing level ``$\hbox{\vkw;nofol;} + 1$''. For
example, if the injection (ground) level is placed at 0 (1000)
g/cm$^2$, the directive \kwbf;ObservingLevels 19; will set 19
observing levels placed at depths 50, 100, 150, \ldots, 950 g/cm$^2$.
%
\idl OutputListing
\sdbf;OutputLis;ting; Brief Brief
\ddir Hidden output data items are not printed in the output summary
file unless \kwbf;OutputListing Full; is specified.
%
\idl PerShowerData
\stx{\akw;PerShower;Data\ ; \vkw;option;}
\dft{\kw;PerShowerData\ ; is equivalent to \kw;PerShowerData Full;}
\dftp{\kw;PerShowerData Brief\ ; is assumed in case of missing
specification.}
\sdir Directive to control the amount of individual shower data to be
stored after each shower is completed. \vkw;option; is a character string
that can take any one of the following values: \akw;N;one;, \akw;B;rief;
or \akw;F;ull;. When \kwbf;None; is specified, no individual shower data
is saved. The \kwbf;Brief; level implies saving global parameters such as
the depth of shower maximum
$X_{\mathrm{max}}$\index{Xmax@$X_{\mathrm{max}}$}, for example; and the
\kwbf;Full; level is the \kwbf;Brief; level plus all the single shower
tables (see page \pageref{IDL0:ExportPerShower}).
%
\idl PhotoNuclear
\sdswitch;PhotoNuc;lear; On On
\shdir Switch to include/exclude the inelastic collisions gamma-air
 nucleus (photonuclear reactions\index{photonuclear reactions}) from
 the gamma ray propagating
 algorithms. The collisions are {\bf enabled} by default. Disabling
 them may lead to non realistic air shower simulations.

 This directive belongs to the model-dependent IDL instruction set and
may be changed or not implemented in future versions of AIRES.
\clearpage
\idl PrimaryAzimAngle
\stx{\akw;PrimaryA;zimAngle\ ; \vkw;minang; [ \vkw;maxang; ]
     [ \aop{\akw;M;agnetic;\asep\akw;G;eographic;} ]}
\dft{\kw;PrimaryAzimAngle 0 deg Magnetic ; if the zenith angle is fixed;
\kw;PrimaryAzimAngle 0 deg 360 deg Magnetic ; otherwise}
\dftp{(see \kw;PrimaryZenAngle;).}
\sdir Primary azimuth angle. The angle for each shower is selected
with uniform probability distribution in the interval
$[\hbox{\vkw;minang;}, \hbox{\vkw;maxang;}]$.  If the angle
\vkw;maxang; is not specified, it is taken equal to \vkw;minang;
(fixed azimuth angle). The \kwbf;Geographic;\index{geographic
azimuth}\index{magnetic azimuth} keyword indicates that the specified
azimuth is measured with respect to the {\em geographic\/} north,
positive for eastwards directions; in this case the azimuth angle used
by AIRES is obtained applying equation (\ref{eq:geophi}). If no
keyword or the \kwbf;Magnetic; keyword is specified, then the origin
for the azimuths is the {\em magnetic\/} north, and the given angles
are interpreted accordingly with the orientation of the AIRES
coordinate system\index{AIRES coordinate system} defined in section
\ppref{S:coordinates}.
%
\idl PrimaryEnergy
\stx{\akw;PrimaryE;nergy\ ; \vkw;minener; [ \vkw;maxener; [ \vkw;gamma; ] ]}
\dft{None.}
\sdir Energy of primary. If only \vkw;minener; is specified then all
primaries have a fixed energy equal to this parameter. Otherwise the
energy will be sampled from the interval $[E_{\mathrm{min}},
E_{\mathrm{max}}] = [\hbox{\vkw;minener;}, \hbox{\vkw;maxener;}]$ with
the probability distribution\index{primary energy spectrum} of
equation (\ref{eq:gammadist}) with exponent $\gamma$ optionally
specified by \vkw;gamma;.

The primary energy must be larger than $500$ MeV and less than
$3\times 10^{12}$ GeV ($3 \times 10^{21}$ eV). There are no
restrictions on $\gamma$. If not specified it is set to 1.7.
%
\idl PrimaryParticle
\stx{\akw;Primary;Particle\ ; \vkw;particle; [ \vkw;weight; ]}
\dft{None. This directive is always required.}
\sdir Primary particle specification. \vkw;particle; is the particle
name.  \kwbf;Proton;, \kwbf;Iron;, \kwbf;Fe\^{}56;, etc. are valid
particle names. Special particle names\index{special primary
particles} defined by means of directive \idlbfx;AddSpecialParticle;
can also be used with this instruction. If more than one
\kwbf;PrimaryParticle; directive appear within the input instructions,
then the primary particles will be selected at random among the
different specified particle kinds, with probabilities proportional to
the weights specified in the corresponding \vkw;weight; fields. If
\vkw;weight; is not specified, then the particle weight is taken as 1.
\clearpage
\idl PrimaryZenAngle
\stx{\akw;PrimaryZ;enAngle\ ; \vkw;minang; [ \vkw;maxang; [
\aop{\kw;S;\asep\kw;SC;\asep\kw;CS;} ] ]}
\dft{\kw;PrimaryZenAngle 0 deg;}
\sdir Primary zenith angle, $\Theta$. If only \vkw;minang; is
specified, then the zenith angle is fixed and equal to this value, and
the default for the azimuth angle will be 0. Otherwise the zenith
angle for each shower is selected randomly within the interval
$[\hbox{\vkw;minang;}, \hbox{\vkw;maxang;}]$, with the {\em sine\/}
probability distribution of equation (\ref{eq:sindist}), which is
proportional to $\sin\Theta$ (default or \kwbf;S; specification), or
the {\em sine-cosine\/} probability distribution of equation
(\ref{eq:sincosdist}), which is proportional to $\sin\Theta\cos\Theta$
(\kwbf;SC; or \kwbf;CS; specifications). In this case the default for
the azimuth angle is \kwbf;PrimaryAzimAngle 0 deg 360 deg;. Both
\vkw;minang; and \vkw;maxang; must belong to the interval
$[0^\circ,90^\circ)$.
%
\idl PrintTables
\stx{\akw;Print;Tables\ ; \vkw;mincode; [ \vkw;maxcode; ] [
\akw;O;ptions; \vkw;optstring; ]}
\stxp{\akw;Print;Tables\ ; \akw;Cl;ear;}
\dft{No tables are printed by default.}
\ddir Tables whose codes range from \vkw;mincode; to \vkw;maxcode; are
selected for being displayed in the summary output file.  If
\vkw;maxcode; is not specified, it is taken equal to
\vkw;mincode;. The table codes are integers. A complete list of
available tables (more than 180) is placed in appendix
\ppref{A:outables}, or can be obtained with directives \kwbf;Help
tables; and/or \idlbfx;TableIndex;. The \kwbf;Clear; option permits
clearing the list of printed tables, thus overriding all the previous
\kwbf;PrintTables; directives. \vkw;opstring; is a string of
characters to set available options: \kwbf;n; suppress plotting minimum
(\kw;<;) and maximum (\kw;>;) characters; \kwbf;m; include minimum and
maximum plots in the tables; \kwbf;M; do not insert character plots,
make a completely numerical table instead; \kwbf;S; (\kwbf;R;) use
standard deviations (RMS errors of the means) to plot error bars;
\kwbf;r; (\kwbf;d;) normal (density) lateral distributions; \kwbf;L;
(\kwbf;l;) distributions normalized as $d/d\log_{10}$ ($d/d\ln$). The
default options are: \kwbf;nSr;.
%
\idl Prompt
\sdswitch;Prom;pt; On Off
\ddir Turns prompting on/off. This directive is meaningful only in
interactive sessions.
%
\idl PropagatePrimary
\sdswitch;PropagateP;rimary; On On
\shdir This directive controls the initial propagation of the primary. If
the \kwbf;On; option is selected (the default), then the primary is
normally advanced before the first interaction takes place, and therefore
the first interaction altitude will be variable. Otherwise the first
interaction will be forced to occur at the injection altitude.

This directive is ignored for showers initiated by special
primaries\index{special primary particles} (see section
\ref{S:specialprim}).
%
\idl RandomSeed
\stx{\akw;RandomS;eed\ ; \vkw;seed;}
\stxp{\akw;RandomS;eed\ ; \akw;Get;From\ ; \vkw;idfile;}
\dft{\kw;RandomSeed 0.0;}
\sdir This directive sets the random number generator\index{random
  number generator} seed. \vkw;seed; is a real number. If it belongs
to the interval $(0,1)$ then the seed is effectively taken as the
given number. Otherwise it is evaluated internally (using the system
clock). The alternative syntax with the keyword \kwbf;GetFrom; allows
extracting the random generator seed from an already existing internal
dump file\index{internal dump file}. This is most useful to
reproducing a previous simulation repeating the original random number
simulator configuration.

%
\idl RecordObsLevels
\stx{\akw;RecordOb;sLevels\ ; [ \akw;Not;; ]
                               [ \vkw;lev1; [ \vkw;lev2; [ \vkw;step;
                                   ] ] ] }
\stxp{\akw;RecordOb;sLevels\ ; [ \akw;Not;; ]
                              \aop{\akw;All;; \asep
                                   \akw;All;;/\vkw;step; \asep
                                   \akw;None;; }}
\dft{\kw;RecordObsLevels All;}
\sdir Directive to mark a certain subset of the defined observing
levels for inclusion (or exclusion) in the set of levels that are
included in the longitudinal tracking compressed particle file. The
integer variables \vkw;lev1; \vkw;lev2; and \vkw;step; are the
arguments of a FORTRAN do loop which starts at \vkw;lev1;, ends at
\vkw;lev2; advancing in steps of \vkw;step;. The keyword \kwbf;Not;
indicates that the corresponding levels must be {\em excluded\/} for
being recorded in the file. If \vkw;lev2; and/or \vkw;step; are not
indicated they default to \vkw;lev1; and 1
respectively. \kwbf;RecordObsLevels All;/\vkw;step; is a short form
for \kwbf;RecordObsLevels; 1 $N_o$ \vkw;step;, where $N_o$ is the
number of defined observing levels. \kwbf;RecordObsLevels All; is
equivalent to \kwbf;RecordObsLevels All/1; while \kwbf;RecordObsLevels
None; can be used in place of \kwbf;RecordObsLevels Not All;. This
directive can be repeatedly used within an input instruction stream to
mark or unmark arbitrary subsets of observing levels, as explained in
page \pageref{P:recobslev}.
%
\idl RecordSpecPrimaries
\sdswitch;RecordSpecP;rimaries; On On
\sdir Directive to enable or disable recording in the compressed
output files\index{compressed output files} the list of special
primary particles\index{special primary particles} injected at the
beginning of each shower.
\clearpage
\idl Remark
\stx{\akw;Rem;ark\ ; \vkw;string;}
\stxp{\akw;Rem;ark\ ; \kw;\&;\vkw;label;}
\stxp{{\it First line of remarks.}}
\stxp{{\it \ldots}}
\stxp{{\it Last line of remarks.}}
\stxp{\kw;\&;\vkw;label;}
\sdir Remarks directive. Each time this directive appears in the input
data stream, the corresponding remark string(s) are appended to the
remarks text. All the entered remarks will be printed in the log and
summary files, and stored in different output data files. There is no
limit in the number of remark lines, but every line cannot be longer
than 75 characters. Notice that the \idlidl;Echo; directive can also
be used if what is needed is to type a message to standard output
while the AIRES input file(s) are being processed.
%
\idl ResamplingRatio
\stx{\akw;Resamp;lingRatio\ ; \vkw;rsratio;}
\dft{\kw;ResamplingRatio 10;}
\sdir This directive sets the variable $s_r$ used in the resampling
algorithm\index{resampling algorithm} defined in
section \ppref{def:rsa}. \vkw;rsratio; is a real number that must be greater
or equal than 1.
%
\idl RLimsFile
\stx{\akw;RLimsF;ile\ ; \vkw;filext\ ; \vkw;rmin\ ;  \vkw;rmax;}
\dft{\kw;RLimsFile\ ; \vkw;any\_file\ ; \kw;250 m 12 km;}
\sdir This directive defines the lateral limits for the compressed
data file\index{compressed output files} whose extension is
\vkw;filext;. For the ground particle file, \vkw;rmin; and \vkw;rmax;
define, together with the resampling ratio\index{resampling algorithm}
that is controlled by the IDL instruction \idlbfx;ResamplingRatio; the
radial limits of the zone where the particles are going to be saved
(see page \pageref{def:rsa}). In the case of longitudinal tracking
particle files, those parameters define the inclusion zone at ground
level. At an arbitrary altitude, the particles are included
accordingly with the rules explained in section \ppref{S:ciocustom}.
%
\idl RLimsTables
\stx{\akw;RLimsT;ables\ ; \vkw;rmin\ ; \vkw;rmax;}
\dft{\kw;RLimsTables 50 m 2 km;}
\sdir This directive defines the radial interval to use in the lateral
distribution\index{lateral distributions} tables (histograms). Each
lateral distribution histogram consists of 40 logarithmic bins starting
with \vkw;rmin; (lower radius of bin 1) and ending with \vkw;rmax; (upper
radius of bin 40).
\clearpage
\idl RunsPerProcess
\stx{\akw;RunsPerP;rocess\ ; \aop{\vkw;number;\asep\akw;Inf;inite;}}
\dft{\kw;RunsPerProcess Infinite;}
\ddir Number of runs within a process (see also \kwbf;MaxCpuTimePerRun; and
\kwbf;ShowersPerRun;).
%
\idl SaveInFile
\stx{\akw;Save;InFile\ ; \vkw;filext\ ; \vkw;particle1; [ \vkw;particle2; ]
\ldots}
\dft{\kw;SaveInFile\ ; \kw;grdpcles All;}
\dftp{\kw;SaveInFile\ ; \kw;lgtpcles None;}
\sdir This directive allows to control the particles being saved in
the compressed file\index{compressed output files} whose extension is
\vkw;filext; (see directive \kwbf;RLimsFile;). \vkw;particle1;,
\vkw;particle2;, \ldots, are valid particle or particle group
names. This directive, together with \kwbf;SaveNotInFile; are useful
to save output file space in certain circumstances.
%
\idl SaveNotInFile
\stx{\akw;SaveNot;InFile\ ; \vkw;filext\ ; \vkw;particle1; [ \vkw;particle2; ]
\ldots}
\sdir The syntax of this directive is similar to \kwbf;SaveInFile;, and its
meaning is opposite (\kwbf;SaveInFile; \vkw;filext; \kwbf;None; is equivalent
to \kwbf;SaveNotInFile; \vkw;filext; \kwbf;All;).
%
\idl SeparateShowers
\stx{\akw;SeparateS;howers\ ; \aop{\kw;Off;\asep\vkw;number;}}
\dft{\kw;SeparateShowers Off;}
\sdir In a task involving more than one shower, the compressed output
files\index{compressed output files} can be split into several pieces
each one storing the data corresponding to \vkw;number; showers. In
particular, \kwbf;SeparateShowers 1; generates one compressed file per
shower while \kwbf;SeparateShowers Off; disables file splitting.
%
\idl SetGlobal
\stx{\akw;SetGl;obal\ ; [ \aop{\akw;Dyn;amic;\asep\akw;Sta;tic;} ] 
     \vkw;varname; \vkw;value;}
\dft{No environmental variables are imported by default.}
\ddir Setting global variables\index{global variables}. The variable
 \vkw;varname; is set to the string \vkw;value;. If the variable was
 already set, then its old setting is superseded. The defined
 variables can either be used within the IDL input stream or passed to
 the compressed output files or special primary modules\index{special
   primary particles}. The
 \kwbf;Dynamic; qualifier (default) indicates the {\em dynamic\/}
 character of the corresponding variable. This means that the value
 currently passed to the external modules is modified each time AIRES
 is invoked for a given task. On the other hand, \kwbf;Static;
 variables are set at the first invocation of AIRES; and further
 settings have no effect.
%
\idl SetTimeAtInjection
\sdswitch;SetTime;AtInjection; On On
\shdir Directive to set whether the time count for each shower is
started at the moment of
injecting the primary particle (\kwbf;On;) or at its first interaction
(\kwbf;Off;).

This directive is ignored for showers initiated by special
primaries\index{special primary particles} (see section
\ref{S:specialprim}).
%
\idl SetTopAtInjection
\sdswitch;SetTop;AtInjection; On On
\ddir When this switch is enabled, the top surface of the shower
bounding box is set accordingly with the atmospheric depth of the
primary injection point. Otherwise it is set accordingly with the
depth of the first primary interaction.
%
\idl Shell
\stx{\akw;Sh;ell\ ; \vkw;shell\_instruction;}
\stxp{\akw;Sh;ell\ ; \vkw;shell\_instruction; \kw;<<;\  \vkw;input\_file;}
\stxp{\akw;Sh;ell\ ; \vkw;shell\_instruction; \kw;<<;\  \kw;\&;\vkw;label;}
\stxp{{\it\ . . . \ \ (lines with input data)}}
\stxp{\kw;\&;\vkw;label;}
\ddir Executing a given shell instruction. The string
\vkw;shell\_instruction; is passed to the operating system to have it
executed within the current shell interpreter. It is optionally
possible to specify a input data file or, alternatively, place such
input data as a here-document\index{here-document} delimited by a
label \kw;\&;\vkw;label;.
%
\idl ShowersPerRun
\stx{\akw;ShowersP;erRun\ ; \aop{\vkw;number;\asep\akw;Inf;inite;}}
\dft{\kw;ShowersPerRun Infinite;}
\ddir Maximum number of showers in a run (see also
\kwbf;MaxCpuTimePerRun; and \kwbf;RunsPerProcess;). Notice that this
parameter is related with the computer environment only and does not
affect the total number of showers that define a task (see
\kwbf;TotalShowers;).
\clearpage
\idl Site
\stx{\akw;Site;\ ; \vkw;name;}
\dft{\kw;Site Site00;}
\sdir The \kwbf;Site; directive specify the geographical location that
define the environment (latitude, longitude and altitude) where the
simulations take place. \vkw;name; is a string identifying the
selected site. It must either be one of the predefined sites of the
AIRES site library\index{AIRES site library}, listed in table
\ppref{TAB:sitelib}, or have been previously defined by means of the
\idlbfx;AddSite; directive.
%
\idl Skip
\stx{\akw;Ski;p\ ; \kw;\&;\vkw;label;}
\ddir Instruction to skip part of an input data stream. All directives
placed after the \kwbf;Skip; statement and before \kwbf;\&;\vkw;label; are
skipped. Notice that this is not a ``go to'' statement: It is only
possible to skip forwards, never backwards.
%
\idl SpecialParticLog
\stx{\akw;SpecialParticL;og\ ; \vkw;lvl;}
\dft{\kw;SpecialParticLog\ ; is equivalent to \kw;SpecialParticLog 1;}
\dftp{\kw;SpecialParticLog 0\ ; is assumed in case of missing
specification.}
\ddir Controlling the amount of data related with special primary
particles\index{special primary particles} to be saved in the
corresponding log file\index{log file}. \vkw;lvl; is an integer
parameter that can take the following values:
\begin{center}
\begin{tabular}{rl}
0 & No information written in the log file. \\
1 & Messages before and after invoking the external module. \\
2 & Level 1 plus detailed list of valid primaries.\\
\end{tabular}
\end{center}
%
\idl SPMaxFieldsToAdd
\stx{\akw;SPMaxField;sToAdd\ ; \vkw;mxdynfields;}
\dft{\kw;SPMaxFieldsToAdd 0;}
\sdir Sets the maximum number of fields that can be dynamically added
to the AIRES compressed output files\index{compressed output
  files!dynamically added fields}.
%
\idl StackInformation
\sdswitch;StackI;nformation; On Off
\ddir Directive to instruct AIRES to print detailed stack usage
information in the summary output file.
%
\idl StopOnError
\sdswitch;StopOnErr;or; On Off
\ddir Directive to enable or disable the {\em\bfseries
  ``stop-on-error''\/} condition. When this condition is in effect,
the severity of every error that could happen while parsing the AIRES
IDL directives is set to the maximum level (fatal error). In consequence,
the simulation program will always abort if there are errors within
the input instructions. This directive is useful as a security tool to
allow execution of simulations only if the input files do not contain
instructions with errors. Consider also using it together with
\idlbfx;ForceModel;.
%
\idl Summary
\sdswitch;Sum;mary; On On
\ddir Directive to enable or disable the output summary.
%
\idl TableIndex
\sdswitch;TableI;ndex; On Off
\ddir Directive to instruct AIRES to print a table index
in the summary output file.
%
\idl TaskName
\stx{\akw;Task;Name\ ; [ \akw;Append;; ] \vkw;taskname; [ \vkw;taskversion; ]}
\dft{\kw;TaskName GIVE\_ME\_A\_NAME\_PLEASE;}
\ddir Task name assignment. \vkw;taskname; is a character string which
identifies the current task. If its length is greater than 64 characters,
it will be truncated to the first 64 characters. \vkw;taskversion; is an
optional integer between 0 (default) and 999. If \vkw;taskversion; is not
zero, the effective task name is
\vkw;taskname;\kwbf;\_;\vkw;taskversion;. If the keyword \kwbf;Append; is
used, then \vkw;taskname; is appended to the existing task name
string. The task name is used to set the file names of all output files.
%
\idl ThinningEnergy
\stx{\akw;Thin;ningEnergy\ ; \aop{\vkw;energy;\asep\vkw;number;
\ \akw;R;elative;}}
\dft{\kw;ThinningEnergy 1.0e-4 Relative;}
\sdir Thinning\index{thinning} energy. It can be expressed either as
an absolute energy or as a real (positive) number with the keyword
\kwbf;Relative; (In this case the thinning energy is the primary
energy times the specified number).
%
\idl ThinningWFactor
\stx{\akw;ThinningWF;actor\ ; \vkw;number;}
\dft{\kw;ThinningWFactor 12;}
\shdir Thinning\index{thinning} weight factor\index{statistical weight
factor}. This instruction permits setting the weight factor $W_f$ of
equation (\ref{eq:Wfactor}).
%
\idl TotalShowers
\stx{\akw;TotalSh;owers\ ; \vkw;nofshowers;}
\dft{None. This directive is always required.}
\ddir Total number of showers. \vkw;nofshowers; is a positive integer in
the range $[1, 759375]$ defining the number of showers to be simulated in
the current task. Notice that this is a dynamic parameter, that is, it can
be modified (either enlarged or reduced) during the simulations.
%
\idl Trace
\sdswitch;Trace;; On Off
\ddir Directive to enable or disable input data tracing. If enabled
(\kwbf;On;) then trace information about the directives being processed by
the IDL parser is written into the standard output channel. This
directive is useful to debug IDL input data sets.
%
\idl TSSFile
\sdswitch;TSS;File; On Off
\ddir If \kwbf;TSSFile On; is specified, then a task summary script
file\index{task summary script file} will be generated upon task
completion. The task summary script file (TSS) is a plain text file
containing information about the main parameters of the simulation, in
the format \kwbf;Keyword; = \vkw;value;, suitable for processing with
other programs.

\end{description}
}

%
\chapter{Output data table index}
\label{A:outables}\index{output data tables}

We list here all the tables defined in AIRES
{\currairesversion}. These tables can be processed using directives
\idlbfx;PrintTables; and/or \idlbfx;ExportTables;\index{exported data
files} (see chapter \ref{C:steering}).
\bigskip
%
\definecolor{tabcolorb}{cmyk}{1.0,0.0,0.0,0.1}
\def\clra{\color{black}}
\def\clrb{\color{black}}
\def\clrc{\color{red}}
\def\tcolorswap{\let\clrx=\clra \global\let\clra=\clrb \global\let\clrb=\clrx}
\def\tcolorswapskip{\noalign{\smallskip}\tcolorswap}
{\openup-1pt\frenchspacing
\def\tabnewpage{\noalign{\clearpage}
&{\clrc\itbf Code\ }&{\clrc\itbf Table name}\cr
\noalign{\medskip}}
\tabskip=1em plus 1fil
\halign to\textwidth{\hfil{\itshape #\/}\tabskip=1em&\hfil
{\bfseries #}&#\hfil\tabskip=1em
plus 1fil\cr
&{\clrc\itbf Code\ }&{\clrc\itbf Table name}\cr
\noalign{\medskip}\clra\index{system and environment tables}
   1 &\clrb 0100 & Atmospheric profile.\cr
\tcolorswapskip\clra\index{longitudinal development}
   2 &\clrb 1001 & Longitudinal development: Gamma rays.\cr\clra
   3 &\clrb 1005 & Longitudinal development: Electrons.\cr\clra
   4 &\clrb 1006 & Longitudinal development: Positrons.\cr\clra
   5 &\clrb 1007 & Longitudinal development: Muons ($+$).\cr\clra
   6 &\clrb 1008 & Longitudinal development: Muons ($-$).\cr\clra
   7 &\clrb 1011 & Longitudinal development: Pions ($+$).\cr\clra
   8 &\clrb 1012 & Longitudinal development: Pions ($-$).\cr\clra
   9 &\clrb 1013 & Longitudinal development: Kaons ($+$).\cr\clra
  10 &\clrb 1014 & Longitudinal development: Kaons ($-$).\cr\clra
  11 &\clrb 1021 & Longitudinal development: Neutrons.\cr\clra
  12 &\clrb 1022 & Longitudinal development: Protons.\cr\clra
  13 &\clrb 1023 & Longitudinal development: Antiprotons.\cr\clra
  14 &\clrb 1041 & Longitudinal development: Nuclei.\cr\clra
  15 &\clrb 1091 & Longitudinal development: Other charged pcles.\cr\clra
  16 &\clrb 1092 & Longitudinal development: Other neutral pcles.\cr\clra
  17 &\clrb 1205 & Longitudinal development: e$+$ and e$-$\cr\clra
  18 &\clrb 1207 & Longitudinal development: mu$+$ and mu$-$\cr\clra
  19 &\clrb 1211 & Longitudinal development: pi$+$ and pi$-$\cr\clra
  20 &\clrb 1213 & Longitudinal development: K$+$ and K$-$\cr\clra
  21 &\clrb 1291 & Longitudinal development: All charged particles.\cr\clra
  22 &\clrb 1292 & Longitudinal development: All neutral particles.\cr\clra
  23 &\clrb 1293 & Longitudinal development: All particles.\cr
\tabnewpage
\tcolorswapskip\clra\index{unweighted distributions}
  24 &\clrb 1301 & Unweighted longit. development: Gamma rays.\cr\clra
  25 &\clrb 1305 & Unweighted longit. development: Electrons.\cr\clra
  26 &\clrb 1306 & Unweighted longit. development: Positrons.\cr\clra
  27 &\clrb 1307 & Unweighted longit. development: Muons ($+$).\cr\clra
  28 &\clrb 1308 & Unweighted longit. development: Muons ($-$).\cr\clra
  29 &\clrb 1311 & Unweighted longit. development: Pions ($+$).\cr\clra
  30 &\clrb 1312 & Unweighted longit. development: Pions ($-$).\cr\clra
  31 &\clrb 1313 & Unweighted longit. development: Kaons ($+$).\cr\clra
  32 &\clrb 1314 & Unweighted longit. development: Kaons ($-$).\cr\clra
  33 &\clrb 1321 & Unweighted longit. development: Neutrons.\cr\clra
  34 &\clrb 1322 & Unweighted longit. development: Protons.\cr\clra
  35 &\clrb 1323 & Unweighted longit. development: Antiprotons.\cr\clra
  36 &\clrb 1341 & Unweighted longit. development: Nuclei.\cr\clra
  37 &\clrb 1391 & Unweighted longit. development: Other charged
  pcles.\cr\clra
  38 &\clrb 1392 & Unweighted longit. development: Other neutral
  pcles.\cr\clra
  39 &\clrb 1405 & Unweighted longit. development: e$+$ and e$-$\cr\clra
  40 &\clrb 1407 & Unweighted longit. development: mu$+$ and mu$-$\cr\clra
  41 &\clrb 1411 & Unweighted longit. development: pi$+$ and pi$-$\cr\clra
  42 &\clrb 1413 & Unweighted longit. development: K$+$ and K$-$\cr\clra
  43 &\clrb 1491 & Unweighted longit. development: All charged
  particles.\cr\clra
  44 &\clrb 1492 & Unweighted longit. development: All neutral
  particles.\cr\clra
  45 &\clrb 1493 & Unweighted longit. development: All particles.\cr
\tcolorswapskip\clra
  46 &\clrb 1501 & Longitudinal development: Energy of gamma rays.%
\index{longitudinal development!in energy}\cr\clra
  47 &\clrb 1505 & Longitudinal development: Energy of electrons.\cr\clra
  48 &\clrb 1506 & Longitudinal development: Energy of positrons.\cr\clra
  49 &\clrb 1507 & Longitudinal development: Energy of muons ($+$).\cr\clra
  50 &\clrb 1508 & Longitudinal development: Energy of muons ($-$).\cr\clra
  51 &\clrb 1511 & Longitudinal development: Energy of pions ($+$).\cr\clra
  52 &\clrb 1512 & Longitudinal development: Energy of pions ($-$).\cr\clra
  53 &\clrb 1513 & Longitudinal development: Energy of kaons ($+$).\cr\clra
  54 &\clrb 1514 & Longitudinal development: Energy of kaons ($-$).\cr\clra
  55 &\clrb 1521 & Longitudinal development: Energy of neutrons.\cr\clra
  56 &\clrb 1522 & Longitudinal development: Energy of protons.\cr\clra
  57 &\clrb 1523 & Longitudinal development: Energy of antiprotons.\cr\clra
  58 &\clrb 1541 & Longitudinal development: Energy of nuclei.\cr\clra
  59 &\clrb 1591 & Longitudinal development: Energy of other charged
 particles.\cr\clra
  60 &\clrb 1592 & Longitudinal development: Energy of other neutral
 particles.\cr\clra
  61 &\clrb 1705 & Longitudinal development: Energy of e$+$ and e$-$\cr\clra 
  62 &\clrb 1707 & Longitudinal development: Energy of mu$+$ and mu$-$\cr\clra
  63 &\clrb 1711 & Longitudinal development: Energy of pi$+$ and pi$-$\cr\clra
  64 &\clrb 1713 & Longitudinal development: Energy of K$+$ and K$-$\cr
\tabnewpage\clra
  65 &\clrb 1791 & Longitudinal development: Energy of all charged
 particles.\cr\clra
  66 &\clrb 1792 & Longitudinal development: Energy of all neutral
 particles.\cr\clra
  67 &\clrb 1793 & Longitudinal development: Energy of all particles.\cr
\tcolorswapskip\clra\index{lateral distributions}
  68 &\clrb 2001 & Lateral distribution: Gamma rays.\cr\clra
  69 &\clrb 2005 & Lateral distribution: Electrons.\cr\clra
  70 &\clrb 2006 & Lateral distribution: Positrons.\cr\clra
  71 &\clrb 2007 & Lateral distribution: Muons ($+$).\cr\clra
  72 &\clrb 2008 & Lateral distribution: Muons ($-$).\cr\clra
  73 &\clrb 2011 & Lateral distribution: Pions ($+$).\cr\clra
  74 &\clrb 2012 & Lateral distribution: Pions ($-$).\cr\clra
  75 &\clrb 2013 & Lateral distribution: Kaons ($+$).\cr\clra
  76 &\clrb 2014 & Lateral distribution: Kaons ($-$).\cr\clra
  77 &\clrb 2021 & Lateral distribution: Neutrons.\cr\clra
  78 &\clrb 2022 & Lateral distribution: Protons.\cr\clra
  79 &\clrb 2023 & Lateral distribution: Antiprotons.\cr\clra
  80 &\clrb 2041 & Lateral distribution: Nuclei.\cr\clra
  81 &\clrb 2091 & Lateral distribution: Other charged pcles.\cr\clra
  82 &\clrb 2092 & Lateral distribution: Other neutral pcles.\cr\clra
  83 &\clrb 2205 & Lateral distribution: e$+$ and e$-$ \cr\clra
  84 &\clrb 2207 & Lateral distribution: mu$+$ and mu$-$\cr\clra
  85 &\clrb 2211 & Lateral distribution: pi$+$ and pi$-$\cr\clra
  86 &\clrb 2213 & Lateral distribution: K$+$ and K$-$ \cr\clra
  87 &\clrb 2291 & Lateral distribution: All charged particles.\cr\clra
  88 &\clrb 2292 & Lateral distribution: All neutral particles.\cr\clra
  89 &\clrb 2293 & Lateral distribution: All particles.\cr
\tcolorswapskip\clra\index{unweighted distributions}
  90 &\clrb 2301 & Unweighted lateral distribution: Gamma rays.\cr\clra
  91 &\clrb 2305 & Unweighted lateral distribution: Electrons.\cr\clra
  92 &\clrb 2306 & Unweighted lateral distribution: Positrons.\cr\clra
  93 &\clrb 2307 & Unweighted lateral distribution: Muons ($+$).\cr\clra
  94 &\clrb 2308 & Unweighted lateral distribution: Muons ($-$).\cr\clra
  95 &\clrb 2311 & Unweighted lateral distribution: Pions ($+$).\cr\clra
  96 &\clrb 2312 & Unweighted lateral distribution: Pions ($-$).\cr\clra
  97 &\clrb 2313 & Unweighted lateral distribution: Kaons ($+$).\cr\clra
  98 &\clrb 2314 & Unweighted lateral distribution: Kaons ($-$).\cr\clra
  99 &\clrb 2321 & Unweighted lateral distribution: Neutrons.\cr\clra
 100 &\clrb 2322 & Unweighted lateral distribution: Protons.\cr\clra
 101 &\clrb 2323 & Unweighted lateral distribution: Antiprotons.\cr\clra
 102 &\clrb 2341 & Unweighted lateral distribution: Nuclei.\cr\clra
 103 &\clrb 2391 & Unweighted lateral distribution: Other charged
  pcles.\cr\clra
 104 &\clrb 2392 & Unweighted lateral distribution: Other neutral
  pcles.\cr\clra
 105 &\clrb 2405 & Unweighted lateral distribution: e$+$ and e$-$\cr
\tabnewpage\clra
 106 &\clrb 2407 & Unweighted lateral distribution: mu$+$ and mu$-$\cr\clra
 107 &\clrb 2411 & Unweighted lateral distribution: pi$+$ and pi$-$\cr\clra
 108 &\clrb 2413 & Unweighted lateral distribution: K$+$ and K$-$\cr
\clra
 109 &\clrb 2491 & Unweighted lateral distribution: All charged
  particles.\cr\clra
 110 &\clrb 2492 & Unweighted lateral distribution: All neutral
  particles.\cr\clra
 111 &\clrb 2493 & Unweighted lateral distribution: All particles.\cr
\tcolorswapskip\clra\index{energy distributions}
 112 &\clrb 2501 & Energy distribution at ground: Gamma rays.\cr\clra
 113 &\clrb 2505 & Energy distribution at ground: Electrons.\cr\clra
 114 &\clrb 2506 & Energy distribution at ground: Positrons.\cr\clra
 115 &\clrb 2507 & Energy distribution at ground: Muons ($+$).\cr\clra
 116 &\clrb 2508 & Energy distribution at ground: Muons ($-$).\cr\clra
 117 &\clrb 2511 & Energy distribution at ground: Pions ($+$).\cr\clra
 118 &\clrb 2512 & Energy distribution at ground: Pions ($-$).\cr\clra
 119 &\clrb 2513 & Energy distribution at ground: Kaons ($+$).\cr\clra
 120 &\clrb 2514 & Energy distribution at ground: Kaons ($-$).\cr\clra
 121 &\clrb 2521 & Energy distribution at ground: Neutrons.\cr\clra
 122 &\clrb 2522 & Energy distribution at ground: Protons.\cr\clra
 123 &\clrb 2523 & Energy distribution at ground: Antiprotons.\cr\clra
 124 &\clrb 2541 & Energy distribution at ground: Nuclei.\cr\clra
 125 &\clrb 2591 & Energy distribution at ground: Other charged pcles.\cr\clra
 126 &\clrb 2592 & Energy distribution at ground: Other neutral pcles.\cr\clra
 127 &\clrb 2705 & Energy distribution at ground: e$+$ and e$-$ \cr\clra
 128 &\clrb 2707 & Energy distribution at ground: mu$+$ and mu$-$ \cr\clra
 129 &\clrb 2711 & Energy distribution at ground: pi$+$ and pi$-$ \cr\clra
 130 &\clrb 2713 & Energy distribution at ground: K$+$ and K$-$\cr\clra
 131 &\clrb 2791 & Energy distribution at ground: All charged
  particles.\cr\clra
 132 &\clrb 2792 & Energy distribution at ground: All neutral
  particles.\cr\clra 
 133 &\clrb 2793 & Energy distribution at ground: All particles.\cr
\tcolorswapskip\clra\index{unweighted distributions}
 134 &\clrb 2801 & Unweighted energy distribution: Gamma rays.\cr\clra
 135 &\clrb 2805 & Unweighted energy distribution: Electrons.\cr\clra
 136 &\clrb 2806 & Unweighted energy distribution: Positrons.\cr\clra
 137 &\clrb 2807 & Unweighted energy distribution: Muons ($+$).\cr\clra
 138 &\clrb 2808 & Unweighted energy distribution: Muons ($-$).\cr\clra
 139 &\clrb 2811 & Unweighted energy distribution: Pions ($+$).\cr\clra
 140 &\clrb 2812 & Unweighted energy distribution: Pions ($-$).\cr\clra
 141 &\clrb 2813 & Unweighted energy distribution: Kaons ($+$).\cr\clra
 142 &\clrb 2814 & Unweighted energy distribution: Kaons ($-$).\cr\clra
 143 &\clrb 2821 & Unweighted energy distribution: Neutrons.\cr\clra
 144 &\clrb 2822 & Unweighted energy distribution: Protons.\cr\clra
 145 &\clrb 2823 & Unweighted energy distribution: Antiprotons.\cr\clra
 146 &\clrb 2841 & Unweighted energy distribution: Nuclei.\cr
\tabnewpage\clra
 147 &\clrb 2891 & Unweighted energy distribution: Other charged
  pcles.\cr\clra
 148 &\clrb 2892 & Unweighted energy distribution: Other neutral
  pcles.\cr\clra
 149 &\clrb 2905 & Unweighted energy distribution: e$+$ and e$-$\cr\clra
 150 &\clrb 2907 & Unweighted energy distribution: mu$+$ and mu$-$\cr\clra
 151 &\clrb 2911 & Unweighted energy distribution: pi$+$ and pi$-$\cr\clra
 152 &\clrb 2913 & Unweighted energy distribution: K$+$ and K$-$\cr\clra
 153 &\clrb 2991 & Unweighted energy distribution: All charged
  particles.\cr\clra
 154 &\clrb 2992 & Unweighted energy distribution: All neutral
  particles.\cr\clra
 155 &\clrb 2993 & Unweighted energy distribution: All particles.\cr
\tcolorswapskip\clra\index{time distributions}
 156 &\clrb 3001 & Mean arrival time distribution: Gamma rays.\cr\clra
 157 &\clrb 3005 & Mean arrival time distribution: Electrons and
 positrons.\cr\clra
 158 &\clrb 3007 & Mean arrival time distribution: Muons.\cr\clra
 159 &\clrb 3091 & Mean arrival time distribution: Other charged
 pcles.\cr\clra
 160 &\clrb 3092 & Mean arrival time distribution: Other neutral
 pcles.\cr\clra
 161 &\clrb 3291 & Mean arrival time distribution: All charged
 particles.\cr\clra
 162 &\clrb 3292 & Mean arrival time distribution: All neutral
 particles.\cr\clra
 163 &\clrb 3293 & Mean arrival time distribution: All particles.\cr
\tcolorswapskip\clra
 164 &\clrb 5001 & Number and energy of ground gammas versus shower
 number.\cr\clra
 165 &\clrb 5005 & Number and energy of ground e$-$ versus shower
 number.\cr\clra
 166 &\clrb 5006 & Number and energy of ground e$+$ versus shower
 number.\cr\clra
 167 &\clrb 5007 & Number and energy of ground mu$+$ versus shower
 number.\cr\clra
 168 &\clrb 5008 & Number and energy of ground mu$-$ versus shower
 number.\cr\clra
 169 &\clrb 5011 & Number and energy of ground pi$+$ versus shower
 number.\cr\clra
 170 &\clrb 5012 & Number and energy of ground pi$-$ versus shower
 number.\cr\clra
 171 &\clrb 5013 & Number and energy of ground K$+$ versus shower
 number.\cr\clra
 172 &\clrb 5014 & Number and energy of ground K$-$ versus shower
 number.\cr\clra
 173 &\clrb 5021 & Number and energy of ground neutrons versus shower
 number.\cr\clra   
 174 &\clrb 5022 & Number and energy of ground protons versus shower
 number.\cr\clra
 175 &\clrb 5023 & Number and energy of ground pbar versus shower
 number.\cr\clra
 176 &\clrb 5041 & Number and energy of ground nuclei versus shower
 number.\cr\clra
 177 &\clrb 5091 & Number and energy of other grd. ch. pcles. versus shower
number.\cr\clra
 178 &\clrb 5092 & Number and energy of other grd. nt. pcles. versus shower
 number.\cr\clra
 179 &\clrb 5205 & Number and energy of ground e$+$ and e$-$ versus shower
number.\cr\clra 
 180 &\clrb 5207 & Number and energy of ground mu$+$ and mu$-$ versus
shower number.\cr\clra 
 181 &\clrb 5211 & Number and energy of ground pi$+$ and pi$-$ versus
shower number.\cr\clra 
 182 &\clrb 5213 & Number and energy of ground K$+$ and K$-$ versus shower
number.\cr\clra 
 183 &\clrb 5291 & Number and energy of ground ch. pcles. versus shower
 number.\cr\clra
 184 &\clrb 5292 & Number and energy of ground nt. pcles. versus shower
 number.\cr\clra
 185 &\clrb 5293 & Number and energy of all ground particles versus shower
 number.\cr
\tabnewpage\clra
\index{Xmax@$X_{\mathrm{max}}$}\index{Nmax@$N_{\mathrm{max}}$}
 186 &\clrb 5501 & Xmax and Nmax (charged particles) versus shower
number.\cr\clra 
 187 &\clrb 5511 & First interact. depth and primary energy versus shower
number\index{depth of first interaction}.\cr\clra
 188 &\clrb 5513 & Zenith and azimuth angles versus shower number.\cr
\tcolorswapskip\clra\index{longitudinal development!created particles}
 189 &\clrb  6001 & Number of created particles: Gamma rays.\cr\clra
 190 &\clrb  6005 & Number of created particles: Electrons.\cr\clra
 191 &\clrb  6006 & Number of created particles: Positrons.\cr\clra
 192 &\clrb  6007 & Number of created particles: Muons (+).\cr\clra
 193 &\clrb  6008 & Number of created particles: Muons (-).\cr\clra
 194 &\clrb  6011 & Number of created particles: Pions (+).\cr\clra
 195 &\clrb  6012 & Number of created particles: Pions (-).\cr\clra
 196 &\clrb  6013 & Number of created particles: Kaons (+).\cr\clra
 197 &\clrb  6014 & Number of created particles: Kaons (-).\cr\clra
 198 &\clrb  6021 & Number of created particles: Neutrons.\cr\clra
 199 &\clrb  6022 & Number of created particles: Protons.\cr\clra
 200 &\clrb  6023 & Number of created particles: Antiprotons.\cr\clra
 201 &\clrb  6041 & Number of created particles: Nuclei.\cr\clra
 202 &\clrb  6091 & Number of created particles: Other charged pcles.\cr\clra
 203 &\clrb  6092 & Number of created particles: Other neutral pcles.\cr\clra
 204 &\clrb  6205 & Number of created particles: e+ and e-\cr\clra
 205 &\clrb  6207 & Number of created particles: mu+ and mu-\cr\clra
 206 &\clrb  6211 & Number of created particles: pi+ and pi-\cr\clra
 207 &\clrb  6213 & Number of created particles: K+ and K-\cr\clra
 208 &\clrb  6291 & Number of created particles: All charged particles.\cr\clra
 209 &\clrb  6292 & Number of created particles: All neutral particles.\cr\clra
 210 &\clrb  6293 & Number of created particles: All particles.\cr\clra
 211 &\clrb  6296 & Number of created particles: All neutrinos.\cr
\tcolorswapskip\clra\index{unweighted distributions}
 212 &\clrb  6301 & Number of created entries: Gamma rays.\cr\clra
 213 &\clrb  6305 & Number of created entries: Electrons.\cr\clra
 214 &\clrb  6306 & Number of created entries: Positrons.\cr\clra
 215 &\clrb  6307 & Number of created entries: Muons (+).\cr\clra
 216 &\clrb  6308 & Number of created entries: Muons (-).\cr\clra
 217 &\clrb  6311 & Number of created entries: Pions (+).\cr\clra
 218 &\clrb  6312 & Number of created entries: Pions (-).\cr\clra
 219 &\clrb  6313 & Number of created entries: Kaons (+).\cr\clra
 220 &\clrb  6314 & Number of created entries: Kaons (-).\cr\clra
 221 &\clrb  6321 & Number of created entries: Neutrons.\cr\clra
 222 &\clrb  6322 & Number of created entries: Protons.\cr\clra
 223 &\clrb  6323 & Number of created entries: Antiprotons.\cr\clra
 224 &\clrb  6341 & Number of created entries: Nuclei.\cr\clra
 225 &\clrb  6391 & Number of created entries: Other charged pcles.\cr\clra
 226 &\clrb  6392 & Number of created entries: Other neutral pcles.\cr
\tabnewpage\clra
 227 &\clrb  6405 & Number of created entries: e+ and e-\cr\clra
 228 &\clrb  6407 & Number of created entries: mu+ and mu-\cr\clra
 229 &\clrb  6411 & Number of created entries: pi+ and pi-\cr\clra
 230 &\clrb  6413 & Number of created entries: K+ and K-\cr\clra
 231 &\clrb  6491 & Number of created entries: All charged particles.\cr\clra
 232 &\clrb  6492 & Number of created entries: All neutral particles.\cr\clra
 233 &\clrb  6493 & Number of created entries: All particles.\cr\clra
 234 &\clrb  6496 & Number of created entries: All neutrinos.\cr
\tcolorswapskip\clra\index{longitudinal development!energy of created
 particles}
 235 &\clrb  6501 & Energy of created particles: Gamma rays.\cr\clra
 236 &\clrb  6505 & Energy of created particles: Electrons.\cr\clra
 237 &\clrb  6506 & Energy of created particles: Positrons.\cr\clra
 238 &\clrb  6507 & Energy of created particles: Muons (+).\cr\clra
 239 &\clrb  6508 & Energy of created particles: Muons (-).\cr\clra
 240 &\clrb  6511 & Energy of created particles: Pions (+).\cr\clra
 241 &\clrb  6512 & Energy of created particles: Pions (-).\cr\clra
 242 &\clrb  6513 & Energy of created particles: Kaons (+).\cr\clra
 243 &\clrb  6514 & Energy of created particles: Kaons (-).\cr\clra
 244 &\clrb  6521 & Energy of created particles: Neutrons.\cr\clra
 245 &\clrb  6522 & Energy of created particles: Protons.\cr\clra
 246 &\clrb  6523 & Energy of created particles: Antiprotons.\cr\clra
 247 &\clrb  6541 & Energy of created particles: Nuclei.\cr\clra
 248 &\clrb  6591 & Energy of created particles: Other charged pcles.\cr\clra
 249 &\clrb  6592 & Energy of created particles: Other neutral pcles.\cr\clra
 250 &\clrb  6705 & Energy of created particles: e+ and e-\cr\clra
 251 &\clrb  6707 & Energy of created particles: mu+ and mu-\cr\clra
 252 &\clrb  6711 & Energy of created particles: pi+ and pi-\cr\clra
 253 &\clrb  6713 & Energy of created particles: K+ and K-\cr\clra
 254 &\clrb  6791 & Energy of created particles: All charged particles.\cr\clra
 255 &\clrb  6792 & Energy of created particles: All neutral particles.\cr\clra
 256 &\clrb  6793 & Energy of created particles: All particles.\cr\clra
 257 &\clrb  6796 & Energy of created particles: All neutrinos.\cr
\tcolorswapskip\clra
 258 &\clrb 7001 & Longitudinal development:
                    Low energy gamma rays.
\index{longitudinal development!low energy particles}\cr\clra
 259 &\clrb 7005 & Longitudinal development: Low energy electrons.\cr\clra
 260 &\clrb 7006 & Longitudinal development: Low energy positrons.\cr\clra
 261 &\clrb 7007 & Longitudinal development: Low energy muons (+).\cr\clra
 262 &\clrb 7008 & Longitudinal development: Low energy muons (-).\cr\clra
 263 &\clrb 7091 & Longitudinal development:
                    Other charged low egy. pcles.\cr\clra
 264 &\clrb 7092 & Longitudinal development:
                    Other neutral low egy. pcles.\cr\clra
 265 &\clrb 7205 & Longitudinal development: Low energy e+ and e-\cr\clra
 266 &\clrb 7207 & Longitudinal development: Low energy mu+ and mu-\cr\clra
 267 &\clrb 7291 & Longitudinal development:
                   All low energy charged pcles.\cr
\tabnewpage\clra
 268 &\clrb 7292 & Longitudinal development:
                   All low energy neutral pcles.\cr\clra
 269 &\clrb 7293 & Longitudinal development: All low energy pcles.\cr
\tcolorswapskip\clra
 270 &\clrb 7301 & Unweighted longit. devel.: Low energy gamma rays.\cr\clra
 271 &\clrb 7305 & Unweighted longit. devel.: Low energy electrons.\cr\clra
 272 &\clrb 7306 & Unweighted longit. devel.: Low energy positrons.\cr\clra
 273 &\clrb 7307 & Unweighted longit. devel.: Low energy muons (+).\cr\clra
 274 &\clrb 7308 & Unweighted longit. devel.: Low energy muons (-).\cr\clra
 275 &\clrb 7391 & Unweighted longit. devel.:
                   Other charged low egy. pcles.\cr\clra
 276 &\clrb 7392 & Unweighted longit. devel.:
                   Other neutral low egy. pcles.\cr\clra
 277 &\clrb 7405 & Unweighted longit. devel.: Low energy e+ and e-\cr\clra
 278 &\clrb 7407 & Unweighted longit. devel.: Low energy mu+ and mu-\cr\clra
 279 &\clrb 7491 & Unweighted longit. devel.:
                   All low energy charged pcles.\cr\clra
 280 &\clrb 7492 & Unweighted longit. devel.:
                   All low energy neutral pcles.\cr\clra
 281 &\clrb 7493 & Unweighted longit. devel.: All low energy pcles.\cr
\tcolorswapskip\clra
 282 &\clrb 7501 & Longitudinal development:
                    Energy of low egy. gamma rays.\cr\clra
 283 &\clrb 7505 & Longitudinal development:
                    Energy of low egy. electrons.\cr\clra
 284 &\clrb 7506 & Longitudinal development:
                    Energy of low egy. positrons.\cr\clra
 285 &\clrb 7507 & Longitudinal development:
                    Energy of low egy. muons(+).\cr\clra
 286 &\clrb 7508 & Longitudinal development:
                    Energy of low egy. muons(-).\cr\clra
 287 &\clrb 7591 & Longitudinal development:
                    Egy. of other charged low egy. pcles.\cr\clra
 288 &\clrb 7592 & Longitudinal development:
                    Egy. of other neutral low egy. pcles.\cr\clra
 289 &\clrb 7705 & Longitudinal development:
                    Energy of low egy. e+ and e-\cr\clra
 290 &\clrb 7707 & Longitudinal development:
                    Energy of low egy. mu+ and mu-\cr\clra
 291 &\clrb 7791 & Longitudinal development:
                    Egy. of all low egy. charged pcles.\cr\clra
 292 &\clrb 7792 & Longitudinal development:
                    Egy. of all low egy. neutral pcles.\cr\clra
 293 &\clrb 7793 & Longitudinal development:
                    Egy. of all low energy pcles.\cr
\tcolorswapskip\clra
 294 &\clrb 7801 & Longitudinal development:
                    Energy deposited by gamma rays.
\index{longitudinal development!deposited energy}\cr\clra
 295 &\clrb 7805 & Longitudinal development:
                    Energy deposited by electrons.\cr\clra
 296 &\clrb 7806 & Longitudinal development:
                    Energy deposited by positrons.\cr\clra
 297 &\clrb 7807 & Longitudinal development:
                    Energy deposited by muons (+).\cr\clra
 298 &\clrb 7808 & Longitudinal development:
                    Energy deposited by muons (-).\cr\clra
 299 &\clrb 7891 & Longitudinal development:
                    Egy. deposited by other charged pcles.\cr\clra
 300 &\clrb 7892 & Longitudinal development:
                    Egy. deposited by other neutral pcles.\cr\clra
 301 &\clrb 7905 & Longitudinal development:
                    Energy deposited by e+ and e-\cr\clra
 302 &\clrb 7907 & Longitudinal development:
                    Energy deposited by mu+ and mu-\cr\clra
 303 &\clrb 7991 & Longitudinal development:
                    Egy. deposited by all charged pcles.\cr\clra
 304 &\clrb 7992 & Longitudinal development:
                    Egy. deposited by all neutral pcles.\cr\clra
 305 &\clrb 7993 & Longitudinal development: Energy deposited by all pcles.\cr
}}

%
\chapter{The AIRES object library}\index{AIRES object library}
\label{A:aireslib}

The AIRES object library is a collection of modules that are useful in
several applications, including (but not limited to) special primary
modules\index{special primary particles} (see section
\ref{S:specialprim}), and output file processing, particularly
compressed files\index{compressed output files} generated by the AIRES
compressed i/o unit (CIO), and many other analysis procedures.

There is an on-line reference\index{AIRES object library!on-line
  reference} for the AIRES library that can be accessed at the
following link:
\kwdisplay;aires.fisica.unlp.edu.ar/doc/aireslibref;
The information currently available on the on-line library reference
covers all the user-callable library modules and is permanently updated.
\section{C/C++ interface}\index{AIRES object library!C/C++ interface}

The modules of the AIRES object library are callable from a C/C++
program. In general the calling statement is similar to the FORTRAN
one, taking into account that all arguments are passed {\em by
reference.\/} That means that the actual arguments must be pointers to
the corresponding data items. 

This requirement is made evident when describing the different
routines by placing an ampersand (\kwbf;\&;) before the corresponding
arguments. The experienced C programmer will understand, however, that
this character is not required in actual calling statements
containing pointer variables as arguments. The following example
illustrates this point:
\kwexample{%
{\tt int *channel, *vrb, *irc;}\\
{\tt int recnumber;}\\
{\tt int crogotorec();}\\
\kwt;. . .;\\
\kwt;if (;crogotorec(channel, \&recnumber, vrb, irc)\kwt;) \{ . . .;
}
All the arguments of \libbfx;crogotorec; are defined as pointers,
except \kwbf;recnumber; which is declared as an integer variable. The
\kwbf;\&; placed before this argument ensures that this variable be
passed by reference to the called routine.

In general, all the FORTRAN routines of the library can be directly
called from a C program. In a few cases it was necessary to write
special C routines, which were named appending a ``c'' to the original
FORTRAN name, as in the case of \libbfx;opencrofile; that must be
called \kwbf;opencrofilec; from a C program (see page
\pageref{rou0:opencrofile}).

It is also worthwhile mentioning that some FORTRAN compilers do place
an underscore (\kw;\_;) after the names of the routines. In such cases
this character must be manually appended to all the routines used
within the C program, excluding, of course, all the special C routines
of the previous paragraph.

\section{List of most frequently used library modules.}

In this appendix we list the definitions of the most frequently used
routines, alphabetically ordered. At each case the FORTRAN as well as
the C calling statements are placed.

\def\routb #1
{\item[{\color{airesblue}\large\bf #1}]\label{rou0:#1}%
  \vrule depth2.4ex width0pt
\index{AIRES object library!#1@\kw;#1;|bfpage{ }}%
\index{#1@\kw;#1;|see{AIRES object library.}}%
\hfil\break}
\def\rout #1
{\clearpage\routb #1
}
\def\cend{;}
\def\fcall #1(#2){\makebox[5.7em][l]{FORTRAN}
\kw;call #1;\if.#2\else%
\kw;(;\strut\parbox[t]{7.5cm}{\raggedright\kw;#2);\strut}\fi\hfil\break}
\def\ccall #1(#2){\makebox[5.7em][l]{C}
\kw;#1;\if.#2\kw;\cend;\else%
\kw;(;\strut\parbox[t]{8.3cm}{\raggedright\kw;#2)\cend;\strut}\fi\hfil\break}
\def\ffun #1 = #2(#3){\makebox[5.7em][l]{FORTRAN}
\if.#3\kw;#1 = #2();\strut\else\kw;#1 = #2(;\strut%
\parbox[t]{6.5cm}{\raggedright\kw;#3);\strut}\fi\hfil\break}
\def\cfun #1 = #2(#3){\makebox[5.7em][l]{C}
\if.#3\kw;#1 = #2();\strut;\else\kw;#1 = #2(;\strut%
\parbox[t]{6.5cm}{\raggedright\kw;#3)\cend;\strut}\fi\hfil\break}
\def\dh{\vrule height 4.5ex width0pt\relax}
\def\barlist{\vrule height4ex width0pt{\bf Arguments:}
\begin{description}}
\def\earlist{\end{description}}
\def\argu[#1; #2]{\item[\kwbf;#1;] ({\itshape\bfseries #2\/}) }
\def\retval[#1]#2{\begin{description}
\item[Returned value:]
\vrule height3ex width0pt({\itshape\bfseries #1\/}) #2
\end{description}}
\def\arguchannel{\argu[channel; Input, integer]
Variable that uniquely identifies the
I/O channel assigned to the corresponding file. This variable must be
already set by means of routine \kwbf;opencrofile;.}
\def\arguvrb{\argu[vrb; Input, integer] Verbosity control.
                       If \kwbf;vrb; is
                       zero or negative then no error or informative
                       messages are printed; error conditions are
                       communicated to the calling program via the
                       return code. If \kwbf;vrb; is positive error messages
                       will be printed: $\kwbf;vrb; = 1$ means that messages
                       will be printed even with successful operations.
                       $\kwbf;vrb; = 2,3$ means that only error messages will
                       be printed. $\kwbf;vrb; > 3$ is similar to
                       $\kwbf;vrb; = 3$, but
                       with the additional action of stopping the
                       program if a fatal error takes place.}
\def\arguouflag{\argu[ouflag; Integer, input] Logical output unit(s)
                       selection flag. See routine \kwbf;croheaderinfo;.}
\begin{description}
%

\rout  atmodelinit
\fcall atmodelinit(modidstr, modparstr, vrb, modname, rc)
\ccall atmodelinitc(\&modidstr, \&modparstr, \&vrb, \&modname, \&rc)
\dh

     Initialization of the atmospheric model and specifying its
     parameters if necessary.  This routine is automatically called
     every time a compressed output file\index{compressed output
     files} is opened to ensure by default that the atmospheric
     model\index{atmospheric model} used in the analysis is the same
     as the one in effect during the simulations that originated the
     compressed data file. For this reason it is very unlikely that a
     standard analysis application need manual invocation of
     \kwbf;atmodelinit;.

\barlist
\argu[modidstr; Input, character string] Label to switch among
                      atmospheric models. Maximum length is 16
                      characters.
              Current available options are:
\begin{description}
             \item[Linsley.] Linsley's standard model.
             \item[SouthPoleAvg.]     South pole average atmosphere.
             \item[LSouthPole.]       Linsley model for the South pole.
             \item[MalargueAvg.]      Malargue site annual average atmosphere.
             \item[GAMMA.]            J. C. Moreno's model.
             \item[Isothermic.]       Isothermic atmosphere
             \item[Homogeneous.]      Constant density atmosphere.
\end{description}
\argu[modparstr; Input, character string] String containing
                      model parameters.
\argu[vrb; Input, integer] Verbosity control. If vrb is
                      zero or negative then no error/informative
                      messages are printed; error conditions are
                      communicated to the calling program via the
                      return code. If vrb is positive error messages
                      will be printed: vrb = 1 means that messages
                      will be printed even with successful operations.
                      vrb = 2,3 means that only error messages will
                      be printed. vrb > 3 is similar to vrb = 3, but
                      with the additional action of stopping the
                      program if a fatal error takes place.
\argu[modname; Output, character string] A name for the
                      atmospheric model, to be typed somewhere.
                      Maximum length: 42 characters.
\argu[rc; Output, integer] Return code. Zero means
                      successful return.
\earlist
%

\rout atmosinit
\fcall atmosinit(modlabel, atmosname)
\ccall atmosinitc(\&modlabel, \&atmosname)
\dh

     Initialization of the atmospheric parameters.
Obsolete routine maintained just to guarantee backwards compatibility
of the AIRES system.
Presently, this routine initializes the standard Linsley atmosphere,
 regardless of the model label used.

\barlist
\argu[modlabel; Input, integer] Label to switch among
      atmospheric models. No longer used.
\argu[atmosname; Output, character string] A name for the
                      atmospheric model.
                      Maximum length: 42 characters.
\earlist
%

\routb adstydepth, adstymdepth
\ffun  dsty = adstydepth(Xvert, atlayer)
\cfun  dsty = adstydepth(\&Xvert, \&atlayer)
\ffun  dsty = adstymdepth(Xvert, atlayer)
\cfun  dsty = adstymdepth(\&Xvert, \&atlayer)
\dh

     Local density at a given depth. Multilayer atmospheric
     model\index{atmospheric model}.
     The atmospheric depth is measured in \gcmsq, and the density in
     g/cm$^3$ (\kwbf;adstydepth;) or m$^{-1}$\gcmsq (\kwbf;adstydepth;).

\barlist
\argu[Xvert; Input, double precision] Vertical atmospheric depth (in
\gcmsq) of the corresponding point. Must be positive.
\argu[atlayer; Output, integer] Atmospheric layer corresponding to the
depth \kwbf;Xvert;. This parameter depends on the selected atmospheric
model.\index{atmospheric model}
\earlist
\retval[Double precision]{The local density in g/cm$^3$
           (\kwbf;adstydepth;) or m$^{-1}$\gcmsq (\kwbf;adstydepth;).}
%

\rout  cioclose
\fcall cioclose(.)
\ccall cioclose(.)
\dh
Closing {\em all} the currently already opened CIO files.
%

\routb cioclose1
\fcall cioclose1(channel)
\ccall cioclose1(\&channel)
\dh
Closing an already opened CIO file.

\barlist
\arguchannel
\earlist
%

\rout ciorinit
\fcall ciorinit(inilevel, codsys, vrb, irc)
\ccall ciorinit(\&inilevel, \&codsys, \&vrb, \&irc)
\dh
Initializing the AIRES compressed I/O system for reading data. This
routine {\em must\/} be invoked at the beginning of every program
using the compressed I/O system routines.

\barlist
\argu[inilevel; Input, integer] Initialization switch. If inilevel is
zero or negative, all
                       needed
                       initialization routines are called. If positive
                       only the CIO system is initialized (The other
                       routines must be called within the invoking
                       program, before calling ciorinit:
$\kwbf;inilevel; = 1$ means complete
                       cio initialization, while $\kwbf;inilevel; > 1$
implies only particle coding initialization. This last case allows
changing the particle coding system\index{particle codes}
at any moment during a CIO processing session.
\argu[codsys; Input, integer] Particle coding system
                       identification. This variable permits selecting
                       among several particle coding systems supported
                       by AIRES (see table \ref{TAB:codsys}). The
                       menu of available systems is the following:
                       \begin{enumerate}
                         \item[0]  AIRES internal coding
                          system\index{AIRES particle codes}.
                         \item[1]  AIRES internal coding for elementary
                            particles and decimal nuclear notation
                            ($code=A + 100 * Z$).
                         \item[4]  Particle Data Group coding
system \cite{PDG} extended with decimal nuclear notation.\footnote{For
 nuclei the notation: $code = A + 100000 * Z$, is used.}
            \item[5]  CORSIKA program particle coding
                      system\index{CORSIKA!particle codes} \cite{CORSIKA}.
                         \item[6]  GEANT particle coding
                         system\index{GEANT!particle codes} \cite{GEANT}.
                         \item[8]  SIBYLL particle coding
                         system\index{SIBYLL!particle codes}
                         \cite{SIBYLL23c}, extended with decimal nuclear
                         notation.
                         \item[9]  MOCCA\index{MOCCA!particle codes}
                        style particle coding
                        system, extended with decimal nuclear notation.
                         \item[--]  Any other value is equivalent to
                            $\kwbf;codsys; = 1$.
                       \end{enumerate}
      \arguvrb
      \argu[irc; Output, integer] Return code. 0 means
                       successful return. 1 means that an invalid
                       particle coding system was specified by codsys
                       (in this case the default coding system is used).
\earlist
%

\rout ciorshutdown
\fcall ciorshutdown(.)
\ccall ciorshutdown(.)
\dh
      Terminating (in an ordered fashion) a compressed file analysis
      session. This routine should be invoked at the end of every CIO
      processing program.
%

\routb clockrandom
\ffun  r = clockrandom(.)
\cfun  r = clockrandom(.)
\dh

This function invokes the AIRES elementary random number
generator\index{random number generator!elementary without seed} and
returns a pseudo-random number uniformly distributed
in the interval $(0,1)$, generated with the current clock and CPU
usage lectures. No initialization is needed before using this random
number generator.

{\bfseries WARNING: This function is not to be used as a high quality
random number generator.}
This routine is intended only for some special applications like
generating a single random seed, for example.

     Multiple calls may eventually return correlated numbers if there
     is no enough time between invocations. Nevertheless, a sequence
     of different numbers passes direct 1d and 2d chi-square tests,
     ensuring a minimum quality for the generated numbers.

\retval[Double precision]{The uniform pseudo-random number.}
%

\rout crofieldindex
\ffun idx = crofieldindex(channel, rectype, fieldname, vrb, datype,
                            irc)
\cfun idx = crofieldindex(\&channel, \&rectype, \&fieldname, \&vrb, \&datype,
                          \&irc)
\dh

     Returning the index corresponding to a given field within a
     compressed file record. It is convenient to use this routine to
     set integer variables, and use them to manage the data returned by
     \kwbf;getcrorecord;, as explained in section \ppref{S:usinglib}.

\barlist
\arguchannel
\argu[rectype; Input, integer] Record type (0 for default
                 record type).
\argu[fieldname; Input, character string] First characters of
                 field name (enough characters must be provided
                 to make an unambiguous specification).
\arguvrb
\argu[datype; Output, integer] The data type that corresponds
                 to the specified field: 1 for integer data,
                 2 for date-time data, and 3 for real data.
\argu[irc; Output, integer] Return code. 0 means
                   successful return.
\earlist
\retval[Integer]{The field index. Zero if there was an error.}
%

\rout crofileinfo
\fcall crofileinfo(channel, ouflag, vrb, irc)
\ccall crofileinfo(\&channel, \&ouflag, \&vrb, \&irc)
\dh

Printing information about the records of an already opened compressed
file. This routine retrieves information about the complete record
structure of the corresponding file: How many record types are
defined, and for each record type the number of fields and a list of
their names and relative logical positions. The ordering in the list
of fields is equal to the ordering of data in the integer and real
arrays returned by routine \kwbf;getcrorecord; when reading a record
of the same type.

\barlist
\arguchannel
\arguouflag
\arguvrb
\argu[irc; Output, integer] Return code. 0 means
                   successful return.
\earlist
%

\rout crofileversion
\ffun ivers = crofileversion(channel)
\cfun ivers = crofileversion(\&channel)
\dh

     Returning the AIRES version used to write an already opened
     compressed file.

\barlist
\arguchannel
\earlist
\retval[Integer]{The corresponding version in integer
                 format (for example the number 01040200
                 for version 1.4.2,
                 01040201 for version 1.4.2a, etc.). If the file
                 is not opened or if there is an error, then the
                 return value is negative.}
%

\rout crogotorec
\ffun okflag = crogotorec(channel, recnumber, vrb, irc)
\cfun okflag = crogotorec(\&channel, \&recnumber, \&vrb, \&irc)
\dh

Positioning the file after a given record. This routine, used in
connection with \kwbf;crorecnumber;, allows emulating direct access to
compressed files. Notice however that a completely random access
regime with very large files may eventually imply longer processing
times.

\barlist
\arguchannel
\argu[recnumber; Input, integer] The record number. A negative
                      value is taken as zero.\\
                      If $\kwbf;recnumber; \le 0$, the
                      return code is always set to zero for
                      successful operations (Notice that in this case
the file will be positioned at the beginning of the data records).
\arguvrb
\argu[irc; Output, integer] Return code. The meanings of the different
values that can be returned are as explained for routine
\kwbf;getcrorecord;. When the return code is a record type, it
corresponds to the record type of the {\em last\/} scanned record.
\earlist

\retval[Logical]{True if the positioning was successfully
done. False otherwise.}
%

\rout croheaderinfo
\fcall croheaderinfo(ouflag, vrb, irc)
\ccall croheaderinfo(\&ouflag, \&vrb, \&irc)
\dh
     Printing a summary of the information contained in the header
     of the {\em most recently opened\/} compressed file.

\barlist
\argu[ouflag; Input, integer] Logical output unit(s)
                   selection flag: 0 or negative means FORTRAN unit 6 only,
                   1 means unit 7 only, 2 means both units 6 and 7,
                   3 means unit 8 only, $\kwbf;ouflag; > 8$ means unit
                   \kwbf;ouflag; only. FORTRAN unit 6 corresponds to
                   the  standard output channel.
\arguvrb
\argu[irc; Output, integer] Return code. 0 means
                   successful return.
\earlist
%

\rout croinputdata0
\fcall croinputdata0(intdata, realdata, shprimcode, shprimwt)
\ccall croinputdata0(\&intdata[1], \&realdata[1],
                     \&shprimcode[1], \&shprimwt[1])
\dh

     Copying into arrays some header data items corresponding to
     the {\em most recently opened\/} compressed file. Notice that
     some additional input parameters must be retrieved using routines
     \kwbf;getinpreal;, \kwbf;getinpint; or \kwbf;getinpswitch; (see
     pages \pageref{rou0:getinpint}--\pageref{rou0:getinpswitch}).

\barlist

\argu[intdata; Output, integer, array(*)] Integer data array.
                  The calling program must provide enough space
                  for it. The following list describes the
                  different data items:
\begin{list}{\bf no default}%
{\setlength{\itemsep}{0ex}%
\setlength{\leftmargin}{4.0em}%
\setlength{\labelwidth}{3.4em}%
\setlength{\labelsep}{0.6em}}
             \item[ 1] Number of different primary particles.
             \item[ 2-4] Reserved for future use.
             \item[ 5] Primary energy distribution: 0 fixed
                       energy; 1 varying energy.
             \item[ 6] Zenith angle distribution: 0, fixed angle;
                       1, {\em sine\/} distribution (equation
                       (\ref{eq:sindist})) ; 2, {\em sine-cosine\/}
                       distribution (equation (\ref{eq:sincosdist})). 
             \item[ 7] Azimuth angle distribution: 0 (10), fixed angle
                       (geographic azimuth)\index{geographic azimuth};
                       1 (11), varying angle (geographic azimuths).
             \item[ 8] Number of observing levels.
             \item[ 9] Atmospheric model\index{atmospheric model}
                       label (See page \pageref{IDL0:Atmosphere}).
             \item[10-14] Reserved for future use.
             \item[15] First shower number.
\end{list}
\argu[realdata; Output, double precision, array(*)] Real data
                  array. The calling program must provide enough
                  space for it. The following list describes the
                  different data items:
\begin{list}{\bf no default}%
{\setlength{\itemsep}{0ex}%
\setlength{\leftmargin}{4.0em}%
\setlength{\labelwidth}{3.4em}%
\setlength{\labelsep}{0.6em}}
             \item[ 1] Minimum primary energy (GeV).
             \item[ 2] Maximum primary energy (GeV).
             \item[ 3] Exponent $\gamma$ of energy distribution
                       (equation (\ref{eq:gammadist})).
             \item[ 4] Minimum zenith angle (deg).
             \item[ 5] Maximum zenith angle (deg).
             \item[ 6] Minimum azimuth angle (deg).
             \item[ 7] Maximum azimuth angle (deg).
             \item[ 8] Thinning\index{thinning} energy
                       parameter.\footnote{The thinning energy
                       parameter, $t_p$, must be interpreted as
                       follows: When positive, it gives the absolute
                       thinning energy in GeV. Otherwise it indicates
                       a relative thinning specification, being
                       $E_{\mathrm{th}} = |t_p|E_{\mathrm{primary}}$.}
             \item[ 9] Injection altitude (m).\footnote{Measured
                       vertically, starting from the intersection
                       point between the sea level and the line that
                       goes form the Earth's center to the particle
                       injection point, i.e., $z_v$ in figure
                       \ref{FIG:coor1}.}
             \item[10] Injection depth (\gcmsq).
             \item[11] Ground altitude (m).
             \item[12] Ground depth (\gcmsq).
             \item[13-14] Reserved for future use.
             \item[15] Altitude of first observing level (m).
             \item[16] Vertical depth of first observing level (\gcmsq).
             \item[17] Altitude of last observing level (m).
             \item[18] Vertical depth of last observing level (\gcmsq).
             \item[19] Distance between consecutive observing
                       levels in \gcmsq.
             \item[20] Site latitude (deg).
             \item[21] Site longitude (deg).
             \item[22] Geomagnetic field\index{geomagnetic field}
                       strength, F, (nT).
             \item[23] Local geomagnetic inclination, I, (deg).
             \item[24] Local geomagnetic declination, D, (deg).
             \item[25] Amplitude of random fluctuation of magnetic
                       field\footnote{The magnetic fluctuation
                       parameter, $f_p$, must be interpreted
                       as follows: When positive, it gives the
                       absolute fluctuation in nT. Otherwise it
                       indicates a relative fluctuation specification,
                       being $\Delta B = |f_p|\,\mathrm{F}$.}.
             \item[26-29] Reserved for future use.
             \item[30] Minimum lateral distance used for ground
                       particle histograms (m).
             \item[31] Maximum lateral distance used for ground
                       particle histograms (m).
             \item[32] Minimum energy used for histograms (GeV).
             \item[33] Maximum energy used for histograms (GeV).
             \item[34-35] Reserved for future use.
             \item[36] Minimum radial distance parameter for the
                       {\em most recently opened\/} compressed file (m).
             \item[37] Maximum radial distance parameter for the
                       {\em most recently opened\/} compressed file (m).
\end{list}
\argu[shprimcode; Output, integer, array(*)] For $i$ from 1 to
                  $\kwbf;intdata;(1)$, $\kwbf;shprimcode;(i)$ gives the
                  corresponding primary particle code. The
                  coding system\index{particle codes} used is the one
                  defined when
                  starting the cio system.
\argu[shprimwt; Output, double precision, array(*)] For $i$ from
                  1 to $\kwbf;intdata;(1)$, $\kwbf;shprimwt;(i)$ gives the
                  corresponding primary particle weight. This
                  weight is 1 in the single primary case.
\earlist
%

\rout crooldata
\fcall crooldata(vrb, nobslev, olzv, oldepth, irc)
\ccall crooldata(\&vrb, \&nobslev, \&olzv[1], \&oldepth[1], \&irc)
\dh

 Calculating observing levels information from data contained in
 a compressed data file header.

     Since the header data is of global nature, the data used
     by this routine corresponds to the {\em most recently opened\/}
     compressed file.

\barlist
\arguvrb
\argu[nobslev; Output, integer] The number of observing
                  levels.
\argu[olzv; Output, double precision, array(*)] Altitudes
                  (in m) of the corresponding observing levels, from
                  1 to \kwbf;nobslev;. The calling program must ensure
                  that there is enough space for this array.
\argu[oldepth; Output, double precision, array(*)] Vertical atmospheric
                  depth (in \gcmsq) of the corresponding
                  observing
                  levels, from 1 to \kwbf;nobslev;. The calling
                  program must ensure that there is enough space
                  for this array.
\argu[irc; Output, integer] Return code. 0 means
                  successful return.
\earlist
%

\rout croreccount
\fcall croreccount(channel, vrb, nrtype, nrec, irc)
\ccall croreccount(\&channel, \&vrb, \&nrtype[0], \&nrec, \&irc)
\dh

     Counting the records of a compressed file starting from the
     first non-read record. Once the file was scanned, the
     corresponding I/O channel is left in ``end of file''
     status.

\barlist
\arguchannel
\arguvrb
\argu[nrtype; Output, integer] The highest record type
                 defined for the file (record types range from
                 zero to \kwbf;nrtype;).
\argu[nrec; Output, integer, array(0:\/\kw;nrtype;)] For each
                 record type, the number of records found.
                 No check is made to ensure that the length of the
array is enough to store all the data items.
\argu[irc; Output, integer] Return code. 0 means
                   successful return.
\earlist
%
%

\rout crorecfind
\ffun okflag = crorecfind(channel, intype, vrb, infield1, rectype)
\cfun okflag = crorecfind(\&channel, \&intype, \&vrb, \&infield1, \&rectype)
\dh

Reading records until getting a specified record type. The compressed
file associated with \kwbf;channel; is scanned until a record of type
\kwbf;intype; is found.

\barlist
\arguchannel
\argu[intype; Input, integer] Record type to find.
\arguvrb
\argu[infield1; Output, integer] If \kwbf;intype; is zero, this
variable contains the current value of the first integer field of the
last scanned record (which will be, in general, a particle code).
Otherwise it is set to zero.
\argu[rectype; Output, integer] Last scanned record type and return code.
This argument contains the same information as argument \kwbf;irc; of
routine \kwbf;getcrorecord;. Notice that in the case of successful
return, \kwbf;rectype; is equal to \kwbf;intype;.
\earlist
\retval[Logical]{True if the last record was successfully
read. False otherwise (End of file or I/O error).}
%
%

\rout crorecinfo
\fcall crorecninfo(channel, poskey, ouflag, vrb, irc)
\ccall crorecninfo(\&channel, \&poskey, \&ouflag, \&vrb, \&irc)
\dh

Printing information about the total number of records within an
already opened compressed file. The file is scanned starting after the
last record already read to count the number of records of each type
that were written into it.

\barlist
\arguchannel
\argu[poskey; Input, integer] Positioning key. This parameter
                  allows to control the file positioning after
                  returning from this routine: If zero or negative
                  the file remains positioned at the ``end of file''
                  point, if 1 at the beginning of data, and if
                  greater than 1, at the position found before the
                  call (This last option may eventually imply a
                  significant increase in
                  processing time for very large files).
\arguouflag
\arguvrb
\argu[irc; Output, integer] Return code. 0 means
                   successful return.
\earlist
%

\rout crorecnumber
\ffun recno = crorecnumber(channel, vrb, irc)
\cfun recno = crorecnumber(\&channel, \&vrb, \&irc)
\dh

This function returns the current record number corresponding to an
already opened compressed file.

\barlist
\arguchannel
\arguvrb
\argu[irc; Output, integer] Return code. 0 means
                   successful return.
\earlist

\retval[Integer]{The record number. If the file is not
  ready (closed or end of file), then $-1$ is returned.}
%

\rout crorecstrut
\fcall crorecstruct(channel, nrtype, nintf, nrealf, irc)
\ccall crorecstruct(\&channel, \&nrtype, \&nintf, \&nrealf, \&irc)
\dh

     Getting information about the records of an already opened
     compressed file.

\barlist
\arguchannel
\argu[nrtype; Output, integer] The highest record type
                 defined for the file (record types range from
                 zero to \kwbf;nrtype;).
\argu[nintf; Output, integer, array(0:\/\kw;nrtype;)] Number
                 of integer fields contained at each record type,
                 for record types from zero to \kwbf;nrtype;.
                 No check is made to ensure that the length of the
array is enough to store all the data items.
\argu[nrealf; Output, integer, array(0:\/\kw;nrtype;)] Number
                 of real fields contained at each record type,
                 for record types from zero to \kwbf;nrtype;.
                 No check is made to ensure that the length of the
array is enough to store all the data items.
\argu[irc; Output, integer] Return code. 0 means
                   successful return.
\earlist
%

\rout crorewind
\fcall crorewind(channel, vrb, irc)
\ccall crorewind(\&channel, \&vrb, \&irc)
\dh

     ``Rewinding''\index{rewinding compressed files} an already opened
compressed file. The file is positioned just before the first data
record. In other words, the file is system rewound and its header is
re-scanned so the file pointer remains located at the beginning of the
record data stream.

\barlist
\arguchannel
\arguvrb
\argu[irc; Output, integer] Return code. 0 means
                      successful return.
\earlist
%

\rout crospcode
\ffun isspecial = crospcode(pcode, splabel)
\cfun isspecial = crospcode(\&pcode, \&splabel)
\dh

This logical function determines whether or not a given particle code
corresponds to a special primary particle\index{special primary
particles}.

\barlist
\argu[pcode; Input, integer] The particle code to check.
\argu[splabel; Output, integer] Label associated to the
                      special particle, or zero if the code does not
                      corresponds to a special particle. This variable
                      is useful for further use with other library
routines, and should not be set by the calling program.
\earlist

\retval[Logical]{True if the input code corresponds to a special
  primary particle. False otherwise.}
%

\rout crospmodinfo
\fcall crospmodinfo(spname, spmodu, spml, sppars, sppl, irc)
\ccall crospmodinfoc(\&spname, \&spmodu, \&spml, \&sppars, \&sppl,
\&irc)
\dh

    Retrieving information about the external module associated to a
    already defined special particle\index{special primary particles}.
    When this routine is used to retrieve information stored in a
    compressed file, the data returned correspond to the {\em most
    recently opened\/} compressed file.

\barlist
\argu[intdata; Input, string] The name of the special
                      particle.
\argu[spmodu; Output, string] The name of the
                      associated module. The calling program must
                      provide enough space for this string.
\argu[spmodu; Output, integer] Length of string \kwbf;spmodu;.
\argu[sppars; Output, string] String containing the
                      parameters passed to the module. The calling
                      program must provide enough space for this
                      string.
\argu[sppl; Output, integer] Length of string \kwbf;sppars;.
\earlist
%

\rout crospnames
\fcall crospnames(nspp, spname)
\ccall crospnamesc(\&nspp, \&spname[1])
\dh

Retrieving the names of the currently defined special
particles\index{special primary particles}.
    When this routine is used to retrieve information stored in a
    compressed file, the data returned correspond to the {\em most
    recently opened\/} compressed file.

\barlist
\argu[nspp; Input, integer] The number of special particles defined.
\argu[spname; Output, string, array(*)]  Array
                containing the names of the defined particles.
                The calling program must provide enough
                space for this array, and its elements (maximum 16
                characters each).
\earlist
%

\rout crotaskid
\fcall crotaskid(taskname, tasknamelen, taskversion, startdate)
\ccall crotaskidc(\&taskname, \&tasknamelen, \&taskversion, \&startdate)
\dh

Getting task name and starting date for the task corresponding to the {\em
most recently opened\/} compressed file.

\barlist
\argu[taskname; Output, string] The task name. The calling program must
ensure there is enough space to store the string.
\argu[tasknamelen; Output, integer] Length of task name.
\argu[tasknameversion; Output, integer] Task version.
\argu[startdate; Output, string] Task starting date in the format
``dd/Mmm/yyyy~hh:mm:ss'' (20 characters).
\earlist
%

\rout dati
\fcall dati(datistr)
\ccall dati(\&datistr)
\dh

     Current date and time in the format dd/Mmm/yyyy hh:mm:ss

\barlist
\argu[datistr; Output, character string] The string containing
                      the current date and time.
\earlist
%

\rout depthfromz
\ffun Xvert = depthfromz(z, atlayer)
\cfun Xvert = depthfromz(\&z, \&atlayer)
\dh

     Atmospheric depth from vertical altitude. Multilayer atmospheric
     model.\index{atmospheric model}.

\barlist
\argu[z; Input, double precision] The vertical altitude in meters
                       above sea level.
\argu[atlayer; Output, integer] Atmospheric layer corresponding to the
depth \kwbf;Xvert;. This parameter depends on the selected atmospheric
model.\index{atmospheric model}
\earlist
\retval[Double precision]{The vertical depth in \gcmsq}
%

\rout dumpfileversion
\ffun ivers = dumpfileversion(.)
\cfun ivers = dumpfileversion(.)
\dh

     Returning the AIRES version associated with the dump file that was
     {\em most recently read in\/} (this can be done using routine
     \libbfx;loadumpfile;).

\retval[Integer]{The corresponding version in integer
                 format (for example the number 01040200
                 for version 1.4.2,
                 01040201 for version 1.4.2a, etc.). If there is an
                 error, then the return value is negative.}
%

\rout dumpfileversiono
\ffun ivers = dumpfileversiono(.)
\cfun ivers = dumpfileversiono(.)
\dh

     Returning the AIRES version used to write for the first time (the
     original version) the dump file that was {\em most recently read
     in\/} (this can be done using routine \libbfx;loadumpfile;).

\retval[Integer]{The corresponding version in integer
                 format (for example the number 01040200
                 for version 1.4.2,
                 01040201 for version 1.4.2a, etc.). If there is an
                 error, then the return value is negative.}
%

\rout dumpinputdata0
\fcall dumpinputdata0(intdata, realdata)
\ccall dumpinputdata0(\&intdata[1], \&realdata[1])
\dh

     Copying into arrays some global input data parameters stored in the
     dump file that was {\em most recently read in\/} (this can be
     done using routine \libbfx;loadumpfile;), that are
     not returned by \libbfx;croinputdata0;.

\barlist
\argu[intdata; Output, integer, array(*)] Integer data array.
                  The calling program must provide enough space
                  for it. The following list describes the
                  different data items:
\begin{list}{\bf no default}%
{\setlength{\itemsep}{0ex}%
\setlength{\leftmargin}{4.0em}%
\setlength{\labelwidth}{3.4em}%
\setlength{\labelsep}{0.6em}}
             \item[ 1] Total number of showers.
             \item[ 2] Number of completed showers.
             \item[ 3] First shower number.
             \item[ 4-9] Reserved for future use.
             \item[10] Separate showers integer parameter.
\end{list}
\argu[realdata; Output, double precision, array(*)] Real data
                  array. The calling program must provide enough
                  space for it. The following list describes the
                  different data items:
\begin{list}{\bf no default}%
{\setlength{\itemsep}{0ex}%
\setlength{\leftmargin}{4.0em}%
\setlength{\labelwidth}{3.4em}%
\setlength{\labelsep}{0.6em}}
             \item[ 1-  ] Reserved for future use.
\end{list}
\earlist
%

\rout fitghf
\fcall fitghf(bodata0, eodata0, depths, nallch, weights, ws,
              minnmax, nminratio, bodataeff, eodataeff,
              nmax, xmax, x0, lambda, sqsum, irc)
\ccall fitghf(\&bodata0, \&eodata0, \&depths[1], \&nallch[1],
              \&weights[1], \&ws,
              \&minnmax, \&nminratio, \&bodataeff, \&eodataeff,
              \&nmax, \&xmax, \&x0, \&lambda, \&sqsum, \&irc)
\dh

     Performing a 4-parameter nonlinear least squares fit to evaluate
     the parameters $N_{\mathrm{max}}$, $X_{\mathrm{max}}$, $X_0$ and
     $\lambda$ of the Gaisser-Hillas function\index{Gaisser-Hillas
     function!fitting} of equation (\ref{eq:ghf4p}). The fit is done
     using the Levenberg-Mardquardt algorithm, as implemented in the
     public domain software library Netlib\index{external
     packages}\index{Netlib} \cite{netlib}.

\barlist
\argu[bodata0, eodata0; Input, integer] Positive integer parameters
                      defining the number of data points to use
                      in the fit.
\argu[depths; Input, double precision, array(eodata0)] Depths
                      of the observing levels used in the fit. Only
                      the range (\kwbf;bodata0;:\kwbf;eodata0;) is used.
\argu[nallch; Input, double precision, array(eodata0)] Number
                      of charged particles crossing the different
                      levels. Only the range
                      (\kwbf;bodata0;:\kwbf;eodata0;) is used.
\argu[weights; Input, double precision, array(eodata0)]
                      Positive weights to be assigned to each one of
                      the data points. Only the range
                      (\kwbf;bodata0;:\kwbf;eodata0;) is used.
\argu[ws; Input, integer] If $\kwbf;ws; = 2$, the weights are
                      evaluated internally (proportionally to the
                      square root of the number of particles).
                      If $\kwbf;ws; = 1$ they must be provided as input data.
                      If $\kwbf;ws; = 2$ the array \kwbf;weights; is not used.
\argu[minnmax; Input, double precision] Threshold value for
                      the maximum number of particles in the input
                      data set. The fit is not performed if the
                      maximum number of particles is below this
                      parameter. If \kwbf;minnmax; is negative, it is
                      taken as zero.
\argu[nminratio; Input, double precision] Positive parameter
                      used to determine the end of the data set.
                      Must be equal or greater than 5. Once the
                      maximum of the data set is found. the points
                      located after this maximum up to the point
                      where the number of charged particles is less
                      than the maximum divided \kwbf;nminratio;. The
                      remaining part of the data is not taken into
                      account in the fit. A similar analysis is
                      performed with the points located before the
                      maximum. The recommended value is 100. A very
                      large value will enforce inclusion of all the
                      data set.
\argu[bodataeff, eodataeff; Output, integer] The actual range of data
                      points used in the fit.
\argu[nmax; Output, double precision] Estimated number of
                      charged particles at the shower maximum
                      (parameter $N_{\mathrm{max}}$). If no
                      fit was possible, then the value coming from a
                      direct estimation from the input data is returned.
\argu[xmax; Output, double precision] Fitted position of
                      the shower maximum, $X_{\mathrm{max}}$, in
                      \gcmsq. If no fit was
                      possible, then the value coming from a direct
                      estimation from the input data is returned.
\argu[x0; Output, double precision] Fitted position of
                      the point where the Gaisser-Hillas
                      function\index{Gaisser-Hillas function} is zero
                      (parameter $X_0$), expressed in \gcmsq.
\argu[lambda; Output, double precision] Fitted parameter
                      $\lambda$, in \gcmsq.
\argu[sqsum; Output, double precision] The resulting
                      normalized sum of squares:
\begin{equation}                                   \label{eq:nsofsq}
S= \frac{1}{N\,N_{\mathrm{max}}}
\sum_{i=1}^N \frac{\left[N^{(I)}(i)-N^{(GH)}(i)\right]^2}{N^{(GH)}(i)},
\end{equation}
where $N$ is the number of data points used in the fit, and $N^{(I)}$
($N^{(GH)}$) represent the set of particle numbers given as input
(returned from equation \ref{eq:ghf4p} for the corresponding depths).
\argu[irc; Output, integer] Return code. Zero means that
                      the fit was successfully completed.
\earlist
%

%
%

%
%

%
%

\rout getcrorecord
\ffun okflag = getcrorecord(channel, intfields, realfields, altrec,
                             vrb, irc)
\cfun okflag = getcrorecord(\&channel, \&intfields[1],
                             \&realfields[1], \&altrec,
                             \&vrb, \&irc)
\dh
 Reading a record from a compressed data file already opened. This
routine can be used to read records from {\em every} kind of
compressed file: The routine automatically processes the records
without needing any user-level specification beyond file identity
(parameter \kwbf;channel;). The logical returned value (here assigned
to logical variable \kwbf;okflag;) permits determining whether or not
the read operation was successful. The characteristics of the read
record are informed via the return code (\kwbf;irc;), and the arrays
\kwbf;intfields; and \kwbf;realfields; contain the corresponding data
items. Their contents depend on the file being processed and on the
record type. The auxiliary routines \kwbf;crofileinfo; and
\kwbf;crofieldindex; are useful to process adequately the returned
data at each case.

\barlist
\arguchannel
\argu[intfields; Output, integer, array(*)] Integer fields
                  of the last read record. This includes the non-scaled
                  integer quantities and (in the last positions) the date-time
                  specification(s), if any. The calling
                  program must provide enough space for this
                  array (The minimum dimension is the maximum
                  number of fields that can appear in a record
                  plus 1). Positions beyond the last integer
                  fields are used as scratch working space. The
                  meaning of each data item within this array varies
                  with the class of file processed and with the record
                  type (see also argument \kwbf;irc; and routine
                  \kwbf;crofileinfo;).
\argu[realfields; Output, double precision, array(*)] Real
                  fields of the record. The calling program must
                  provide enough space for this array. The
                  meaning of each data item within this array varies
                  with the class of file processed and with the record
                  type (see also argument \kwbf;irc; and routine
                  \kwbf;crofileinfo;).
\argu[altrec; Output, logical] True if the corresponding record type
                  is positive (alternative record type)
                  False if the record type is zero (default record
                  type).
\arguvrb
\argu[irc; Output, integer] Return code. 0 means that a record with
                  zero (default) record type was successfully read.
                  $i$ ($i > 0$) means that an alternative record of
                  type $i$ was successfully read. $-1$ means that an
                  end-of-file condition was got from the
                  corresponding file.  Any other value indicates
                  a reading error (\kwbf;irc; equals the system return
                  code plus 10000).
\earlist
\retval[Logical]{True if a record was successfully read.
False otherwise (End of file or I/O error).}
%

\rout getcrorectype
\ffun okflag = getcrorectype(channel, vrb, infield1, rectype)
\cfun okflag = getcrorectype(\&channel, \&vrb, \&infield1, \&rectype)
\dh

Getting the record type of the record which is located next to the
last read record of the compressed file identified by argument
\kwbf;channel;.

The action of this routine consists in reading the first part of the
record to obtain the record type, and then skip the remaining part to
position the file at the end of the corresponding record. The use of
this routine is recommended whenever only the record type is needed, since
it is faster than \kwbf;getcrorecord;. When additional data of an
already scanned record is required, routine \libbfx;regetcrorecord; can
be used to re-scan the last processed one.

\barlist
\arguchannel
\arguvrb
\argu[infield1; Output, integer] If \kwbf;rectype; is zero, this
variable contains the current value of the first integer field of the
record (which is, in general, a particle code). Otherwise it is set to
zero.
\argu[rectype; Output, integer] Record type and return code. This
argument contains the same information as argument \kwbf;irc; of
routine \kwbf;getcrorecord;.
\earlist
\retval[Logical]{True if a record was successfully read.
False otherwise (End of file or I/O error).}
%

\rout getglobal
\fcall getglobal(gvname, sdynsw, gvval, valen)
\ccall getglobalc(\&gvname, \&sdynsw, \&gvval, \&valen)
\dh

     Getting the current value of an already defined global
    varible\index{global variables}.
    When this routine is used to retrieve information stored in a
    compressed file, the data returned correspond to the {\em most
    recently opened\/} compressed file.

\barlist
\argu[gvname; Input, string] Name of global variable.
\argu[sdynsw; Output, integer] Type of variable: 1 dynamic, 2 static, 0
if the variable is undefined.
\argu[gvval; Output, string] The string currently
                      assigned to the variable. The calling program
                      must ensure enough space to store the string.
\argu[valen; Output, integer] Length of \kwbf;gvval;. \kwbf;valen; is
negative for undefined variables.
\earlist
%

%
%

\rout getinpint
\ffun value = getinpint(dirname)
\cfun value = getinpintc(\&dirname)
\dh

     Getting the current value for an {\bf integer} (static) input
     parameter corresponding to the {\em most recently opened\/}
     compressed file. This routine is used to get from the current file's
     header those integer input parameters not returned by routine
     \kwbf;croinputdata0; (see page \pageref{rou0:croinputdata0}).

\barlist
\argu[dirname; Input, string] Name of the IDL directive associated with
the parameter (can be abbreviated accordingly with the rules described
in appendix \ref{A:idlref}).
\earlist
\retval[integer]{The current setting for the corresponding
parameter. In case of error the returned value is undefined.}
%

\rout getinpreal
\ffun value = getinpreal(dirname)
\cfun value = getinprealc(\&dirname)
\dh

     Getting the current value for a {\bf real} (static) input parameter
     corresponding to the {\em most recently opened\/} compressed
     file. This routine is used to get from the current file's header
     those real input parameters not returned by routine
     \kwbf;croinputdata0; (see page \pageref{rou0:croinputdata0}).

\barlist
\argu[dirname; Input, string] Name of the IDL directive associated with
the parameter (can be abbreviated accordingly with the rules described
in appendix \ref{A:idlref}).
\earlist
\retval[double precision]{The current setting for the corresponding
parameter. In case of error the returned value is undefined.}
%

\rout getinpstring
\fcall getinpstring(dirname, value, slen)
\ccall getinpstringc(\&dirname, \&value, \&slen)
\dh

     Getting the current value for an input (static) {\bf character string}
     corresponding to the {\em most recently opened\/} compressed
     file.

\barlist
\argu[dirname; Input, string] Name of the IDL directive associated with
the parameter (can be abbreviated accordingly with the rules described
in appendix \ref{A:idlref}).
\argu[value; Output, string] The current parameter value. The calling
program must ensure that there is enough space to store the string.
\argu[slen; Output, integer] Length of the current parameter value. On
error, \kwbf;slen; is negative.
\earlist
%

\rout getinpswitch
\ffun value = getinpswitch(dirname)
\cfun value = getinpswitchc(\&dirname)
\dh

     Getting the current value for an input (static) {\bf logical switch}
     corresponding to the {\em most recently opened\/} compressed
     file. This routine is used to get from the current file's header
     those logical input parameters not returned by routine
     \kwbf;croinputdata0; (see page \pageref{rou0:croinputdata0}).

\barlist
\argu[dirname; Input, string] Name of the IDL directive associated with
the parameter (can be abbreviated accordingly with the rules described
in appendix \ref{A:idlref}).
\earlist
\retval[Logical]{The current setting for the corresponding
parameter. In case of error the returned value is undefined.}
%

%
%

%
%

\rout getlgtinit
\fcall getlgtinit(channel, vrb, irc)
\ccall getlgtinit(\&channel, \&vrb, \&irc)
\dh
 
      Initializing internal data needed to process records from
      compressed longitudinal particle tracking files by means of
      routine \libbfx;getlgtrecord; and related ones.
      This routine should be called immediately after opening
      the corresponding compressed file.

\barlist
\arguchannel
\arguvrb
\argu[irc; Output, integer] Return code. 0 means
                      successful return.
\earlist
%

\rout getlgtrecord
\ffun okflag = getlgtrecord(channel, currol, updown,
                            intfields, realfields, altrec,
                             vrb, irc)
\cfun okflag = getlgtrecord(\&channel, \&currol, \&updown,
                            \&intfields[1],
                             \&realfields[1], \&altrec,
                             \&vrb, \&irc)
\dh

 Reading a record from a compressed longitudinal particle tracking
 file and returning the read data in a ``level per level'' basis.
 This routine invokes \libbfx;getcrorecord; to get a record from the
 corresponding compressed file when it is necessary, and must be used
 jointly with \libbfx;getlgtinit;.

\barlist
\arguchannel
\argu[currol; Output, integer] Observing level crossed by the particle.
\argu[updown; Output, integer] Up-down indicator: 1 if the
                  particle is going upwards, $-1$ otherwise.
\argu[intfields; Output, integer, array(*)] Integer fields
                  of the last read record. This includes the non-scaled
                  integer quantities and (in the last positions) the date-time
                  specification(s), if any. The calling
                  program must provide enough space for this
                  array (The minimum dimension is the maximum
                  number of fields that can appear in a record
                  plus 1). Positions beyond the last integer
                  fields are used as scratch working space. The
                  meaning of each data item within this array varies
                  with the class of file processed and with the record
                  type (see also argument \kwbf;irc; and routine
                  \kwbf;crofileinfo;).
\argu[realfields; Output, double precision, array(*)] Real
                  fields of the record. The calling program must
                  provide enough space for this array. The
                  meaning of each data item within this array varies
                  with the class of file processed and with the record
                  type (see also argument \kwbf;irc; and routine
                  \kwbf;crofileinfo;).
\argu[altrec; Output, logical] True if the corresponding record type
                  is positive (alternative record type)
                  False if the record type is zero (default record
                  type).
\arguvrb
\argu[irc; Output, integer] Return code. 0 means that a record with
                  zero (default) record type was successfully read.
                  $i$ ($i > 0$) means that an alternative record of
                  type $i$ was successfully read. $-1$ means that an
                  end-of-file condition was got from the
                  corresponding file.  Any other value indicates
                  a reading error (\kwbf;irc; equals the system return
                  code plus 10000).
\earlist
\retval[Logical]{True if a record was successfully read.
False otherwise (End of file or I/O error).}

\rout ghfpars
\fcall ghfpars(nmax, xmax, x0, lambda, vrb, irc)
\ccall ghfpars(\&nmax, \&xmax, \&x0, \&lambda, \&vrb, \&irc)
\dh

     Setting the internal quantities needed to work with the
     Gaisser-Hillas function\index{Gaisser-Hillas function} (equation
     (\ref{eq:ghf4p})) related routines.

\barlist
\argu[nmax;   Input, double precision] Parameter $N_{\mathrm{max}}$ of
                      equation \ref{eq:ghf4p}.
\argu[xmax;   Input, double precision] Parameter $X_{\mathrm{max}}$ of
                      equation \ref{eq:ghf4p}.
\argu[x0;     Input, double precision] Parameter $X_{0}$ of
                      equation \ref{eq:ghf4p}.
\argu[lambda; Input, double precision] Parameter $\lambda$ of
                      equation \ref{eq:ghf4p}.
\argu[irc; Output, integer] Return code. 0 means
                   successful return.
\arguvrb
\earlist
%

\rout ghfin
\ffun x = ghfin(np, prepost)
\cfun x = ghfin(\&np, \&prepost)
\dh

     Numerical evaluation of the inverse of the Gaisser-Hillas
     function\index{Gaisser-Hillas function!inverse of} (equation
     (\ref{eq:ghf4p})) for a given number of particles
     \kwbf;np;. The four parameters $N_{\mathrm{max}}$,
     $X_{\mathrm{max}}$, $X_{0}$, and $\lambda$ must be specified
     previously by means of \libbfx;ghfpars;.

\barlist
\argu[np; Input, double precision] The number of
                      particles. If $\kwbf;np; < 0$ or $\kwbf;np; >
                      N_{\mathrm{max}}$, the
                      result is a large negative number.
\argu[prepost; Input, integer] Integer parameter labeling
                      which of the two abscissas $X$ has to be
                      returned: If prepost is less or equal to 0 then
                      $X < X_{\mathrm{max}}$; 
                      otherwise $X > X_{\mathrm{max}}$.
                      Notice that the inverse of the Gaisser-Hillas
                      function is bi-valuated.
\earlist
\retval[Double precision]{The value of the inverse Gaisser-Hillas
function, expressed in {\gcmsq}, that is,
\kwbf;x; such that \kwbf;np; = \kwbf;ghfx(x);.}
%

\rout ghfx
\ffun np = ghfx(x)
\cfun np = ghfx(\&x)
\dh

     Evaluating the Gaisser-Hillas function\index{Gaisser-Hillas
     function} (equation (\ref{eq:ghf4p})) for a given depth
     \kwbf;x;. The four parameters $N_{\mathrm{max}}$,
     $X_{\mathrm{max}}$, $X_{0}$, and $\lambda$ must be specified
     previously by means of \libbfx;ghfpars;.

\barlist
\argu[x; Input, double precision] Atmospheric depth in {\gcmsq}.
\earlist
\retval[Double precision]{The value of the function at the specified
                          \kwbf;x;.}
%

\rout grandom
\ffun r = grandom(.)
\cfun r = grandom(.)
\dh

This function invokes the AIRES random number generator\index{random
number generator} and returns a pseudo-random number with normal
Gaussian distribution (zero mean and unit standard deviation). It is
necessary to initialize the random series calling \libbfx;raninit;
before using this function.

\retval[Double precision]{The Gaussian pseudo-random number.}
%

%
%

\rout idlcheck
\ffun ikey = idlcheck(dirname)
\cfun ikey = idlcheckc(\&dirname)
\dh

     Checking a string to see if it matches any of the IDL
     instructions currently defined, that is, the ones corresponding
     to the {\em most recently opened\/} compressed
     file.

\barlist
\argu[dirname; Input, string] Name of the IDL directive to be checked
(can be abbreviated accordingly with the rules described
in appendix \ref{A:idlref}).
\earlist
\retval[Integer]{If an error occurs, then the returned
            value will be negative. Other return values are the
            following:
            \begin{enumerate}
            \item[0] The string does not match any of the currently
                     valid IDL instructions.
            \item[1] The string matches a directive belonging to the
                     ``basic'' instruction set with no parameter(s)
                     associated with it, for example \kwbf;Help;.
            \item[2] The string matches a directive belonging to the
                     ``basic'' instruction set. If there is a parameter
                     associated with the directive, then it can be
                     obtained by means of routine \libbfx;croinputdata0;.
            \item[4] The directive corresponds to a real input
                     parameter. The parameter can be retrieved by
                     means of function \libbfx;getinpreal;.
            \item[6] The directive corresponds to an integer input
                     parameter. The parameter can be retrieved by
                     means of function \libbfx;getinpint;.
            \item[8] The directive corresponds to a logical input
                     parameter. The parameter can be retrieved by
                     means of function \libbfx;getinpswitch;.
           \item[10] The directive correspond to a string input
                     parameter. The parameter can be retrieved by
                     means of routine \libbfx;getinpstring;.
            \end{enumerate}}
%

%
%

%
%

\rout loadumpfile
\fcall loadumpfile(wdir, taskname, vrb, irc)
\ccall loadumpfilec(\&wdir, \&taskname, \&vrb, \&irc)
\dh

  Reading the dump file\index{internal dump file} associated with a
  given task, and copying into internal variables all the information
  contained within it.

\barlist
\argu[wdir; Input, character string] The name of the directory
                      where the file is placed. It defaults to the
                      current directory when blank.
\argu[taskname; Input, character string] Task name, or dump file name.
\arguvrb
\argu[irc; Output, integer] Return code. 0 means
                 successful return. 1 means successful return,
                 but the dump file was not created using the
                 same AIRES version. 8 means that no dump file
                 (in the sequence \kwbf;taskname;, \kwbf;taskname.adf;,
                 \kwbf;taskname.idf;) exists. 12 means invalid file
                 name. Other return codes
                 come from the adf or idf read routines.
\earlist
%

\rout nuclcode
\ffun ncode = nuclcode(z, n, irc)
\cfun ncode = nuclcode(\&z, \&n, \&irc)
\dh

     This routine returns the AIRES code of a nucleus of $Z$ protons
     and $N$ neutrons, as defined in page \pageref{eq:nuclcode}.

\barlist
\argu[z; Input, integer] The number of protons in the nucleus.
\argu[n; Input, integer] The number of neutrons in the nucleus.
\argu[irc; Output, integer] Return code. 0 means that a valid pair of
input parameters $(Z,N)$ was successfully processed. 3 means that the
nucleus cannot be specified with the AIRES system. 5 means that either
$Z$ or $N$ are out of allowed ranges.
\earlist
\retval[Integer]{The nucleus code of equation (\ref{eq:nuclcode}).}
%

\rout nucldecode
\fcall nucldecode(ncode, z, n, a)
\ccall nucldecode(\&ncode, \&z, \&n, \&a)
\dh

     This routine returns the charge, neutron and mass numbers
     corresponding to a given AIRES nuclear code (see page
     \pageref{eq:nuclcode}).

\barlist
\argu[ncode; Input, integer] The AIRES nuclear code of equation
(\ref{eq:nuclcode}).
\argu[z; Output, integer] The number of protons in the nucleus.
\argu[n; Output, integer] The number of neutrons in the nucleus.
\argu[a; Output, integer] The mass number.
\earlist
%

\rout olcoord
\fcall olcoord(nobslev, olzv, groundz, injz, zenith, azimuth,
               xaxis, yaxis, zaxis, tshift, mx, my, irc)
\ccall olcoord(\&nobslev, \&olzv[1], \&groundz, \&injz, \&zenith, \&azimuth,
               \&xaxis[1], \&yaxis[1], \&zaxis[1], \&tshift[1],
               \&mx[1], \&my[1], \&irc)
\dh

 This routine evaluates the coordinates of the intersections of
 observing level surfaces with the shower axis,
 $(x_{0i},y_{0i},z_{0i})$, $i=1,\ldots, N_o$, the corresponding
 time shifts, $t_{0i}$, and the coefficients, $m_{xi}$, $m_{yi}$, of
 the plane tangent to the surface at the intersection point:
\begin{equation}                                  \label{eq:oltanpl}
             z - z_{0i} = m_{xi} (x - x_{0i}) + m_{yi} (y - y_{0i}),
             \quad
             i = 1,\ldots, N_o .
\end{equation}

\barlist
\argu[nobslev; Input, integer] The number of observing
                  levels ($N_o$).
\argu[olzv; Input, double precision, array(\/\kw;nobslev;)]
                  Altitudes (in m) of the corresponding observing
                  levels.
\argu[groundz; Input, double precision] Ground altitude (in m).
\argu[injz; Input, double precision] Injection altitude
                  (in m).
\argu[zenith; Input, double precision] Shower zenith angle (deg).
\argu[azimuth; Input, double precision] Shower azimuth angle (deg).
\argu[xaxis, yaxis,
 zaxis; Output, double precision, array(\/\kw;nobslev;)]
                  Respectively $x_{0i}$, $y_{0i}$ and $z_{0i}$,
                  $i=1,\ldots, N_o$, coordinates (in m) of the intersection
                  points between the observing level surfaces and
                  the shower axis.
\argu[tshift; Output, double precision, array(\/\kw;nobslev;)]
                  Observing levels time shifts, $t_{0i}$,
                  $i=1,\ldots,N_o$, (in ns), that is,
                  the amount of time a particle moving at the speed of
                  light
                  needs to go from the shower injection point to
                  corresponding intersection point $(x_{0i}, y_{0i},
                  z_{0i})$.
\argu[mx, my; Output, double precision, array(\/\kw;nobslev;)]
                  Coefficients of the planes which are tangent to the
                  observing levels and pass by the corresponding
                  intersection points.
\argu[irc; Output, integer] Return code. Zero means
                  successful return.
\earlist
%

\rout olcrossed
\fcall olcrossed(olkey, updown, firstol, lastol)
\ccall olcrossed(\&olkey, \&updown, \&firstol, \&lastol)
\dh

 This routine reconstructs the information contained in the {\em
 crossed observing levels key\index{crossed observing levels key},\/}
 one of the data items saved at each particle record in any
 longitudinal tracking compressed file.

 This
 key encodes the first and last crossed observing
 observing levels and the direction of motion.
 The encoding formula defined in equation (\ref{eq:olcrossed}), where
 $L$, $i_f$ and $i_l$ correspond to \kwbf;olkey;, \kwbf;firstol; and
 \kwbf;lastol;, respectively.

 The routine returns all the variables of the right hand side of
 equation (\ref{eq:olcrossed}). The variable associated to $s_{\mathrm{ud}}$,
 \kwbf;updown; is set in a
 slightly different way: It is be set to 1 when the particle
 goes upwards, and to $-1$ otherwise.

\barlist
\argu[olkey; Input, integer] Key with information about the
                  crossed observing levels.
\argu[updown; Output, integer] Up-down indicator: 1 if the
                  particle is going upwards, $-1$ otherwise.
\argu[firstol; Output, integer] First observing level crossed
                  ($1 \le \kwbf;firstol; \le 510$).
\argu[lastol; Output, integer] Last observing level crossed
                  ($1 \le \kwbf;lastol; \le 510$).
\earlist
%

\rout olcrossedu
\fcall olcrossedu(olkey, ux, uy, uz, firstol, lastol)
\ccall olcrossedu(\&olkey, \&ux, \&uy, \&uz, \&firstol, \&lastol)
\dh

      This routine is similar to \kwbf;olcrossed;, but retrieves the
      information about the particle's direction of motion (up or
      down) in the form of an unitary vector.

\barlist
\argu[olkey; Input, integer] Key with information about the
                  crossed observing levels (See routine \kwbf;olcrossed;).
\argu[ux, uy; Input, double precision] $x$ and $y$ components of
                  the unitary vector marking the particle's
                  direction of motion.
\argu[uz; Output, double precision] $z$ component of the
                  direction of motion. Positive means upwards motion.
\argu[firstol; Output, integer] First observing level crossed
                  ($1 \le \kwbf;firstol; \le 510$).
\argu[lastol; Output, integer] Last observing level crossed
                  ($1 \le \kwbf;lastol; \le 510$).
\earlist
%

\rout olsavemarked
\ffun ismarked = olsavemarked(obslev, vrb, irc)
\cfun ismarked = olsavemarked(\&obslev, \&vrb, \&irc)
\dh

     Logical function returning ``true'' if an observing level is
     marked to be saved into longitudinal files, ``false'' otherwise.
     An arbitrary subset of the defined observing levels can be
     selected for inclusion into the longitudinal compressed files
     (see page \pageref{IDL0:RecordObsLevels}); this function allows
     to determine if a given observing level was or not marked at the
     moment of performing the simulations that generated the
     corresponding compressed file.

\barlist
\argu[obslev; Input, integer] The number of observing level.
                      If it is out of range the returned value will
                      always be ``false''.
\arguvrb
\argu[irc; Output, integer] Return code. 0 means
                   successful return.
\earlist
\retval[Logical]{``true'' if the level is marked for file recording,
                 ``false'' otherwise.}
%

\rout olv2slant
\fcall olv2slant(nobslev, olxv, Xv0, zendis, zen1, zen2, groundz, olxs)
\ccall olv2slant(\&nobslev, \&olxv[1], \&Xv0, \&zendis, \&zen1, \&zen2,
                 \&groundz, \&olxs[1])
\dh

     Evaluating the slant depths\index{atmospheric depth!slant} of a set
     of observing levels. The slant depths are calculated along an axis
     starting at altitude \kwbf;zground;, for the ``segment'' that ends at
     vertical depth \kwbf;Xv0; ($\hbox{\kwbf;Xv0;} = 0$ is the top of the
     atmosphere). The integer variable \kwbf;zendis; allows to select
     among fixed, {\em sine\/} and {\em sine-cosine\/} zenith angle
     distributions (see section \ref{S:zenazimdist}).

\barlist
\argu[nobslev; Input, integer] The number of observing
                  levels ($N_o$).
\argu[olxv; Input, double precision, array(\/\kw;nobslev;)]
                  Vertical atmospheric depths (in \gcmsq) of the
                  corresponding observing levels.
\argu[Xv0; Input, double precision] Vertical atmospheric depth (in
        \gcmsq) of the point marking the end of the integration
        path. If \kwbf;Xv0; is zero, then the end of the integration path is
        the top of the atmosphere.
\argu[zendis; Input, integer] Zenith angle distribution switch: 0 -- fixed
zenith angle, 1 -- sine distribution, 2 -- sine-cosine distribution.
\argu[zen1, zen2; Input, double precision] Minimum and maximum zenith
angles (degrees). If \kwbf;zendis; is 0, then \kwbf;zen2; is not used and
\kwbf;zen1; gives the corresponding fixed zenith angle.
\argu[groundz; Input, double precision] Ground altitude (in m).
\argu[olxs; Output, double precision, array(\/\kw;nobslev;)]
                  Slant atmospheric depths (in \gcmsq) of the
                  corresponding observing levels.
\earlist
%

\rout opencrofile
\fcall opencrofile(wdir, filename, header1, logbase, vrb, channel, irc)
\ccall opencrofilec(\&wdir, \&filename, \&header1, \&logbase, \&vrb,
                    \&channel, \&irc)
\dh
       Opening a CIO file for reading. This routine performs both the
       system open operation and file header processing and checking.

\barlist
\argu[wdir; Input, character string] The name of the directory
                      where the file is placed. It defaults to the
                      current directory when blank.
\argu[filename; Input, character string] The name of the file to
                    open.
\argu[header1; Input, integer] Integer switch to select
                  reading (greater than or equal to 0) or skipping
      (less than 0) the first
                  part of the header.
\argu[logbase; Input, integer] Variable to control the
                  logarithmically scaled fields of the file
                  records. If \kwbf;logbase; is less than 2, then
                  the returned logarithms will be natural
                  logarithms. Otherwise base \kwbf;logbase; will be
                  returned (decimal ones if $\kwbf;logbase; = 10$).
\arguvrb
\argu[channel; Output, integer] File identification. This
                  variable should not be changed by the calling
                  program. It must be used as a parameter of
                  the reading and closing routines in order to
                  specify the corresponding file.
\argu[irc; Output, integer] Return code. 0 means
                  successful return. 1 means successful return
                  obtained with a file that was written with a
                  previous AIRES version. 10 means that the file
                  could be opened normally, but that it seems not
                  to be a valid AIRES compressed data file, or is
                  a corrupted file; 12 invalid file header; 14
                  not enough size in some of the internal arrays;
                  16 format incompatibilities. 20: too many
                  compressed files already opened. $300 < \kwbf;irc; < 400$
                  indicates a version incompatibility (when
                  processing files written with other AIRES
                  version) or invalid version field (corrupt
                  header). Any other value indicates an opening /
                  header-reading error (\kwbf;irc; equals the system
                  return code plus 10000).
\earlist
%

%
%

%
%

%
%

%
%

\rout raninit
\fcall raninit(seed)
\ccall raninit(\&seed)
\dh
    Initialization of the uniform pseudo-random number
    generator\index{random number generator}. This routine {\em must\/} be
    called before the first invocation of \libbfx;grandom;,
    \libbfx;urandom;, or \libbfx;urandomt;.

\barlist
\argu[seed; Input, double precision] Seed to initialize the random
series. If \kwbf;seed; does not belong to the interval $(0,1)$, then
the seed actually used for initialization is internally generated
using the elementary generator \libbfx;clockrandom;.
\earlist
%

\rout regetcrorecord
\ffun okflag = regetcrorecord(channel, intfields, realfields, altrec,
                              vrb, irc)
\cfun okflag = regetcrorecord(\&channel, \&intfields[1], \&realfields[1],
                              \&altrec, \&vrb, \&irc)
\dh

     Re-reading the current record. The input and output parameters of
     this routine are equivalent to the respective arguments of
     routine \kwbf;getcrorecord;. The difference between this routine
     and the mentioned one is that \kwbf;regetcrorecord; re-scans the
     last read record instead of advancing across the input file.
     \kwbf;regetcrorecord; is thought to be used jointly with
     \kwbf;getcrorectype;, \kwbf;crorecfind; and other related
     procedures.

\barlist
\arguchannel
\argu[intfields; Output, integer, array(*)] Integer fields
                      of the record.
For a complete description of
this argument see routine \kwbf;getcrorecord;
\argu[realfields; Output, double precision, array(*)] Real fields
                      of the record.
For a complete description of
this argument see routine \kwbf;getcrorecord;
\argu[altrec; Output, logical] Alternative/default record type label.
See \kwbf;getcrorecord;
\arguvrb
\argu[irc; Output, integer] Return code.
For a complete description of
this argument see routine \kwbf;getcrorecord;
\earlist

\retval[Logical]{True if a record was successfully re-read.
False otherwise (EOF or I/O error).}
%

\rout sp1stint
\fcall sp1stint(csys, x1, y1, z1, irc)
\ccall sp1stint(\&csys, \&x1, \&y1, \&z1, \&irc)
\dh
  
     Setting manually the position of the first
     interaction\index{depth of first interaction}. When using special
     primary particles processed by external modules which may inject
     more that a single primary, AIRES cannot determine automatically
     the point where the first interaction takes place, and will take
     it as equal to the injection point unless it is set explicitly
     using \kwbf;sp1stint;.
     This routine should be used only within modules designed to
     process special primaries\index{special primary particles}, and
     following the guidelines of section \ppref{S:specialprim}.

\barlist
\argu[csys; Input, integer] Parameter labeling the coordinate system
used. $\kwbf;csys; = 0$ selects the AIRES coordinate
system\index{AIRES coordinate system}. $\kwbf;csys; = 1$ selects the
shower axis-injection point system\index{shower axis-injection point
coordinate system} defined in section \ref{S:specialprim}.
\argu[x1, y1, z1; Input, double precision] Coordinates of the
first interaction point with respect to the chosen coordinate system (in
meters).
\argu[irc; Output, integer] Return code. 0 means normal
                       return.
\earlist
%

\rout spaddnull
\fcall spaddnull(pener, pwt, irc)
\ccall spaddnull(\&pener, \&pwt, \&irc)
\dh
  
     Adding a {\em null\/} (unphysical) particle to the list of primaries
     to be passed
     from the external module to the main simulation program. This
     ``particle'' will not be propagated, but its energy will be added to
     the unphysical particle counter included in the shower energy balance.
     This routine should be used only within modules designed to
     process special primaries\index{special primary particles}, and
     following the guidelines of section \ppref{S:specialprim}.

\barlist
\argu[pener; Input, double precision] Energy (GeV).
\argu[pwt; Input, double precision] Null particle weight. Must be equal or
greater than one.
\argu[irc; Output, integer] Return code. 0 means normal
                       return.
\earlist
%

\rout spaddp0
\fcall spaddp0(pcode, pener, csys, ux, uy, uz, pwt, irc)
\ccall spaddp0(\&pcode, \&pener, \&csys, \&ux, \&uy, \&uz, \&pwt,
               \&irc)
\dh
  
     Adding a primary particle to the list of primaries to be passed
     from the external module to the main simulation program.
     This routine should be used only within modules designed to
     process special primaries\index{special primary particles}, and
     following the guidelines of section \ppref{S:specialprim}.

\barlist
\argu[pcode; Input, integer] Particle code, accordingly with the AIRES
coding system\index{AIRES particle codes} described in section
\ppref{S:acodes}.
\argu[pener; Input, double precision] Kinetic energy (GeV).
\argu[csys; Input, integer] Parameter labeling the coordinate system
used. $\kwbf;csys; = 0$ selects the AIRES coordinate
system\index{AIRES coordinate system}. $\kwbf;csys; = 1$ selects the
shower axis-injection point system\index{shower axis-injection point
coordinate system} defined in section \ref{S:specialprim}.
\argu[ux, uy, uz; Input, double precision] Direction of motion with
respect to the chosen coordinate system. The vector $(\kwbf;ux;,
\kwbf;uy;, \kwbf;uz;)$ does not need to be normalized.
\argu[pwt; Input, double precision] Particle weight. Must be equal or
greater than one.
\argu[irc; Output, integer] Return code; can be one of the following:
\begin{enumerate}
\item[0] The particle was successfully added.
\item[8] Negative kinetic energy.
\item[9] Particle weight less than 1.
\item[10] The direction of motion is a null vector.
\item[11] Invalid coordinate system specification.
\end{enumerate}
\earlist
%

\rout spaddpn
\fcall spaddpn(n, pcode, pener, csys, ldu, uxyz, pwt, irc)
\ccall spaddpn(\&n, \&pcode, \&pener, \&csys, \&ldu, \&uxyz[1][1], \&pwt,
               \&irc)
\dh
  
     Adding a set of \kwbf;n; primary particles to the list of
     primaries to be passed
     from the external module to the main simulation program.
     This routine should be used only within modules designed to
     process special primaries\index{special primary particles}, and
     following the guidelines of section \ppref{S:specialprim}.

\barlist
\argu[n; Input, integer] The number of particles to add to the list.
\argu[pcode; Input, integer, array(n)] Particle codes, accordingly
with the AIRES
coding system\index{AIRES particle codes} described in section
\ppref{S:acodes}.
\argu[pener; Input, double precision, array(n)] Kinetic energies (GeV).
\argu[csys; Input, integer] Parameter labeling the coordinate system
used. $\kwbf;csys; = 0$ selects the AIRES coordinate
system\index{AIRES coordinate system}. $\kwbf;csys; = 1$ selects the
shower axis-injection point system\index{shower axis-injection point
coordinate system} defined in section \ref{S:specialprim}.
\argu[ldu; Input, integer] Leading dimension of array \kwbf;uxyz;;
must be equal or greater than 3.
\argu[uxyz; Input, double precision, array(ldu, n)\footnote{If \kwbf;uxyz; is
defined in a C environment, then its two dimensions should be swapped,
i.e., {\kw;double uxyz[n][ldu];}.}]
Directions of motion with
respect to the chosen coordinate system. The vectors $(\kwbf;uxyz;(1,i),
\kwbf;uxyz;(2,i), \kwbf;uxyz(3, i);)$, $i=1,\ldots,n$, do not need to
be normalized.
\argu[pwt; Input, double precision, array(n)] Particle weights. The
weights must be equal or greater than one.
\argu[irc; Output, integer] Return code. 0 means normal
                       return.
\earlist
%

\rout speiend
\fcall speiend(retcode)
\ccall speiend(\&retcode)
\dh
  
     Closing the interface for the special primary
     particle\index{special primary particles} external process. This
     routine should be used only within modules designed to process
     special primaries, and following the guidelines of section
     \ppref{S:specialprim}.

\barlist
\argu[retcode; Input, integer] Return code to pass to the
                       main simulation program. $\kwbf;retcode; = 0$ means
                       normal return. If \kwbf;retcode; is not zero, a
                       message will be printed and saved in the log
                       file (extension \kwbf;.lgf;). $0 <
                       |\kwbf;retcode;| < 10$,
                       $10 \le |\kwbf;retcode;| < 20$, $20 \le
                       |\kwbf;retcode;| < 30$, and
                       $|\kwbf;retcode;| \ge 30$ correspond, respectively, to
                       information, warning, error and fatal messages.
\earlist
%

\rout speigetmodname
\fcall speigetmodname(mn, mnlen, mnfull, mnfullen)
\ccall speigetmodnamec(\&mn, \&mnlen, \&mnfull, \&mnfullen)
\dh
  
     Getting the name of the module invoked by the simulation program,
     that is, the one specified in the definition of the corresponding
     special particle.
     This routine should be used only within modules designed to
     process special primaries\index{special primary particles}, and
     following the guidelines of section \ppref{S:specialprim}.

\barlist
\argu[mn; Output, string ] Name of external module. The calling
program must ensure there is enough space to store the string.
\argu[mnlen; Output, integer] Length of external module name.
\argu[mnfull; Output, string ] Full name of external module (Will be
different of \kwbf;mn; if the module was placed within one of the
directories specified with the \idlbfx;InputPath; directive. The
calling program must ensure there is enough space to store the
string.
\argu[mnfullen; Output, integer] Length of full external module name.
\earlist
%

\rout speigetpars
\fcall speigetpars(parstring, pstrlen)
\ccall speigetparsc(\&parstring, \&pstrlen)
\dh
  
     Getting the parameter string specified in the IDL instruction
     that defines the corresponding special particle.
     This routine should be used only within modules designed to
     process special primaries\index{special primary particles}, and
     following the guidelines of section \ppref{S:specialprim}.

\barlist
\argu[parstring; Output, string ] Parameter string. The calling
program must ensure there is enough space to store the string.
\argu[pstrlen; Output, integer] Length of parameter string. Zero if
there are no parameters.
\earlist
%

\rout speimv
\fcall speimv(mvnew, mvold)
\ccall speimv(\&mvnew, \&mvold)
\dh
  
     Setting and/or getting the external macro version.
     This routine should be used only within modules designed to
     process special primaries\index{special primary particles}, and
     following the guidelines of section \ppref{S:specialprim}.

\barlist
\argu[mvnew; Input, integer] Macro version number. Must be an integer
in the range $[1, 759375]$. If \kwbf;mvnew; is zero, then the macro
version is not set.
\argu[mvold; Output, integer] Macro version number effective at the
moment of invoking the routine. This variable will be set to zero in
the first call to \kwbf;speimv;.
\earlist
%

\rout spinjpoint
\fcall spinjpoint(csys, x0, y0, z0, tsw, t0beta, irc)
\ccall spinjpoint(\&csys, \&x0, \&y0, \&z0, \&tsw, \&t0beta, \&irc)
\dh
  
     Setting the current injection point for primary particles.
     This routine should be used only within modules designed to
     process special primaries\index{special primary particles}, and
     following the guidelines of section \ppref{S:specialprim}.

\barlist
\argu[csys; Input, integer] Parameter labeling the coordinate system
used. $\kwbf;csys; = 0$ selects the AIRES coordinate
system\index{AIRES coordinate system}. $\kwbf;csys; = 1$ selects the
shower axis-injection point system\index{shower axis-injection point
coordinate system} defined in section \ref{S:specialprim}.
\argu[x0, y0, z0; Input, double precision] Coordinates of the
injection point with respect to the chosen coordinate system (in
meters).
\argu[tsw; Input, integer] Injection time switch. If
                       \kwbf;tsw; is zero then \kwbf;t0beta; is an
absolute injection
                       time; if \kwbf;tsw; is 1, then the injection time is
                       set as the time employed by a particle whose
                       speed is $\kwbf;t0beta; \times c$ to go from
the original
                       injection point to the intersection point of
                       the shower axis with the plane orthogonal to
                       that axis and containing the point $(\kwbf;x0;,
\kwbf;y0;, \kwbf;z0;)$.
\argu[t0beta; Input, double precision] The meaning of this
                       argument depends on the current value of \kwbf;tsw;.
                       It can be the absolute injection time (ns)
                       (time at original injection is taken as zero);
                       or the speed of a particle divided by $c$.
\argu[irc; Output, integer] Return code. 0 means normal
                       return.
\earlist
%

\rout speistart
\fcall speistart(showerno, primener, injpos, xvinj, zground, xvground,
                 dgroundinj, uprim)
\ccall speistart(\&showerno, \&primener, \&injpos[1], \&xvinj,
                 \&zground, \&xvground, \&dgroundinj, \&uprim[1])
\dh
  
     Starting the interface for the special primary
     particle\index{special primary particles} external process. This
     routine should be used only within modules designed to process
     special primaries, and following the guidelines of section
     \ppref{S:specialprim}.

\barlist
\argu[showerno; Output, integer] Current shower number.
\argu[primener; Output, double precision] Primary energy (GeV).
\argu[injpos; Output, double precision, array(3)] Position of the
initial injection point with respect to the AIRES coordinate
system\index{AIRES coordinate system} (in meters).
\argu[xvinj; Output, double precision] Vertical atmospheric depth of
the injection point (in {\gcmsq}).
\argu[zground; Output, double precision] Altitude og ground level (in
m.a.s.l.).
\argu[xvground; Output, double precision] Vertical atmospheric depth of
the ground surface (in {\gcmsq}).
\argu[dgroundinj; Output, double precision] Distance from the
injection point to the intersection between the shower axis and the
ground surface (in meters).
\argu[uprim; Output, double precision, array(3)] Unitary vector in the
direction of the straight line going from the injection point towards
the intersection between the shower axis and the ground plane.
\earlist
%

\rout speitask
\fcall speitask(taskn, tasklen, tver)
\ccall speitaskc(\&taskn, \&tasklen, \&tver)
\dh
  
     Getting the current task name and version.
     This routine should be used only within modules designed to
     process special primaries\index{special primary particles}, and
     following the guidelines of section \ppref{S:specialprim}.

\barlist
\argu[taskn; Output, string ] Task name. The calling program must
ensure there is enough space to store the string.
\argu[tasklen; Output, integer] Length of task name.
\argu[tver; Output, integer] Task name version.
\earlist
%

\rout spnshowers
\fcall spnshowers(totsh, firstsh, lastsh)
\ccall spnshowers(\&totsh, \&firstsh, \&lastsh)
\dh
  
     Getting the current values of the first and last shower, and
     total number of showers.
     This routine should be used only within modules designed to
     process special primaries\index{special primary particles}, and
     following the guidelines of section \ppref{S:specialprim}.

\barlist
\argu[totsh; Output, integer] Total number of showers for the current
task.
\argu[firstsh; Output, integer] Number of first shower.
\argu[lasttsh; Output, integer] Number of last shower.
\earlist
%

\rout sprimname
\fcall sprimname(pname, pnamelen)
\ccall sprimnamec(\&pname, \&pnamelen)
\dh
  
     Getting the name of the special primary particle specified in the
     corresponding IDL instruction.
     This routine should be used only within modules designed to
     process special primaries\index{special primary particles}, and
     following the guidelines of section \ppref{S:specialprim}.

\barlist
\argu[pname; Output, string] The name of the special particle. The
calling program must ensure there is enough space to store the string.
\argu[pnamelen; Output, integer] Length of particle name.
\earlist
%

%
%

\rout thisairesversion
\ffun iavers = thisairesversion(.)
\cfun iavers = thisairesversion(.)
\dh

Returning the current version of the AIRES library.

\retval[Integer]{The corresponding version in integer
                 format (for example the number 01040200
                 for version 1.4.2,
                 01040201 for version 1.4.2a, etc.).}
%

\rout urandom
\ffun r = urandom(.)
\cfun r = urandom(.)
\dh

This function invokes the AIRES random number generator\index{random
number generator} and returns a pseudo-random number uniformly distributed
in the interval $[0,1)$. It is necessary to initialize the random series
calling \libbfx;raninit; before using this function.

\retval[Double precision]{The uniform pseudo-random number.}
%

\routb urandomt
\ffun r = urandomt(threshold)
\cfun r = urandomt(threshold)
\dh

This function invokes the AIRES random number generator\index{random
number generator} and returns a pseudo-random number uniformly distributed
in the interval $[t,1)$, where $t$ is a specified threshold ($0\le t<1$).
It is necessary to initialize the random series calling \libbfx;raninit;
before using this function.

\barlist
\argu[threshold; Input, double precision] The threshold $t$.
\earlist

\retval[Double precision]{The uniform pseudo-random number.}
%

\rout xslant
\ffun X = xslant(Xvert, Xv0, cozenith, zground)
\cfun X = xslant(\&Xvert, \&Xv0, \&cozenith, \&zground)
\dh

Converting vertical atmospheric depths into slant atmospheric
depths\index{atmospheric depth!slant}. This routine evaluates the
slanted path (in \gcmsq) of equation (\ref{eq:Xsdef}), starting
(ending) at the point whose {\em vertical\/} depth is \kwbf;Xvert;
(\kwbf;Xv0;). The inclination of the integration path is controlled by
parameters \kwbf;cozenith; (cosine of the zenith angle) and
\kwbf;zground; (altitude, in meters, of the intersection between the
oblique axis and the $z$-axis), as illustrated in figure
\ppref{FIG:coor1}.

\barlist
\argu[Xvert; Input, double precision] Vertical atmospheric depth (in
\gcmsq) of the point marking the beginning of the integration
path. Must be positive.
\argu[Xv0; Input, double precision] Vertical atmospheric depth (in
\gcmsq) of the point marking the end of the integration
path. If \kwbf;Xv0; is zero, then the end of the integration path is
the top of the atmosphere. If \kwbf;Xv0; corresponds to a point
located below the point corresponding to \kwbf;Xvert; ($\kwbf;Xv0; >
\kwbf;Xvert;$), then the returned slant depth will be negative.
\argu[cozenith; Input, double precision] Cosine of the zenith angle
$\Theta$ (see figure \ref{FIG:coor1}) corresponding to the integration
line. Must be in the range $(0,1]$.
\argu[zground; Input, double precision] $z$-coordinate (in meters) of
the intersection point between the oblique axis and the $z$-axis,
which is normally coincident with the ``ground altitude''.
\earlist

\retval[Double precision]{The slant atmospheric depth in \gcmsq; or
zero in case of error or invalid argument.}
%

%
%

\rout zfromdepth
\ffun z = zfromdepth(Xvert, atlayer)
\cfun z = zfromdepth(\&Xvert, \&atlayer)
\dh

     Vertical altitude from atmospheric depth. Multilayer atmospheric
     model.\index{atmospheric model}.

\barlist
\argu[Xvert; Input, double precision] Vertical atmospheric depth (in
\gcmsq) of the corresponding point. Must be positive.
\argu[atlayer; Output, integer] Atmospheric layer corresponding to the
depth \kwbf;Xvert;. This parameter depends on the selected atmospheric
model.\index{atmospheric model}
\earlist
\retval[Double precision]{The vertical altitude in m.a.s.l.}
%


\end{description}

\clearpage
%
%
%
\typeout{References.}
{\frenchspacing
\def\journal#1#2#3#4{{\em #1,\/} {\bf #2}, #3 (#4)}
}
\clearpage
%
\typeout{Index.}
{\openup-1pt\small
\addcontentsline{toc}{chapter}{\protect\numberline{ }%
{\large\bfseries Index}}%
\def\previndexmessage{Page numbers in {\bfseries boldface} represent
the definition or the main source of information about whatever is
being indexed.\par\bigskip}
\raggedright
\input UsersManual.ind
}
\clearpage
%
%
\markboth {{\scshape NOTES}}{{\scshape NOTES}}%
{\parindent=0pt\baselineskip=0pt
\newcounter{iline}
\definecolor{rulegray}{cmyk}{0,0,0,0.3}
\def\linerule{{\color{rulegray}\hrule height 0.4pt width \textwidth}}
\def\oneline{\vfil\linerule}
\def\onepage{\setcounter{iline}{25}
\linerule
\loop\oneline\ifnum\value{iline}>0\addtocounter{iline}{-1}\repeat
\eject}
\onepage
\onepage
\onepage
\ifodd\value{page}\else\onepage\fi
}
\cleardoublepage

\begin{thebibliography}{99}
\addcontentsline{toc}{chapter}{\protect\numberline{ }%
{\large\bfseries References}}
%
\bibitem{MOCCA} A. M. Hillas,
 \journal{Nucl. Phys. B (Proc. Suppl.)}{52B}{29}{1997};
 A. M. Hillas, \journal{Proc. 19th ICRC (La Jolla)}{1}{155}{1985}.
%
\bibitem{Aires120} S. J. Sciutto, {\em AIRES: A minimum document,\/}
Auger technical note GAP-97-029 (1997).
%
\bibitem{airmoc} M. T. Dova and S. J. Sciutto, {\em Air Shower
Simulations: Comparison Between AIRES and MOCCA,\/} Auger technical
note GAP-97-053 (1997).
%
\bibitem{splithin} A. M. Hillas, {\em Proc. of the Paris Workshop on
Cascade simulations,\/} J. Linsley and A. M. Hillas (eds.), p 39
(1981).
%
\bibitem{slthin} M. Kobal, A. Filip\v{c}i\v{c} and D. Zavrtanik, Auger
technical notes GAP-98-001 and GAP-98-058 (1998).
%
\bibitem{EPOSLHC}
T. Pierog, Iu. Karpenko, J.M. Katzy, E. Yatsenko, K. Werner,
\journal{Phys. Rev. C}{92}{034906}{2015}.
%
\bibitem{QGSJET2R4}
S. S. Ostapchenko, \journal{Phys. Rev. D}{83}{014018}{2011}.
%
\bibitem{QGSJET2R3}
S. Ostapchenko, \journal{Nucl. Phys. B (Proc. Suppl.)}{151}{143}{2006}.
%
\bibitem{QGSJETold} N. N. Kalmykov and S. S. Ostapchenko,
           \journal{Yad. Fiz.}{56}{105}{1993};
           \journal{Phys. At. Nucl.}{56}{(3) 346}{1993};
           N. N. Kalmykov, S. S. Ostapchenko and A. I. Pavlov,
           \journal{Bull. Russ. Acad. Sci. (Physics)}{58}{1966}{1994}.
%
\bibitem{SIBYLL23c}
 Eun-Joo Ahn {\em et al.,\/} \journal{Phys. Rev. D}{80}{094003}{2009};
 F. Riehn {\em et al.,\/} \journal{Proc. 35th Int. Cosmic Ray Conf.
       (Bexco, Busan, Korea)}{}{cont. 301}{2017}.
%
\bibitem{SIBYLL23}
 Eun-Joo Ahn {\em et al.,\/} \journal{Phys. Rev. D}{80}{094003}{2009};
 F. Riehn {\em et al.,\/} \journal{Proc. 35th Int. Cosmic Ray Conf.
       (The Hague, The Netherlands)}{}{cont. 1313}{2015}.
%
\bibitem{SIBYLL21} R. Engel, T. K. Gaisser, T. Stanev,
\journal{Proc. 26th ICRC (Utah)}{1}{415}{1999}.
%
\bibitem{SIBYLLold} R. T. Fletcher, T. K. Gaisser, P. Lipari and T.
     Stanev, {\em Phys. Rev. D,\/} {\bf 50}, 5710 (1994);
     J. Engel, T. K. Gaisser, P. Lipari and T. Stanev,
     {\em Phys. Rev. D,\/} {\bf 46}, 5013 (1992).
%
\bibitem{IGRF} The data, software and documentation related with the
International Geomagnetic Reference Field are distributed by the {\em
National Geophysical Data Center,\/} Boulder (CO), USA, and can be
obtained electronically at the following Web address: {\ttbf
www.ngdc.noaa.gov/IAGA/vmod}. 
%
\bibitem{netlib} NETLIB is a public collection of mathematical software,
 papers, and databases, that can be accessed through Internet, at the
 World Wide Web address {\ttbf www.netlib.org}.
%
\bibitem{PAW} CERN Program library Long Writeup {\bf Q121} (1995).
%
\bibitem{Aires142} S. J. Sciutto, {\em AIRES users guide and reference
manual ,\/} version 1.4.2,
Auger technical note GAP-98-032 (1998). 
%
\bibitem{usatm76} National Aerospace Administration (NASA), National
Oceanic and Atmospheric Administration (NOAA) and US Air Force, {\em
US standard atmosphere 1976,\/} NASA technical report NASA-TM-X-74335,
NOAA technical report NOAA-S/T-76-1562 (1976).
%
\bibitem{usatmCRC} R. C. Weast (editor), {\em CRC Handbook of Chemistry
and Physics, 61st edition,\/} pp F206 -- F213, CRC Press, Boca Raton
(FL, USA) (1981).
%
\bibitem{rossi} B. Rossi, {\em High-energy particles,\/}
Prentice-Hall, New Jersey (USA) (1956).
%
\bibitem{Linsley1} We were not able to find official references related with
Linsley's standard atmosphere model. References \cite{MOCCA,CORSIKA} contain
information about parameterization data.
%
\bibitem{geomag1} A. Cillis and S. J. Sciutto,
\journal{J. Phys. G}{26}{309-321}{2000}.
%
\bibitem{mubrem} A. Cillis and S. J. Sciutto,
\journal{Phys. Rev. D}{64}{013010}{2001}.
%
\bibitem{LPMigdal} A. B. Migdal,
\journal{Phys. Rev.}{103}{1811}{1956}.
%
\bibitem{LPMpaper} A. Cillis, C. A. Garc\'{\i}a Canal, H. Fanchiotti
and S. J. Sciutto, {\em Phys. Rev. D,\/} {\bf 59}, 113012 (1999).
%
\bibitem{LPMHeck} D. Heck, private communication.
%
\bibitem{gaisserbk} T. K. Gaisser, {\em Cosmic Rays and Particle Physics,\/}
Cambridge University Press, Cambridge (1992).
%
\bibitem{papercpc} S. J. Sciutto, in preparation.
%
\bibitem{LPM} L. D. Landau and I. Ya. Pomeranchuk,
\journal{Dokl. Akad. Nauk SSSR}{92}{535, 735}{1953}.
%
\bibitem{KleinRev} S. Klein, preprint hep-ph/9820442 (1998).
%
\bibitem{rwnbtests} I. Vattulainen, T. Ala-Nissila, K. Kankaala, {\em
Phys. Rev. Lett.\/} {\bf 73}, 2513 (1994).
%
\bibitem{inclinedgap} S. J. Sciutto, in preparation.
%
\bibitem{berezin1} V. S. Berezinski\u\i, {\em et. al.,\/} V. L. Ginzburg
(editor), {\em Astrophysics of cosmic rays,\/} North-Holland (1990).
%
\bibitem{gaihfun} T. K. Gaisser and A. M Hillas,
\journal{Proc. 15th ICRC (Plovdiv)}{8}{353}{1977}.%
%
\bibitem{CORSIKA} D. Heck, J. Knapp, J.N.  Capdevielle, G. Schatz, and
       T. Thouw, {\em Forschungszentrum Karlsruhe,\/} Report FZKA 6019
       (1998).
%
\bibitem{pierresample} P. Billoir, private communication.
%
\bibitem{PDG} Particle Data Group, M. Tanabashi {\em et al.\/}
  (Particle Data Group), 
\journal{Phys. Rev. D}{98}{030001}{2018}. 
%
\bibitem{GEANT} CERN Program library Long Writeup {\bf W5013} (1994).
%
\bibitem{PDG00} Particle Data Group, D. E. Groom {\em et. al.,\/} {\em The
European Physical Journal,\/} {\bf C15}, 1 (2000); website: {\ttbf
www-pdg.lbl.gov}.
%
\bibitem{JuanCruzAtm} J. C. Moreno, S. J. Sciutto,
\journal{Eur. Phys. J. Plus}{128}{104}{2013}.
%
\bibitem{ZHAireS} J. Alvarez-Mu\~niz, W. R. Carvalho, A. Romero-Wolf,
  M. Tueros, E. Zas,
\journal{Phys. Rev. D}{86}{123007}{2012}.
%
\end{thebibliography}
\end{document}